\documentclass[final,3p,times,twocolumn]{elsarticle}

\usepackage{amsmath,amssymb,amsfonts}
\usepackage{graphicx}
\usepackage{textcomp}
\usepackage{xcolor}
\usepackage{ulem}
\usepackage{float}
\usepackage{amssymb}
\usepackage{color}
\usepackage[pagewise]{lineno}
\usepackage{mdwlist}
\usepackage{algorithm}
\usepackage{algorithmicx}
\usepackage{algpseudocode}
\usepackage{amsmath}
\algrenewcommand\textproc{}
\usepackage[colorlinks]{hyperref}
\usepackage{float}
\usepackage{subfigure}

\usepackage{url}
\usepackage{multirow}

\usepackage{balance}
\usepackage{pifont}
\usepackage{amsmath}
\usepackage{array}

\usepackage[pagewise]{lineno}
\usepackage{amsmath}
\newtheorem{definition}{\textbf{Definition}}

\usepackage{mathrsfs}
\usepackage{tikz}

\usepackage{threeparttable}
\usepackage{supertabular,booktabs}
\usepackage{rotating}

\usepackage{supertabular}

\newcounter{DaveCommentCounter}
\newcounter{RHCommentCounter}
\begin{document}

\begin{frontmatter}



\title{Regression Test Case Prioritization by Code Combinations Coverage}

 \author[JS1]{Rubing Huang\corref{cor1}}
 \ead{rbhuang@ujs.edu.cn}
  \author[JS]{Quanjun Zhang}
 \ead{2211708038@stmail.ujs.edu.cn}
 \author[ZH]{Dave Towey}
 \ead{dave.towey@nottingham.edu.cn}
   \author[JS]{Weifeng Sun}
 \ead{2211808031@stmail.ujs.edu.cn}
 \author[JS]{Jinfu Chen}
 \ead{jinfuchen@ujs.edu.cn}

 \address[JS1]{School of Computer Science and Communication Engineering, and Jiangsu Key Laboratory of Security Technology for Industrial Cyberspace, Jiangsu University, Zhenjiang, Jiangsu 212013, P.R. China}
 \address[JS]{School of Computer Science and Communication Engineering, Jiangsu University, Zhenjiang, Jiangsu 212013, P.R. China}
  \address[ZH]{School of Computer Science, The University of Nottingham Ningbo China, Ningbo, Zhejiang 315100, P.R. China.}
 \cortext[cor1]{Corresponding author.}

\begin{abstract}
    Regression test case prioritization (RTCP) aims to improve the rate of fault detection by executing more important test cases as early as possible.
    Various RTCP techniques have been proposed based on different coverage criteria.
    Among them, a majority of techniques leverage code coverage information to guide the prioritization process, with code units being considered individually, and in isolation.
    In this paper, we propose a new coverage criterion, \textit{code combinations coverage}, that combines the concepts of code coverage and combination coverage.
    We apply this coverage criterion to RTCP, as a new prioritization technique, \textit{code combinations coverage based prioritization} (CCCP).
    We report on empirical studies conducted to compare the testing effectiveness and efficiency of CCCP with four popular RTCP techniques:
    \textit{total}, \textit{additional}, \textit{adaptive random}, and \textit{search-based} test prioritization.
    The experimental results show that even when the lowest combination strength is assigned, overall, the CCCP fault detection rates are greater than those of the other four prioritization techniques.
    The CCCP prioritization costs are also found to be comparable to the additional test prioritization technique.
    Moreover, our results also show that when the combination strength is increased, CCCP provides higher fault detection rates than the state-of-the-art, regardless of the levels of code coverage.
\end{abstract}


\end{frontmatter}


\section{Introduction}
\label{intro}
Modern software systems continuously evolve due to the fixing of detected bugs, the adding of new functionalities, and the refactoring of system architecture.
Regression testing is conducted to ensure that the changed source code does not introduce new defects.
However, it can become expensive to run an entire regression test suite because its size naturally increases during software maintenance and evolution:
In an industrial case reported by Rothermel et al.~\cite{Rothermel1999}, for example, the execution time for running the entire test suite could become several weeks.

Regression test case prioritization (RTCP) has become one of the most effective approaches to reduce the overheads in regression testing~\cite{Li2007,Jiang2009,2012Mei,2015Saha,2009Ledru}.
RTCP techniques reorder the execution sequence of regression test cases, aiming to execute those test cases more likely to detect faults (according to some award function) as early as possible~\cite{2016Hao,2014Hao,Zhang2009}.

Traditional RTCP techniques~\cite{Rothermel1999,Zhang2013,2017Wang} usually use code coverage criteria to guide the prioritization process.
Intuitively speaking, a code coverage criterion indicates the percentage of some code units (e.g. statements) covered by a test case.
The expectation is that test cases with higher code coverage value have a greater chance of detecting faults~\cite{Zhu1997}.
Because of this, a goal of maximizing code coverage has been incorporated into various RTCP techniques, including greedy strategies~\cite{Rothermel1999}.
Given a coverage criterion (e.g., method, branch, or statement coverage), the \textit{total} strategy selects the next test case with greatest absolute coverage, whereas the \textit{additional} strategy selects the one with greatest coverage of code units not already covered by the prioritized test cases.
Furthermore, Li et al.~\cite{Li2007} proposed two search-based RTCP techniques (a hill-climbing strategy and a genetic strategy) to explore the search space (the set of all permutations of the test cases) to find a sequence with a better fault detection rate.
Jiang et al.~\cite{Jiang2009} investigated adaptive random techniques \cite{Huang2019} to prioritize test cases using code coverage criteria.
In an attempt to bridge the gap between the two greedy strategies, Zhang et al.~\cite{Zhang2013} proposed a unified approach based on the fault detection probability for each test case (referred to as a \textit{p} value).

In this paper, we propose a new coverage criterion, \textit{code combinations coverage}, that combines the concepts of code coverage~\cite{Zhu1997} and combination coverage~\cite{Nie2011}:
Given a set of regression test cases $T$, each test case is first transferred to an equally-sized tuple.
Each position in this tuple is a binary value representing whether the corresponding item (such as branch, statement, or method) is covered by this test case.
In other words, $T$ is represented by a set of abstract test cases with binary values $T'$.
The code combinations coverage of $T$ is measured by the traditional combination coverage of $T'$.
We apply this new coverage criterion to RTCP, proposing a new prioritization technique:
\textit{code combinations coverage based prioritization} (CCCP).

We conducted empirical studies on 14 versions of four Java programs, and 30 versions of five real-world Unix utility programs.
Our goal was to investigate the testing effectiveness and efficiency of CCCP compared with four widely-used RTCP techniques
---
\textit{total}, \textit{additional}, \textit{adaptive random}, and \textit{search-based} test prioritization.
The results show that when the lowest combination strength is assigned, overall, our approach has better fault detection rates than the other four
test prioritization techniques.
It not only achieves comparable testing efficiency to \textit{additional}, but also requires much less prioritization time than the \textit{adaptive random} and \textit{search-based} techniques.
In addition, while the \textit{code coverage} granularity does not impact on the testing effectiveness of CCCP, the \textit{test case} granularity does significantly impact on it.
Furthermore, when the combination strength is increased, CCCP provides better fault detection rates than all other RTCP techniques, regardless of the level of code coverage.

The main contributions of this paper are:
\begin{itemize}
  \item We propose a new coverage criterion called \textit{code combinations coverage} that combines the concepts of code coverage and combination coverage.
  \item We apply code combinations coverage to RTCP, leading to a new prioritization technique called \textit{code combinations coverage based prioritization} (CCCP).
  \item  We report on empirical studies conducted to investigate the test effectiveness and efficiency of CCCP compared to four widely-used prioritization techniques, and also analyze the impact of code coverage granularity and test case granularity on the effectiveness of CCCP.
  \item We provide some guidelines for how to choose the combination strength and code-coverage level for CCCP, under different testing scenarios.
\end{itemize}

The rest of this paper is organized as follows:
Section \ref{back} presents some background information.
Section \ref{approach} introduces the proposed approach.
Section \ref{exp} presents the research questions, and explains details of the empirical study.
Section \ref{res} provides the detailed results of the study and answers the research questions.
Section \ref{related} discusses some related work, and Section \ref{conclude} concludes this paper, including highlighting some potential future work.

\section{Background}
\label{back}

In this section, we provide some background information about abstract test cases and test case prioritization.

\subsection{Abstract Test Cases}
\label{SEC:ATC}

For the system under test (SUT), there are some parameters $p_1,p_2,\cdots,p_k$ that may influence its performance, such as configuration options, components, and user inputs.
Each parameter $p_i$ can take some discrete values to form the set $V_i$, which is finite.
By selecting a value for each parameter, its combination becomes an \textit{abstract test case} \cite{Grindal2006}.
\begin{definition} \textbf{Abstract Test Case:}
An abstract test case is a discrete test case that can be represented by a $k$-tuple $(v_1,v_2,\cdots,v_k)$, where $v_i~(1\leq i \leq k)$ is a value of a parameter $p_i$ from a finite set $V_i$ (i.e., $v_i \in V_i$).
\end{definition}

Each abstract test case covers some $\lambda$-wise tuples (called $\lambda$\textit{-wise parameter-value combinations} \cite{Zhang2011} or $\lambda$-\textit{wise schemas} \cite{Nie2011}), where $1 \leq \lambda \leq k$.
For example, an abstract test case $tc = (1,3,5,7)$ covers the six 2-wise parameter-value combinations $(1,3)$, $(1,5)$, $(1,7)$, $(3,5)$, $(3,7)$, and $(5,7)$;
and also covers the four 3-wise parameter-value combinations $(1,3,5)$, $(1,3,7)$, $(1,5,7)$, and $(3,5,7)$.
Intuitively speaking, when $\lambda=k$, a $\lambda$-wise parameter-value combination becomes an abstract test case.


For ease of description, we define a function $CombSet(tc,\lambda)$ that returns a set of all $\lambda$-wise parameter-value combinations covered by an abstract test case $tc = (v_1,v_2,\cdots,v_k)$, i.e.,
\begin{equation}
\footnotesize
CombSet(tc, \lambda) = \Big\{\big(v_{i_1}, v_{i_2}, \cdots, v_{i_\lambda}\big)\big| 1 \leq i_1 < i_2 < \cdots < v_\lambda \leq k\Big\}
\end{equation}
Obviously, the size of $CombSet(tc, \lambda)$ is equal to $C\big(k,\lambda\big)$ (the number of $\lambda$-combinations from $k$ elements).
To calculate the $\lambda$-wise parameter-value combinations covered by the set $T$ of abstract test cases, the function $CombSet(T,\lambda)$ is defined as:
\begin{equation}
\footnotesize
CombSet(T, \lambda) = \bigcup_{tc \in T} CombSet(tc,\lambda)
\end{equation}

\subsection{Test Case Prioritization}
\label{back:tcp}

Regression test case prioritization (RTCP)~\cite{Rothermel1999} aims to reorder the test cases to realize a certain goal, such as exposing faults earlier.
RTCP is formally defined as:
\begin{definition}\textbf{Regression Test Case Prioritization:}
Given a regression test suite $T$, $PT$ is the set of its all possible permutations, and $f$ is an object function from $PT$ to real numbers. The problem of RTCP~\cite{Rothermel1999}  is to find $P'\in PT$, such that $\forall P'',P''\in PT (P''\ne P'),f(P')\ge f(P'')$.
\end{definition}

RTCP is an effective means to reduce the cost of regression testing, and has been widely investigated~\cite{Rothermel1999,Li2007,Jiang2009,Zhang2009}, with a large number of studies focusing on the coverage criterion and prioritization algorithms.
Intuitively, code coverage criteria can be regarded as characteristics of the test cases, and many prioritization algorithms have used coverage criteria to guide the prioritization process (such as the greedy strategies~\cite{Rothermel1999}, search-based strategies~\cite{Li2007}, and adaptive random strategies~\cite{Jiang2009}).


\section{Approach}
\label{approach}
In this section, we introduce the details of test case prioritization by code combinations coverage.

\subsection{Greedy Techniques}
\label{back:gt}

There are two widely investigated RTCP strategies:
the \textit{total} greedy strategy and the \textit{additional} greedy strategy.
The \textit{total} strategy selects test cases according to a descending order of code units covered by the test case.
The \textit{additional} strategy also selects test cases according to a descending order, but uses the number of
code units not already covered by previously selected test cases.
According to previous studies~\cite{Luo2016,Luo2019,Henard2016}, although seemingly simple, the greedy strategies (especially \textit{additional}) perform better than most other RTCP techniques in terms of the fault detection rate.
Therefore, in our study, we used a simple greedy algorithm to instantiate the CCCP prioritization function for statement, branch, and method coverage criteria.
As we just want to evaluate the performance of code combinations coverage against traditional code coverage (e.g. statement) and the \textit{additional} strategy has been widely accepted as the most effective prioritization strategy, we thus implemented greedy strategies based on the work of Rothermel et al.~\cite{Rothermel1999} as the control techniques for evaluation of our proposed approaches.

\subsection{Code Combinations Coverage}
\label{Section:codecombination}

Various RTCP approaches, based on different prioritization strategies, have been proposed to reduce regression testing overheads.
Many of these approaches used individual code unit coverage of a test case to guide the prioritization process.
For example, greedy strategies only take the number of covered code units into account, with the code units considered as parameters, or individually, and in isolation.
However, this may lead to a loss of coverage information, and regression testing has traditionally used historical testing information to guide future testing.
Thus, the degree to which the information is used is significant for regression testing, and if we consider the combination between code units, it may be possible to devise strategies to take further advantage of the code coverage information.
In our hypothesis, the code units are related, not isolated, and faults may be triggered by combinations amongst them.
Based on this, we can make use of more accurate testing information than traditional RTCP approaches to guide the prioritization process.

In our work, a code unit is a general term describing one structural code element
---
a statement, branch, or method.
Consider a program $P$ that has $m$ code units (statements, branches, or methods) that form the code unit set $U = \{u_1,u_2, \cdots, u_m\}$, and a regression test set $T$ with $n$ test cases ($T = \{tc_1, tc_2,\cdots, tc_n\}$).
We define a function $isCovered(tc, u)$ to measure whether or not a test case $tc$ covers the code unit $u$, as follows:
\begin{equation}
\footnotesize
\label{EQ:isCovered}
isCovered(tc, u) = \left\{
    \begin{array}{ll}
        1,&{\textit{if }} u \textit{ is covered by } tc, \\
        0,&{\textit{otherwise}}   \\
     \end{array}
     \right.
\end{equation}
As a result, each test case $tc$ can be represented by an $m$-wise binary array through the $convertTest$ function:
$convertTest(tc) = (isCovered(tc,u_1),$ $isCovered(tc,u_2),\cdots,isCovered(tc,u_m))$.
For ease of description, with the increase of $i$ for each $u_i$ in $U$, we make use of the incremental values to describe whether or not $tc$ covers the code unit $u_i$: $isCovered(tc, u_i)$, where $1\leq i \leq m$, defined as:
\begin{equation}
\footnotesize
\label{EQ:isCoveredNew}
isCovered(tc, u_i) = \left\{
    \begin{array}{ll}
        2i-1,&{\textit{if }} u_i \textit{ is covered by } tc, \\
        2i,&{\textit{otherwise}}   \\
     \end{array}
     \right.
\end{equation}
In other words, an odd number represents the situation where the code unit in question is covered by a given test case; and an even number means that it is not covered.
For example, using Equation (\ref{EQ:isCovered}), the values $(1,1,1,0,0)$ for a test case $tc$ would mean that the first three code units are covered by $tc$, but that the last two are not.
Equation (\ref{EQ:isCoveredNew}) would allow this to be represented as $tc=(1,3,5,8,10)$.
In effect, each code unit can be considered a parameter that contains binary parameter values (an odd number and an even number):
the first code unit takes the value 1 or 2; the second code unit takes 3 or 4; and so on.
In other words, each test case becomes an abstract test case (as defined in Section \ref{SEC:ATC}).
The $\lambda$-wise \textit{code combinations coverage} (CCC) value of $tc$ against the test set $T$ is defined as the number of $\lambda$-wise code-unit combinations covered by $tc$ that are not covered by $T$:
\begin{equation}
\label{EQ:ccc}
\footnotesize
CCC(tc, T, \lambda) = |CombSet(convertTest(tc),\lambda) \setminus CombSet(T',\lambda)|
\end{equation}
where $T' = \bigcup_{tc \in T}\{convertTest(tc)\}$.

\begin{algorithm}[!b]
\caption{$\textbf{\textit{calculateCombinations}}(temp\_{cover}, \lambda)$}
\scriptsize
\label{ALG:combSet}
\begin{algorithmic}[1]
    \renewcommand{\algorithmicrequire}{\textbf{Input:}}
    \renewcommand{\algorithmicensure}{\textbf{Output:}}
    \Require $temp\_{cover}$: List of code unit values, $\lambda$: Combination strength
    \Ensure  $Combinations$: Set of $\lambda$-wise code-unit-value combinations
    \State $Combinations  \gets  \emptyset$
    \If{$\lambda$ = 1}
     \For{each $item \in temp\_{cover}$}
      \State $Combinations \gets Combinations \bigcup\{item\}$
     \EndFor
     \State \Return $Combinations$
    \EndIf
    \State $num \gets |temp\_{cover}|$ \Comment{The number of code units}
    \For{each $i$  $(0\leq i \leq num- \lambda)$ }
     \For{each $ item \in$  \Call{\textbf{\textit{Calculate}}}{$temp\_{cover}.sublist(i+1)$, $\lambda-1$}}
      \State $Combinations \gets Combinations \bigcup\{temp\_{cover}.get(i)+item\}$
     \EndFor
    \EndFor
    \State \Return $Combinations$

\end{algorithmic}
\end{algorithm}

\subsection{Code Combinations Coverage based Prioritization}
\label{CCCP}

In our model, we view CCCP as a general strategy that can be applied to different prioritization algorithms using different coverage criteria.
As  greedy strategies are among the most widely-adopted  prioritization strategies~\cite{Rothermel1999,Zhang2013}, and the \textit{additional} greedy strategy is considered to be one of the most effective RTCP approaches~\cite{Jiang2009,Luo2016,Luo2019,Lu2016}, in terms of fault detection rate, we adopted a simple greedy strategy to instantiate the function for the proposed code combinations coverage.

\begin{algorithm}[!t]
\caption{Pseudocode of CCCP}
\scriptsize
\label{ALG:CCCP}
\begin{algorithmic}[1]
    \renewcommand{\algorithmicrequire}{\textbf{Input:}}
    \renewcommand{\algorithmicensure}{\textbf{Output:}}

    \Require $T$: $\{tc_1,tc_2,\cdots,tc_n\}$ is a set of unordered test cases with size $n$, $\lambda$: Combination strength, $U$: $\{u_1,u_2,\cdots,u_m\}$ is a set of $m$ code units for the program $P$
    \Ensure  $S$: Prioritized test cases
    \State $maximum \gets -1$
    \State $k \gets 1$
    \For {each $i~(1 \leq i \leq n)$}
     \State $Selected[i] \gets \textbf{false}$
     \State $num \gets 0$
       \For {each $j~(1 \leq j \leq m)$}
        \State $Cover[i,j] \gets isCovered(tc_i,u_j)$
        \State $temp\_cover[j] \gets Cover[i,j]$
        \If{$Cover[i,j] \% 2 == 1$}
        \State $num \gets num + 1$
        \EndIf
       \EndFor
       \If{$num > maximum$}
       \State $maximum \gets num$
       \State $k \gets i$
       \EndIf
        \State $Combinations[i] \gets$        \Call{\textbf{\textit{calculateCombinations}}}{$temp\_{cover}$, $\lambda$}
        \Comment{Calculate $\lambda$-wise code-unit-value combinations covered by $tc_i$}
        \State $UncoverCombinations \gets UncoverCombinations \bigcup Combinations[i]$
        \State $Selected[k] \gets \textbf{true}$
    \EndFor
    \State $tempCombinations \gets UncoverCombinations$
    \State $S \gets S \succ \langle tc_k \rangle$
        \Comment{Choose a candidate covering the maximum number of code units, and then append it to $S$}
     \State $UncoverCombinations\leftarrow UncoverCombinations \setminus Combinations[k]$
    \State $flag \gets \textbf{false}$
    \While {$|T| \neq |S|$}
        \If{$flag$} \Comment{Restart the process}
            \State $UncoverCombinations \gets tempCombinations$
        \EndIf
        \State $maximum \gets -1$
        \State $k \gets 1$
       \For {each $i~(1 \leq i \leq n)$}
       \If{\textbf{not} $Selected[i]$}
       \State $num \gets |UncoverCombinations \bigcap Combinations[i]|$
       \If{$num > maximum$}
       \State $maximum \gets num$
       \State $k \gets i$
       \EndIf
       \EndIf
       \EndFor
       \If{$maximum == 0$}\Comment{No uncovered code-unit-value combinations}
       \State $flag \gets \textbf{true}$
       \Else
       \State $flag \gets \textbf{false}$
       \EndIf
        \State $S \gets S \succ \langle tc_k \rangle$
        \Comment{Choose the best candidate, and then append it to $S$}
        \State $UncoverCombinations \gets UncoverCombinations \setminus Combinations[k]$
        \State $Selected[k] \gets \textbf{true}$
    \EndWhile
    \State \Return $S$

\end{algorithmic}
\end{algorithm}

Generally speaking, the approach chooses an element from candidates as the next test case such that it covers the largest number of $\lambda$-wise code-unit-value combinations that have not already been covered by previously selected test cases.
Accordingly, the test case with the maximum number of uncovered code-unit-value combinations compared with the already selected test cases $T$ will be selected:
$CCC(tc,T,\lambda)$, from Equation (\ref{EQ:ccc}).
Algorithm \ref{ALG:combSet} provides the detailed $CombSet$ calculation process.
Furthermore, when there are no further new code-unit-value combinations, then all remaining test cases are prioritized according the previous process, in a manner similar to the \textit{additional} greedy strategy.


Algorithm \ref{ALG:CCCP} formally presents the pseudocode of CCCP.
A Boolean array $Selected[i]~(1 \leq i \leq n)$ denotes whether or not test case $tc_i$ has been selected for prioritization;
and another Boolean array $Cover[i, j]~(1\leq i \leq n, 1\leq j \leq m)$ identifies whether or not test case $tc_i$ covers the code unit $u_j$.
Similarly, an array $Combinations[i]$ stores the $\lambda$-wise code-unit-value combinations covered by $tc_i$;
and a set $UncoverCombinations$ stores all uncovered $\lambda$-wise code-unit-value combinations.
In addition, the variable $flag$ indicates whether or not all $\lambda$-wise code-unit-value combinations have been covered.


\begin{figure}[!b]
\graphicspath{{graphs/}}
\centering
    \includegraphics[width=0.48\textwidth]{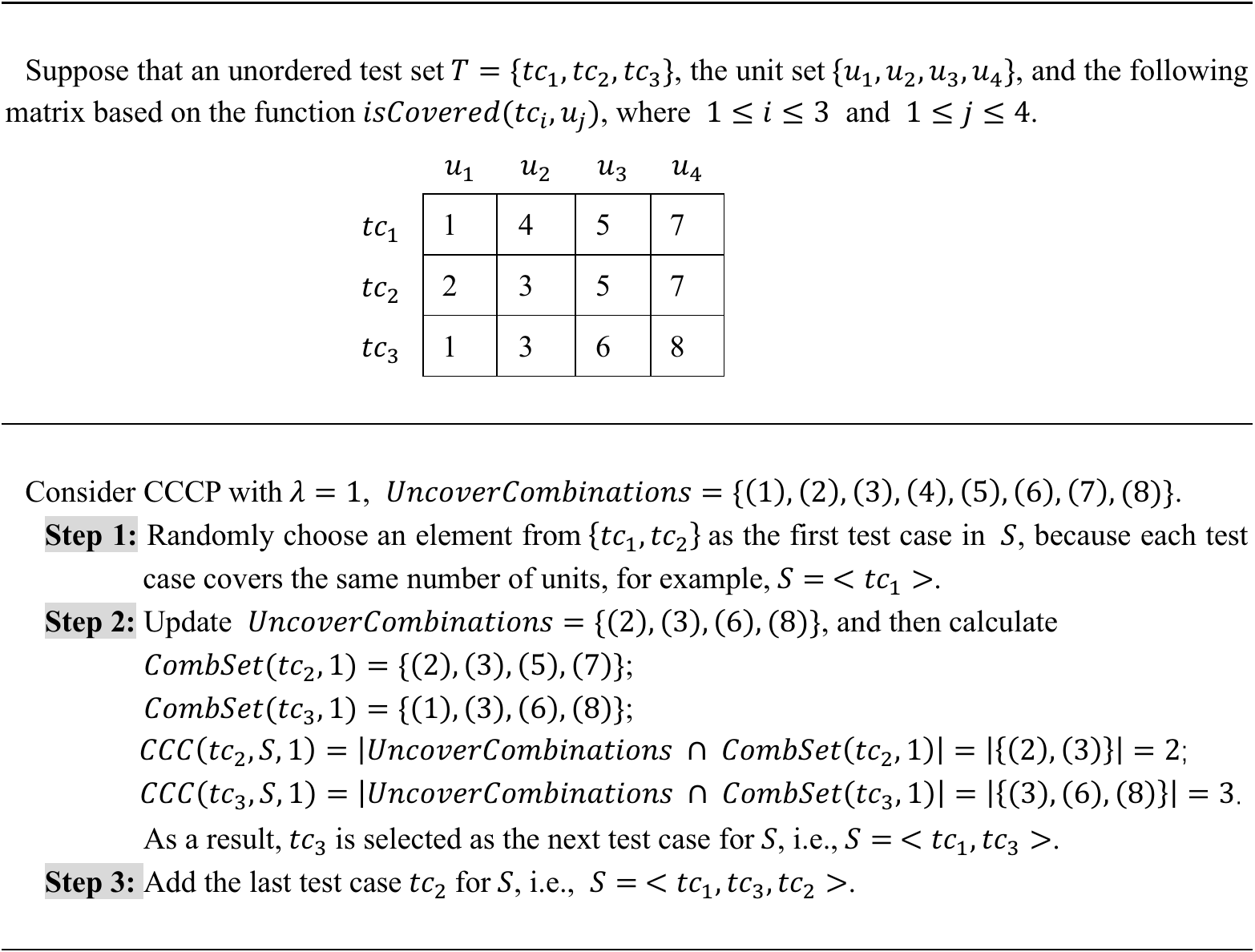}
    \caption{An illustrative example of CCCP}
    \label{FIG:example}
\end{figure}

In Algorithm \ref{ALG:CCCP}, Lines 1--24 perform initialization, and also choose the first test case from the candidates;
while Lines 25--49 prioritize the test cases.
More specifically, since each candidate test case covers the same number of uncovered $\lambda$-wise code-unit-value combinations before prioritization, our approach follows the total and additional test prioritization techniques to choose the first test case:
the one covering the largest number of units (Lines 9--16).
Then, the number of uncovered $\lambda$-wise code-unit-value combinations against previously selected test cases ($S$) is calculated for each remaining test case, and a candidate with the maximum value is selected as the next test case is appended to $S$ (Lines 31--39).
Before choosing the next test case, our approach examines whether or not there are any $\lambda$-wise code-unit-value combinations that are not covered by the test cases in $S$:
If there are not, the remaining candidate test cases are prioritized by restarting the previous process (Lines 26--28).
Once an element is selected as the next test case, our approach updates the set of uncovered $\lambda$-wise code-unit-value combinations (Lines 23 and 46).
This process is repeated until all elements from $T$ have been added to $S$.
Similar to \textit{additional} test prioritization, when facing a tie where more than one test case has the largest number of uncovered code-unit-value combinations, our approach randomly selects one.

To further explain the details of the proposed approach, Figure \ref{FIG:example} illustrates an example of the CCCP process with $\lambda=1$.
Similar to the total and additional test prioritization techniques, CCCP chooses the first test case that covers the largest number of code units (the maximum amount of odd numbers).
Since there are two candidates with the maximum number of code units, $tc_1$ and $tc_2$ (both covering three code units), CCCP randomly chooses one of them (in this case, $tc_1$), and adds it to $S$.
CCCP then updates the set $UncoverCombinations$, and calculates the CCC value for each candidate:
$CCC(tc_2,S,1)=2$ and $CCC(tc_3,S,1)=3$.
Since $tc_3$ has the greater CCC value, it is selected as the next test case, and added to $S$.
In contrast, the \textit{total} prioritization technique would choose $tc_2$ as the second test case, because $tc_2$ covers more code units than $tc_3$; and
the \textit{additional} technique would randomly select one from $tc_2$ and $tc_3$ as the second test case, because both candidates cover the same number of uncovered code units.
Finally, in our approach, the last candidate $tc_2$ is added to $S$, resulting in $S=\langle tc_1,tc_3,tc_2 \rangle$.

\section{Empirical Study}
\label{exp}

In this section, we present our empirical study, including the research questions underlying the study.
We also discuss some independent and dependent variables, and explain the subject programs, test suites, and experimental setup in detail.

\subsection{Research Questions}

Due to space limitations and practical performance constraints (higher $\lambda$ values may result in more substantial running time), we present the evaluation of our proposed approach's performance when $\lambda=1$ and $2$.
Unless explicitly stated, $\lambda=1$ is used as the default value for CCCP.
The empirical study was conducted to answer the following six research questions.

\begin{description}
    \item[RQ1] How does CCCP compare with other RTCP approaches in terms of testing effectiveness measured by the fault detection rates?
    \item[RQ2] How does CCCP compare with other RTCP approaches in terms of testing effectiveness measured by the cost-cognizant fault detection rates?
    \item[RQ3] How does the granularity of code coverage impact the comparative effectiveness of CCCP?
    \item[RQ4] How does the granularity of test cases impact on the comparative effectiveness of CCCP?
    \item[RQ5] How does the efficiency of CCCP compare with other RTCP approaches, in terms of execution time?
    \item[RQ6] How does the use of code combinations coverage with $\lambda=2$ impact on the testing effectiveness of CCCP?
\end{description}

\subsection{Independent Variables}
In this study, we consider the following three independent variables.

\begin{table}[!b]
\centering
\scriptsize
 \caption{Studied RTCP techniques}
  \label{TAB:TCPtechniques}
    \setlength{\tabcolsep}{3.5mm}{
    \begin{tabular}{lll}
     \hline
        \textbf{Mnemonic}  &\textbf{Description} &\textbf{Reference} \\
        \hline
        $\textit{TCP}_\textit{tot}$ &Greedy total test prioritization &\cite{Rothermel1999}\\
        $\textit{TCP}_\textit{add}$ &Greedy additional test prioritization &\cite{Rothermel1999}\\
        $\textit{TCP}_\textit{art}$ &Adaptive random test prioritization &\cite{Jiang2009}\\
        $\textit{TCP}_\textit{search}$ &Search-based test prioritization &\cite{Li2007}\\
        $\textit{TCP}_\textit{ccc}$ &Our proposed CCCP technique &This study\\
        \hline
  \end{tabular}}
\end{table}

\begin{table*}[!t]
\scriptsize
\centering
 \caption{Subject program details}
  \label{TAB:programs}
  \setlength{\tabcolsep}{1.4mm}{
    \begin{tabular}{c|c|r|r|r|r|r|r|r|r|r|c|c}
     \hline
        \multirow{2}*{\textbf{Language}} &\multirow{2}*{\textbf{Program}} &\multirow{2}*{\textbf{Version}} &\multirow{2}*{\textbf{KLoC}} &\multirow{2}*{\textbf{\#Branch}} &\multirow{2}*{\textbf{\#Method}} &\multirow{2}*{\textbf{\#Class}} &\multicolumn{2}{c|}{\textbf{\#Test\_Case}} &\multicolumn{2}{c|}{\textbf{\#Mutant}} &\multicolumn{2}{c}{\textbf{\#Subsuming\_Mutant}} \\\cline{8-13}
        &&&&&&&\textbf{\#T\_Class} &\textbf{\#T\_Method} &\textbf{\#All} &\textbf{\#Detected} &\textbf{\#SM\_Class} &\textbf{\#SM\_Method}\\
     \hline

\multirow{14}*{Java}	
&$ant\_v1$	&v1\_9	&25.80 	&5,240	&2,511	&228	&34 (34)	&137 (135)	&6,498	&1,332	&59	&32	\\
&$ant\_v2$	&1.4	&39.70 	&8,797	&3,836	&342	&52 (52)	&219 (214)	&11,027	&2,677	&90	&47	\\
&$ant\_v3$	&1.4.1	&39.80 	&8,831	&3,845	&342	&52 (52)	&219 (213)	&11,142	&2,661	&92	&47	\\
\cline{2-13}												
&$jmeter\_v1$	&v1\_7\_3	&33.70 	&3,815	&2,919	&334	&26 (21)	&78 (61)	&8,850	&573	&38	&20	\\
&$jmeter\_v2$	&v1\_8   &33.10 	&3,799	&2,838	&319	&29 (24)	&80 (74)	&8,777	&867	&37	&22	\\
&$jmeter\_v3$	&v1\_8\_1 &37.30 	&4,351	&3,445	&373	&33 (27)	&78 (77)	&9,730	&1,667	&47	&25	\\
&$jmeter\_v4$	&v1\_9\_RC1  &38.40 	&4,484	&3,536	&380	&33 (27)	&78 (77)	&10,187	&1,703	&47	&25		\\
&$jmeter\_v5$	&v1\_9\_RC2  &41.10 	&4,888	&3,613	&389	&37 (30)	&97 (83)	&10,459	&1,651	&53	&29		\\
\cline{2-13}												
&$jtopas\_v1$	&0.4	&1.89 	&519	&284	&19	&10 (10)	&126 (126)	&704	&399	&29	&9	\\
&$jtopas\_v2$	&0.5.1	&2.03 	&583	&302	&21	&11 (11)	&128 (128)	&774	&446	&34	&10	\\
&$jtopas\_v3$	&0.6	&5.36 	&1,491	&748	&50	&18 (16)	&209 (207)	&1,906	&1,024	&57	&16	\\
\cline{2-13}												
&$xmlsec\_v1$	&v1\_0\_4    &18.30 	&3,534	&1,627	&179	&15 (15)	&92 (91)	&5,501	&1,198	&32	&12	\\	
&$xmlsec\_v2$	&v1\_0\_5D2  &19.00 	&3,789	&1,629	&180	&15 (15)	&94 (94)	&5,725	&1,204	&33	&12	\\	
&$xmlsec\_v3$	&v1\_0\_71   &16.90 	&3,156	&1,398	&145	&13 (13)	&84 (84)	&3,833	&1,070	&27	&10	\\	
	 \hline												
\multirow{30}*{C}	&$flex\_v0$	&2.4.3	&8.96	&2,005	&138	&--	&\multicolumn{2}{c|}{500}		&--	&--	&\multicolumn{2}{c}{--}		 \\
	&$flex\_v1$	&2.4.7	&9.47	&2,011	&147	&--	&\multicolumn{2}{c|}{500}		&13,873	&6,177	&\multicolumn{2}{c}{32}		 \\
	&$flex\_v2$	&2.5.1	&12.23	&2,656	&162	&--	&\multicolumn{2}{c|}{500}		&14,822	&6,396	&\multicolumn{2}{c}{32}		 \\
	&$flex\_v3$	&2.5.2	&12.25	&2,666	&162	&--	&\multicolumn{2}{c|}{500}		&775	&420	&\multicolumn{2}{c}{20}		 \\
	&$flex\_v4$	&2.5.3	&12.38	&2,678	&162	&--	&\multicolumn{2}{c|}{500}		&14,906	&6,417	&\multicolumn{2}{c}{33}		 \\
	&$flex\_v5$	&2.5.4	&12.37	&2,680	&162	&--	&\multicolumn{2}{c|}{500}		&14,922	&6,418	&\multicolumn{2}{c}{32}		 \\
	\cline{2-13}												
	&$grep\_v0$	&2.0	&8.16	&3,420	&119	&--	&\multicolumn{2}{c|}{144}		&--	&--	&\multicolumn{2}{c}{--}		 \\
	&$grep\_v1$	&2.2	&11.99	&3,511	&104	&--	&\multicolumn{2}{c|}{144}		&23,896	&3,229	&\multicolumn{2}{c}{56}		 \\
	&$grep\_v2$	&2.3	&12.72	&3,631	&109	&--	&\multicolumn{2}{c|}{144}		&24,518	&3,319	&\multicolumn{2}{c}{58}		 \\
	&$grep\_v3$	&2.4	&12.83	&3,709	&113	&--	&\multicolumn{2}{c|}{144}		&17,656	&3,156	&\multicolumn{2}{c}{54}		 \\
	&$grep\_v4$	&2.5	&20.84	&2,531	&102	&--	&\multicolumn{2}{c|}{144}		&17,738	&3,445	&\multicolumn{2}{c}{58}		 \\
	&$grep\_v5$	&2.7	&58.34	&2,980	&109	&--	&\multicolumn{2}{c|}{144}		&17,108	&3492	&\multicolumn{2}{c}{59}		 \\
	\cline{2-13}												
	&$gzip\_v0$	&1.0.7	&4.32	&1,468	&81	&--	&\multicolumn{2}{c|}{156}		&--	&--	&\multicolumn{2}{c}{--}		 \\
	&$gzip\_v1$	&1.1.2	&4.52	&1,490	&81	&--	&\multicolumn{2}{c|}{156}		&7,429	&639	&\multicolumn{2}{c}{8}		 \\
	&$gzip\_v2$	&1.2.2	&5.05	&1,752	&98	&--	&\multicolumn{2}{c|}{156}		&7,599	&659	&\multicolumn{2}{c}{8}		 \\
	&$gzip\_v3$	&1.2.3	&5.06	&1,610	&93	&--	&\multicolumn{2}{c|}{156}		&7,678	&547	&\multicolumn{2}{c}{7}		 \\
	&$gzip\_v4$	&1.2.4	&5.18	&1,663	&93	&--	&\multicolumn{2}{c|}{156}		&7,838	&548	&\multicolumn{2}{c}{7}		 \\
	&$gzip\_v5$	&1.3	&5.68	&1,733	&97	&--	&\multicolumn{2}{c|}{156}		&8,809	&210	&\multicolumn{2}{c}{7}		 \\
	\cline{2-13}												
	&$make\_v0$	&3.75	&17.46	&4,397	&181	&--	&\multicolumn{2}{c|}{111}		&--	&--	&\multicolumn{2}{c}{--}		 \\
	&$make\_v1$	&3.76.1	&18.57	&4,585	&181	&--	&\multicolumn{2}{c|}{111}		&36,262	&5,800	&\multicolumn{2}{c}{37}		 \\
	&$make\_v2$	&3.77	&19.66	&4,784	&190	&--	&\multicolumn{2}{c|}{111}		&38,183	&5,965	&\multicolumn{2}{c}{29}		 \\
	&$make\_v3$	&3.78.1	&20.46	&4,845	&216	&--	&\multicolumn{2}{c|}{111}		&42,281	&6,244	&\multicolumn{2}{c}{28}		 \\
	&$make\_v4$	&3.79	&23.13	&5,413	&239	&--	&\multicolumn{2}{c|}{111}		&48,546	&6,958	&\multicolumn{2}{c}{29}		 \\
	&$make\_v5$	&3.80	&23.40	&5,032	&268	&--	&\multicolumn{2}{c|}{111}		&47,310	&7,049	&\multicolumn{2}{c}{28}		 \\
	\cline{2-13}												
	&$sed\_v0$	&3.01	&7.79	&676	&66	&--	&\multicolumn{2}{c|}{324}		&--	&--	&\multicolumn{2}{c}{--}		 \\
	&$sed\_v1$	&3.02	&7.79	&712	&65	&--	&\multicolumn{2}{c|}{324}		&2,506	&1,009	&\multicolumn{2}{c}{16}		 \\
	&$sed\_v2$	&4.0.6	&18.55	&1,011	&65	&--	&\multicolumn{2}{c|}{324}		&5,947	&1,048	&\multicolumn{2}{c}{18}		 \\
	&$sed\_v3$	&4.0.8	&18.69	&1,017	&66	&--	&\multicolumn{2}{c|}{324}		&5,970	&450	&\multicolumn{2}{c}{18}		 \\
	&$sed\_v4$	&4.1.1	&21.74	&1,141	&70	&--	&\multicolumn{2}{c|}{324}		&6,578	&470	&\multicolumn{2}{c}{19}		 \\
	&$sed\_v$5	&4.2	&26.47	&1,412	&98	&--	&\multicolumn{2}{c|}{324}		&7,761	&628	&\multicolumn{2}{c}{22}		 \\

	 \hline												

  \end{tabular}}
\end{table*}

\subsubsection{Prioritization Techniques}

Since our proposed CCCP technique takes advantage of the information about dynamic coverage and test cases as inputs, for a fair comparison, it is necessary to choose other RTCP techniques that use  similar information to guide the test cases prioritization.
In this study, we selected four such RTCP techniques:
\textit{total test prioritization} \cite{Rothermel1999}, \textit{additional test prioritization} \cite{Rothermel1999}, \textit{adaptive random test prioritization} \cite{Jiang2009}, and \textit{search-based test prioritization} \cite{Li2007}.
The total test prioritization technique prioritizes test cases based on the descending number of code units covered by those tests.
The additional technique, in contrast, greedily chooses each element from candidates such that it covers the largest number of code units not yet covered by the previously selected tests.
The adaptive random technique greedily selects each element from random candidates such that it has the greatest maximum distance from the already selected tests.
Finally, the search-based technique considers all permutations as candidate solutions, and uses a meta-heuristic search algorithm to guide the search for a better test execution order
---
a genetic algorithm was adopted in this study, due to its effectiveness \cite{Li2007}.
These four test prioritization techniques, whose details are presented in Table \ref{TAB:TCPtechniques}, have been widely used in RTCP previous studies \cite{Luo2016,Luo2019,Lu2016}.

\subsubsection{Code Coverage Granularity}
When using code coverage information to support RTCP, the coverage granularity can be considered  a constituent part of the prioritization techniques.
To enable sufficient evaluations, we used structural coverage criteria at the statement-, branch-,  and method-coverage levels.

\subsubsection{Test Case Granularity}
For the subject programs written in Java, we considered an additional factor in the prioritization techniques, the test-case granularity.
Test-case granularity is at either the test-class level, or the test-method level.
For the test-class level, each JUnit test-case class was a test case;
for the test-method level, each test method in a JUnit test-case class was a test case.
In other words, a test case at the test-class level generally involves a number of test cases at the test-method level.
For C subject programs, however, the actual program inputs were the test cases.

\subsection{Dependent Variables}
Because we were examining the fault detection capability, we adopted the widely-used APFD (\textit{average percentage faults detected}) as the evaluation metric for fault detection rates~\cite{Rothermel1999}.
Given a test suite $T$, with $n$ test cases.
If $T'$ is a permutation of $T$, in which $TF_i$ is the position of first test case that reveals the fault $i$, then the APFD value for $T'$ is given by the following equation:
\begin{equation}
\footnotesize
	APFD=1-\frac{\sum_{i=1}^{m}{TF_i}}{n*m}+\frac{1}{2n}
\end{equation}

Although APFD has often been used to evaluate RTCP techniques, it assumes that each test case incurs the same time cost, an assumption that may not hold up in practice.
Elbaum et al. \cite{Elbaum2001}, therefore, introduced an APFD variant,  APFD$\textrm{c}$, a cost-cognizant version of APFD that considers both the fault severity and the test case execution cost.
Similar to APFD, APFD$\textrm{c}$ has also been applied to the evaluation of RTCP techniques, resulting in a more comprehensive evaluation \cite{Epitropakis2015}.
APFD$\textrm{c}$ is defined as:
\begin{equation}
\footnotesize
	APFD_c = \frac{\sum_{i=1}^m(\alpha_i \times (\sum_{j=TF_i}^n \beta_j - \frac{1}{2}\alpha_i))}{\sum_{i=1}^m\alpha_i \times \sum_{i=1}^n\beta_i}
\end{equation}
where $\alpha_1,\alpha_2,\cdots,\alpha_m$ are the severities of the $m$ detected faults,
$\beta_1,\beta_2,\cdots,\beta_n$ are the execution costs of $n$ test cases, and
$TF_i$ has the same meaning as in APFD.
Because of the difficulty involved in estimating fault severity, following previous studies \cite{Epitropakis2015}, we assumed that all faults had the same severity level for this study.
Accordingly, the definition of APFD$\textrm{c}$ can be described as:
\begin{equation}
    \footnotesize
    APFD_c = \frac{\sum_{i=1}^m(\sum_{j=TF_i}^n \beta_j - \frac{1}{2}\beta_{TF_i})}{m \times \sum_{j=1}^n\beta_j}
\end{equation}
Intuitively speaking, prioritized test suites that both find faults faster and require less execution time, will have higher APFD$\textrm{c}$ values.

\subsection{Subject Programs, Test Suites and Faults}
We conducted our study on 14 versions of four Java programs (three versions of $ant$; five versions of $jmeter$; three versions of $jtopas$; and three versions of $xmlsec$) downloaded from the \textit{Software-artifact Infrastructure Repository} (SIR)~\cite{2005Do,sir}, and 30 versions of five real-life Unix utility programs, from the \textit{GNU FTP server}~\cite{gnu}.
Both the Java and C programs have been widely used as benchmarks to evaluate RTCP problems~\cite{Jiang2009,Zhang2013,Henard2016,Eghbali2016}.
Table \ref{TAB:programs} summarizes the subject program details, with Columns 3 to 7 presenting the version,  size,  number of branches,  number of methods, and number of classes (including interfaces), respectively.

Each version of the Java programs has a JUnit test suite that was developed during the program's evolution.
These test suites have two levels of test-case granularity:
the test-class and the test-method.
The numbers of JUnit test cases (including both test-class and test-method levels) are shown in the \textbf{\#Test\_Case} column, as \textbf{\#T\_ Class} and \textbf{\#T\_Method}:
The data is presented as $x~(y)$, where $x$ is the total number of test cases; and
$y$ is the number of test cases that can be successfully executed.
The test suites for the C programs are available from the SIR~\cite{2005Do,sir}:
The number of tests cases in each suite is also shown in the \textbf{\#Test\_Case} column of Table \ref{TAB:programs}.


Because the seeded-in SIR faults were easily detected and small in size, for both C and Java programs, we used mutation faults to evaluate the performance of the different techniques.
Mutation faults have previously been identified as suitable for simulating real program faults~\cite{Andrews2005,Do2005,Just2014}, and have been widely applied to regression test prioritization evaluations~\cite{Rothermel1999,Zhang2013,Luo2016,Luo2019,Henard2016,Lu2016,Elbaum2000}.
Eleven mutation operators from the ``NEW\_DEFAULTS'' group of the PIT mutation tool~\cite{pit} were used to generate mutants for the Java programs.
These operators, whose detailed descriptions can be found on the PIT website \cite{pitop}, were:
\textit{conditionals boundary}, \textit{increments}, \textit{invert negatives}, \textit{math}, \textit{negate conditionals}, \textit{void method calls}, \textit{empty returns}, \textit{false returns}, \textit{null returns}, \textit{primitive returns}, and \textit{true returns}.
We downloaded the mutants from previous RTCP studies \cite{Henard2016} for the C programs, which had been generated using seven mutation operators from Andrews et al.~\cite{Andrews2006}:
\textit{statement deletion}, \textit{unary insertion}, \textit{constant replacement}, \textit{arithmetic operator replacement}, \textit{logical operator replacement}, \textit{bitwise logical operator replacement}, and \textit{relational operator replacement}.


\begin{figure*}[!t]
\graphicspath{{graphs/}}
\centering
    \subfigure[$ant$-statement coverage]
    {
        \includegraphics[width=0.319\textwidth]{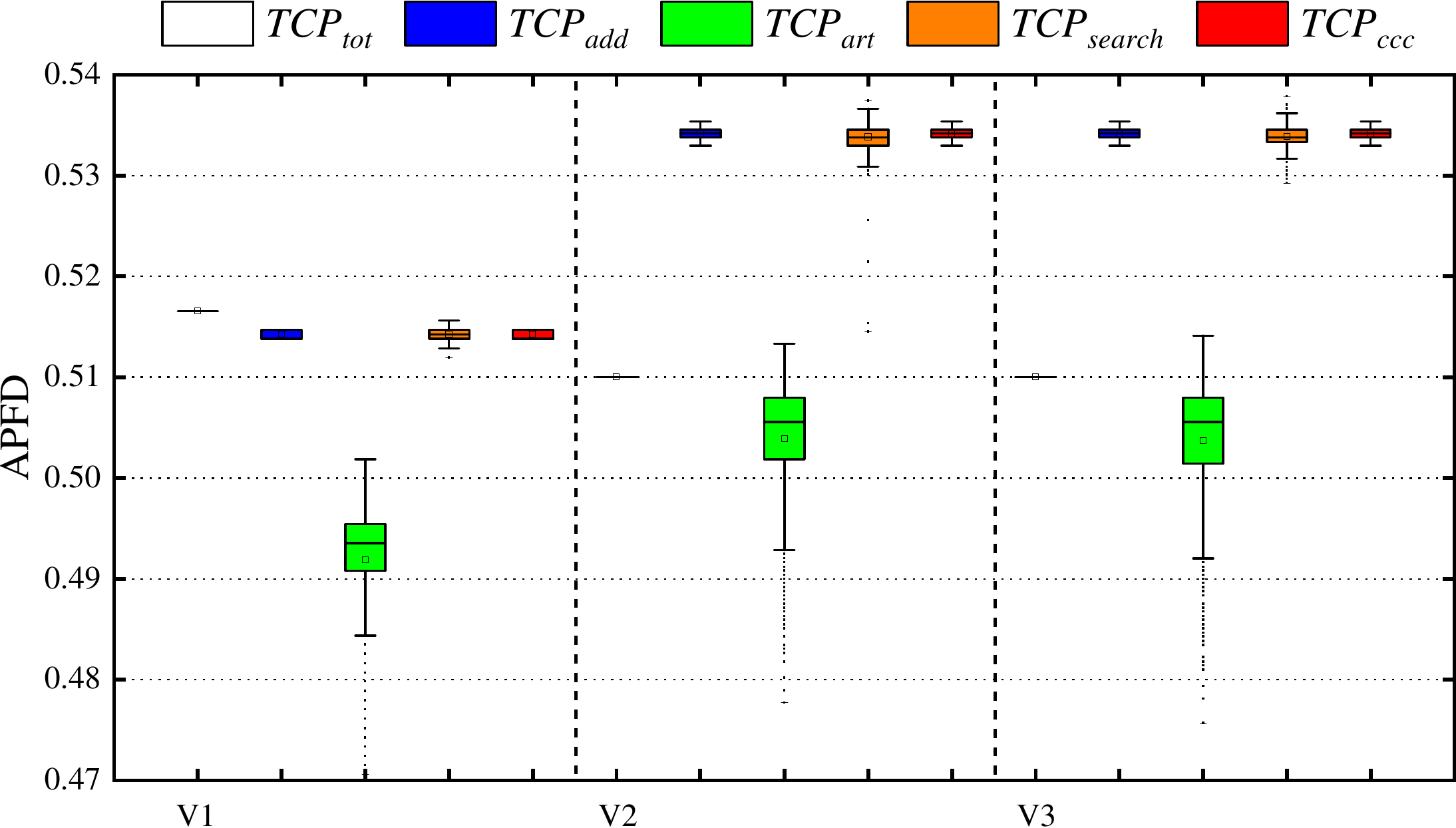}
        \label{apfd_ant_sc}
    }
    \subfigure[$ant$-branch coverage]
    {
        \includegraphics[width=0.319\textwidth]{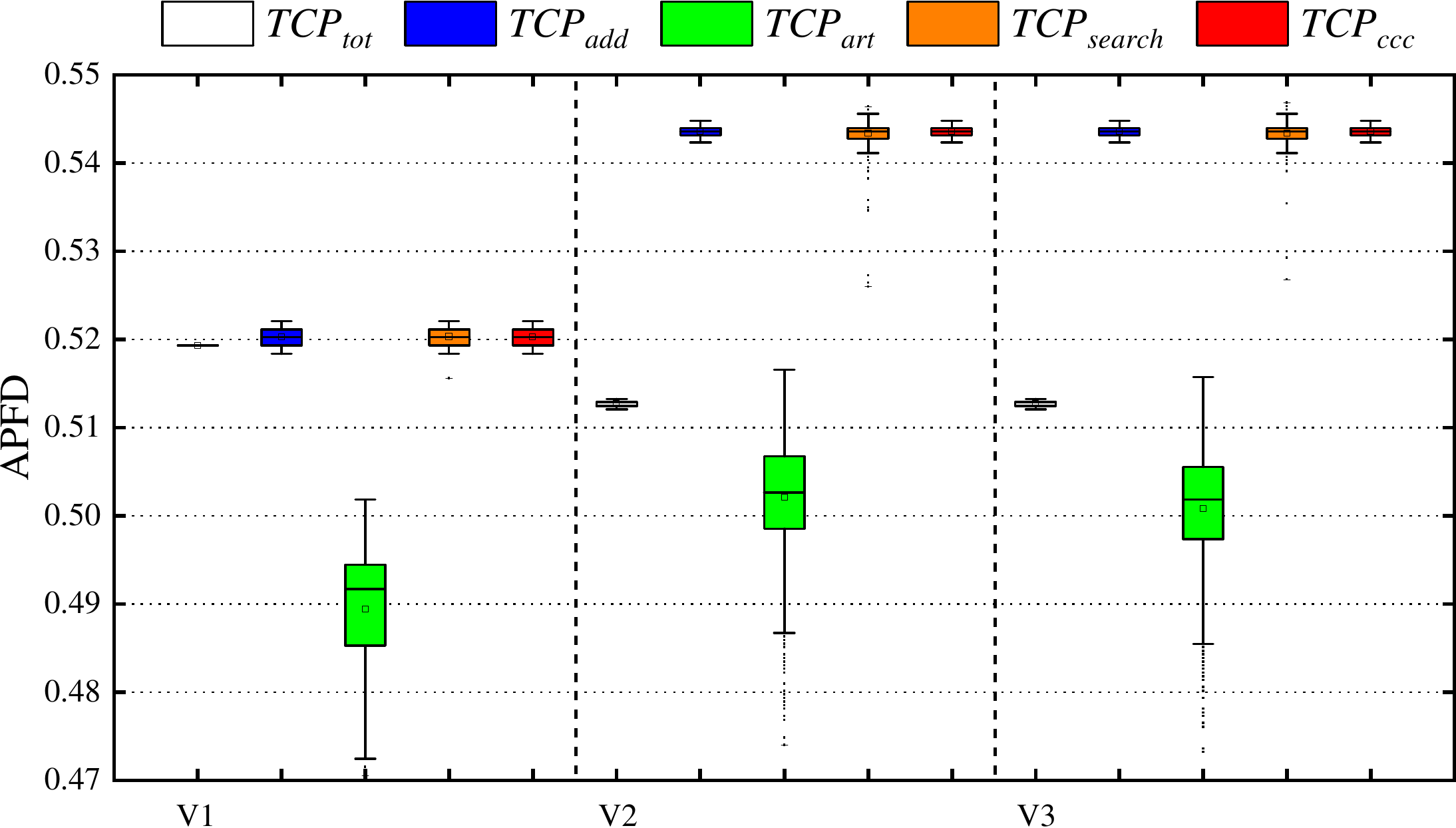}
        \label{apfd_ant_bc}
    }
        \subfigure[$ant$-method coverage]
    {
        \includegraphics[width=0.319\textwidth]{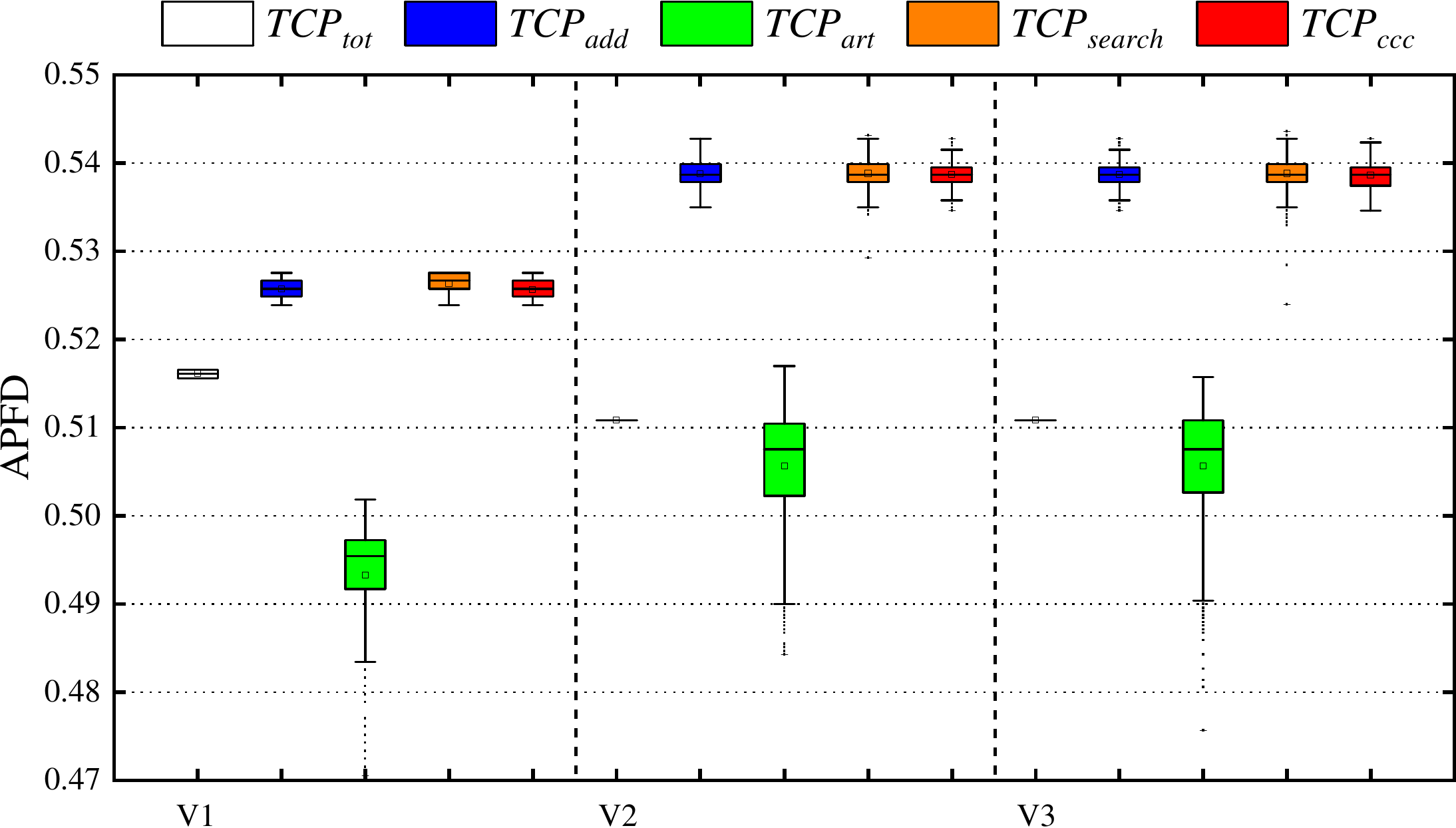}
        \label{apfd_ant_mc}
    }
    \subfigure[$jmeter$-statement coverage]
    {
        \includegraphics[width=0.319\textwidth]{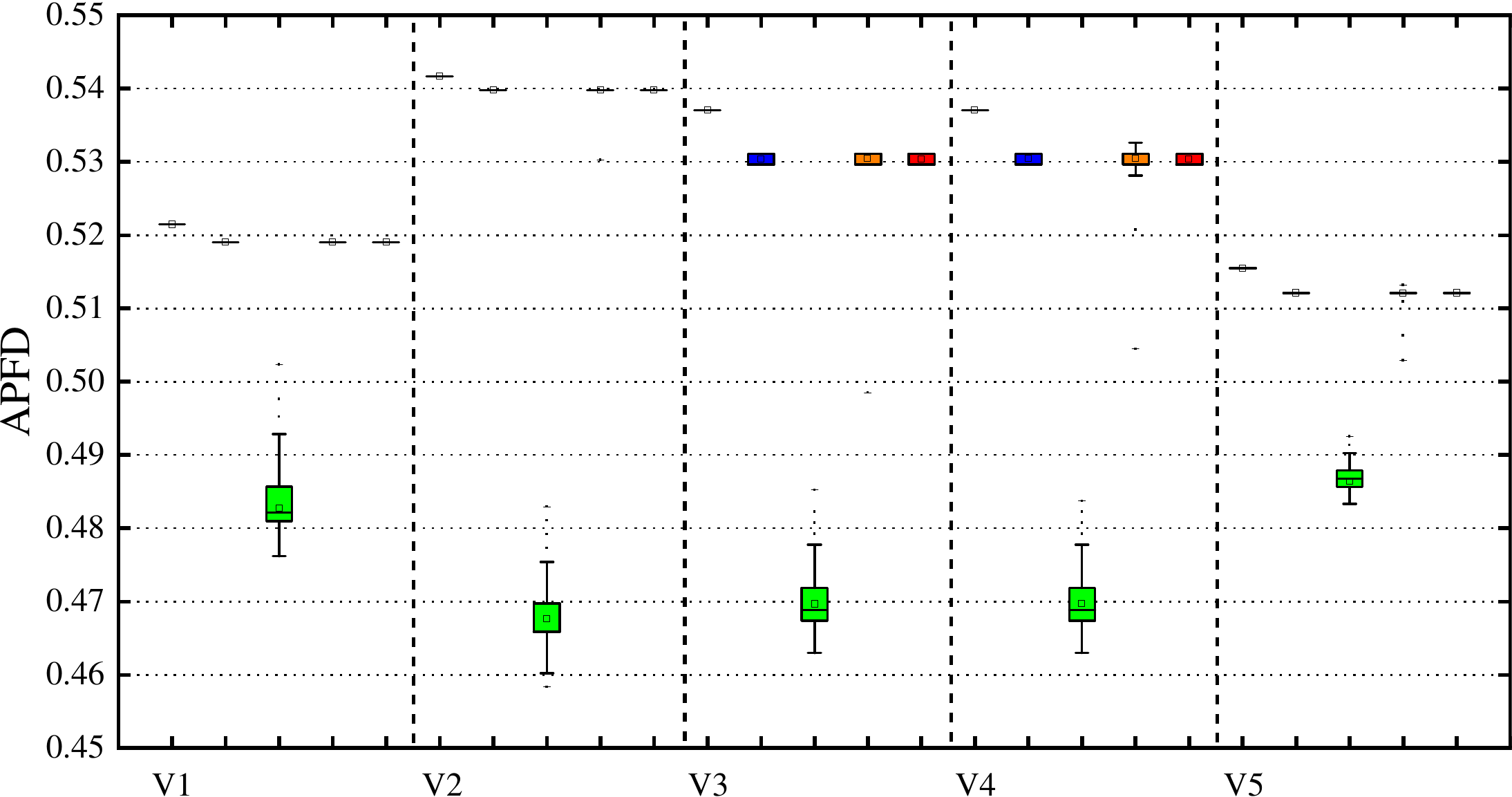}
        \label{apfd_grep_sc}
    }
        \subfigure[$jmeter$-branch coverage]
    {
        \includegraphics[width=0.319\textwidth]{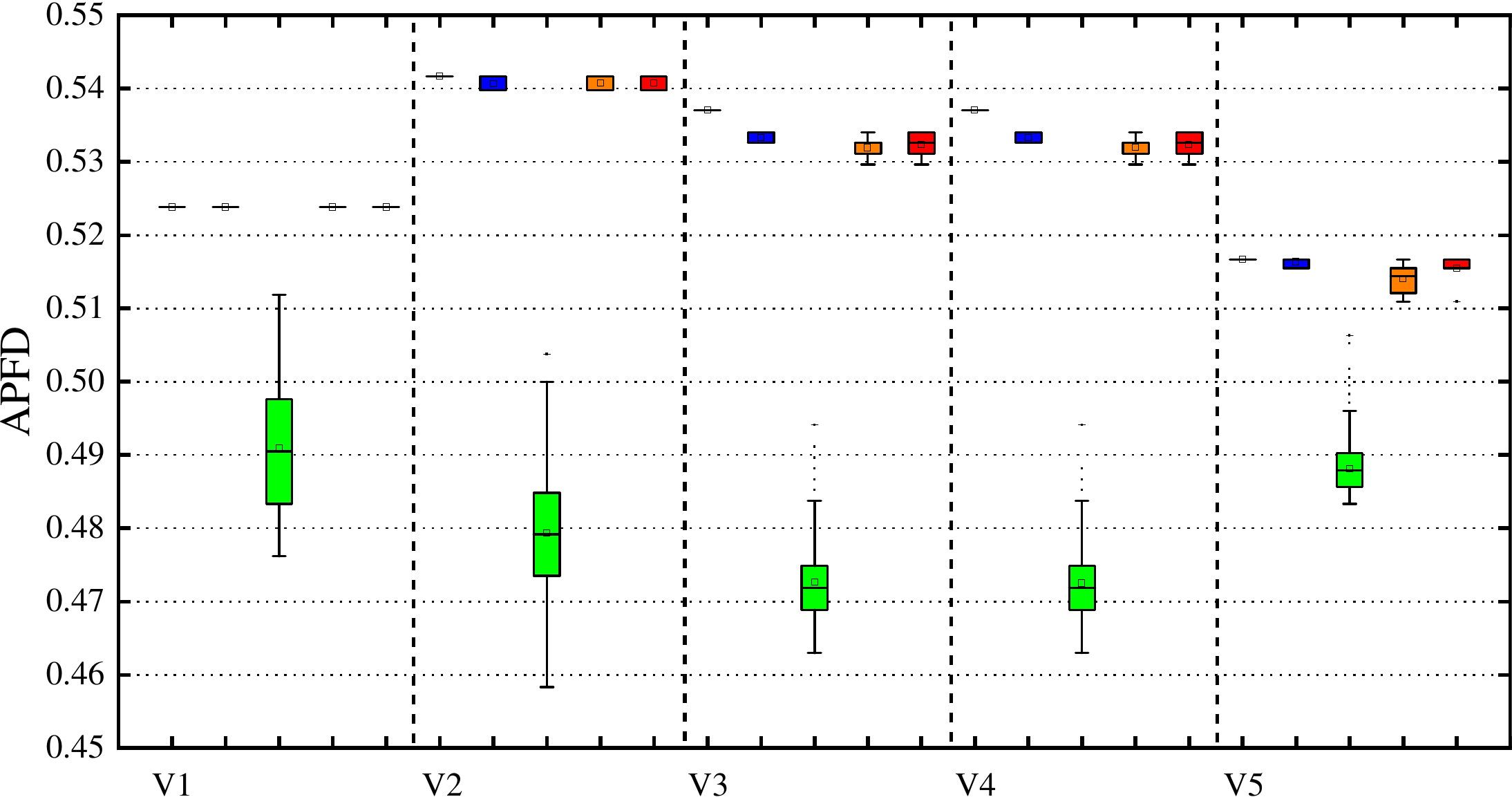}
        \label{apfd_jmeter_bc}
    }
        \subfigure[$jmeter$-method coverage]
    {
        \includegraphics[width=0.319\textwidth]{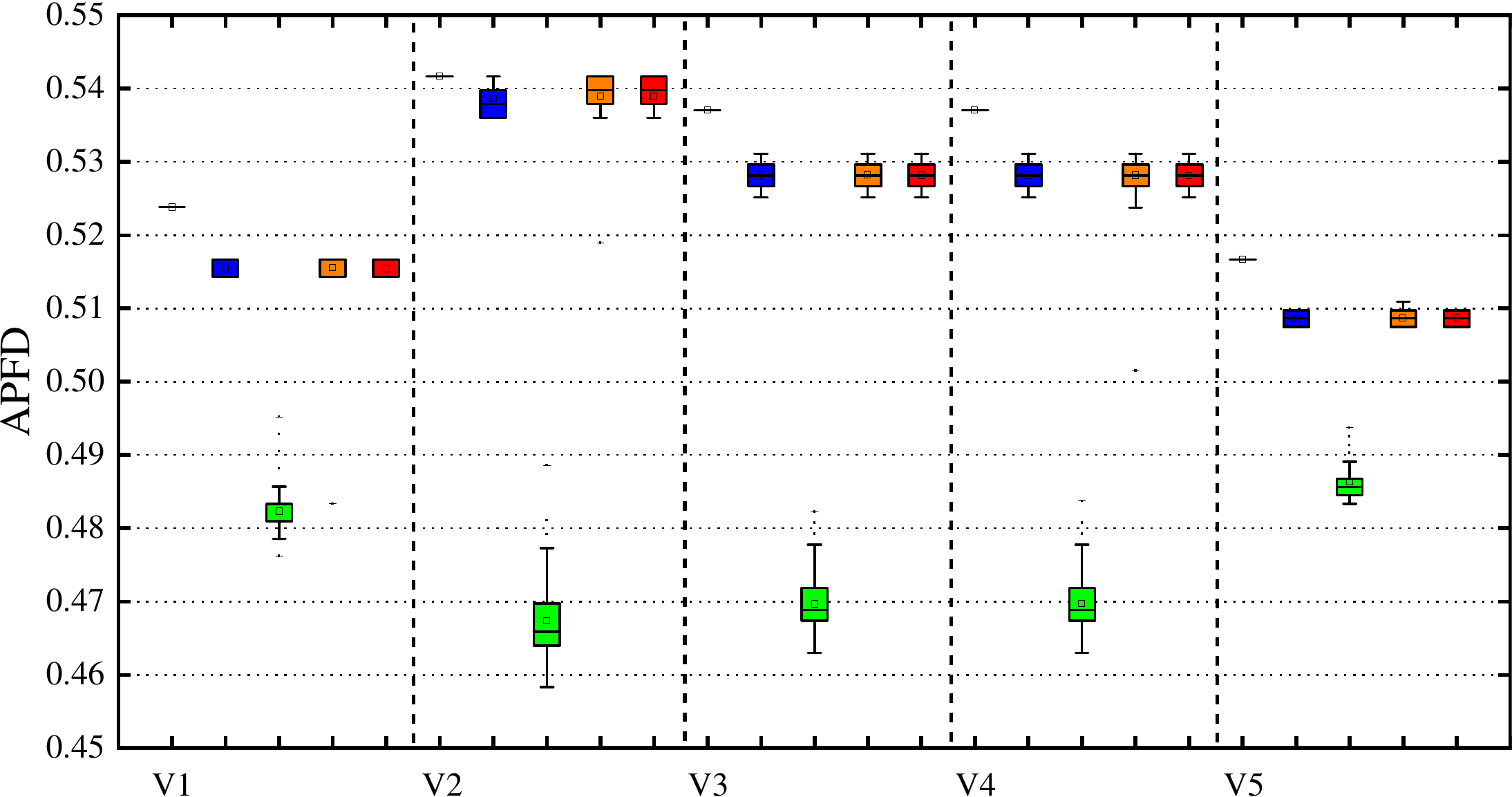}
        \label{apfd_jmeter_mc}
    }
        \subfigure[$jtopas$-statement coverage]
    {
        \includegraphics[width=0.319\textwidth]{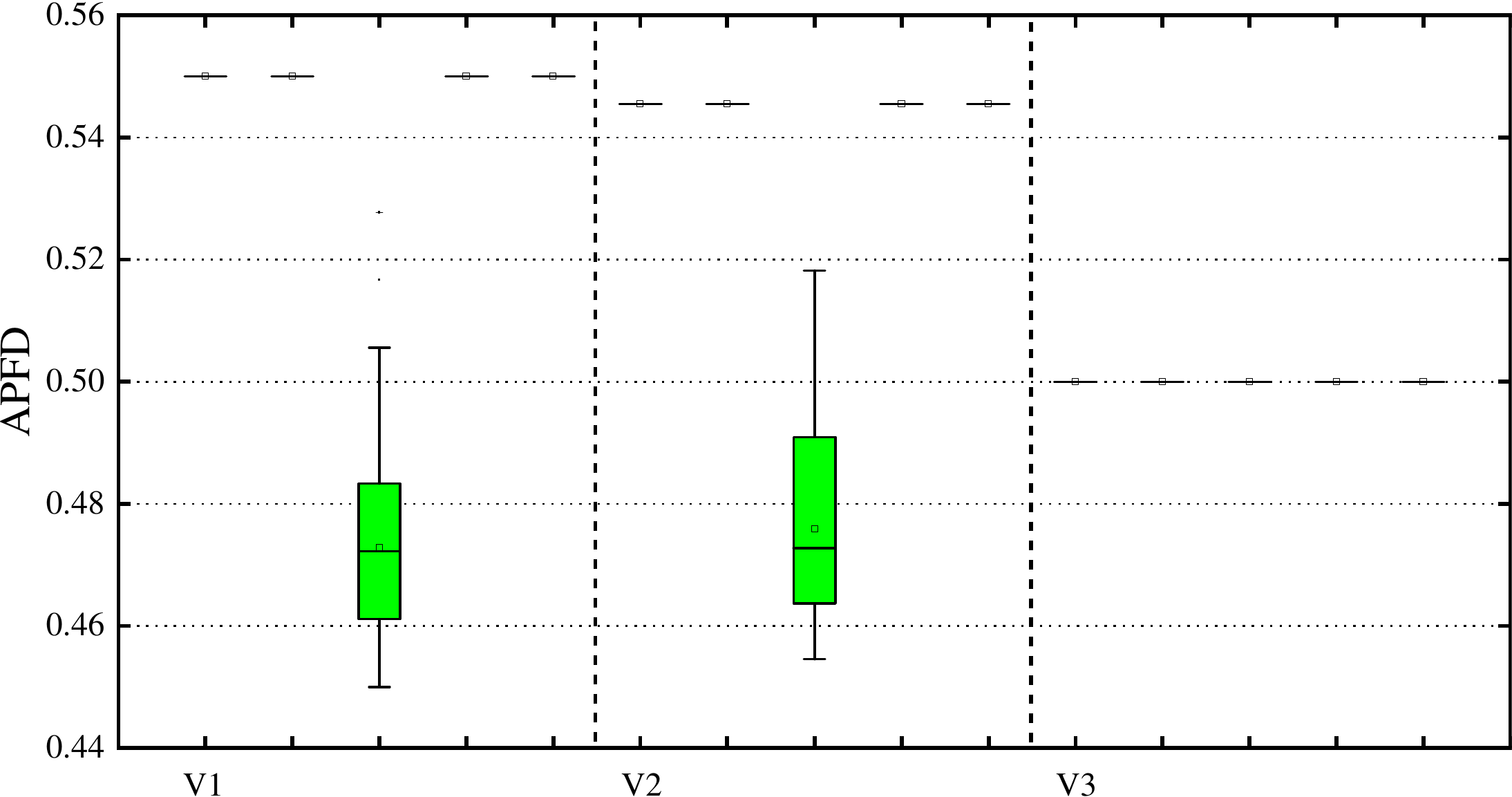}
        \label{apfd_jtopas_sc}
    }
        \subfigure[$jtopas$-branch coverage]
    {
        \includegraphics[width=0.319\textwidth]{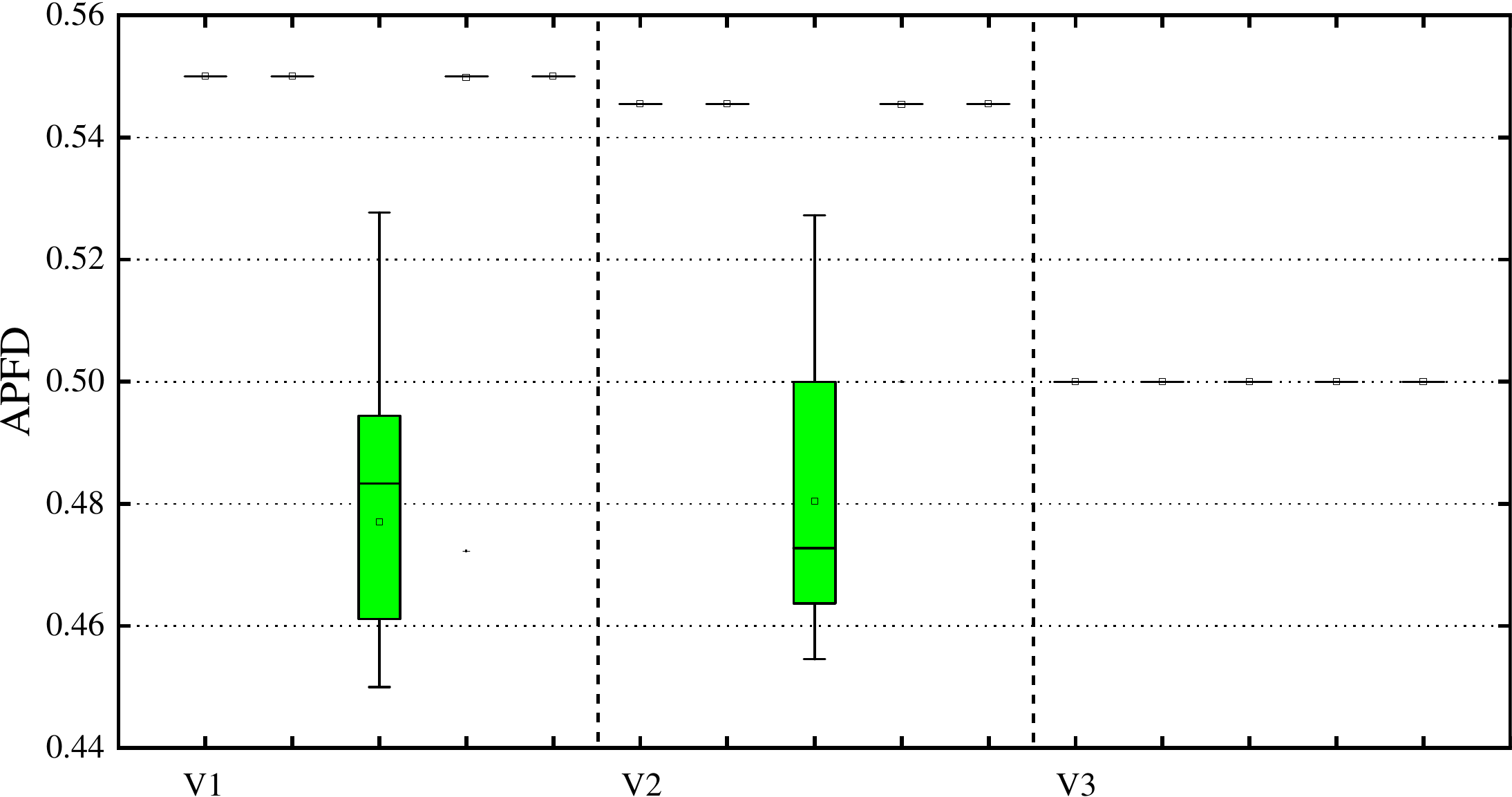}
        \label{apfd_jtopas_bc}
    }
        \subfigure[$jtopas$-method coverage]
    {
        \includegraphics[width=0.319\textwidth]{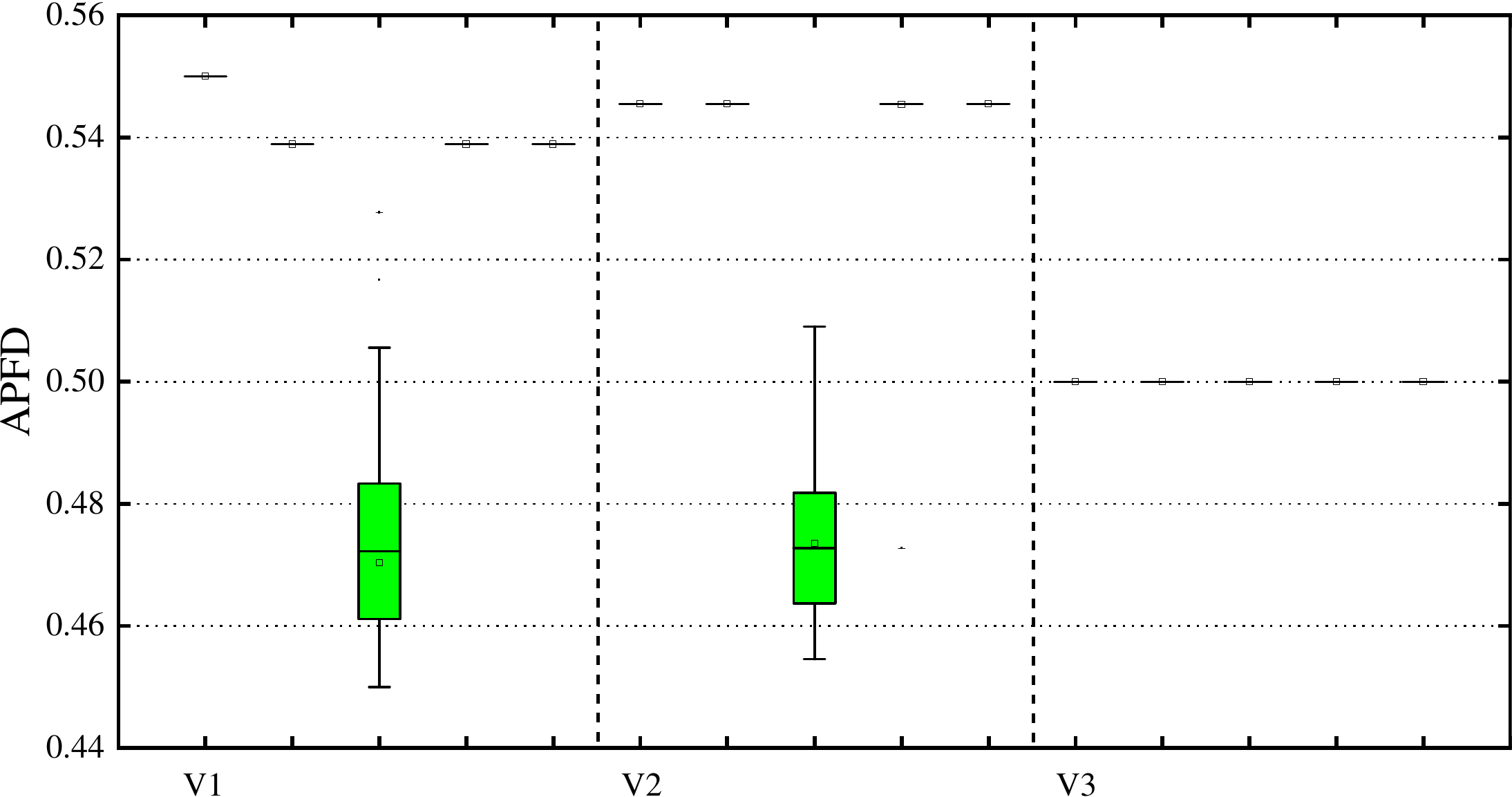}
        \label{apfd_jtopas_mc}
    }2005Do
        \subfigure[$xmlsec$-statement coverage]
    {
        \includegraphics[width=0.319\textwidth]{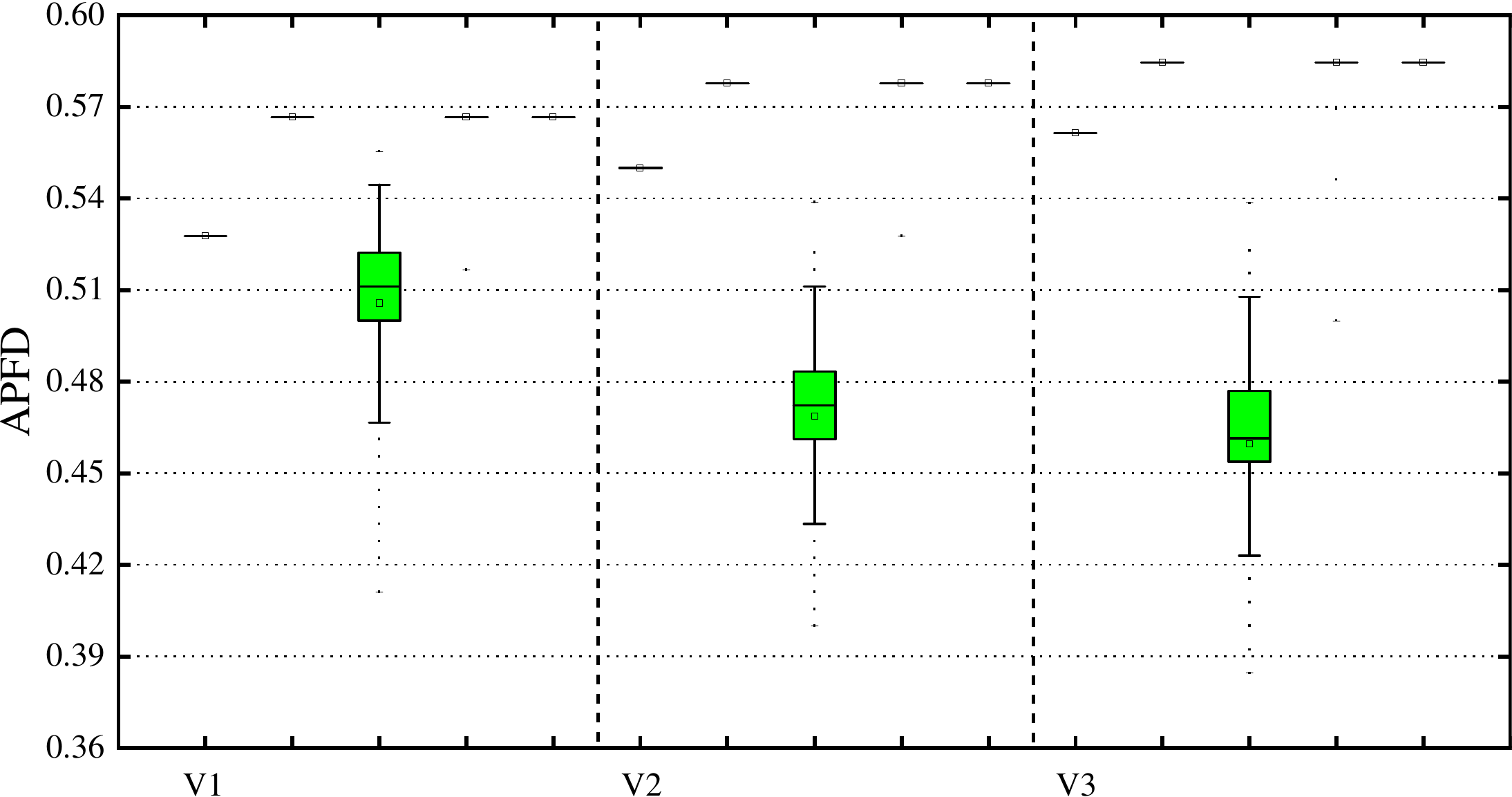}
        \label{apfd_xmlsec_sc}
    }
        \subfigure[$xmlsec$-branch coverage]
    {
        \includegraphics[width=0.319\textwidth]{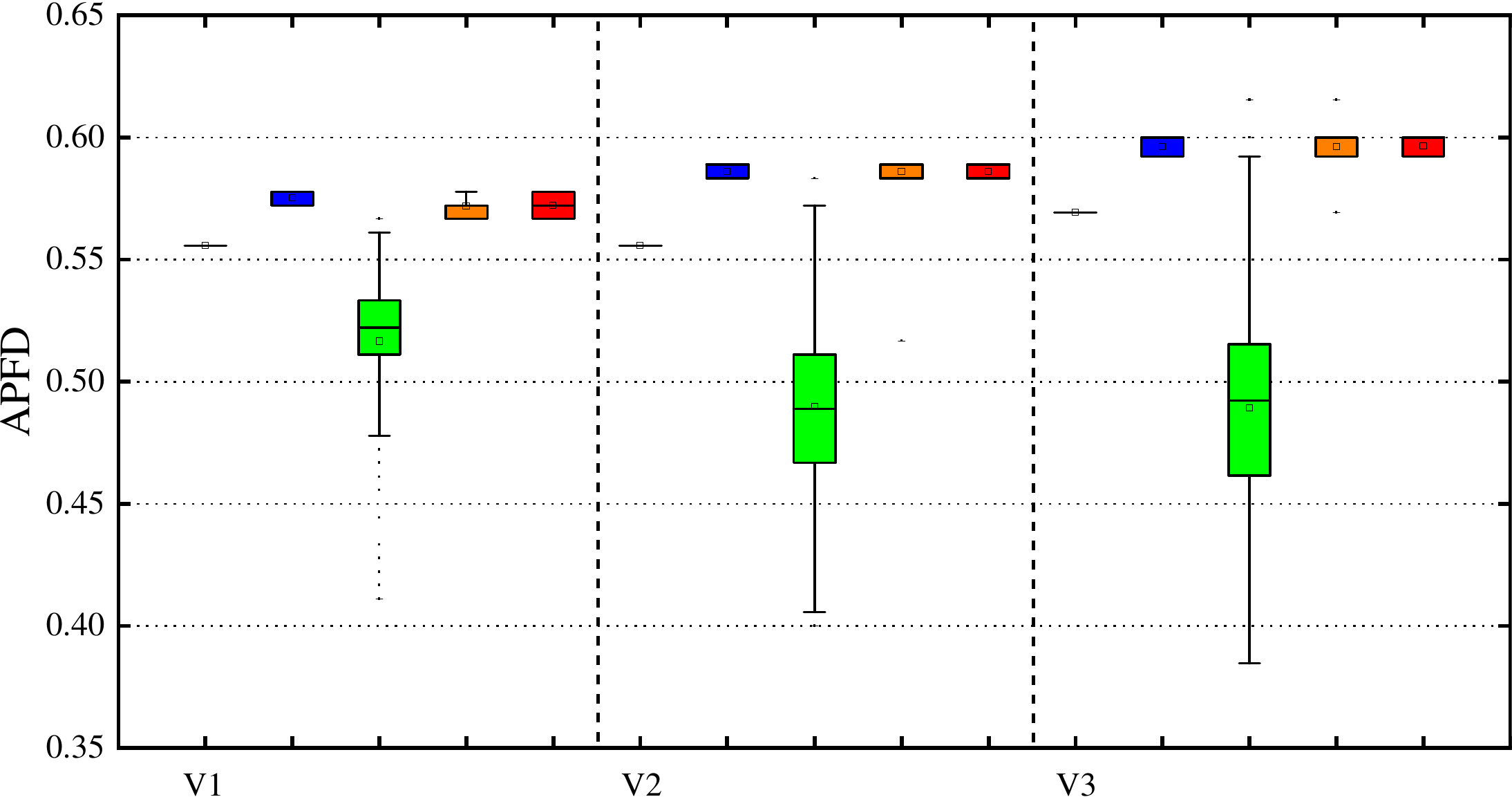}
        \label{apfd_xmlsec_bc}
    }
        \subfigure[$xmlsec$-method coverage]
    {
        \includegraphics[width=0.319\textwidth]{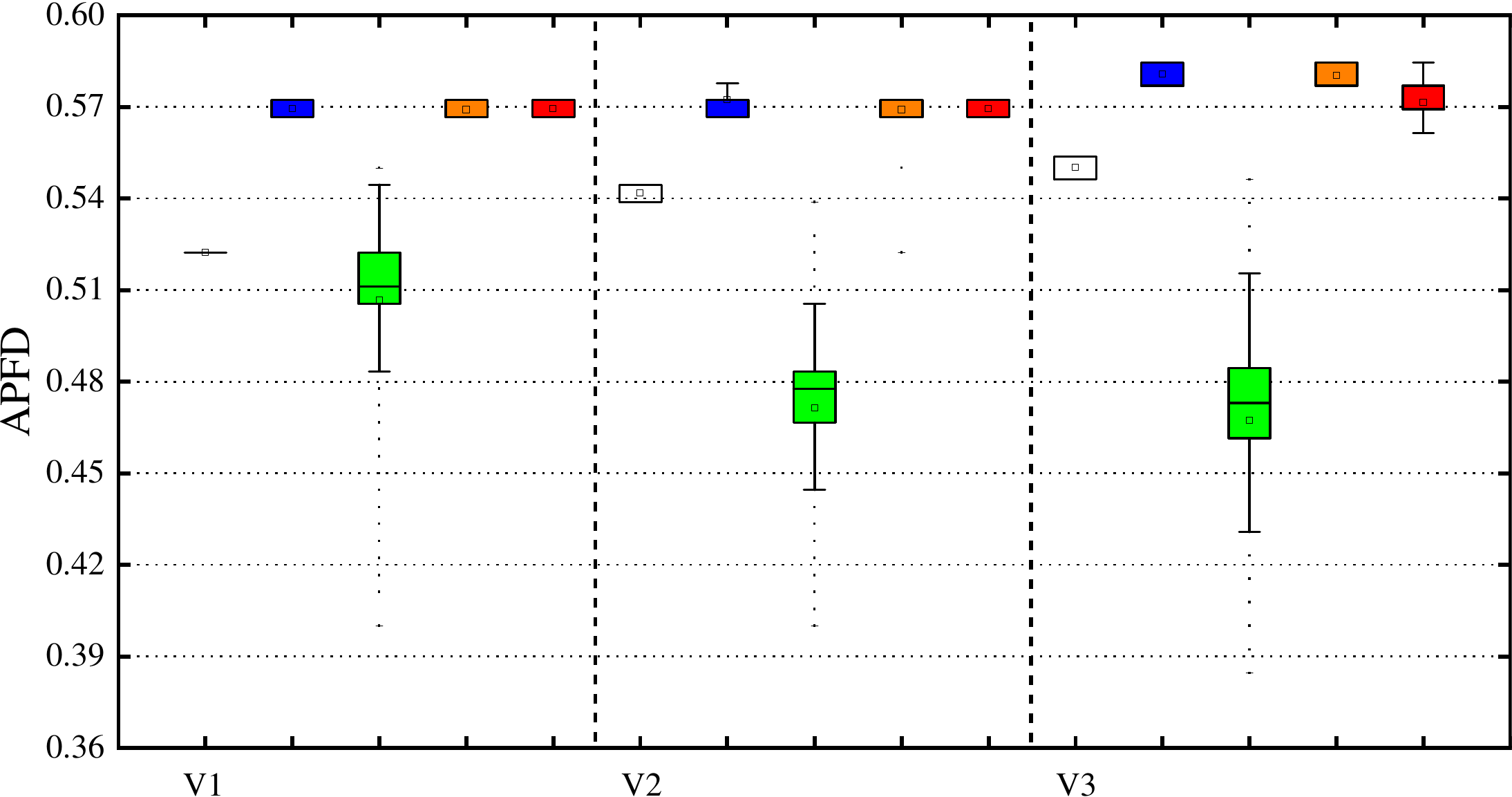}
        \label{apfd_xmlsec_mc}
    }
    \caption{\textbf{Effectiveness:} \textbf{APFD} results for \textbf{Java programs} at the \textbf{test-class level}}
    \label{FIG:apfd-java-c}
\end{figure*}

\textit{Equivalent mutants} \cite{Schuler2009,Papadakis2015} are functionally equivalent versions of the original program, and thus cannot be killed:
no test case applied to both the mutant and the original program could result in different behavior.
In our study, therefore, all equivalent mutants were removed, leaving only those mutants that could be detected by at least one test case.
In Table \ref{TAB:programs}, the \textbf{\#Mutant} column gives the total number of all mutants (\textbf{\#All}), and the (\textbf{\#Detected}) column gives the number of detected mutants.
Although all detected mutants were considered in our study, some mutants, called \textit{duplicated mutants}) \cite{Papadakis2015}, were equivalent to other mutants (but not to the original program).
Similarly, some mutants, called \textit{subsumed mutants} \cite{Just2012,Kaminski2013} were subsumed by others:
If a \textit{subsuming mutant} \cite{Papadakis2016}) is killed, then its subsumed mutants are also killed.
We used the \textit{Subsuming Mutants Identification} (SMI) algorithm \cite{Papadakis2016} to remmove the duplicate and subsuming mutants in our fault set.
SMI first removed duplicate mutants, and then greedily identified the most subsuming mutants
---
those mutants which, when killed, result in the highest number of other mutants being ``collaterally'' killed.
The \textbf{\#Subsuming\_Mutant} column gives the number of subsuming mutants used in our study
(the subsuming faults are classified as either test-class level (\textbf{\#SM\_Class}) or test-method level (\textbf{\#SM\_Method})
for the Java programs).

\subsection{Experimental Setup}

The experiments were conducted on a Linux 4.4.0-170-generic server with two Intel(R) Xeon(R) Platinum 8163 CPUs (2.50 GHz, two cores) and 16 GBs of RAM.

The Java programs were compiled using Java 1.8 \cite{java}.
The coverage information for each program version was obtained using the \texttt{FaultTracer} tool \cite{Zhang2012,Zhang2013a}, which, based on the ASM bytecode manipulation and analysis framework \cite{asm}, uses on-the-fly bytecode instrumentation without any modification of the target program.


There were six versions of each C program:
$P_{V0}$, $P_{V1}$, $P_{V2}$, $P_{V3}$, $P_{V4}$, and $P_{V5}$.
Version $P_{V0}$ was compiled using \texttt{gcc}  5.4.0 \cite{gcc}, and then the coverage information was obtained using the \texttt{gcov} tool~\cite{gcov}, which is one of the \texttt{gcc} utilities.

After collecting the code coverage information, we implemented all RTCP techniques in Java, and applied them to each program version under study, for all coverage criteria.
Because the approaches contain randomness, each execution was repeated 1000 times.
This resulted in, for each testing scenario, 1000 APFD or APFD$_\textrm{c}$ values (1000 orderings) for each RTCP approach.
To test whether there was a statistically significant difference between CCCP and the other RTCP approaches, we performed the unpaired two-tailed Wilcoxon-Mann-Whitney test, at a significance level of $5\%$, following previously reported guidelines for inferential statistical analyses involving randomized algorithms~\cite{Arcuri2014,Gligoric2015}.
To identify which approach was better, we also calculated the effect size, measured by the non-parametric Vargha and Delaney effect size measure~\cite{Vargha2000}, $\hat{\textrm{A}}_{12}$, where $\hat{\textrm{A}}_{12} (X,Y)$ gives probability that the technique $X$ is better than technique $Y$.
The statistical analyses were performed using \texttt{R} \cite{Rstatis}.


\begin{table*}[!t]
\scriptsize
\centering
 \caption{Statistical \textbf{effectiveness} comparisons of  \textbf{APFD} for \textbf{Java programs} at the \textbf{test-class level}.
For a comparison between two methods
$\textit{TCP}_\textit{ccc}$ and $M$,
where
$M \in \{\textit{TCP}_\textit{tot}, \textit{TCP}_\textit{add}, \textit{TCP}_\textit{art}, \textit{TCP}_\textit{search}\}$,
the symbol \ding{52} means that $\textit{TCP}_\textit{ccc}$ is better
($p$-value is less than 0.05, and the effect size $\hat{\textrm{A}}_{12}(\textit{TCP}_\textit{ccc},M)$ is greater than 0.50);
the symbol \ding{54} means that $M$ is better
(the $p$-value is less than 0.05, and $\hat{\textrm{A}}_{12}(\textit{TCP}_\textit{ccc},M)$ is less than 0.50);
and the symbol \ding{109} means that there is no statistically significant difference between them (the $p$-value is greater than 0.05).
}
  \label{TAB:apfd-java-c}
    \setlength{\tabcolsep}{1.8mm}{
    \begin{tabular}{c|cccc|cccc|cccc}
     \hline
       \multirow{2}*{\textbf{Program Name}} &\multicolumn{4}{c|}{\textbf{Statement Coverage}} &\multicolumn{4}{c|}{\textbf{Branch Coverage}} &\multicolumn{4}{c}{\textbf{Method Coverage}} \\
        \cline{2-13}

         &$\textit{TCP}_\textit{tot}$ &$\textit{TCP}_\textit{add}$ &$\textit{TCP}_\textit{art}$ &$\textit{TCP}_\textit{search}$  & $\textit{TCP}_\textit{tot}$ &$\textit{TCP}_\textit{add}$ &$\textit{TCP}_\textit{art}$ &$\textit{TCP}_\textit{search}$ &$\textit{TCP}_\textit{tot}$ & $\textit{TCP}_\textit{add}$ &$\textit{TCP}_\textit{art}$ &$\textit{TCP}_\textit{search}$\\
            \hline

$ant\_v1$	&\ding{54}(0.00)	&\ding{109}(0.49)	&\ding{52}(1.00)	&\ding{109}(0.50)	&\ding{52}(0.76)	&\ding{109}(0.51)	&\ding{52}(1.00)	&\ding{109}(0.49)	&\ding{52}(1.00)	&\ding{109}(0.49)	&\ding{52}(1.00)	&\ding{54}(0.36)	\\
$ant\_v2$	&\ding{52}(1.00)	&\ding{109}(0.51)	&\ding{52}(1.00)	&\ding{52}(0.57)	&\ding{52}(1.00)	&\ding{109}(0.49)	&\ding{52}(1.00)	&\ding{109}(0.52)	&\ding{52}(1.00)	&\ding{109}(0.49)	&\ding{52}(1.00)	&\ding{54}(0.47)	\\
$ant\_v3$	&\ding{52}(1.00)	&\ding{109}(0.50)	&\ding{52}(1.00)	&\ding{52}(0.59)	&\ding{52}(1.00)	&\ding{109}(0.50)	&\ding{52}(1.00)	&\ding{52}(0.54)	&\ding{52}(1.00)	&\ding{109}(0.49)	&\ding{52}(1.00)	&\ding{54}(0.46)	\\\hline
$ant$	&\ding{52}(0.89)	&\ding{109}(0.50)	&\ding{52}(1.00)	&\ding{52}(0.53)	&\ding{52}(0.97)	&\ding{109}(0.50)	&\ding{52}(1.00)	&\ding{109}(0.51)	&\ding{52}(1.00)	&\ding{109}(0.49)	&\ding{52}(1.00)	&\ding{54}(0.47)	\\\hline
$jmeter\_v1$	&\ding{54}(0.00)	&\ding{109}(0.50)	&\ding{52}(1.00)	&\ding{109}(0.50)	&\ding{109}(0.50)	&\ding{109}(0.50)	&\ding{52}(1.00)	&\ding{109}(0.50)	&\ding{54}(0.00)	&\ding{109}(0.50)	&\ding{52}(1.00)	&\ding{54}(0.46)	\\
$jmeter\_v2$	&\ding{54}(0.00)	&\ding{109}(0.50)	&\ding{52}(1.00)	&\ding{109}(0.50)	&\ding{54}(0.24)	&\ding{109}(0.51)	&\ding{52}(1.00)	&\ding{109}(0.50)	&\ding{54}(0.14)	&\ding{52}(0.53)	&\ding{52}(1.00)	&\ding{109}(0.49)	\\
$jmeter\_v3$	&\ding{54}(0.00)	&\ding{109}(0.51)	&\ding{52}(1.00)	&\ding{109}(0.48)	&\ding{54}(0.00)	&\ding{54}(0.34)	&\ding{52}(1.00)	&\ding{52}(0.59)	&\ding{54}(0.00)	&\ding{109}(0.50)	&\ding{52}(1.00)	&\ding{54}(0.47)	\\
$jmeter\_v4$	&\ding{54}(0.00)	&\ding{54}(0.47)	&\ding{52}(1.00)	&\ding{54}(0.46)	&\ding{54}(0.00)	&\ding{54}(0.33)	&\ding{52}(1.00)	&\ding{52}(0.57)	&\ding{54}(0.00)	&\ding{109}(0.49)	&\ding{52}(1.00)	&\ding{109}(0.48)	\\
$jmeter\_v5$	&\ding{54}(0.00)	&\ding{109}(0.50)	&\ding{52}(1.00)	&\ding{109}(0.50)	&\ding{54}(0.32)	&\ding{54}(0.43)	&\ding{52}(1.00)	&\ding{52}(0.72)	&\ding{54}(0.00)	&\ding{109}(0.51)	&\ding{52}(1.00)	&\ding{109}(0.50)	\\\hline
$jmeter$	&\ding{54}(0.36)	&\ding{109}(0.50)	&\ding{52}(1.00)	&\ding{109}(0.50)	&\ding{54}(0.40)	&\ding{54}(0.47)	&\ding{52}(1.00)	&\ding{52}(0.52)	&\ding{54}(0.32)	&\ding{109}(0.50)	&\ding{52}(1.00)	&\ding{109}(0.49)	\\\hline
$jtopas\_v1$	&\ding{109}(0.50)	&\ding{109}(0.50)	&\ding{52}(1.00)	&\ding{109}(0.50)	&\ding{109}(0.50)	&\ding{109}(0.50)	&\ding{52}(1.00)	&\ding{109}(0.50)	&\ding{54}(0.00)	&\ding{109}(0.50)	&\ding{52}(1.00)	&\ding{109}(0.50)	\\
$jtopas\_v2$	&\ding{109}(0.50)	&\ding{109}(0.50)	&\ding{52}(1.00)	&\ding{109}(0.50)	&\ding{109}(0.50)	&\ding{109}(0.50)	&\ding{52}(1.00)	&\ding{109}(0.50)	&\ding{109}(0.50)	&\ding{109}(0.50)	&\ding{52}(1.00)	&\ding{109}(0.50)	\\
$jtopas\_v3$	&\ding{109}(0.50)	&\ding{109}(0.50)	&\ding{109}(0.50)	&\ding{109}(0.50)	&\ding{109}(0.50)	&\ding{109}(0.50)	&\ding{109}(0.50)	&\ding{109}(0.50)	&\ding{109}(0.50)	&\ding{109}(0.50)	&\ding{109}(0.50)	&\ding{109}(0.50)	\\\hline
$jtopas$	&\ding{109}(0.50)	&\ding{109}(0.50)	&\ding{52}(0.93)	&\ding{109}(0.50)	&\ding{109}(0.50)	&\ding{109}(0.50)	&\ding{52}(0.91)	&\ding{109}(0.50)	&\ding{54}(0.33)	&\ding{109}(0.50)	&\ding{52}(0.94)	&\ding{109}(0.50)	\\\hline
$xmlsec\_v1$	&\ding{52}(1.00)	&\ding{109}(0.50)	&\ding{52}(1.00)	&\ding{109}(0.50)	&\ding{52}(1.00)	&\ding{54}(0.31)	&\ding{52}(1.00)	&\ding{109}(0.52)	&\ding{52}(1.00)	&\ding{109}(0.51)	&\ding{52}(1.00)	&\ding{52}(0.54)	\\
$xmlsec\_v2$	&\ding{52}(1.00)	&\ding{109}(0.50)	&\ding{52}(1.00)	&\ding{109}(0.50)	&\ding{52}(1.00)	&\ding{109}(0.50)	&\ding{52}(1.00)	&\ding{109}(0.51)	&\ding{52}(1.00)	&\ding{54}(0.32)	&\ding{52}(1.00)	&\ding{52}(0.54)	\\
$xmlsec\_v3$	&\ding{52}(1.00)	&\ding{109}(0.50)	&\ding{52}(1.00)	&\ding{109}(0.50)	&\ding{52}(1.00)	&\ding{109}(0.52)	&\ding{52}(1.00)	&\ding{109}(0.51)	&\ding{52}(1.00)	&\ding{54}(0.16)	&\ding{52}(1.00)	&\ding{54}(0.18)	\\\hline
$xmlsec$	&\ding{52}(1.00)	&\ding{109}(0.50)	&\ding{52}(1.00)	&\ding{109}(0.50)	&\ding{52}(0.97)	&\ding{54}(0.48)	&\ding{52}(1.00)	&\ding{109}(0.50)	&\ding{52}(1.00)	&\ding{54}(0.31)	&\ding{52}(1.00)	&\ding{54}(0.40)	\\\hline
\textit{\textbf{All Java Programs}}	&\ding{52}(0.58)	&\ding{109}(0.50)	&\ding{52}(0.98)	&\ding{109}(0.50)	&\ding{52}(0.58)	&\ding{109}(0.50)	&\ding{52}(0.96)	&\ding{109}(0.50)	&\ding{52}(0.59)	&\ding{54}(0.49)	&\ding{52}(0.97)	&\ding{54}(0.49)	\\
\hline
\end{tabular}}

\end{table*}

\begin{figure*}[!t]
\graphicspath{{graphs/}}
\centering
    \subfigure[$ant$-statement coverage]
    {
        \includegraphics[width=0.319\textwidth]{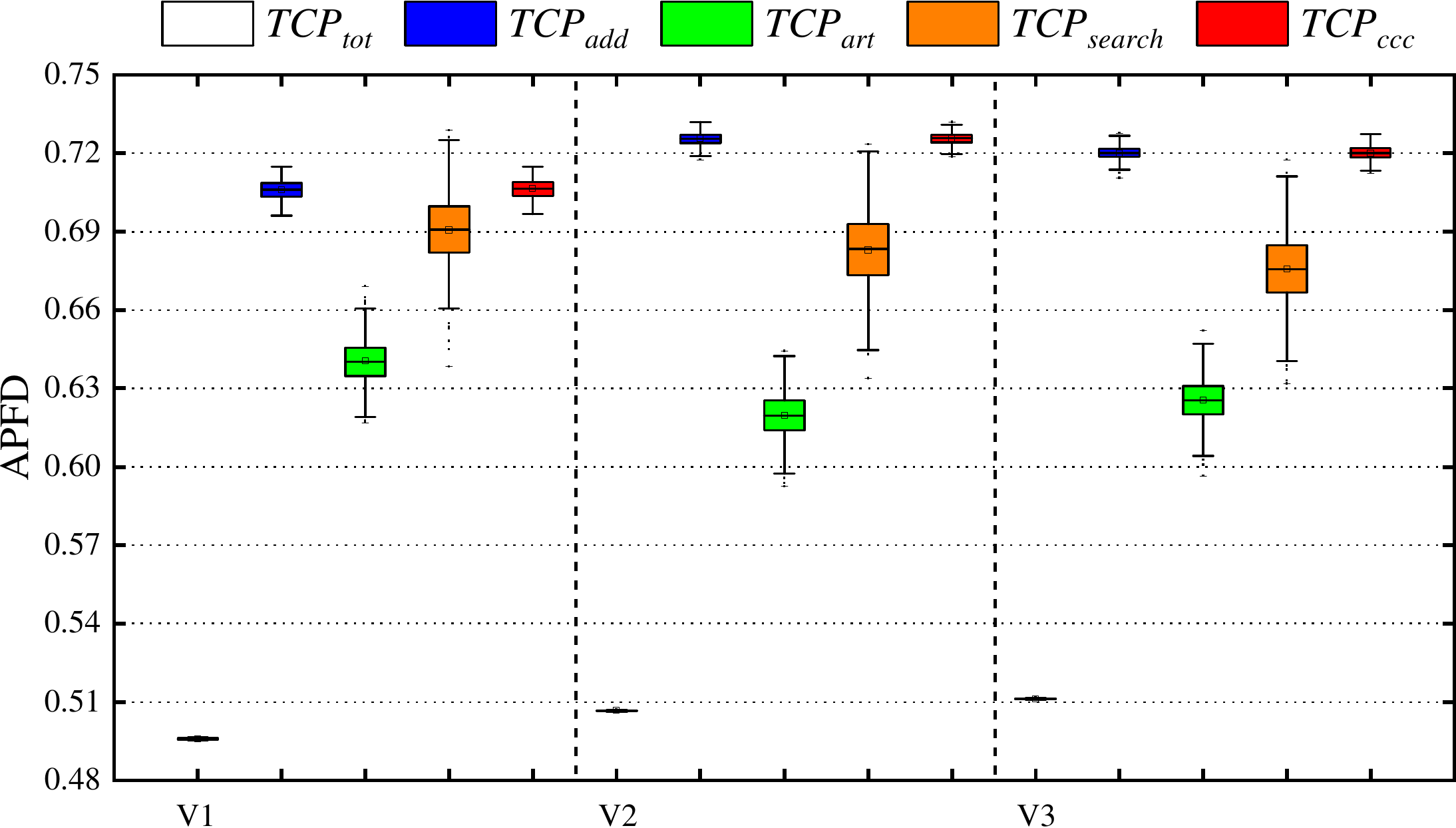}
        \label{apfd_ant_sc}
    }
    \subfigure[$ant$-branch coverage]
    {
        \includegraphics[width=0.319\textwidth]{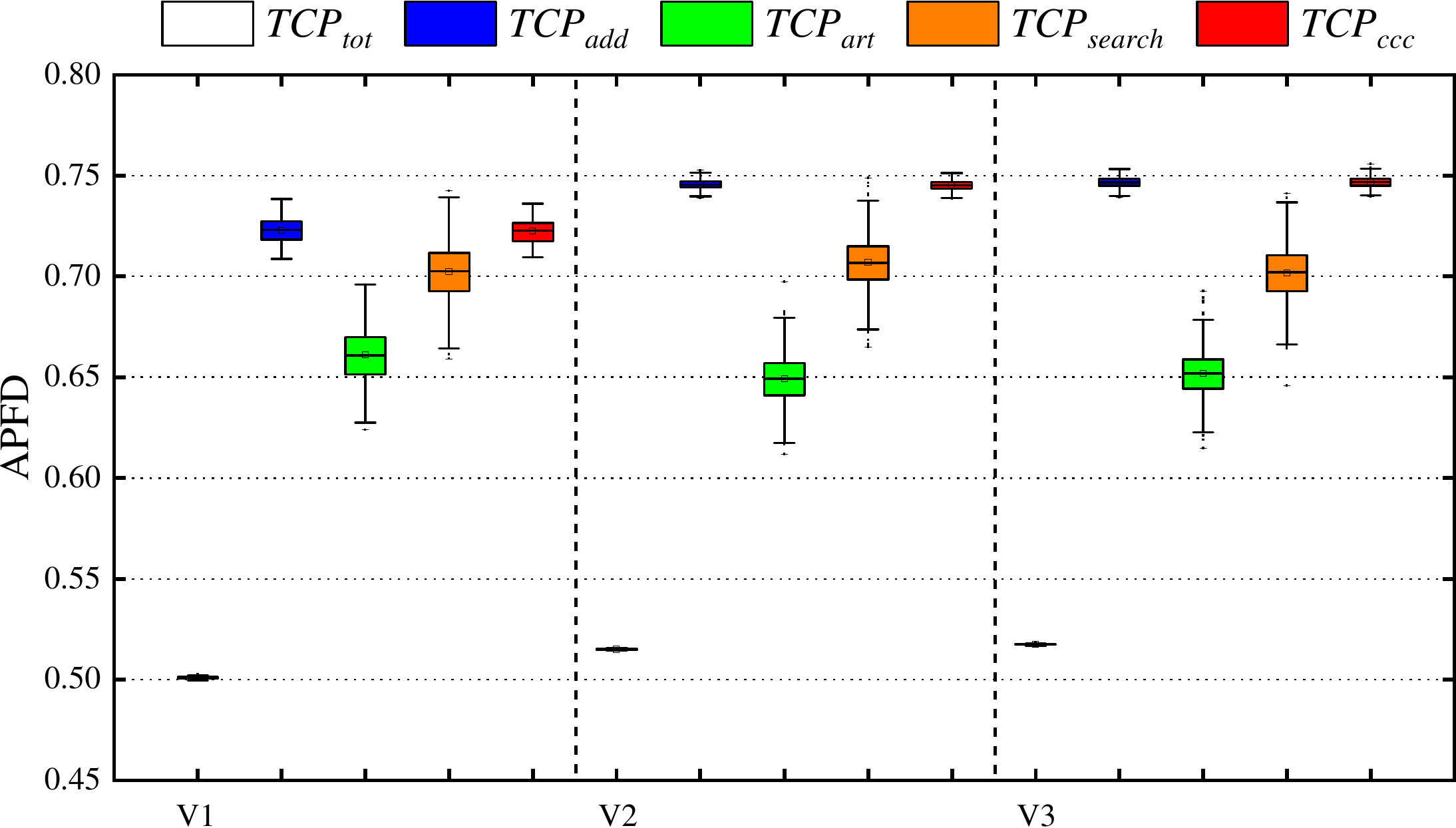}
        \label{apfd_ant_bc}
    }
        \subfigure[$ant$-method coverage]
    {
        \includegraphics[width=0.319\textwidth]{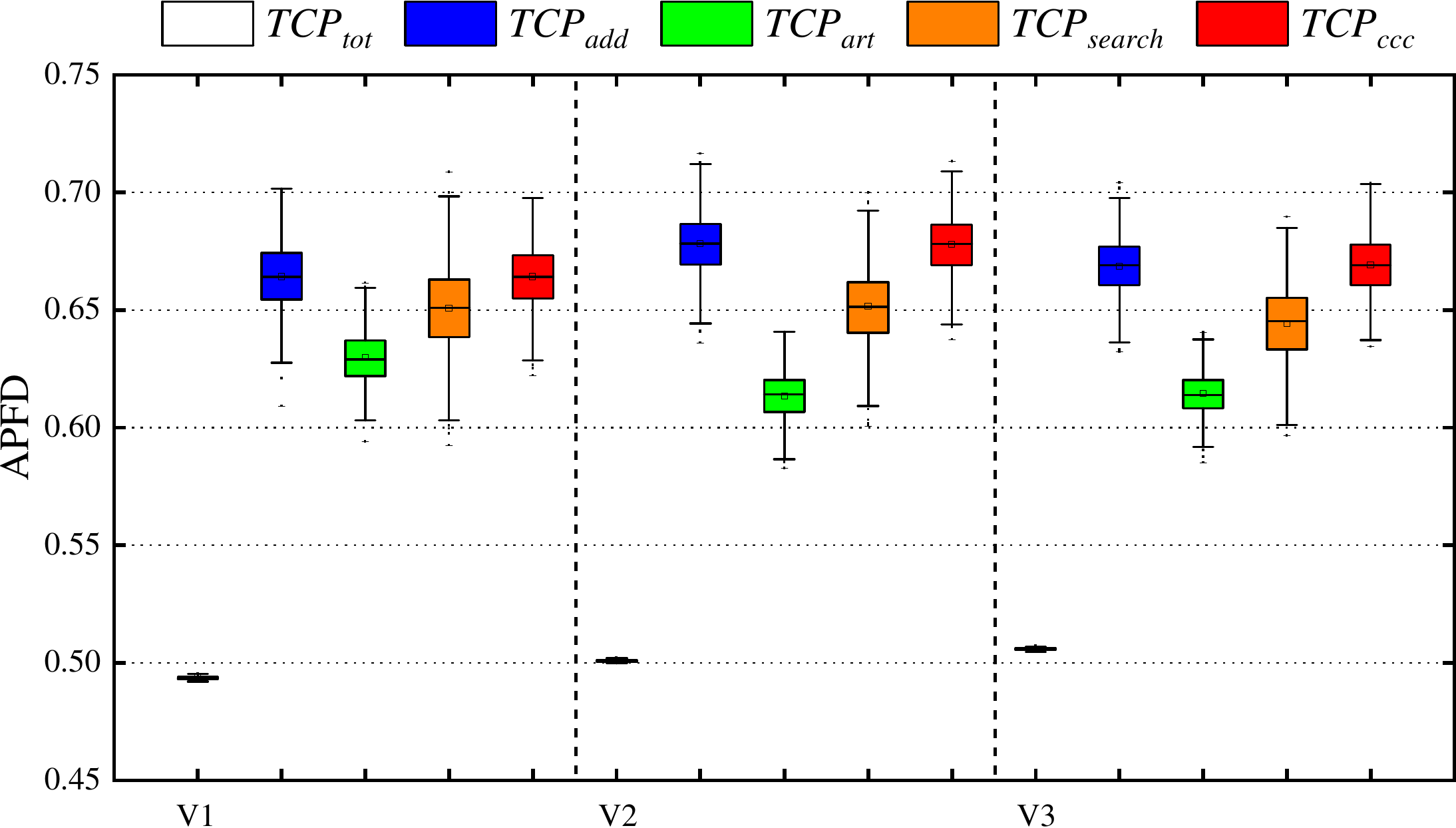}
        \label{apfd_ant_mc}
    }
    \subfigure[$jmeter$-statement coverage]
    {
        \includegraphics[width=0.319\textwidth]{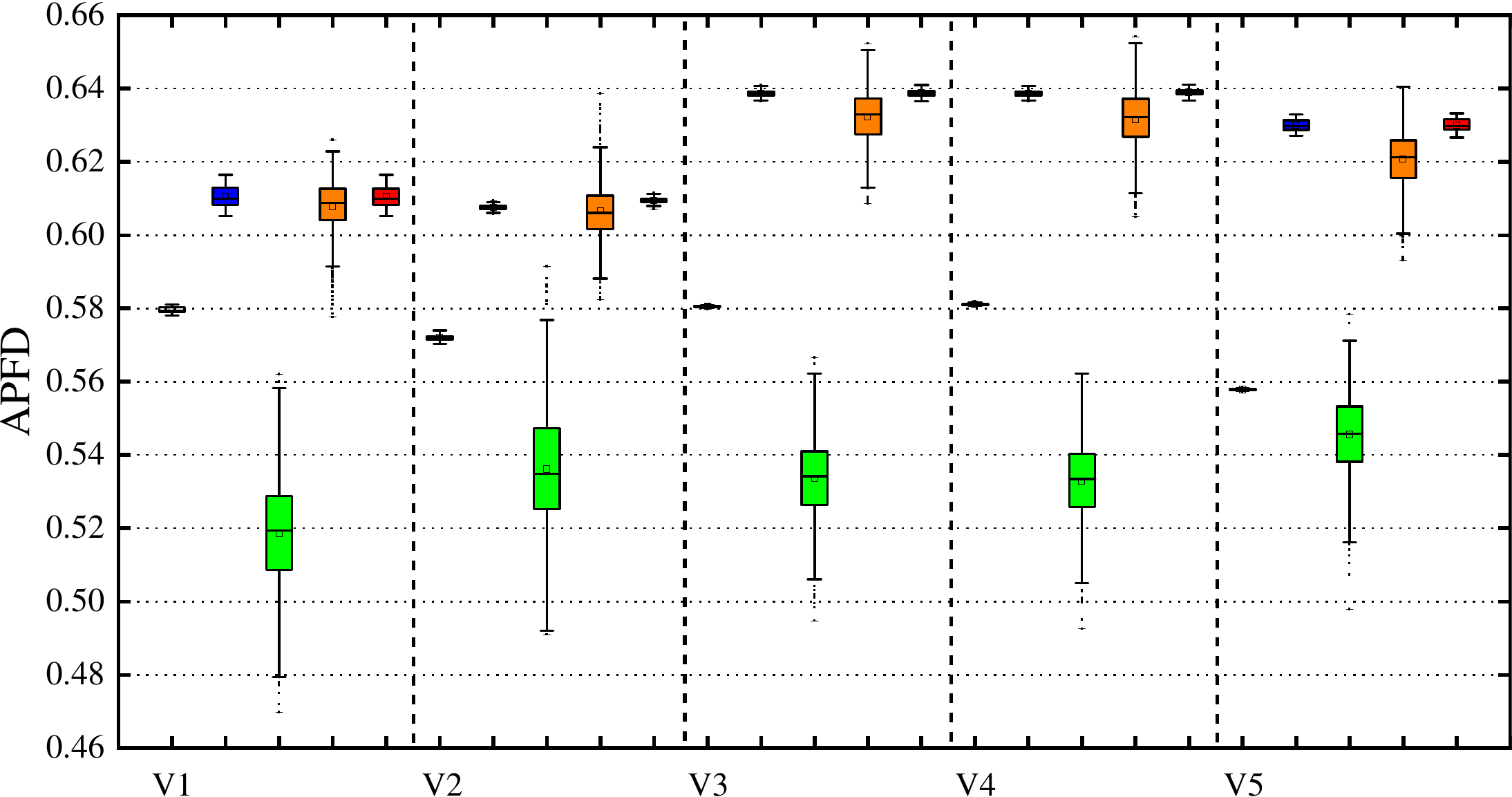}
        \label{apfd_grep_sc}
    }
        \subfigure[$jmeter$-branch coverage]
    {
        \includegraphics[width=0.319\textwidth]{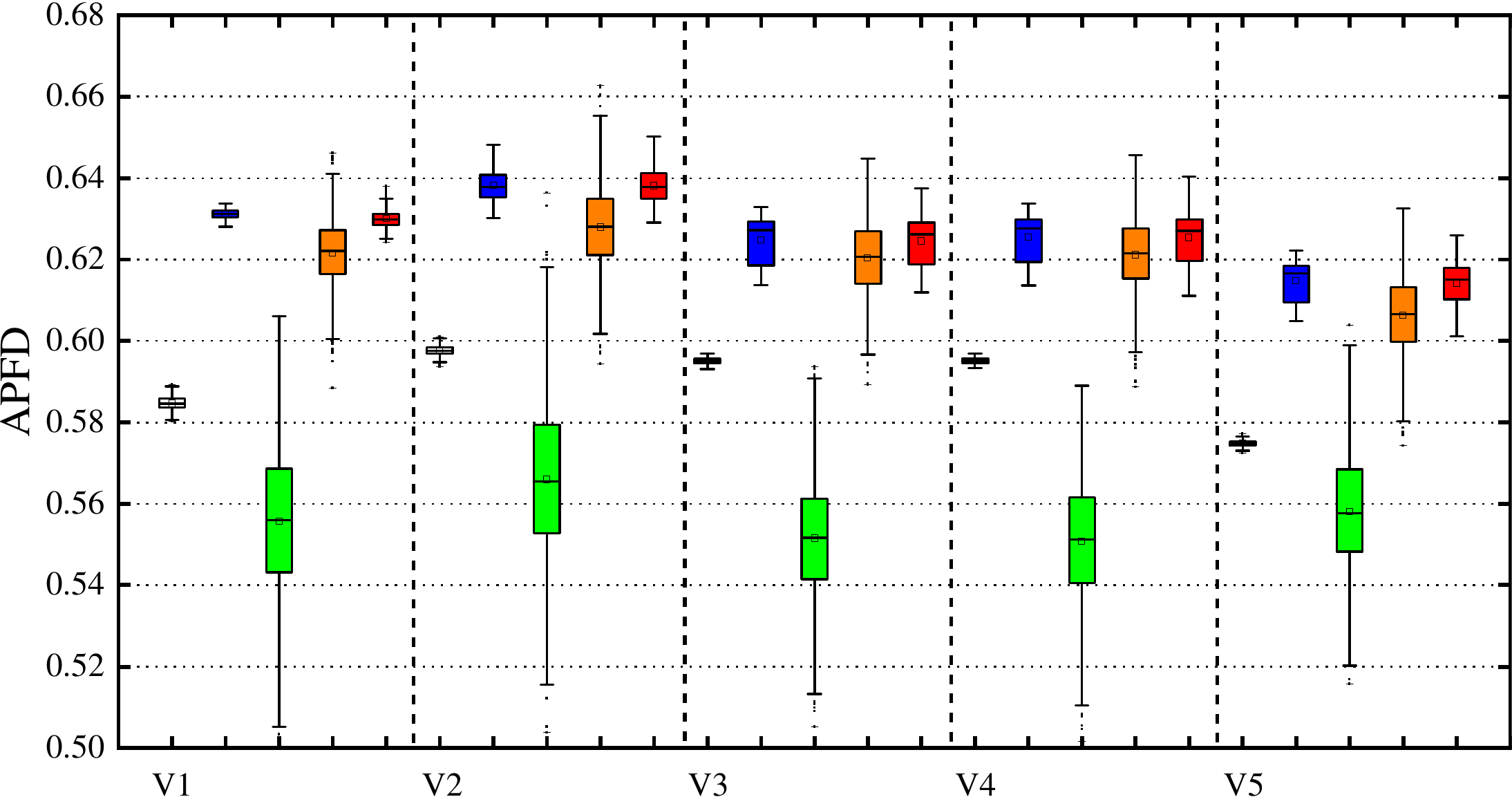}
        \label{apfd_jmeter_bc}
    }
        \subfigure[$jmeter$-method coverage]
    {
        \includegraphics[width=0.319\textwidth]{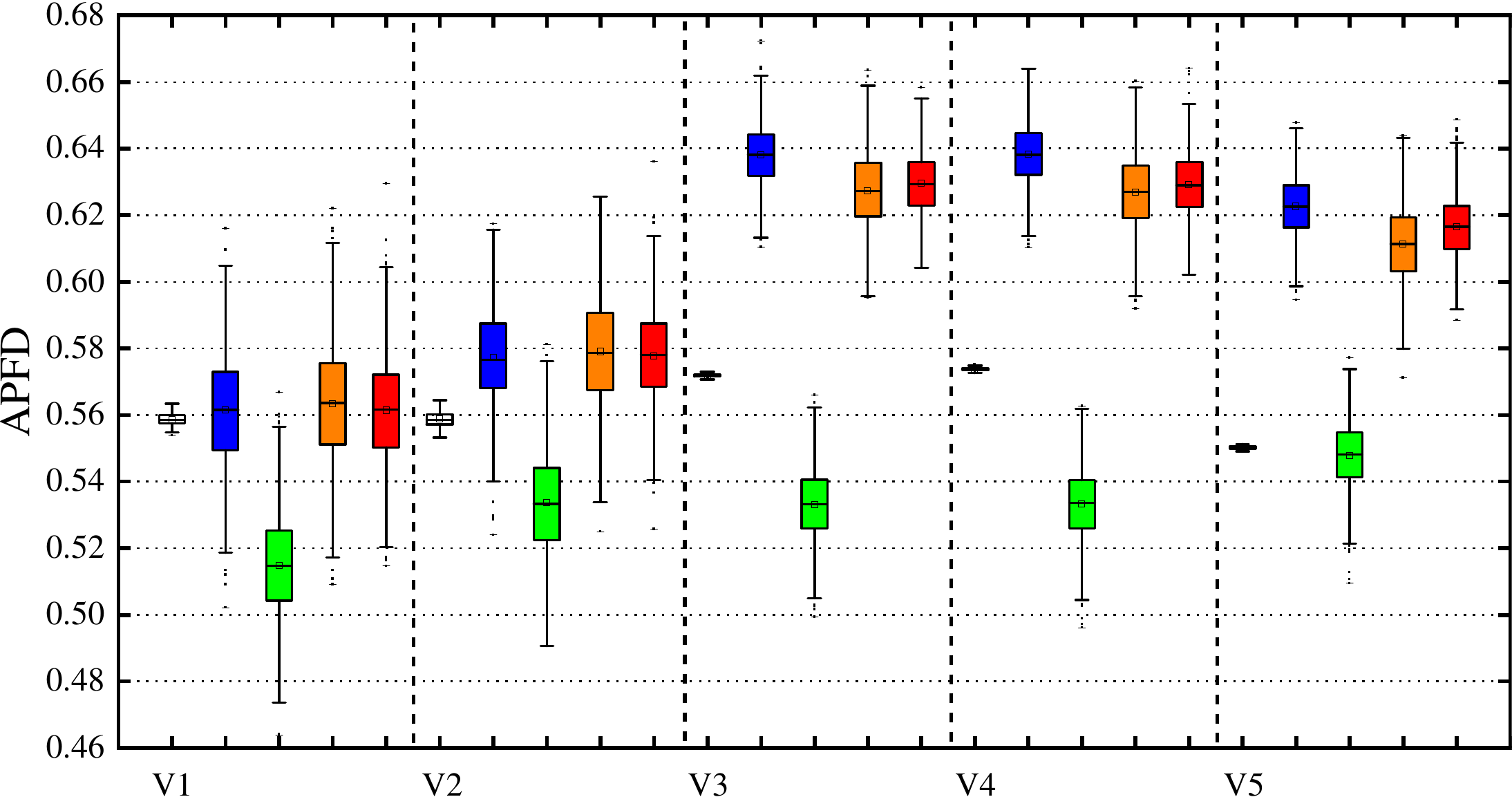}
        \label{apfd_jmeter_mc}
    }
        \subfigure[$jtopas$-statement coverage]
    {
        \includegraphics[width=0.319\textwidth]{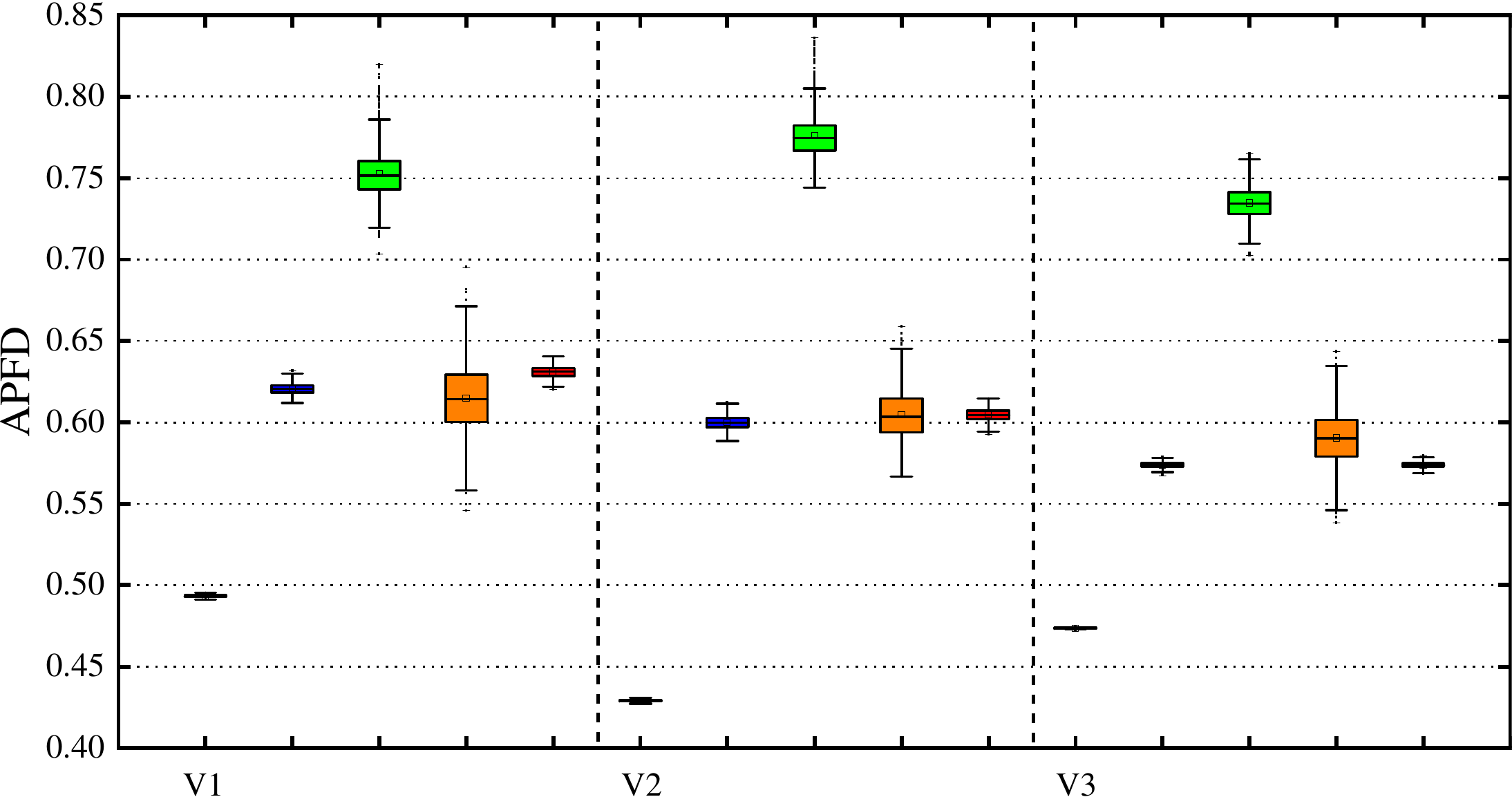}
        \label{apfd_jtopas_sc}
    }
        \subfigure[$jtopas$-branch coverage]
    {
        \includegraphics[width=0.319\textwidth]{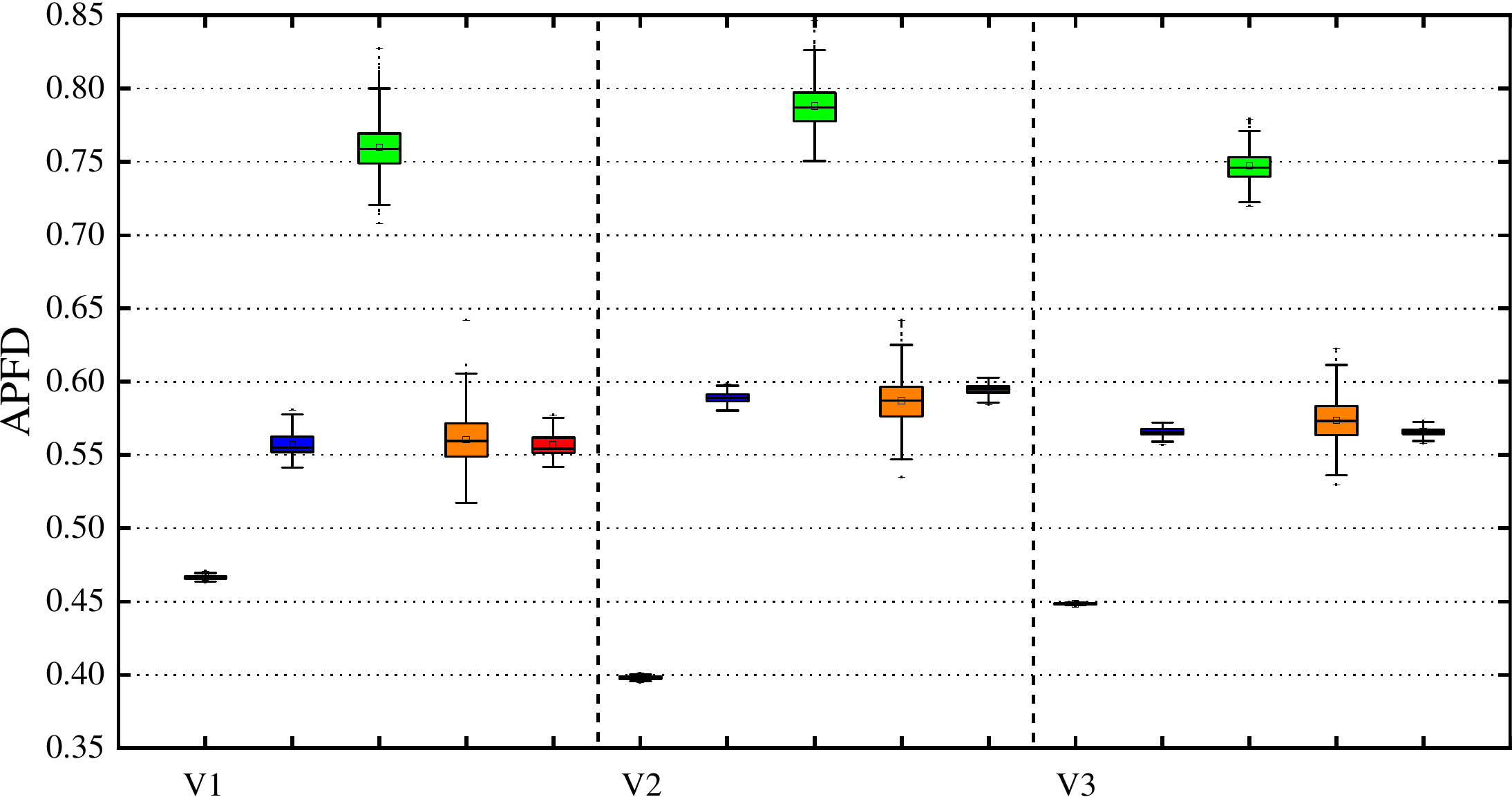}
        \label{apfd_jtopas_bc}
    }
        \subfigure[$jtopas$-method coverage]
    {
        \includegraphics[width=0.319\textwidth]{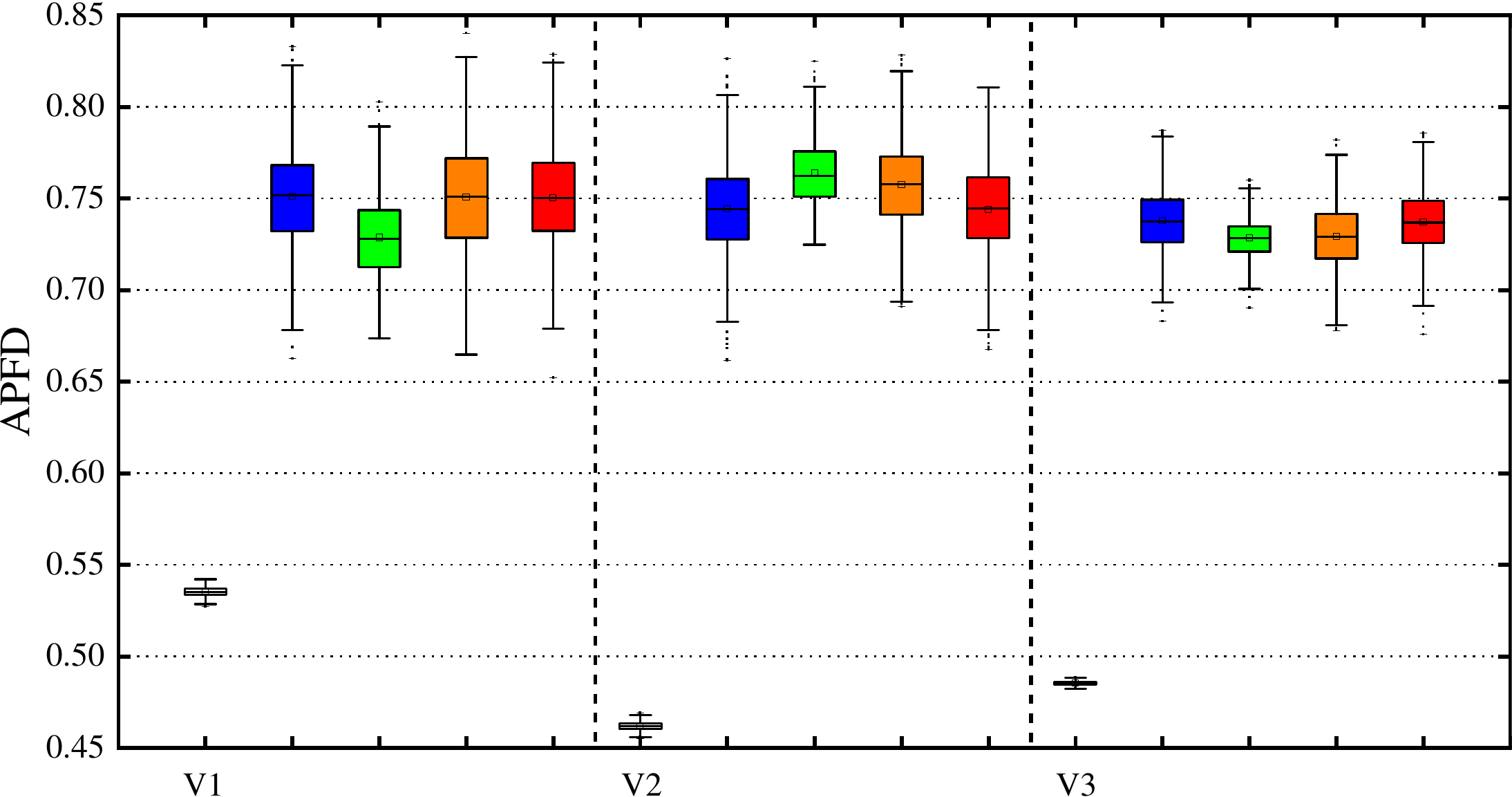}
        \label{apfd_jtopas_mc}
    }
        \subfigure[$xmlsec$-statement coverage]
    {
        \includegraphics[width=0.319\textwidth]{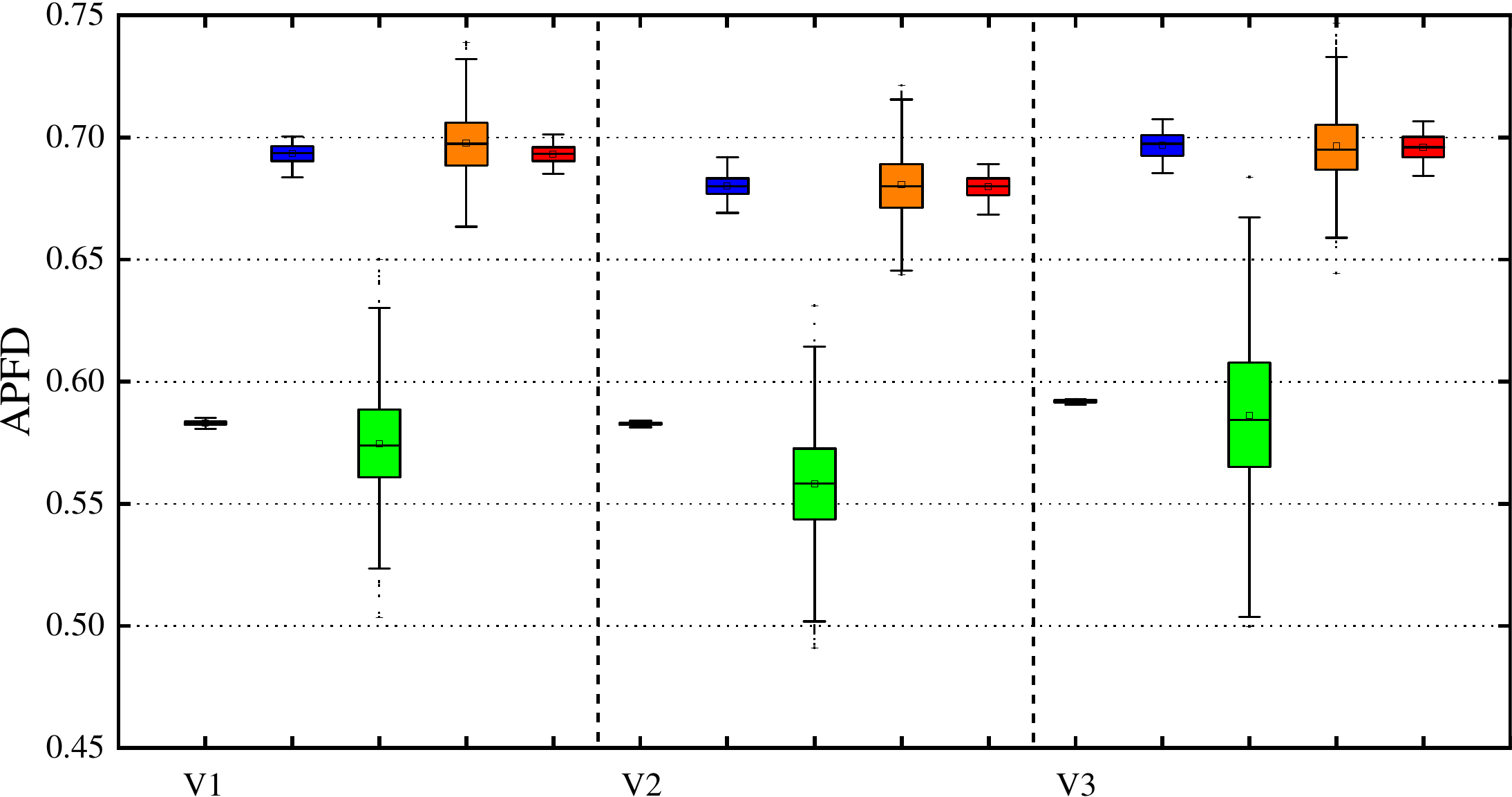}
        \label{apfd_xmlsec_sc}
    }
        \subfigure[$xmlsec$-branch coverage]
    {
        \includegraphics[width=0.319\textwidth]{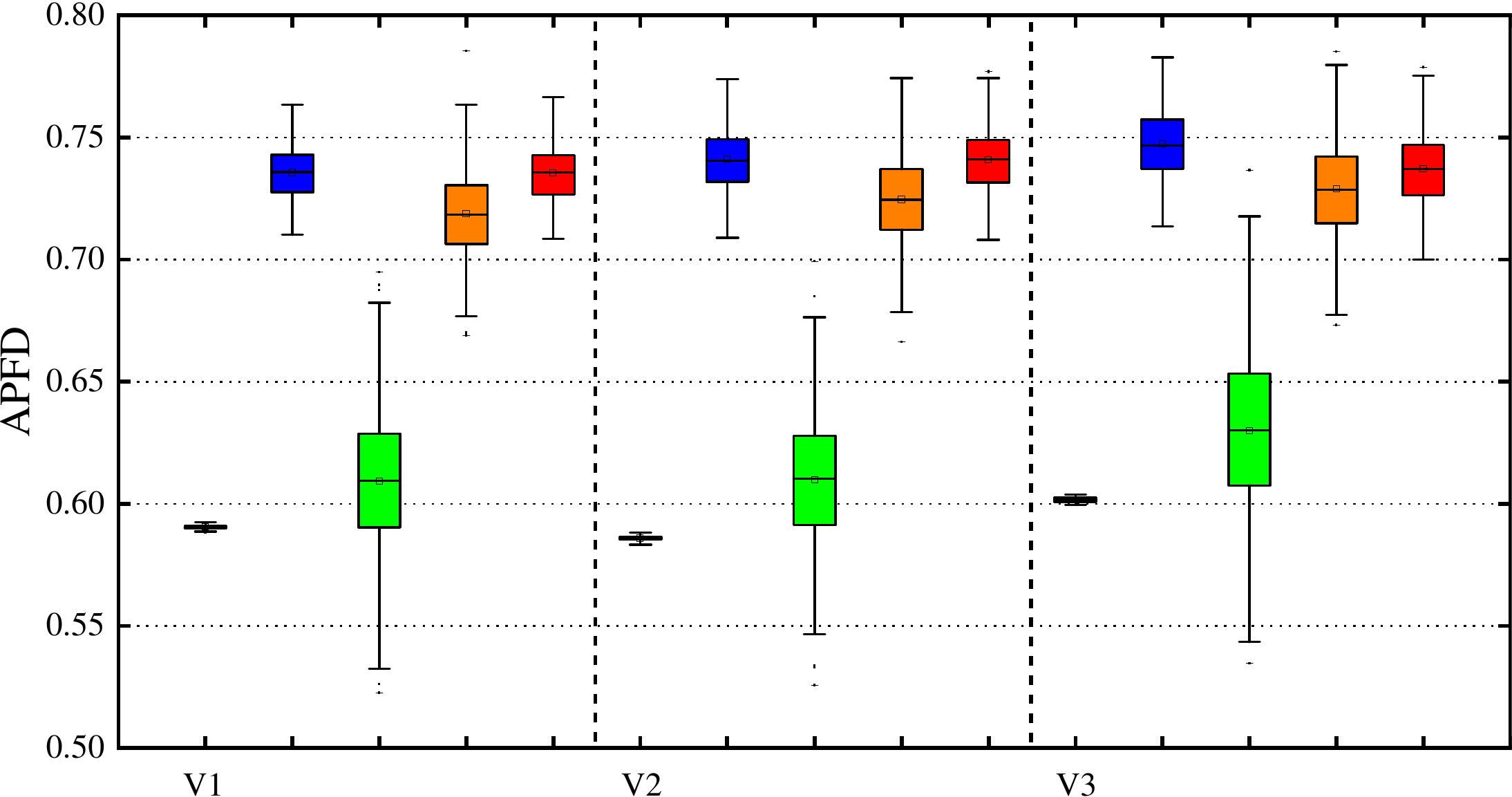}
        \label{apfd_xmlsec_bc}
    }
        \subfigure[$xmlsec$-method coverage]
    {
        \includegraphics[width=0.319\textwidth]{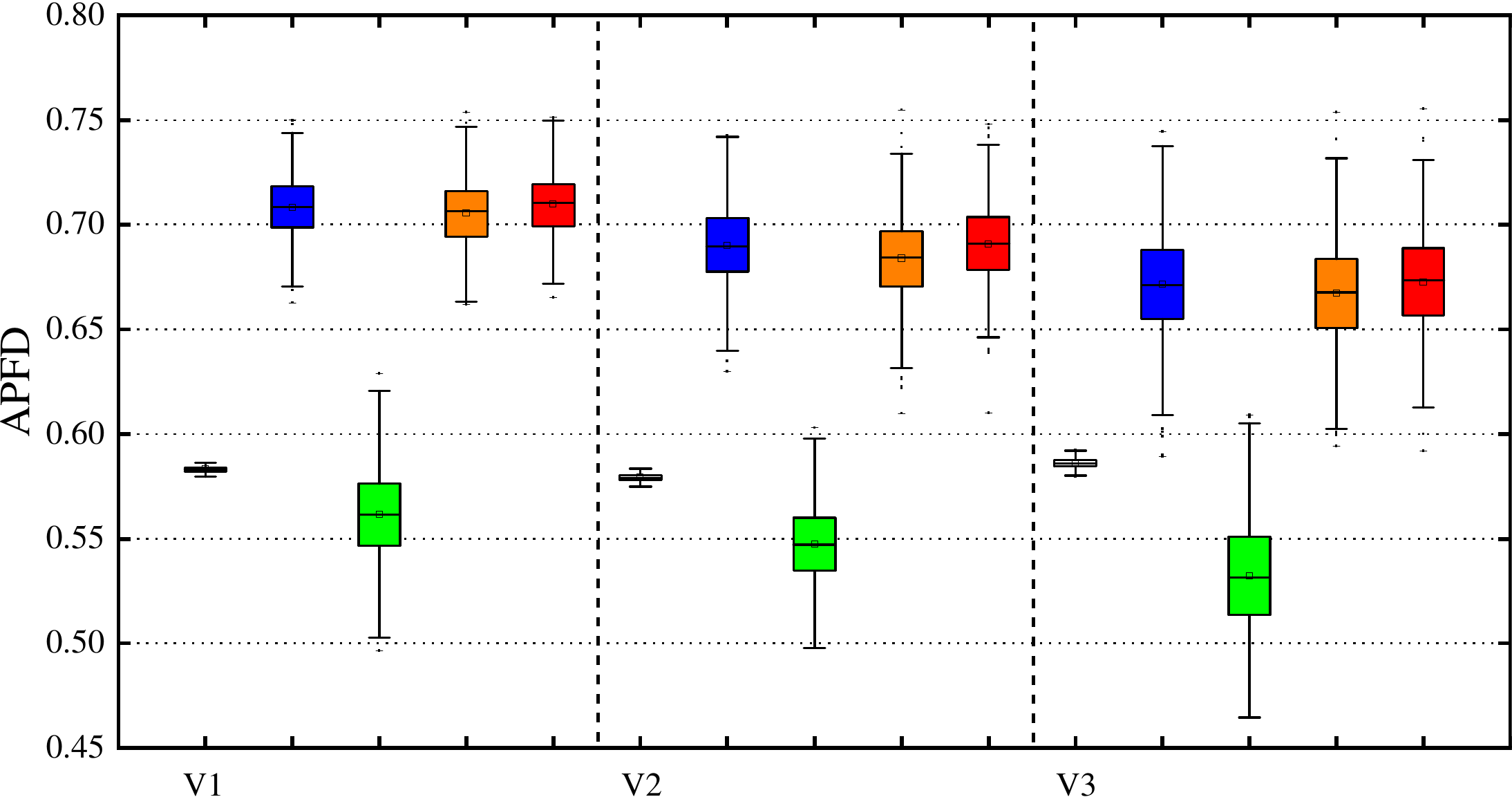}
        \label{apfd_xmlsec_mc}
    }
    \caption{\textbf{Effectiveness:} \textbf{APFD} results for \textbf{Java programs} at the \textbf{test-method level}}
    \label{FIG:apfd-java-m}
\end{figure*}

\begin{table*}[!t]
\scriptsize
\centering
 \caption{Statistical \textbf{effectiveness} comparisons of \textbf{APFD} for \textbf{Java programs} at the \textbf{test-method level}.
or a comparison between two methods
$\textit{TCP}_\textit{ccc}$ and $M$,
where
$M \in \{\textit{TCP}_\textit{tot}, \textit{TCP}_\textit{add}, \textit{TCP}_\textit{art}, \textit{TCP}_\textit{search}\}$,
the symbol \ding{52} means that $\textit{TCP}_\textit{ccc}$ is better
($p$-value is less than 0.05, and the effect size $\hat{\textrm{A}}_{12}(\textit{TCP}_\textit{ccc},M)$ is greater than 0.50);
the symbol \ding{54} means that $M$ is better
(the $p$-value is less than 0.05, and $\hat{\textrm{A}}_{12}(\textit{TCP}_\textit{ccc},M)$ is less than 0.50);
and the symbol \ding{109} means that there is no statistically significant difference between them (the $p$-value is greater than 0.05).
}
  \label{TAB:apfd-java-m}
    \setlength{\tabcolsep}{1.8mm}{
    \begin{tabular}{c|cccc|cccc|cccc}
     \hline
       \multirow{2}*{\textbf{Program Name}} &\multicolumn{4}{c|}{\textbf{Statement Coverage}} &\multicolumn{4}{c|}{\textbf{Branch Coverage}} &\multicolumn{4}{c}{\textbf{Method Coverage}} \\
        \cline{2-13}

         &$\textit{TCP}_\textit{tot}$ &$\textit{TCP}_\textit{add}$ &$\textit{TCP}_\textit{art}$ &$\textit{TCP}_\textit{search}$  & $\textit{TCP}_\textit{tot}$ &$\textit{TCP}_\textit{add}$ &$\textit{TCP}_\textit{art}$ &$\textit{TCP}_\textit{search}$ &$\textit{TCP}_\textit{tot}$ & $\textit{TCP}_\textit{add}$ &$\textit{TCP}_\textit{art}$ &$\textit{TCP}_\textit{search}$\\
            \hline
$ant\_v1$	&\ding{52}(1.00)	&\ding{52}(0.53)	&\ding{52}(1.00)	&\ding{52}(0.88)	&\ding{52}(1.00)	&\ding{109}(0.48)	&\ding{52}(1.00)	&\ding{52}(0.91)	&\ding{52}(1.00)	&\ding{109}(0.50)	&\ding{52}(0.98)	&\ding{52}(0.72)	\\
$ant\_v2$	&\ding{52}(1.00)	&\ding{109}(0.51)	&\ding{52}(1.00)	&\ding{52}(1.00)	&\ding{52}(1.00)	&\ding{54}(0.46)	&\ding{52}(1.00)	&\ding{52}(1.00)	&\ding{52}(1.00)	&\ding{109}(0.50)	&\ding{52}(1.00)	&\ding{52}(0.90)	\\
$ant\_v3$	&\ding{52}(1.00)	&\ding{109}(0.50)	&\ding{52}(1.00)	&\ding{52}(1.00)	&\ding{52}(1.00)	&\ding{109}(0.51)	&\ding{52}(1.00)	&\ding{52}(1.00)	&\ding{52}(1.00)	&\ding{109}(0.51)	&\ding{52}(1.00)	&\ding{52}(0.90)	\\\hline
$ant$	&\ding{52}(1.00)	&\ding{109}(0.51)	&\ding{52}(1.00)	&\ding{52}(0.98)	&\ding{52}(1.00)	&\ding{109}(0.49)	&\ding{52}(1.00)	&\ding{52}(0.97)	&\ding{52}(1.00)	&\ding{109}(0.50)	&\ding{52}(1.00)	&\ding{52}(0.84)	\\\hline
$jmeter\_v1$	&\ding{52}(1.00)	&\ding{109}(0.50)	&\ding{52}(1.00)	&\ding{52}(0.62)	&\ding{52}(1.00)	&\ding{54}(0.30)	&\ding{52}(1.00)	&\ding{52}(0.84)	&\ding{52}(0.56)	&\ding{109}(0.50)	&\ding{52}(0.98)	&\ding{54}(0.47)	\\
$jmeter\_v2$	&\ding{52}(1.00)	&\ding{52}(0.96)	&\ding{52}(1.00)	&\ding{52}(0.68)	&\ding{52}(1.00)	&\ding{109}(0.49)	&\ding{52}(1.00)	&\ding{52}(0.82)	&\ding{52}(0.90)	&\ding{109}(0.51)	&\ding{52}(0.98)	&\ding{109}(0.48)	\\
$jmeter\_v3$	&\ding{52}(1.00)	&\ding{52}(0.55)	&\ding{52}(1.00)	&\ding{52}(0.82)	&\ding{52}(1.00)	&\ding{109}(0.48)	&\ding{52}(1.00)	&\ding{52}(0.64)	&\ding{52}(1.00)	&\ding{54}(0.26)	&\ding{52}(1.00)	&\ding{52}(0.55)	\\
$jmeter\_v4$	&\ding{52}(1.00)	&\ding{52}(0.60)	&\ding{52}(1.00)	&\ding{52}(0.84)	&\ding{52}(1.00)	&\ding{109}(0.49)	&\ding{52}(1.00)	&\ding{52}(0.64)	&\ding{52}(1.00)	&\ding{54}(0.25)	&\ding{52}(1.00)	&\ding{52}(0.56)	\\
$jmeter\_v5$	&\ding{52}(1.00)	&\ding{109}(0.51)	&\ding{52}(1.00)	&\ding{52}(0.89)	&\ding{52}(1.00)	&\ding{54}(0.45)	&\ding{52}(1.00)	&\ding{52}(0.76)	&\ding{52}(1.00)	&\ding{54}(0.32)	&\ding{52}(1.00)	&\ding{52}(0.64)	\\\hline
$jmeter$	&\ding{52}(1.00)	&\ding{52}(0.54)	&\ding{52}(1.00)	&\ding{52}(0.63)	&\ding{52}(1.00)	&\ding{54}(0.48)	&\ding{52}(1.00)	&\ding{52}(0.68)	&\ding{52}(0.86)	&\ding{54}(0.43)	&\ding{52}(0.98)	&\ding{52}(0.52)	\\\hline
$jtopas\_v1$	&\ding{52}(1.00)	&\ding{52}(0.98)	&\ding{54}(0.00)	&\ding{52}(0.77)	&\ding{52}(1.00)	&\ding{54}(0.46)	&\ding{54}(0.00)	&\ding{54}(0.43)	&\ding{52}(1.00)	&\ding{109}(0.49)	&\ding{52}(0.73)	&\ding{109}(0.50)	\\
$jtopas\_v2$	&\ding{52}(1.00)	&\ding{52}(0.79)	&\ding{54}(0.00)	&\ding{52}(0.53)	&\ding{52}(1.00)	&\ding{52}(0.89)	&\ding{54}(0.00)	&\ding{52}(0.70)	&\ding{52}(1.00)	&\ding{109}(0.50)	&\ding{54}(0.27)	&\ding{54}(0.35)	\\
$jtopas\_v3$	&\ding{52}(1.00)	&\ding{109}(0.50)	&\ding{54}(0.00)	&\ding{54}(0.16)	&\ding{52}(1.00)	&\ding{109}(0.49)	&\ding{54}(0.00)	&\ding{54}(0.31)	&\ding{52}(1.00)	&\ding{109}(0.49)	&\ding{52}(0.67)	&\ding{52}(0.62)	\\\hline
$jtopas$	&\ding{52}(1.00)	&\ding{52}(0.59)	&\ding{54}(0.00)	&\ding{109}(0.51)	&\ding{52}(1.00)	&\ding{52}(0.54)	&\ding{54}(0.00)	&\ding{54}(0.47)	&\ding{52}(1.00)	&\ding{109}(0.50)	&\ding{52}(0.55)	&\ding{54}(0.48)	\\\hline
$xmlsec\_v1$	&\ding{52}(1.00)	&\ding{109}(0.49)	&\ding{52}(1.00)	&\ding{54}(0.37)	&\ding{52}(1.00)	&\ding{109}(0.49)	&\ding{52}(1.00)	&\ding{52}(0.78)	&\ding{52}(1.00)	&\ding{52}(0.53)	&\ding{52}(1.00)	&\ding{52}(0.58)	\\
$xmlsec\_v2$	&\ding{52}(1.00)	&\ding{109}(0.49)	&\ding{52}(1.00)	&\ding{109}(0.50)	&\ding{52}(1.00)	&\ding{109}(0.50)	&\ding{52}(1.00)	&\ding{52}(0.77)	&\ding{52}(1.00)	&\ding{109}(0.51)	&\ding{52}(1.00)	&\ding{52}(0.60)	\\
$xmlsec\_v3$	&\ding{52}(1.00)	&\ding{54}(0.46)	&\ding{52}(1.00)	&\ding{109}(0.52)	&\ding{52}(1.00)	&\ding{54}(0.30)	&\ding{52}(1.00)	&\ding{52}(0.63)	&\ding{52}(1.00)	&\ding{109}(0.51)	&\ding{52}(1.00)	&\ding{52}(0.56)	\\\hline
$xmlsec$	&\ding{52}(1.00)	&\ding{109}(0.49)	&\ding{52}(1.00)	&\ding{54}(0.47)	&\ding{52}(1.00)	&\ding{54}(0.43)	&\ding{52}(1.00)	&\ding{52}(0.72)	&\ding{52}(1.00)	&\ding{109}(0.51)	&\ding{52}(1.00)	&\ding{52}(0.56)	\\\hline
\textit{\textbf{All Java Programs}}	&\ding{52}(0.97)	&\ding{52}(0.51)	&\ding{52}(0.70)	&\ding{52}(0.57)	&\ding{52}(0.90)	&\ding{54}(0.49)	&\ding{52}(0.59)	&\ding{52}(0.59)	&\ding{52}(0.96)	&\ding{54}(0.49)	&\ding{52}(0.76)	&\ding{52}(0.54)	\\
\hline

  \end{tabular}}
\end{table*}

\begin{figure*}[!t]
\graphicspath{{graphs/}}
\centering
    \subfigure[$flex$-statement coverage]
    {
        \includegraphics[width=0.319\textwidth]{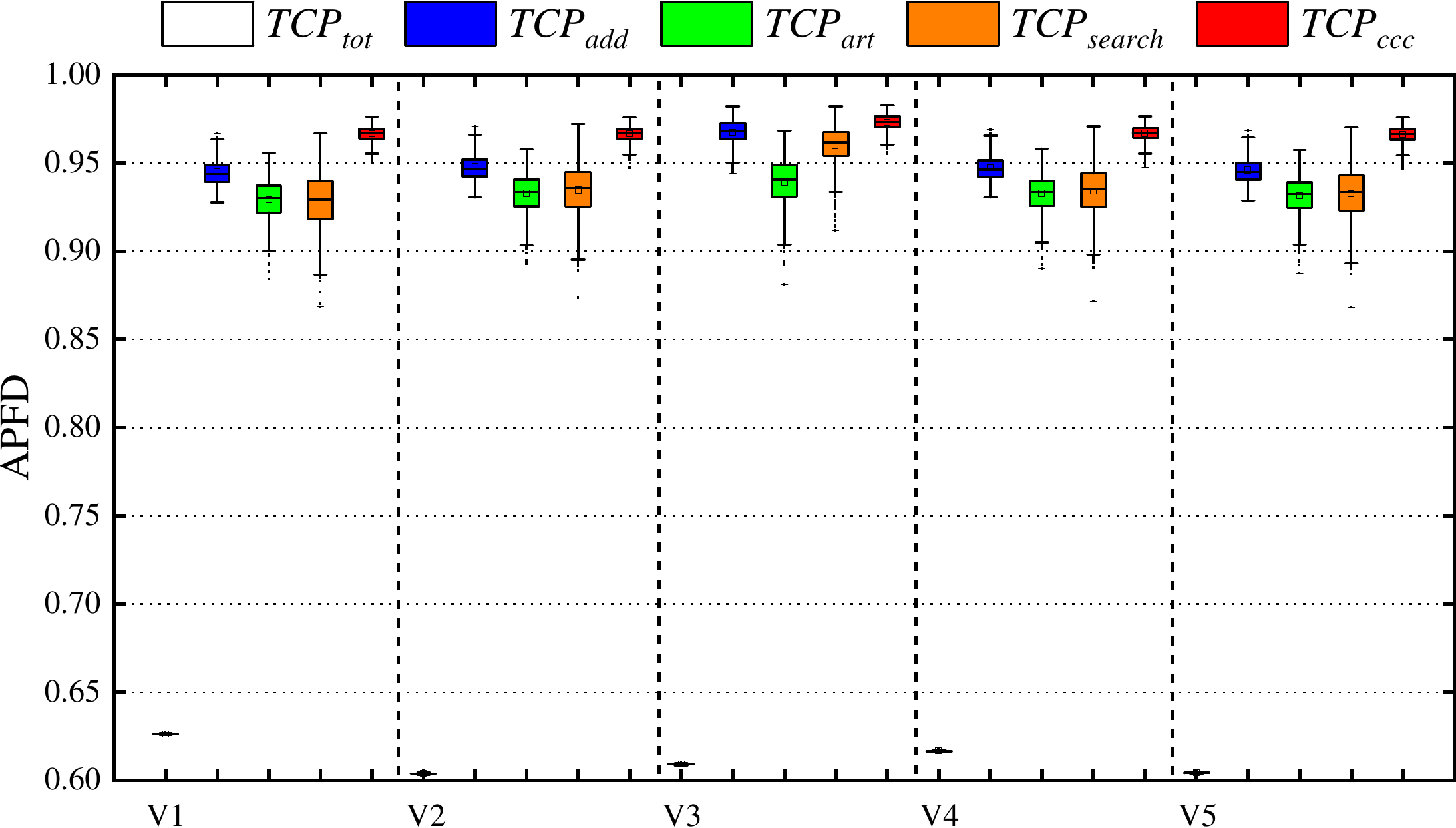}
        \label{apfd_flex_sc}
    }
    \subfigure[$flex$-branch coverage]
    {
        \includegraphics[width=0.319\textwidth]{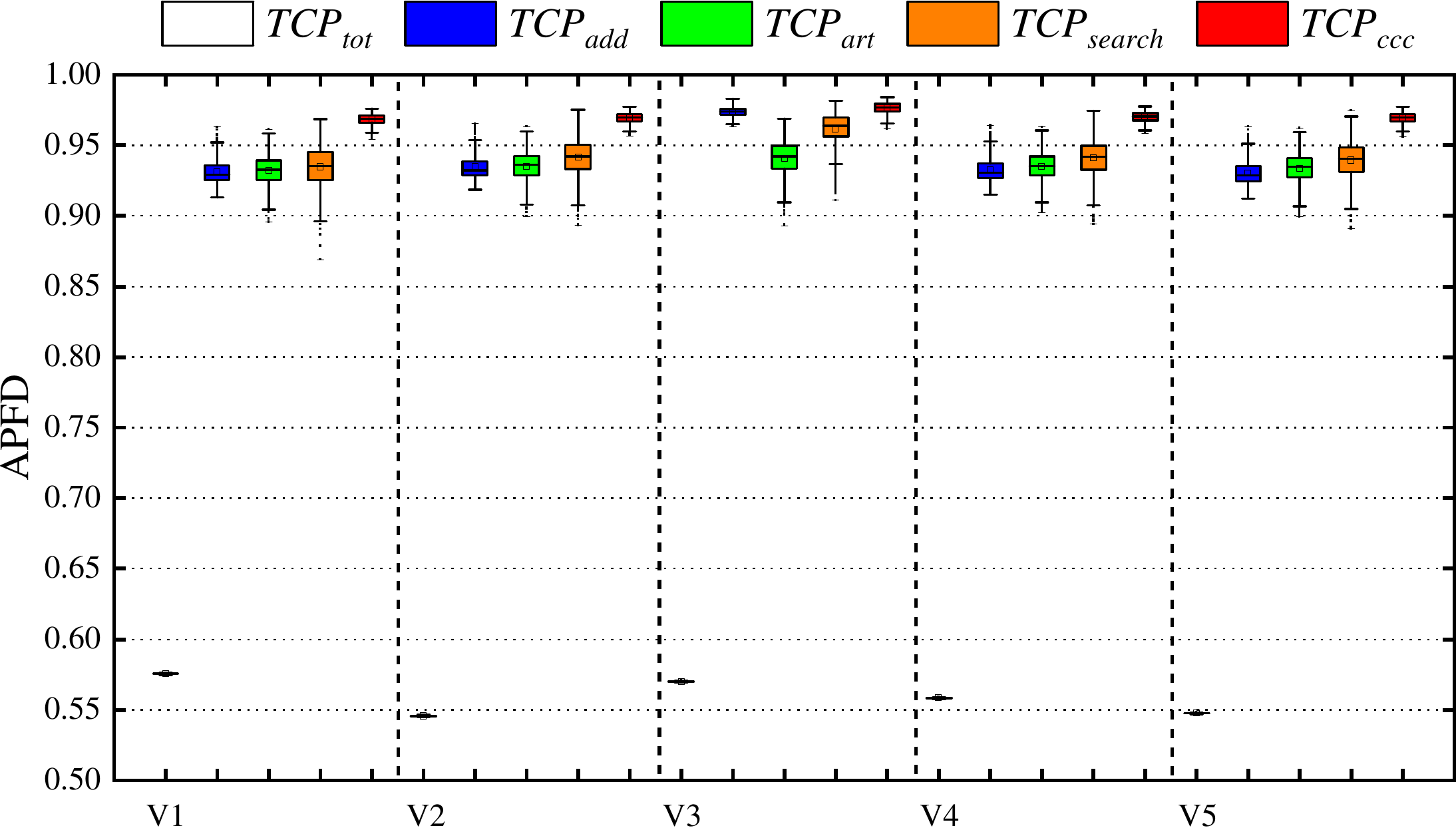}
        \label{apfd_flex_bc}
    }
        \subfigure[$flex$-method coverage]
    {
        \includegraphics[width=0.319\textwidth]{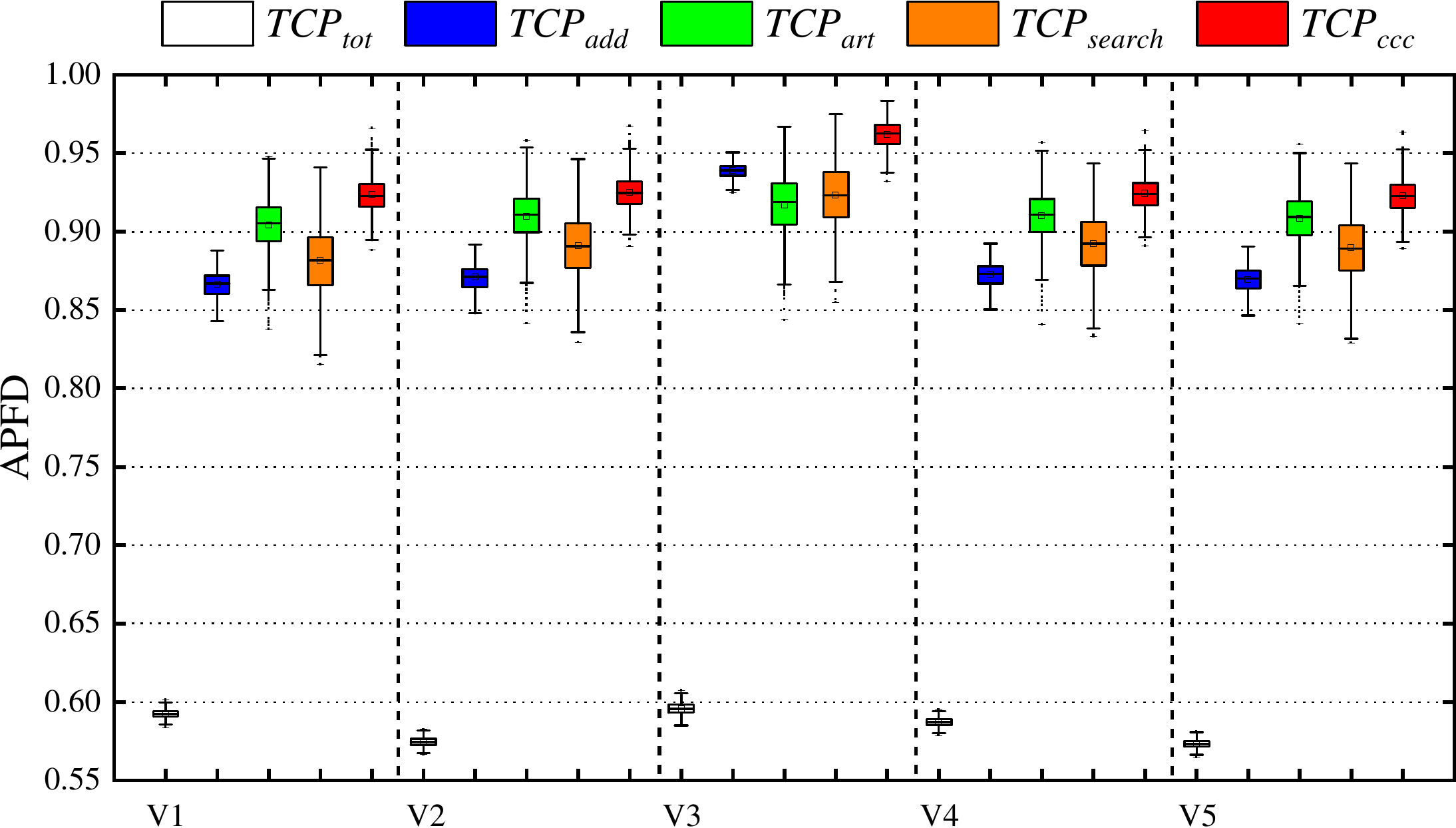}
        \label{apfd_flex_mc}
    }
    \subfigure[$grep$-statement coverage]
    {
        \includegraphics[width=0.319\textwidth]{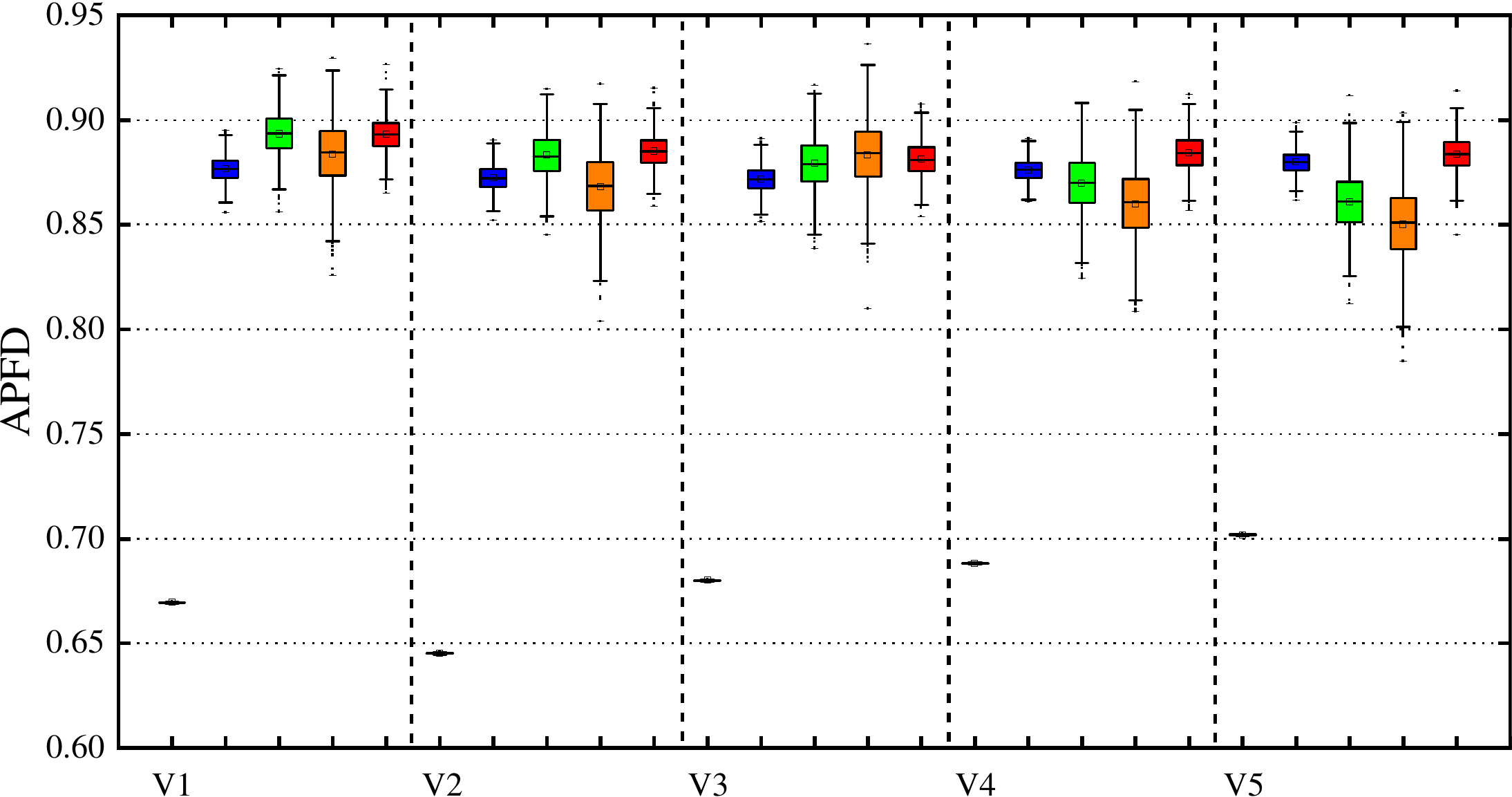}
        \label{apfd_grep_sc}
    }
        \subfigure[$grep$-branch coverage]
    {
        \includegraphics[width=0.319\textwidth]{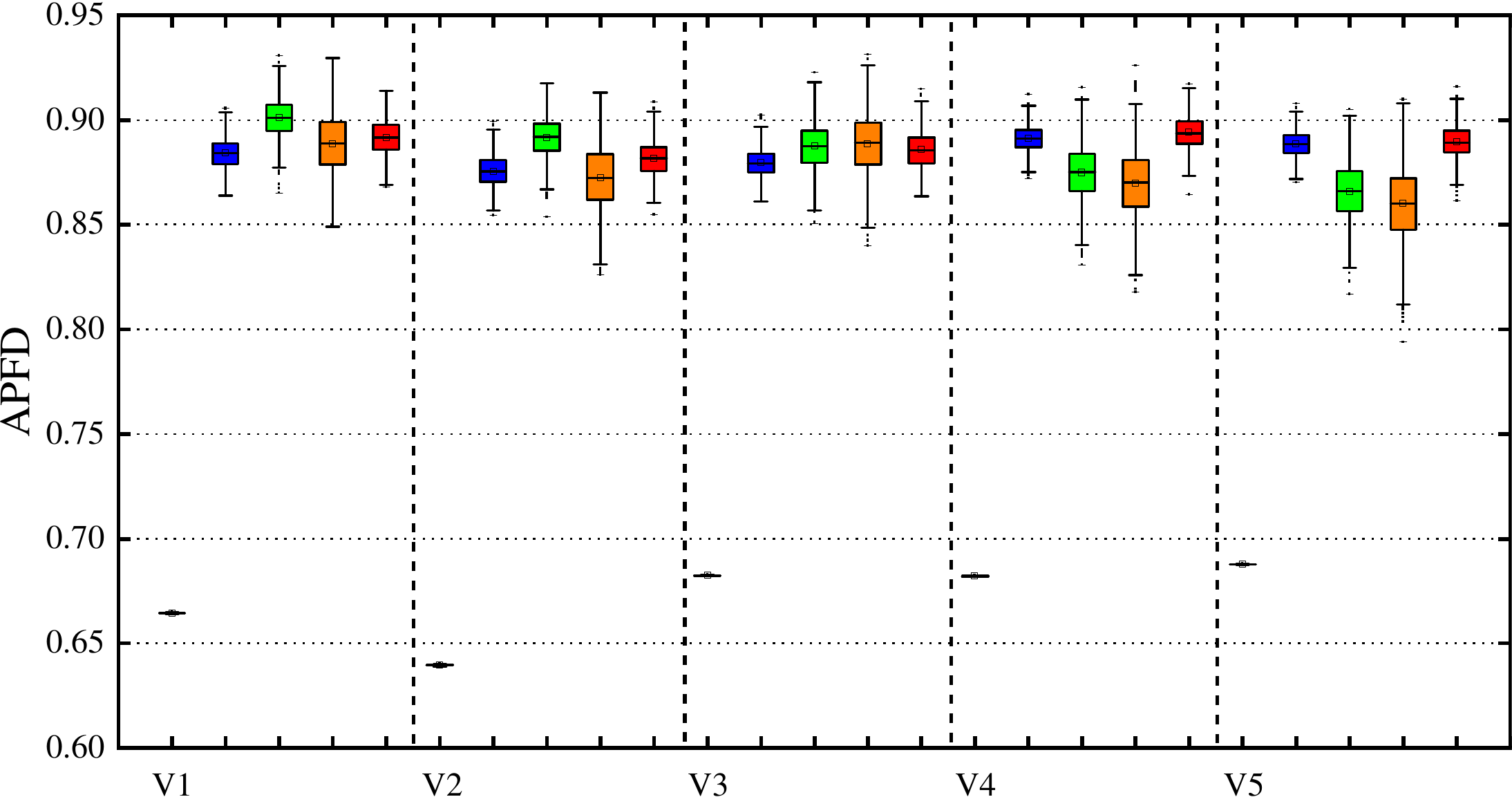}
        \label{apfd_grep_bc}
    }
        \subfigure[$grep$-method coverage]
    {
        \includegraphics[width=0.319\textwidth]{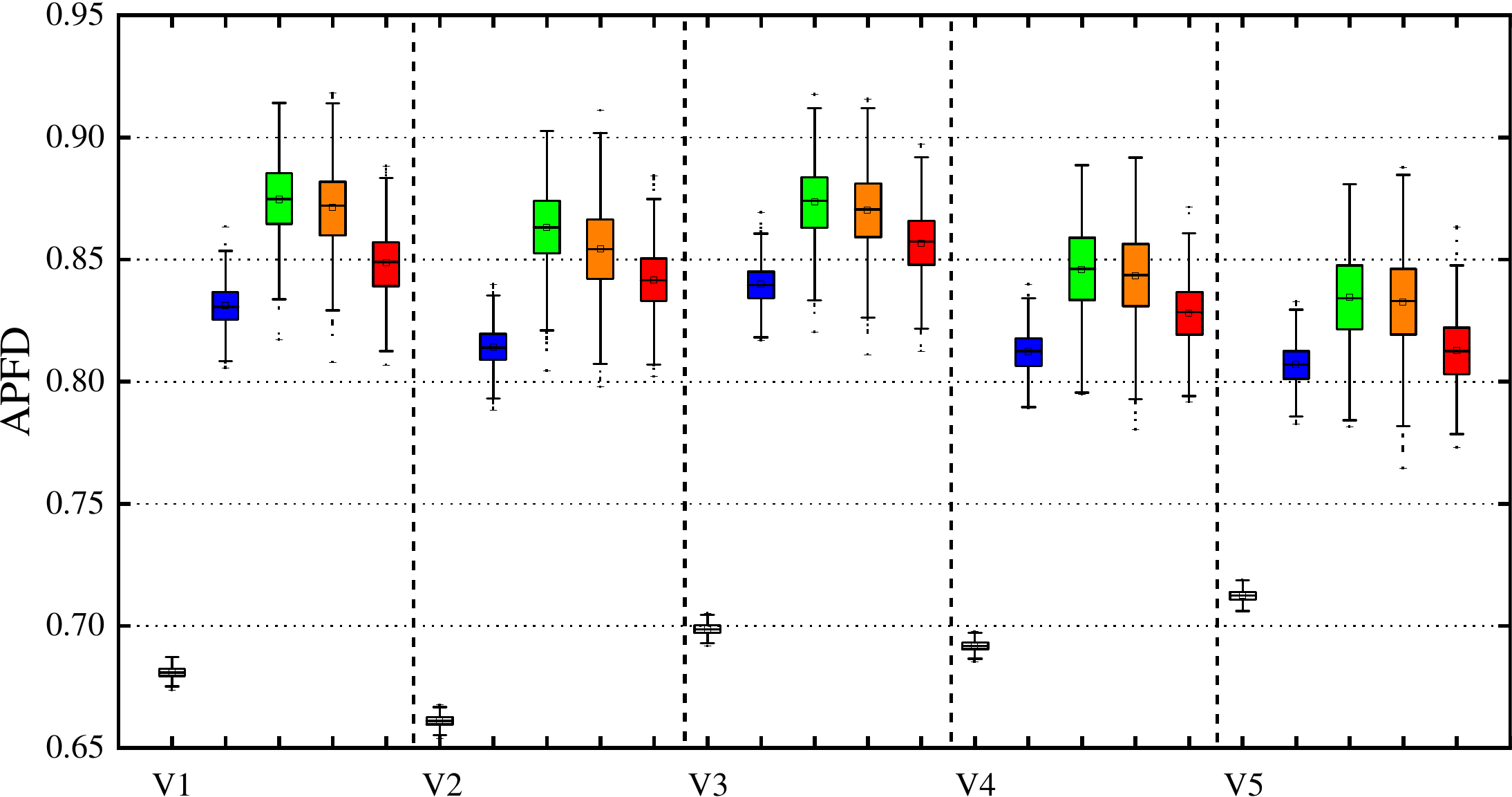}
        \label{apfd_grep_mc}
    }
        \subfigure[$gzip$-statement coverage]
    {
        \includegraphics[width=0.319\textwidth]{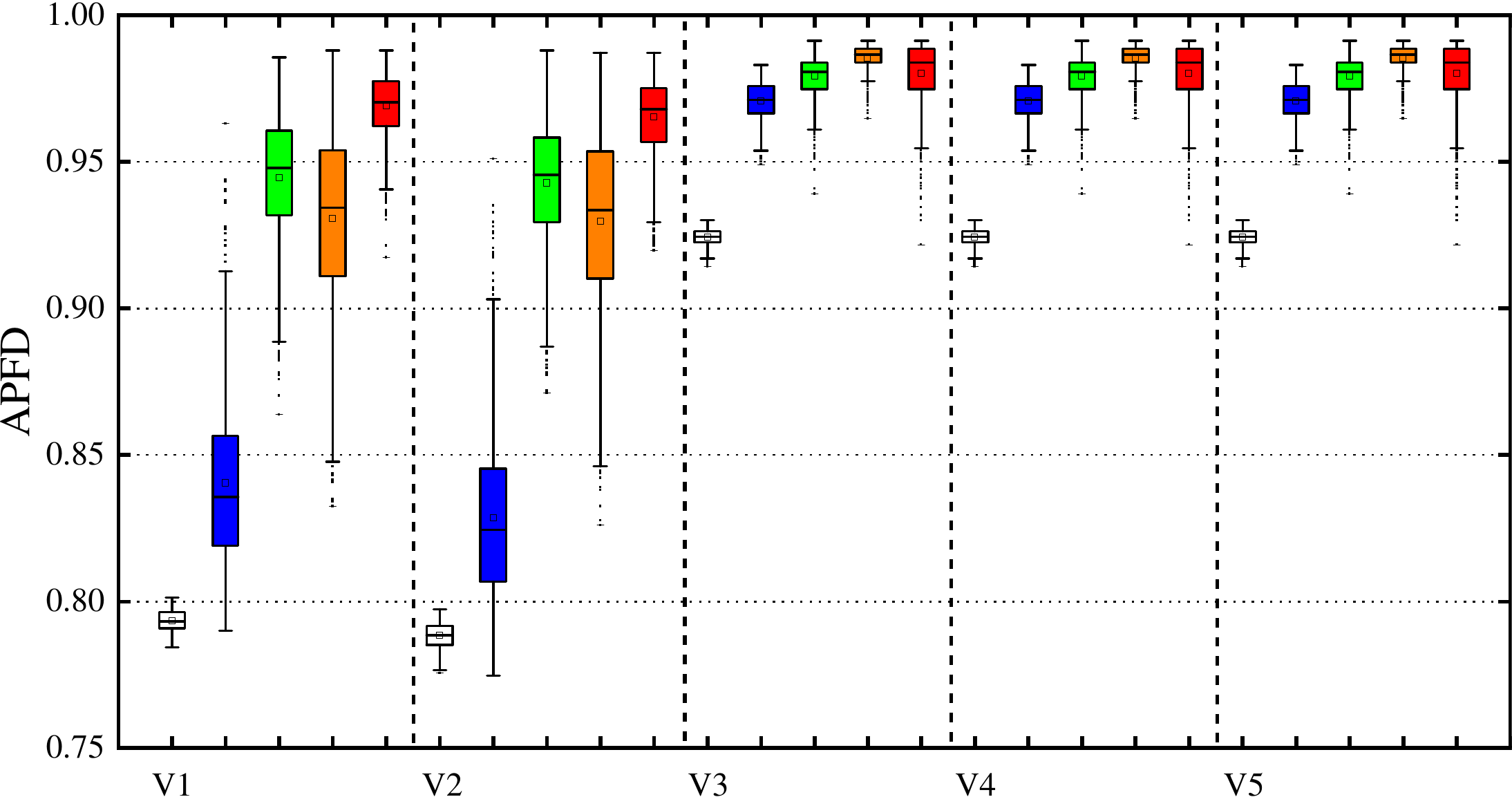}
        \label{apfd_gzip_sc}
    }
        \subfigure[$gzip$-branch coverage]
    {
        \includegraphics[width=0.319\textwidth]{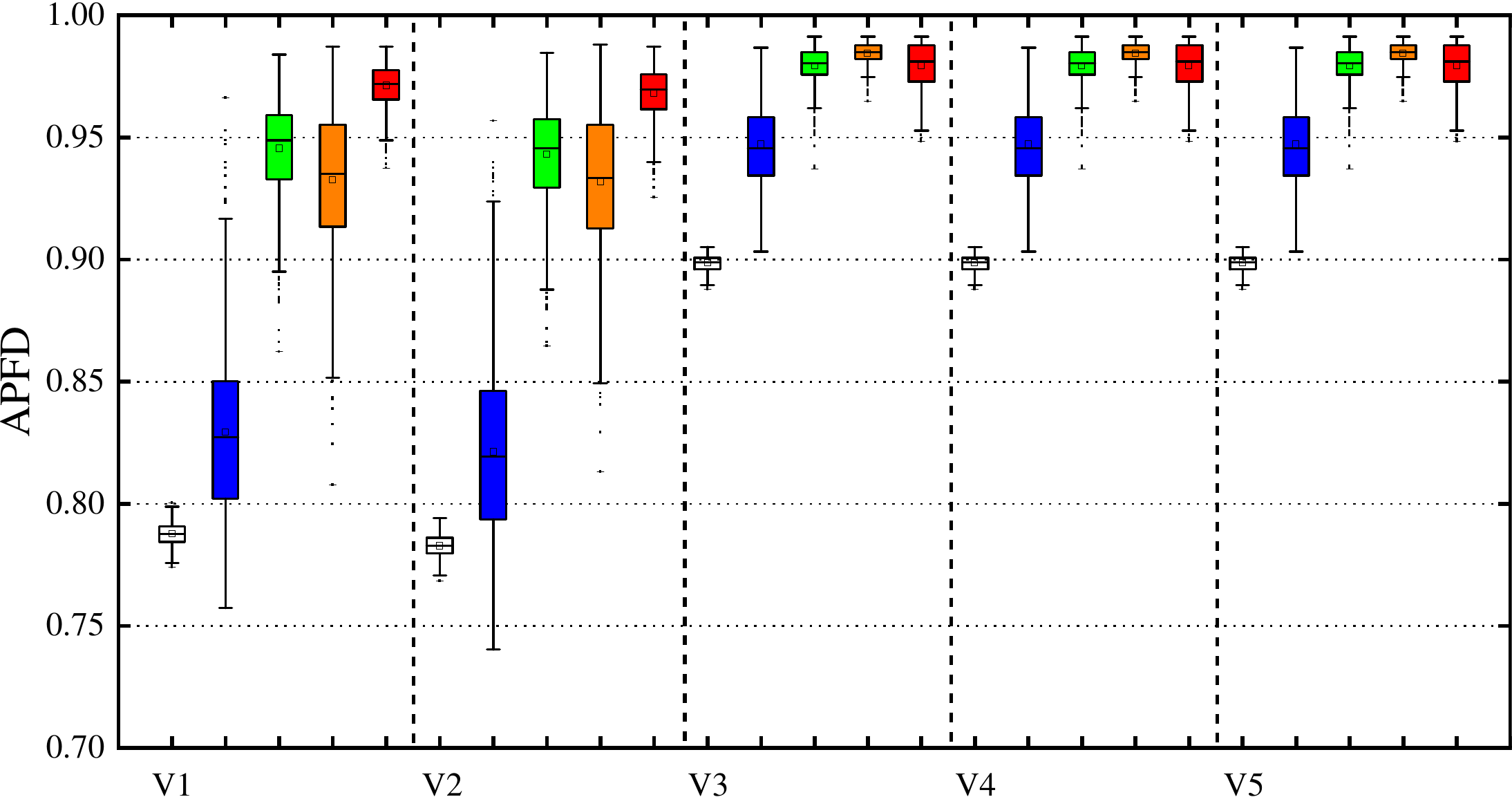}
        \label{apfd_gzip_bc}
    }
        \subfigure[$gzip$-method coverage]
    {
        \includegraphics[width=0.319\textwidth]{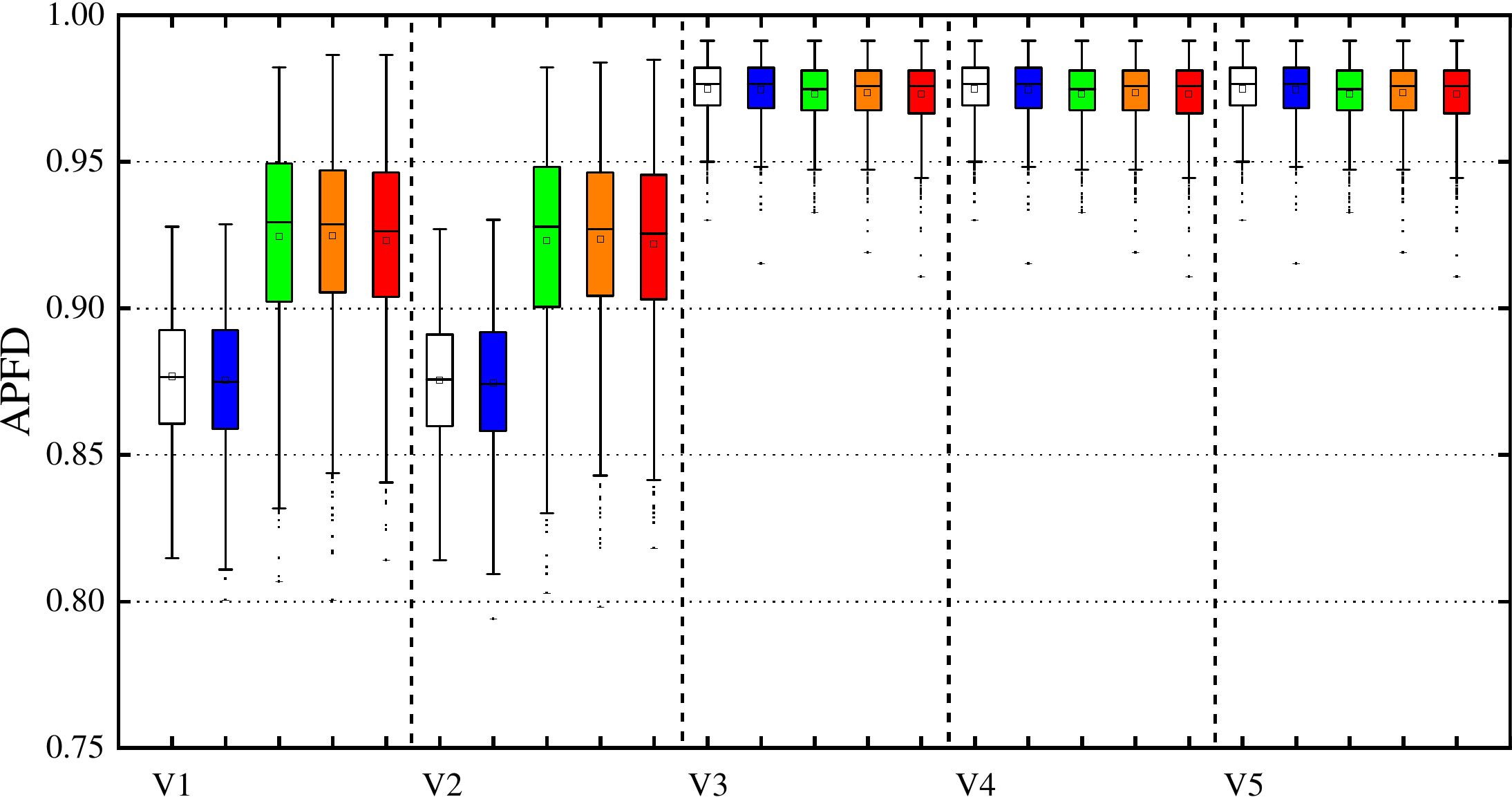}
        \label{apfd_gzip_mc}
    }
        \subfigure[$make$-statement coverage]
    {
        \includegraphics[width=0.319\textwidth]{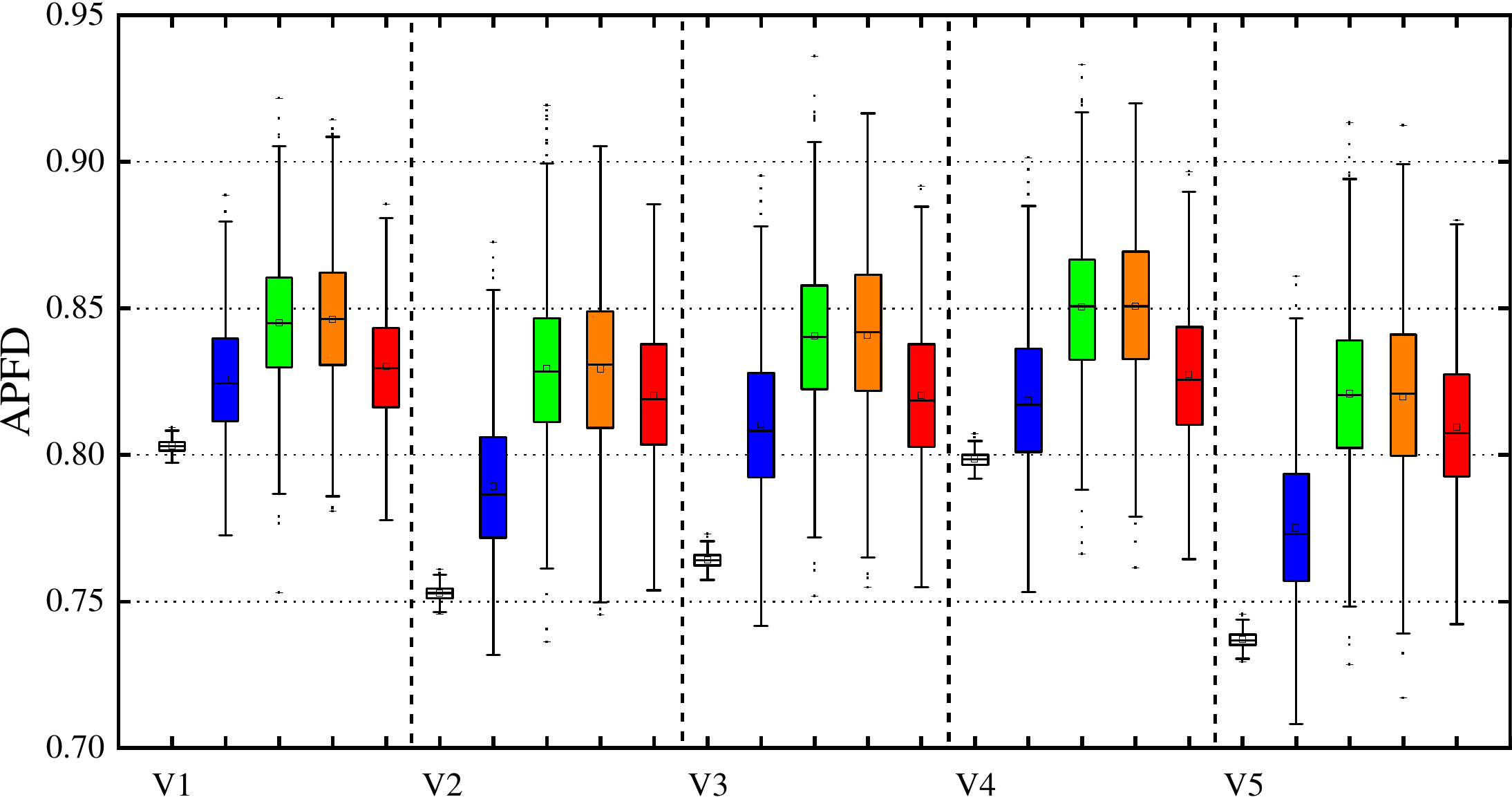}
        \label{apfd_make_sc}
    }
        \subfigure[$make$-branch coverage]
    {
        \includegraphics[width=0.319\textwidth]{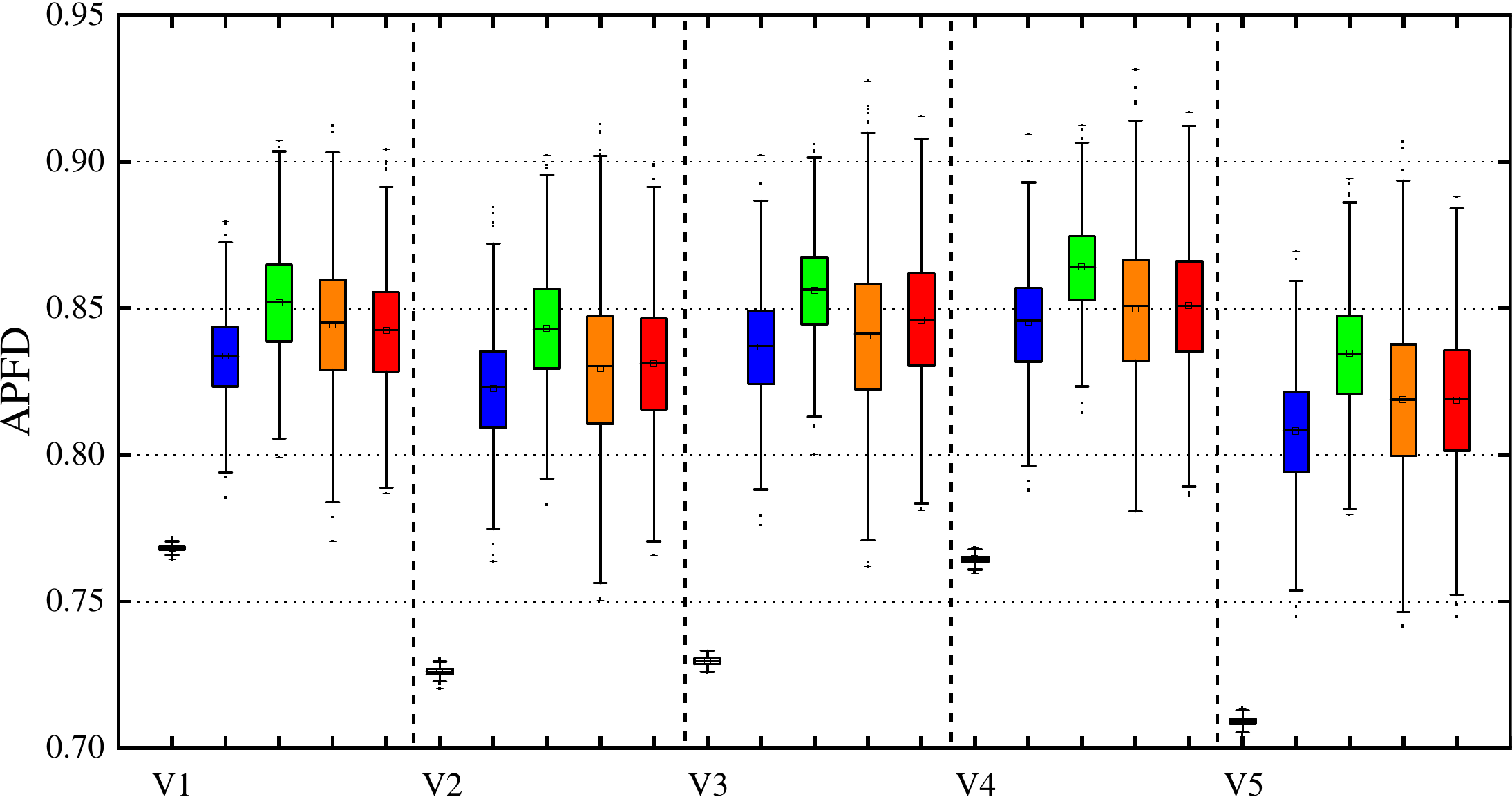}
        \label{apfd_make_bc}
    }
        \subfigure[$make$-method coverage]
    {
        \includegraphics[width=0.319\textwidth]{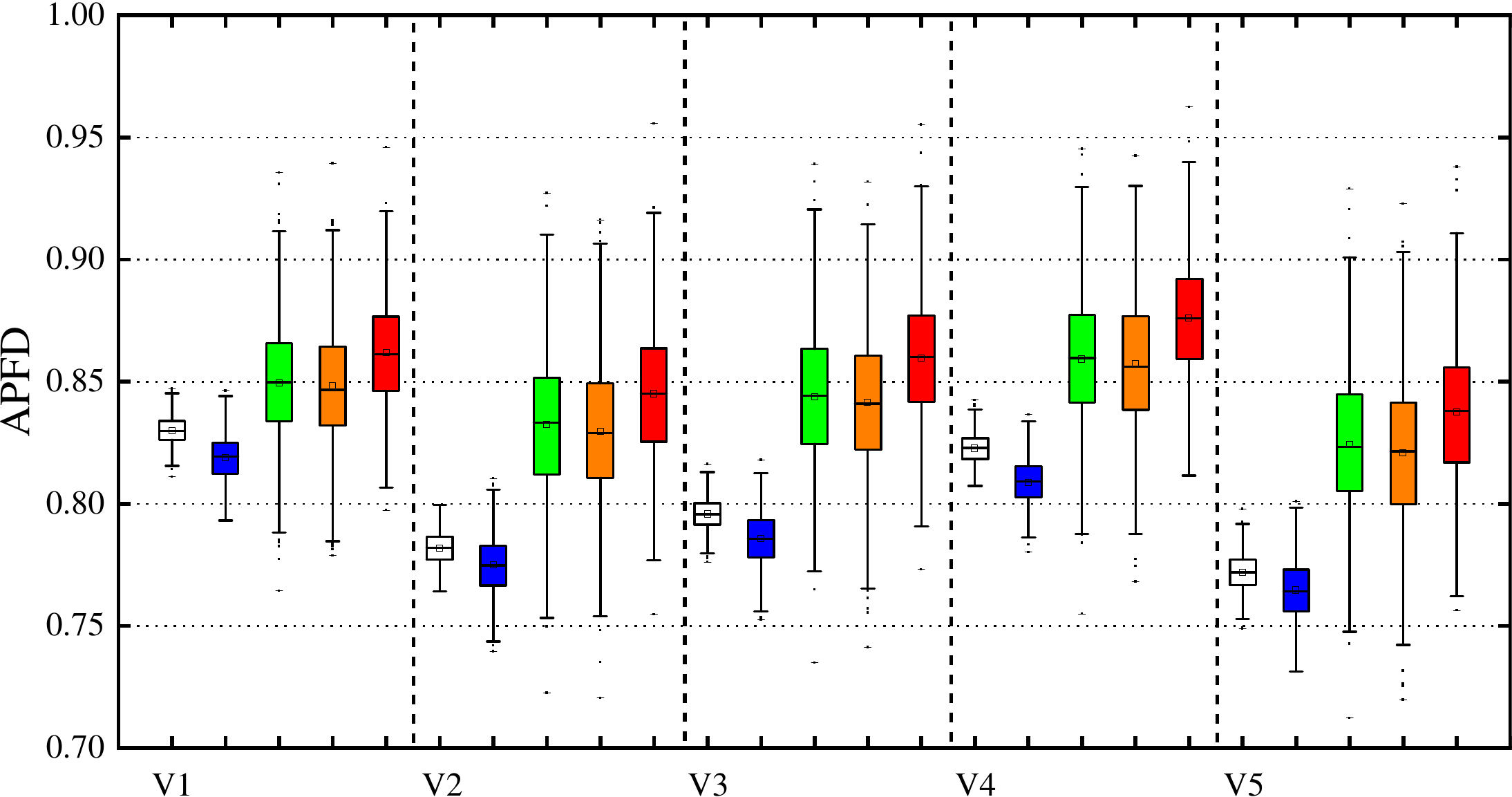}
        \label{apfd_make_mc}
    }
       \subfigure[$sed$-statement coverage]
    {
        \includegraphics[width=0.319\textwidth]{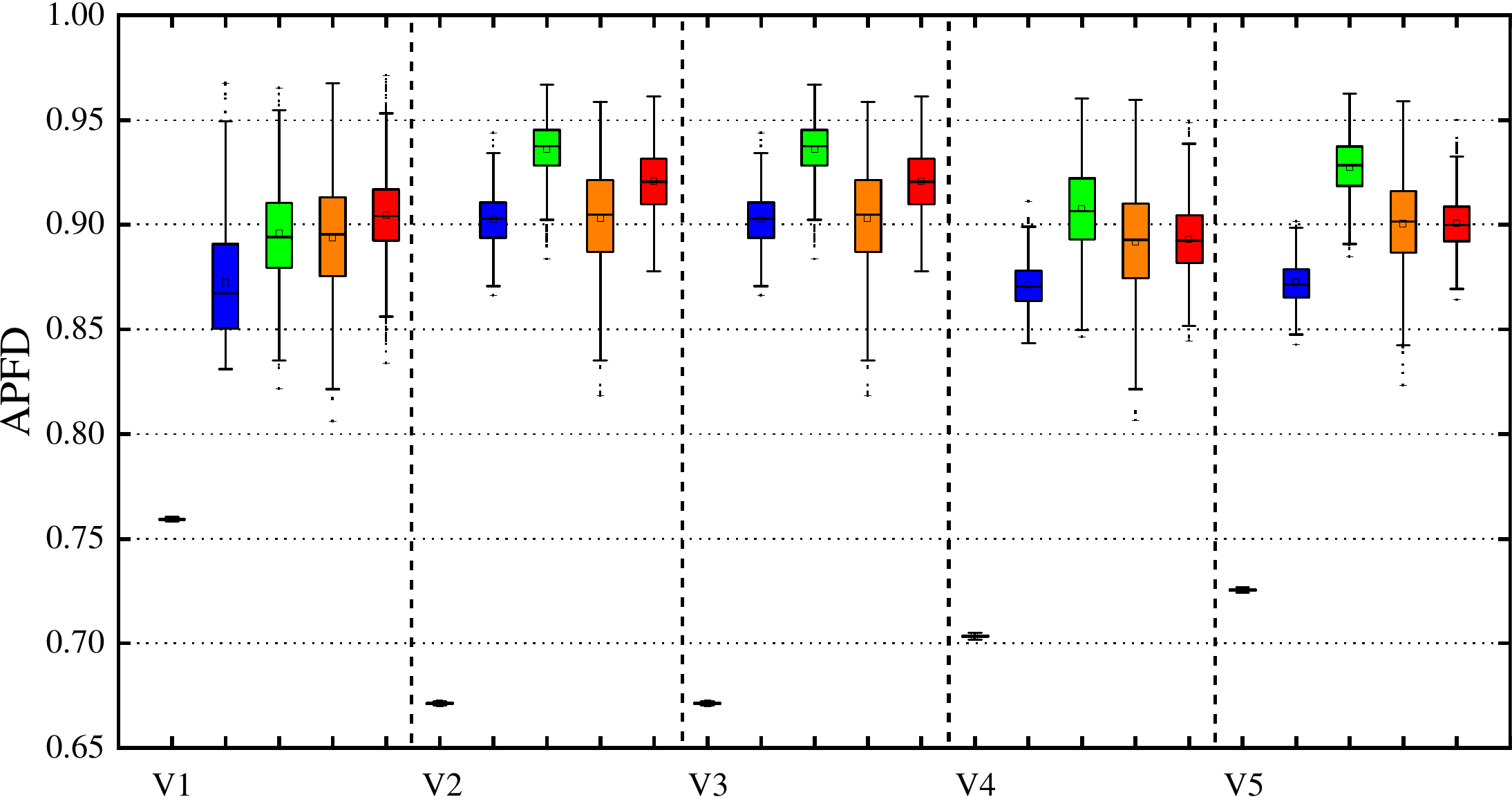}
        \label{apfd_sed_sc}
    }
        \subfigure[$sed$-branch coverage]
    {
        \includegraphics[width=0.319\textwidth]{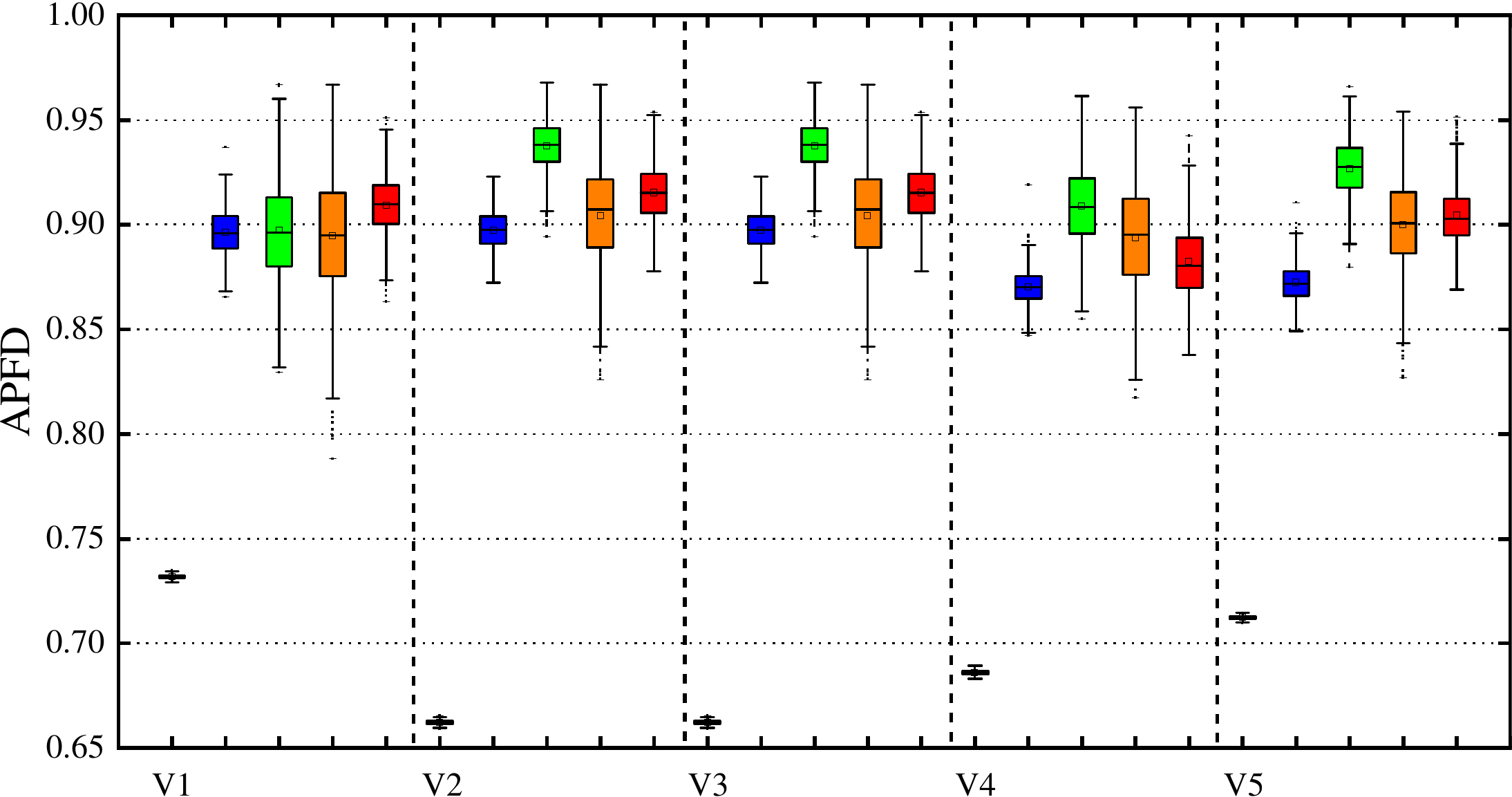}
        \label{apfd_sed_bc}
    }
        \subfigure[$sed$-method coverage]
    {
        \includegraphics[width=0.319\textwidth]{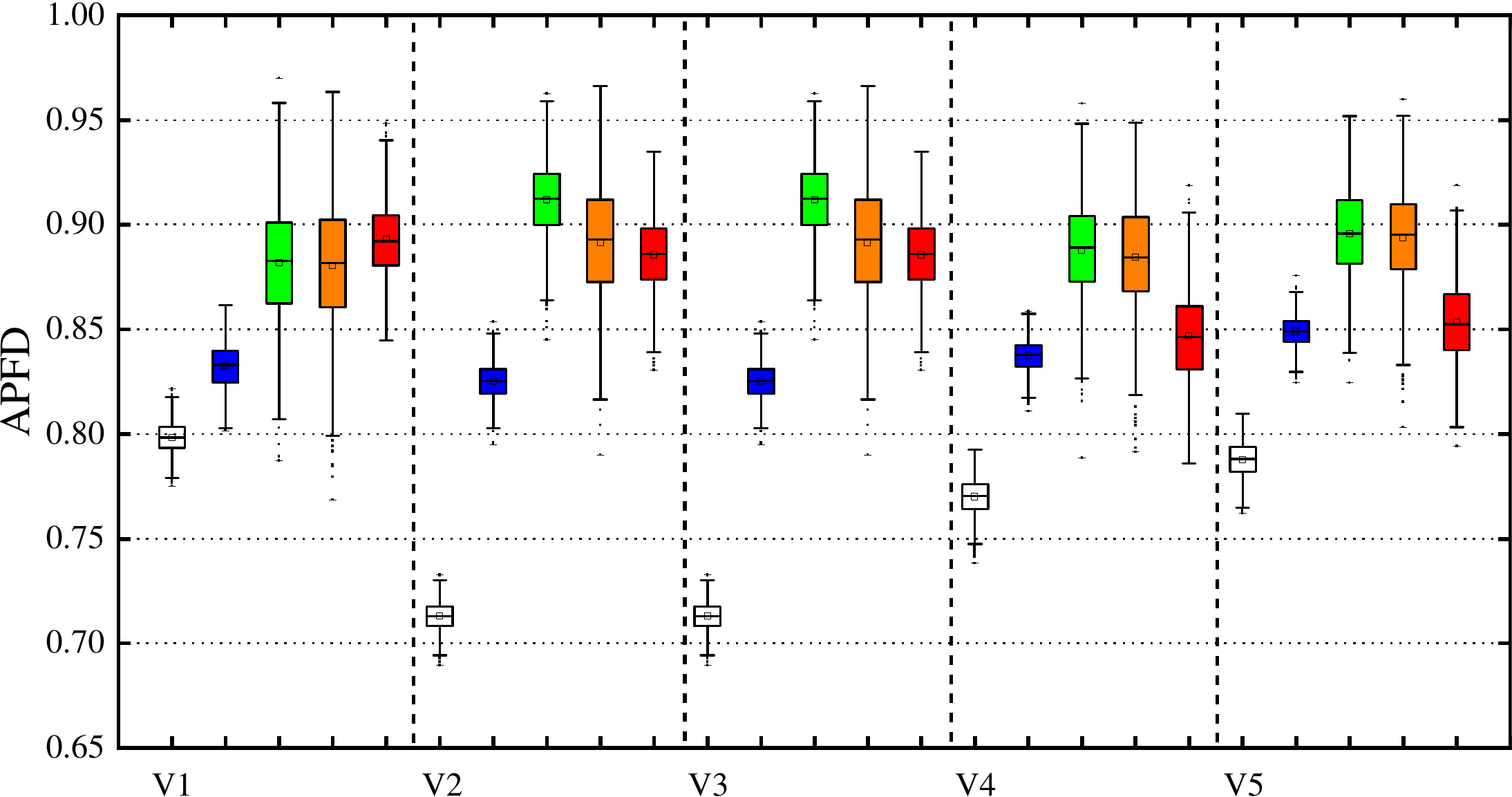}
        \label{apfd_sed_mc}
    }
    \caption{\textbf{Effectiveness:} \textbf{APFD} results for \textbf{C programs}}
    \label{FIG:apfd-c}
\end{figure*}

\begin{table*}[!t]
\scriptsize
\centering
 \caption{Statistical \textbf{effectiveness}  comparisons of \textbf{APFD} for \textbf{C programs}.
For a comparison between two methods
$\textit{TCP}_\textit{ccc}$ and $M$,
where
$M \in \{\textit{TCP}_\textit{tot}, \textit{TCP}_\textit{add}, \textit{TCP}_\textit{art}, \textit{TCP}_\textit{search}\}$,
the symbol \ding{52} means that $\textit{TCP}_\textit{ccc}$ is better
($p$-value is less than 0.05, and the effect size $\hat{\textrm{A}}_{12}(\textit{TCP}_\textit{ccc},M)$ is greater than 0.50);
the symbol \ding{54} means that $M$ is better
(the $p$-value is less than 0.05, and $\hat{\textrm{A}}_{12}(\textit{TCP}_\textit{ccc},M)$ is less than 0.50);
and the symbol \ding{109} means that there is no statistically significant difference between them (the $p$-value is greater than 0.05).
}
  \label{TAB:apfd-c}
    \setlength{\tabcolsep}{1.8mm}{
    \begin{tabular}{c|cccc|cccc|cccc}
     \hline
       \multirow{2}*{\textbf{Program Name}} &\multicolumn{4}{c|}{\textbf{Statement Coverage}} &\multicolumn{4}{c|}{\textbf{Branch Coverage}} &\multicolumn{4}{c}{\textbf{Method Coverage}} \\
        \cline{2-13}

         &$\textit{TCP}_\textit{tot}$ &$\textit{TCP}_\textit{add}$ &$\textit{TCP}_\textit{art}$ &$\textit{TCP}_\textit{search}$  & $\textit{TCP}_\textit{tot}$ &$\textit{TCP}_\textit{add}$ &$\textit{TCP}_\textit{art}$ &$\textit{TCP}_\textit{search}$ &$\textit{TCP}_\textit{tot}$ & $\textit{TCP}_\textit{add}$ &$\textit{TCP}_\textit{art}$ &$\textit{TCP}_\textit{search}$\\
            \hline

$flex\_v1$ &\ding{52}(1.00) &\ding{52}(0.99) &\ding{52}(1.00) &\ding{52}(1.00) &\ding{52}(1.00) &\ding{52}(1.00) &\ding{52}(1.00) &\ding{52}(1.00) &\ding{52}(1.00) &\ding{52}(1.00) &\ding{52}(0.83) &\ding{52}(0.96) \\
$flex\_v2$ &\ding{52}(1.00) &\ding{52}(0.98) &\ding{52}(1.00) &\ding{52}(0.99) &\ding{52}(1.00) &\ding{52}(1.00) &\ding{52}(1.00) &\ding{52}(0.99) &\ding{52}(1.00) &\ding{52}(1.00) &\ding{52}(0.78) &\ding{52}(0.93) \\
$flex\_v3$ &\ding{52}(1.00) &\ding{52}(0.74) &\ding{52}(1.00) &\ding{52}(0.88) &\ding{52}(1.00) &\ding{52}(0.70) &\ding{52}(1.00) &\ding{52}(0.94) &\ding{52}(1.00) &\ding{52}(0.99) &\ding{52}(0.99) &\ding{52}(0.96) \\
$flex\_v4$ &\ding{52}(1.00) &\ding{52}(0.98) &\ding{52}(1.00) &\ding{52}(0.99) &\ding{52}(1.00) &\ding{52}(1.00) &\ding{52}(1.00) &\ding{52}(1.00) &\ding{52}(1.00) &\ding{52}(1.00) &\ding{52}(0.77) &\ding{52}(0.92) \\
$flex\_v5$ &\ding{52}(1.00) &\ding{52}(0.98) &\ding{52}(1.00) &\ding{52}(0.99) &\ding{52}(1.00) &\ding{52}(1.00) &\ding{52}(1.00) &\ding{52}(0.99) &\ding{52}(1.00) &\ding{52}(1.00) &\ding{52}(0.77) &\ding{52}(0.92) \\
\hline
$flex$ &\ding{52}(1.00) &\ding{52}(0.89) &\ding{52}(1.00) &\ding{52}(0.94) &\ding{52}(1.00) &\ding{52}(0.86) &\ding{52}(1.00) &\ding{52}(0.95) &\ding{52}(1.00) &\ding{52}(0.86) &\ding{52}(0.80) &\ding{52}(0.88) \\
\hline

$grep\_v1$ &\ding{52}(1.00) &\ding{52}(0.94) &\ding{109}(0.49) &\ding{52}(0.70) &\ding{52}(1.00) &\ding{52}(0.75) &\ding{54}(0.23) &\ding{52}(0.57) &\ding{52}(1.00) &\ding{52}(0.86) &\ding{54}(0.10) &\ding{54}(0.15) \\
$grep\_v2$ &\ding{52}(1.00) &\ding{52}(0.89) &\ding{52}(0.55) &\ding{52}(0.82) &\ding{52}(1.00) &\ding{52}(0.70) &\ding{54}(0.21) &\ding{52}(0.70) &\ding{52}(1.00) &\ding{52}(0.96) &\ding{54}(0.14) &\ding{54}(0.28) \\
$grep\_v3$ &\ding{52}(1.00) &\ding{52}(0.82) &\ding{52}(0.56) &\ding{54}(0.44) &\ding{52}(1.00) &\ding{52}(0.72) &\ding{54}(0.45) &\ding{54}(0.43) &\ding{52}(1.00) &\ding{52}(0.86) &\ding{54}(0.20) &\ding{54}(0.27) \\
$grep\_v4$ &\ding{52}(1.00) &\ding{52}(0.79) &\ding{52}(0.81) &\ding{52}(0.90) &\ding{52}(1.00) &\ding{52}(0.61) &\ding{52}(0.89) &\ding{52}(0.91) &\ding{52}(1.00) &\ding{52}(0.84) &\ding{54}(0.21) &\ding{54}(0.25) \\
$grep\_v5$ &\ding{52}(1.00) &\ding{52}(0.65) &\ding{52}(0.91) &\ding{52}(0.95) &\ding{52}(1.00) &\ding{52}(0.54) &\ding{52}(0.93) &\ding{52}(0.93) &\ding{52}(1.00) &\ding{52}(0.63) &\ding{54}(0.18) &\ding{54}(0.21) \\
\hline
$grep$ &\ding{52}(1.00) &\ding{52}(0.81) &\ding{52}(0.65) &\ding{52}(0.75) &\ding{52}(1.00) &\ding{52}(0.64) &\ding{52}(0.56) &\ding{52}(0.71) &\ding{52}(1.00) &\ding{52}(0.74) &\ding{54}(0.25) &\ding{54}(0.30) \\
\hline

$gzip\_v1$ &\ding{52}(1.00) &\ding{52}(1.00) &\ding{52}(0.85) &\ding{52}(0.89) &\ding{52}(1.00) &\ding{52}(1.00) &\ding{52}(0.89) &\ding{52}(0.90) &\ding{52}(0.88) &\ding{52}(0.88) &\ding{109}(0.48) &\ding{109}(0.48) \\
$gzip\_v2$ &\ding{52}(1.00) &\ding{52}(1.00) &\ding{52}(0.82) &\ding{52}(0.86) &\ding{52}(1.00) &\ding{52}(1.00) &\ding{52}(0.87) &\ding{52}(0.88) &\ding{52}(0.88) &\ding{52}(0.88) &\ding{109}(0.48) &\ding{109}(0.49) \\
$gzip\_v3$ &\ding{52}(1.00) &\ding{52}(0.80) &\ding{52}(0.60) &\ding{54}(0.39) &\ding{52}(1.00) &\ding{52}(0.95) &\ding{52}(0.53) &\ding{54}(0.36) &\ding{54}(0.46) &\ding{54}(0.46) &\ding{109}(0.50) &\ding{109}(0.49) \\
$gzip\_v4$ &\ding{52}(1.00) &\ding{52}(0.80) &\ding{52}(0.60) &\ding{54}(0.39) &\ding{52}(1.00) &\ding{52}(0.95) &\ding{52}(0.53) &\ding{54}(0.36) &\ding{54}(0.46) &\ding{54}(0.46) &\ding{109}(0.50) &\ding{109}(0.49) \\
$gzip\_v5$ &\ding{52}(1.00) &\ding{52}(0.80) &\ding{52}(0.60) &\ding{54}(0.39) &\ding{52}(1.00) &\ding{52}(0.95) &\ding{52}(0.53) &\ding{54}(0.36) &\ding{54}(0.46) &\ding{54}(0.46) &\ding{109}(0.50) &\ding{109}(0.49) \\
\hline
$gzip$ &\ding{52}(1.00) &\ding{52}(0.79) &\ding{52}(0.62) &\ding{52}(0.52) &\ding{52}(1.00) &\ding{52}(0.95) &\ding{52}(0.61) &\ding{52}(0.52) &\ding{52}(0.55) &\ding{52}(0.56) &\ding{109}(0.50) &\ding{109}(0.49) \\
\hline

$make\_v1$
&\ding{52}(0.92) &\ding{52}(0.56) &\ding{54}(0.31) &\ding{54}(0.30) &\ding{52}(1.00) &\ding{52}(0.64) &\ding{54}(0.36) &\ding{54}(0.47) &\ding{52}(0.92) &\ding{52}(0.97) &\ding{52}(0.64) &\ding{52}(0.66) \\
$make\_v2$ &\ding{52}(1.00) &\ding{52}(0.81) &\ding{54}(0.40) &\ding{54}(0.40) &\ding{52}(1.00) &\ding{52}(0.61) &\ding{54}(0.35) &\ding{109}(0.52) &\ding{52}(0.99) &\ding{52}(0.99) &\ding{52}(0.62) &\ding{52}(0.65) \\
$make\_v3$ &\ding{52}(0.99) &\ding{52}(0.61) &\ding{54}(0.29) &\ding{54}(0.29) &\ding{52}(1.00) &\ding{52}(0.62) &\ding{54}(0.36) &\ding{52}(0.56) &\ding{52}(0.99) &\ding{52}(1.00) &\ding{52}(0.66) &\ding{52}(0.68) \\
$make\_v4$ &\ding{52}(0.89) &\ding{52}(0.60) &\ding{54}(0.26) &\ding{54}(0.26) &\ding{52}(1.00) &\ding{52}(0.58) &\ding{54}(0.32) &\ding{109}(0.51) &\ding{52}(0.99) &\ding{52}(1.00) &\ding{52}(0.68) &\ding{52}(0.69) \\
$make\_v5$ &\ding{52}(1.00) &\ding{52}(0.83) &\ding{54}(0.38) &\ding{54}(0.39) &\ding{52}(1.00) &\ding{52}(0.63) &\ding{54}(0.31) &\ding{109}(0.50) &\ding{52}(0.99) &\ding{52}(0.99) &\ding{52}(0.62) &\ding{52}(0.65) \\
\hline
$make$ &\ding{52}(0.92) &\ding{52}(0.67) &\ding{54}(0.34) &\ding{54}(0.34) &\ding{52}(1.00) &\ding{52}(0.60) &\ding{54}(0.36) &\ding{109}(0.51) &\ding{52}(0.93) &\ding{52}(0.96) &\ding{52}(0.63) &\ding{52}(0.65) \\
\hline
$sed\_v1$ &\ding{52}(1.00) &\ding{52}(0.83) &\ding{52}(0.62) &\ding{52}(0.61) &\ding{52}(1.00) &\ding{52}(0.77) &\ding{52}(0.67) &\ding{52}(0.67) &\ding{52}(1.00) &\ding{52}(1.00) &\ding{52}(0.63) &\ding{52}(0.63) \\
$sed\_v2$ &\ding{52}(1.00) &\ding{52}(0.82) &\ding{54}(0.23) &\ding{52}(0.71) &\ding{52}(1.00) &\ding{52}(0.86) &\ding{54}(0.11) &\ding{52}(0.63) &\ding{52}(1.00) &\ding{52}(1.00) &\ding{54}(0.15) &\ding{54}(0.42) \\
$sed\_v3$ &\ding{52}(1.00) &\ding{52}(0.82) &\ding{54}(0.23) &\ding{52}(0.71) &\ding{52}(1.00) &\ding{52}(0.86) &\ding{54}(0.11) &\ding{52}(0.63) &\ding{52}(1.00) &\ding{52}(1.00) &\ding{54}(0.15) &\ding{54}(0.42) \\
$sed\_v4$ &\ding{52}(1.00) &\ding{52}(0.86) &\ding{54}(0.30) &\ding{109}(0.51) &\ding{52}(1.00) &\ding{52}(0.73) &\ding{54}(0.15) &\ding{54}(0.35) &\ding{52}(1.00) &\ding{52}(0.64) &\ding{54}(0.10) &\ding{54}(0.13) \\
$sed\_v5$ &\ding{52}(1.00) &\ding{52}(0.96) &\ding{54}(0.09) &\ding{109}(0.49) &\ding{52}(1.00) &\ding{52}(0.99) &\ding{54}(0.14) &\ding{52}(0.56) &\ding{52}(1.00) &\ding{52}(0.57) &\ding{54}(0.08) &\ding{54}(0.10) \\
\hline
$sed$ &\ding{52}(1.00) &\ding{52}(0.78) &\ding{54}(0.33) &\ding{52}(0.60) &\ding{52}(1.00) &\ding{52}(0.78) &\ding{54}(0.28) &\ding{52}(0.57) &\ding{52}(1.00) &\ding{52}(0.89) &\ding{54}(0.25) &\ding{54}(0.35) \\
\hline
\textit{A\textbf{ll C Programs}} &\ding{52}(0.93) &\ding{52}(0.64) &\ding{52}(0.54) &\ding{52}(0.56) &\ding{52}(0.95) &\ding{52}(0.65) &\ding{52}(0.53) &\ding{52}(0.56) &\ding{52}(0.84) &\ding{52}(0.70) &\ding{54}(0.48) &\ding{52}(0.52) \\
\hline

  \end{tabular}}
\end{table*}

\begin{table*}[!htpb]
\scriptsize
\centering
 \caption{An analysis of statistical \textbf{effectiveness }results of \textbf{APFD}.
Each cell represents the total times of (\ding{52}), worse (\ding{54}), and (\ding{109}) for corresponding prioritization scenarios described in Tables \ref{TAB:apfd-java-c} to \ref{TAB:apfd-c}.
 }
  \label{TAB:APFDsummary}
  \setlength{\tabcolsep}{0.5mm}{
    \begin{tabular}{c|c|cccc|cccc|cccc|cccc}
     \hline
       \multirow{2}*{\textbf{Language}} &\multirow{2}*{\textbf{Status}}&\multicolumn{4}{c|}{\textbf{Statement Coverage}} &\multicolumn{4}{c|}{\textbf{Branch Coverage}} &\multicolumn{4}{c|}{\textbf{Method Coverage}} &\multicolumn{4}{c}{\textit{\textbf{Sum }}$\sum$ }\\
        \cline{3-18}

         &&$\textit{TCP}_\textit{tot}$ &$\textit{TCP}_\textit{add}$ &$\textit{TCP}_\textit{art}$ &$\textit{TCP}_\textit{search}$  & $\textit{TCP}_\textit{tot}$ &$\textit{TCP}_\textit{add}$ &$\textit{TCP}_\textit{art}$ &$\textit{TCP}_\textit{search}$ &$\textit{TCP}_\textit{tot}$ & $\textit{TCP}_\textit{add}$ &$\textit{TCP}_\textit{art}$ &$\textit{TCP}_\textit{search}$ &$\textit{TCP}_\textit{tot}$ & $\textit{TCP}_\textit{add}$ &$\textit{TCP}_\textit{art}$ &$\textit{TCP}_\textit{search}$\\
            \hline
\multirow{3}*{Java (test-class)}
&\ding{52} &5   &0  &13 &2  &6  &0  &13 &4  &6  &1   &13 &2 &17 &1 &39 &8\\
                            &\ding{54} &6   &1  &0  &1  &4  &4  &0  &0  &6  &2   &0  &6 &16 &7 &0 &7\\
                            &\ding{109}&3   &13 &1  &11 &4  &10 &1  &10 &2  &11  &1  &6 &9 &34 &3 &27\\\hline

\multirow{3}*{Java (test-method)}
                            &\ding{52} &14  &6   &11 &10 &14 &1  &11 &12 &14 &1  &13 &10 &42 &8 &35 &32\\
                            &\ding{54} &0   &1   &3  &2  &0  &5  &3  &2  &0  &3  &1  &2 &0 &9 &7 &6\\
                            &\ding{109}&0   &7   &0  &2  &0  &8  &0  &0  &0  &10 &0  &2 &0 &25 &0 &4\\\hline
\multirow{3}*{C}&\ding{52} &25  &25  &15 &14 &25 &25 &13 &16 &22 &22 &11 &11 &72 &72 &39 &41\\
                            &\ding{54} &0   &0   &9  &9  &0  &0  &12 &6  &3  &3  &9  &9 &3 &3 &30 &24\\
                            &\ding{109}&0   &0   &1  &2  &0  &0  &0  &3  &0  &0  &5  &5 &0 &0 &6 &10\\\hline
\multirow{3}*{\textit{\textbf{Sum}} $\sum$}
                            &\ding{52} &44  &31  &39 &26  &45 &26 &37 &32 &42 &24 &37 &23 &131 &81 &113 &81\\
                            &\ding{54} &6   &2  &12 &12  &4  &9  &15 &13 &9  &8  &10 &17 &19 &19 &37 &42\\
                            &\ding{109}&3   &20   &2  &15  &4  &18 &1  &8  &2  &21 &6  &13 &9 &59 &9 &36\\\hline
  \end{tabular}}
\end{table*}

\begin{figure*}[!t]
\graphicspath{{graphs/}}
\centering
    \subfigure[$ant$-statement coverage]
    {
        \includegraphics[width=0.319\textwidth]{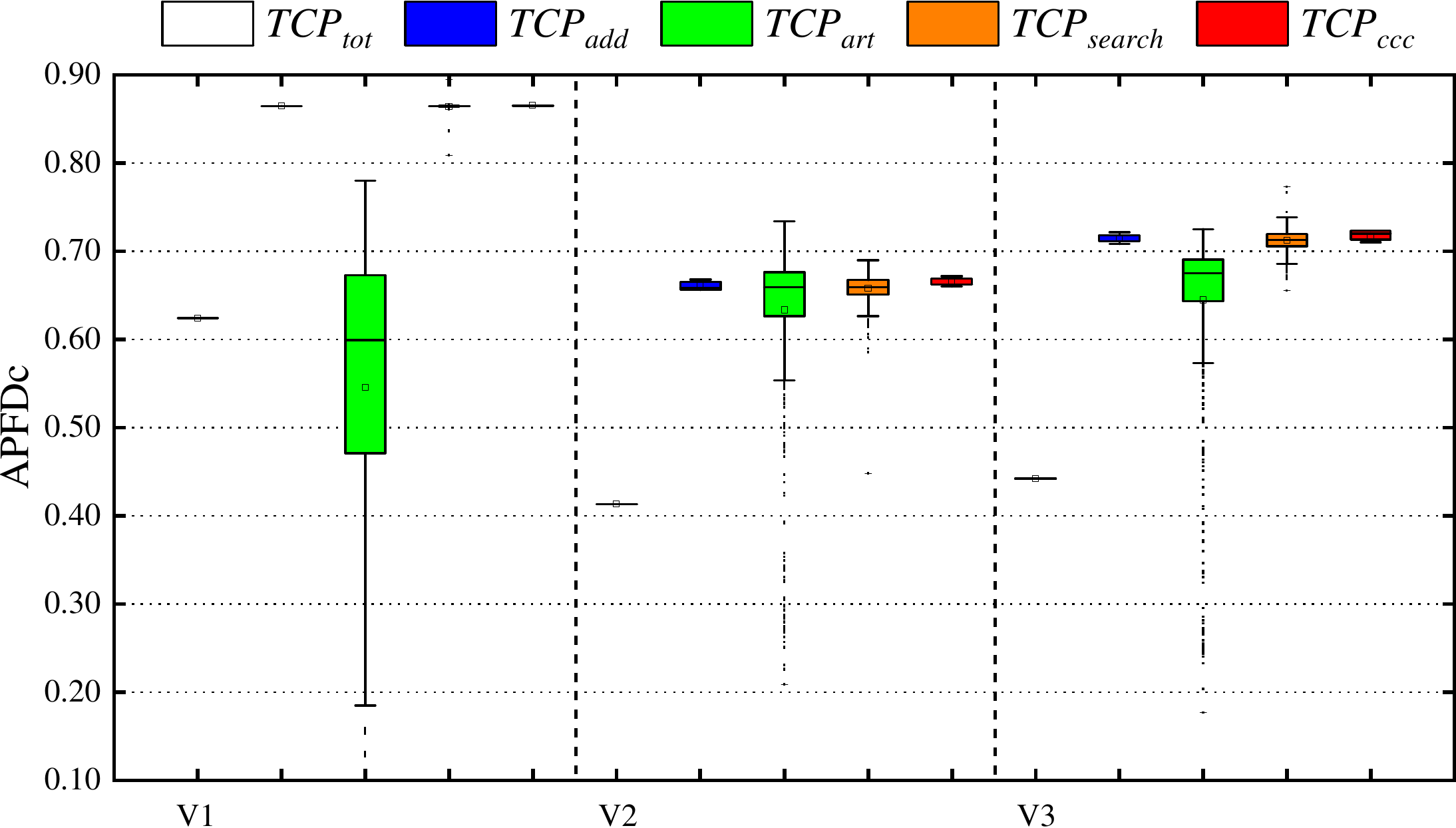}
        \label{apfdc_ant_sc}
    }
    \subfigure[$ant$-branch coverage]
    {
        \includegraphics[width=0.319\textwidth]{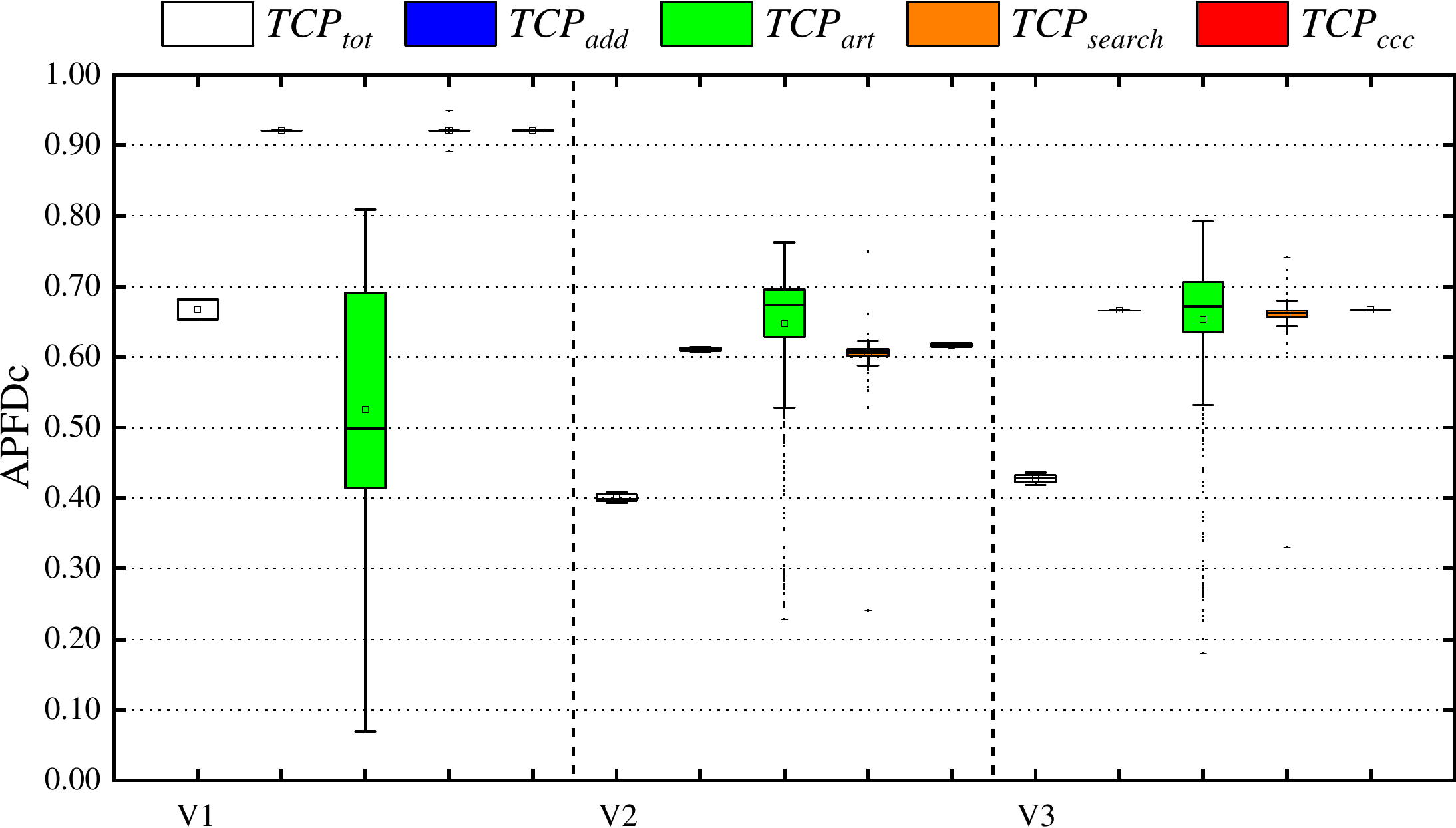}
        \label{apfdc_ant_bc}
    }
        \subfigure[$ant$-method coverage]
    {
        \includegraphics[width=0.319\textwidth]{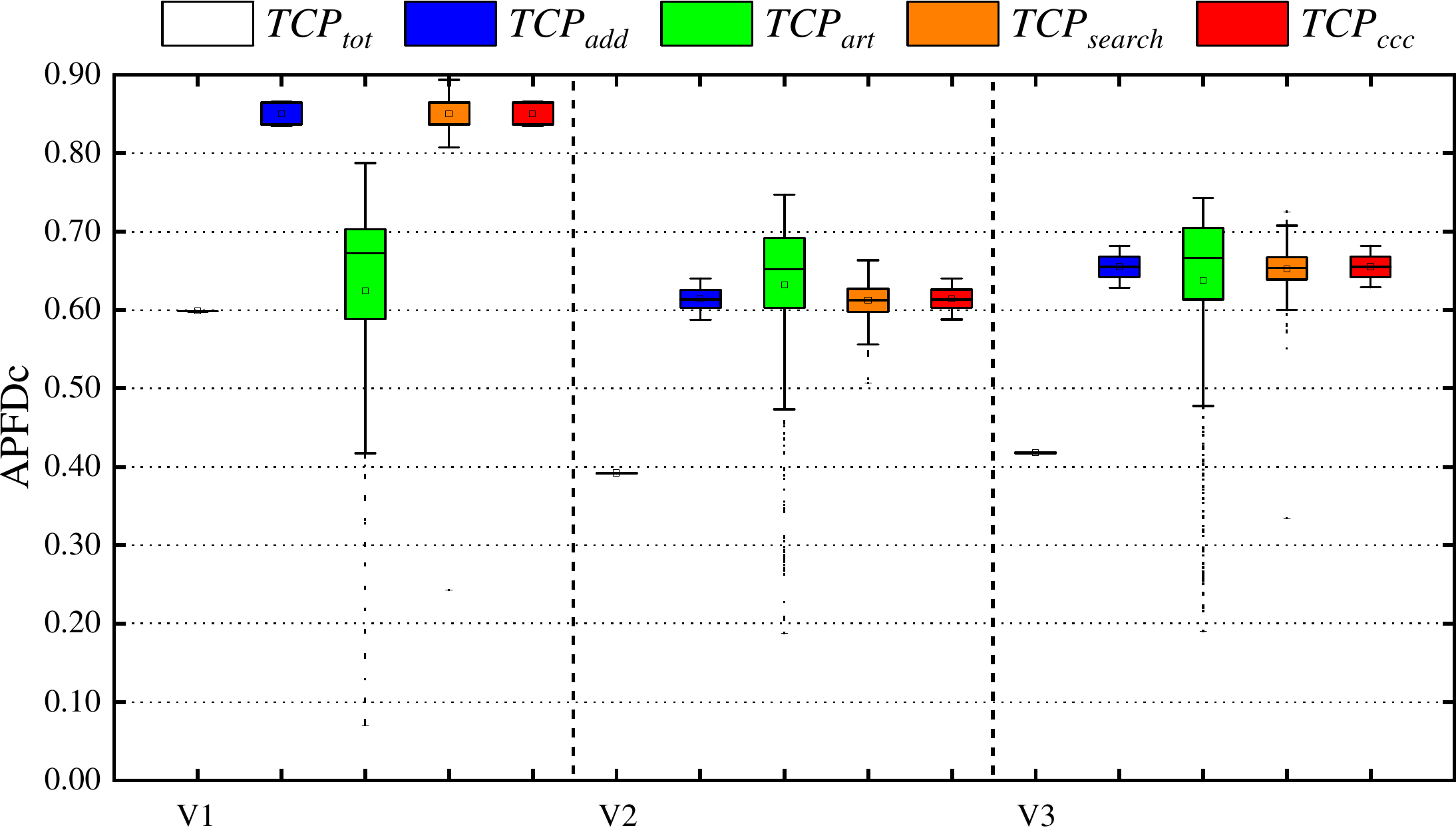}
        \label{apfdc_ant_mc}
    }
    \subfigure[$jmeter$-statement coverage]
    {
        \includegraphics[width=0.319\textwidth]{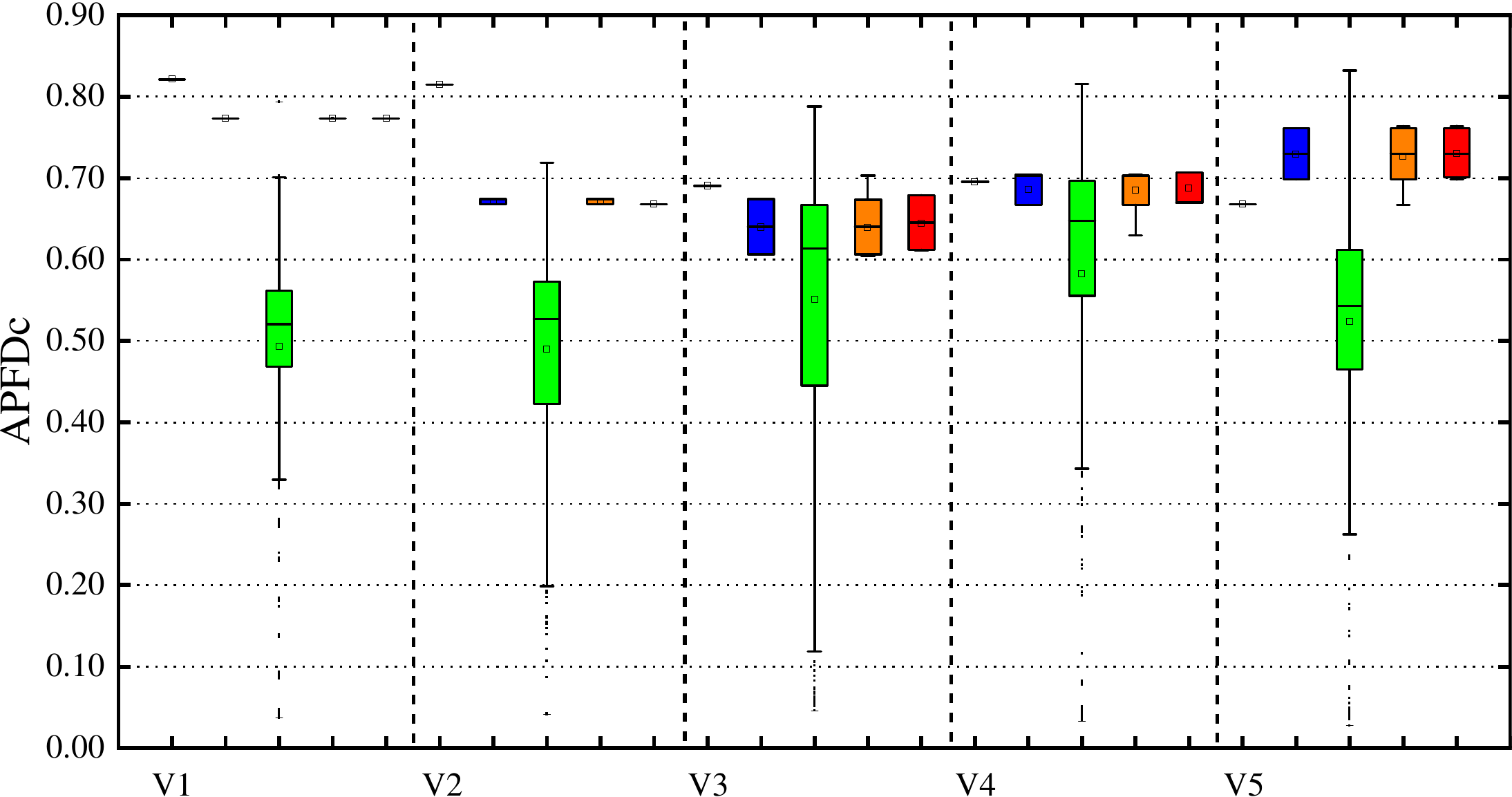}
        \label{apfdc_grep_sc}
    }
        \subfigure[$jmeter$-branch coverage]
    {
        \includegraphics[width=0.319\textwidth]{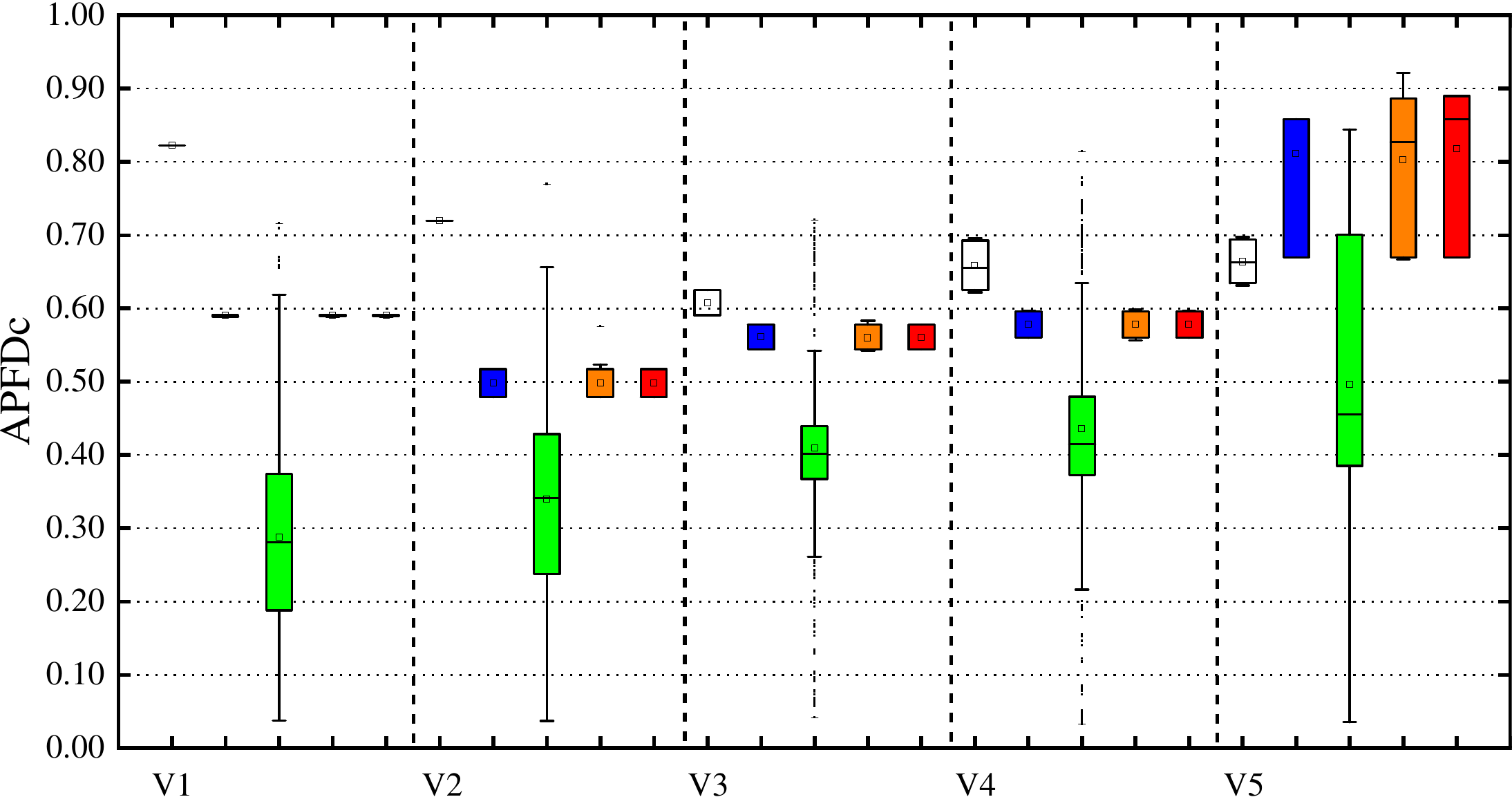}
        \label{apfdc_jmeter_bc}
    }
        \subfigure[$jmeter$-method coverage]
    {
        \includegraphics[width=0.319\textwidth]{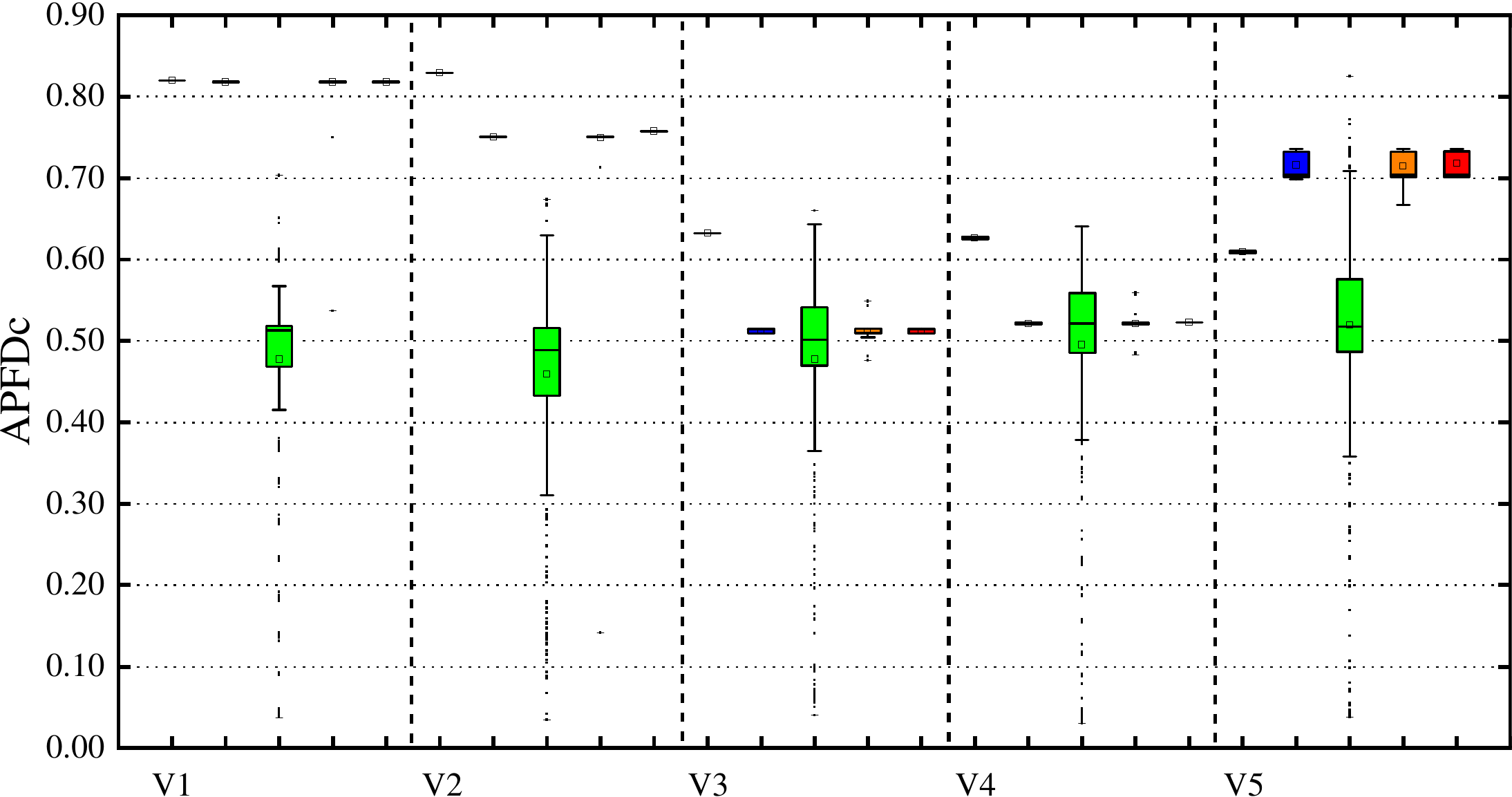}
        \label{apfdc_jmeter_mc}
    }
        \subfigure[$jtopas$-statement coverage]
    {
        \includegraphics[width=0.319\textwidth]{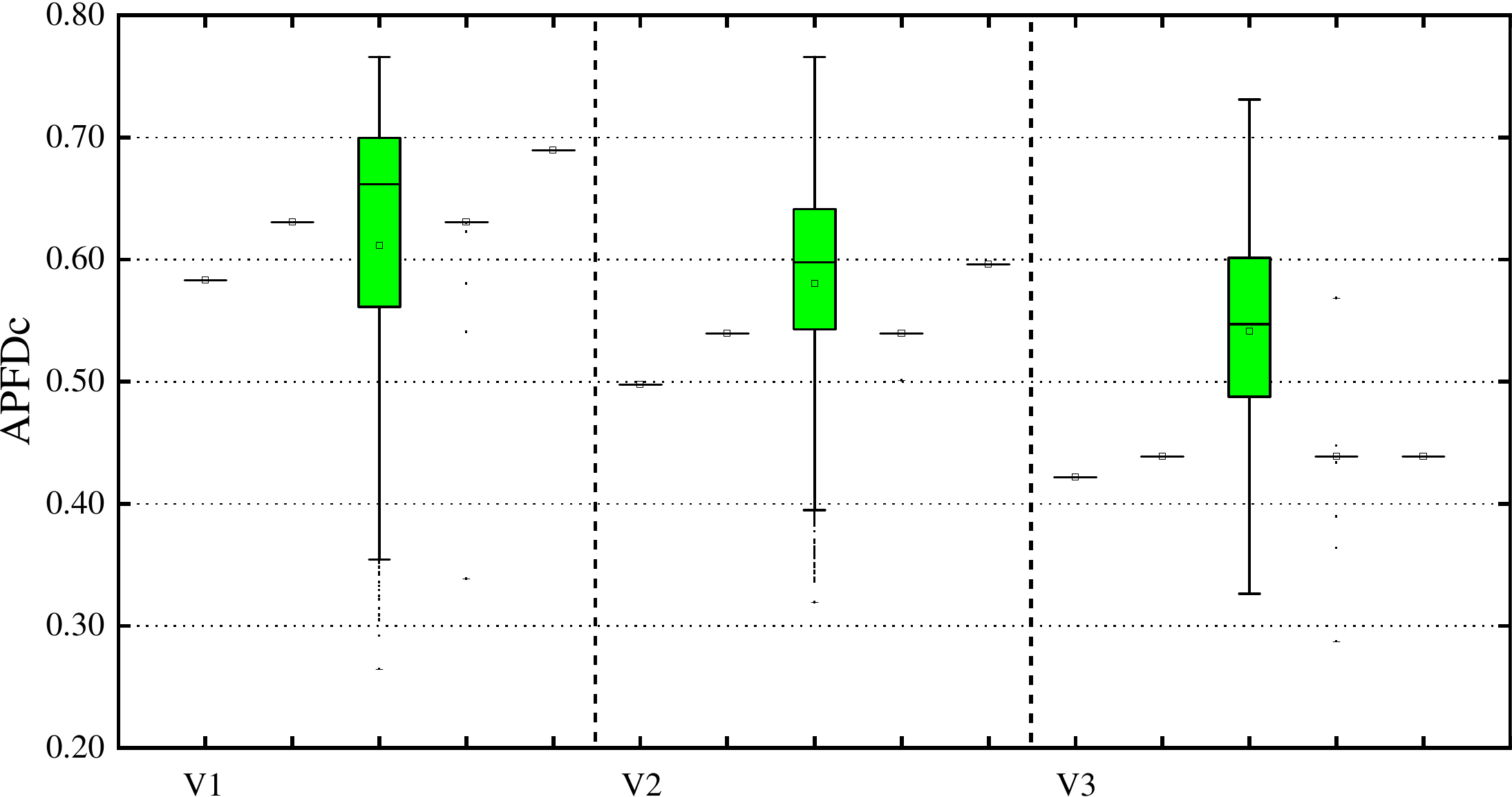}
        \label{apfdc_jtopas_sc}
    }
        \subfigure[$jtopas$-branch coverage]
    {
        \includegraphics[width=0.319\textwidth]{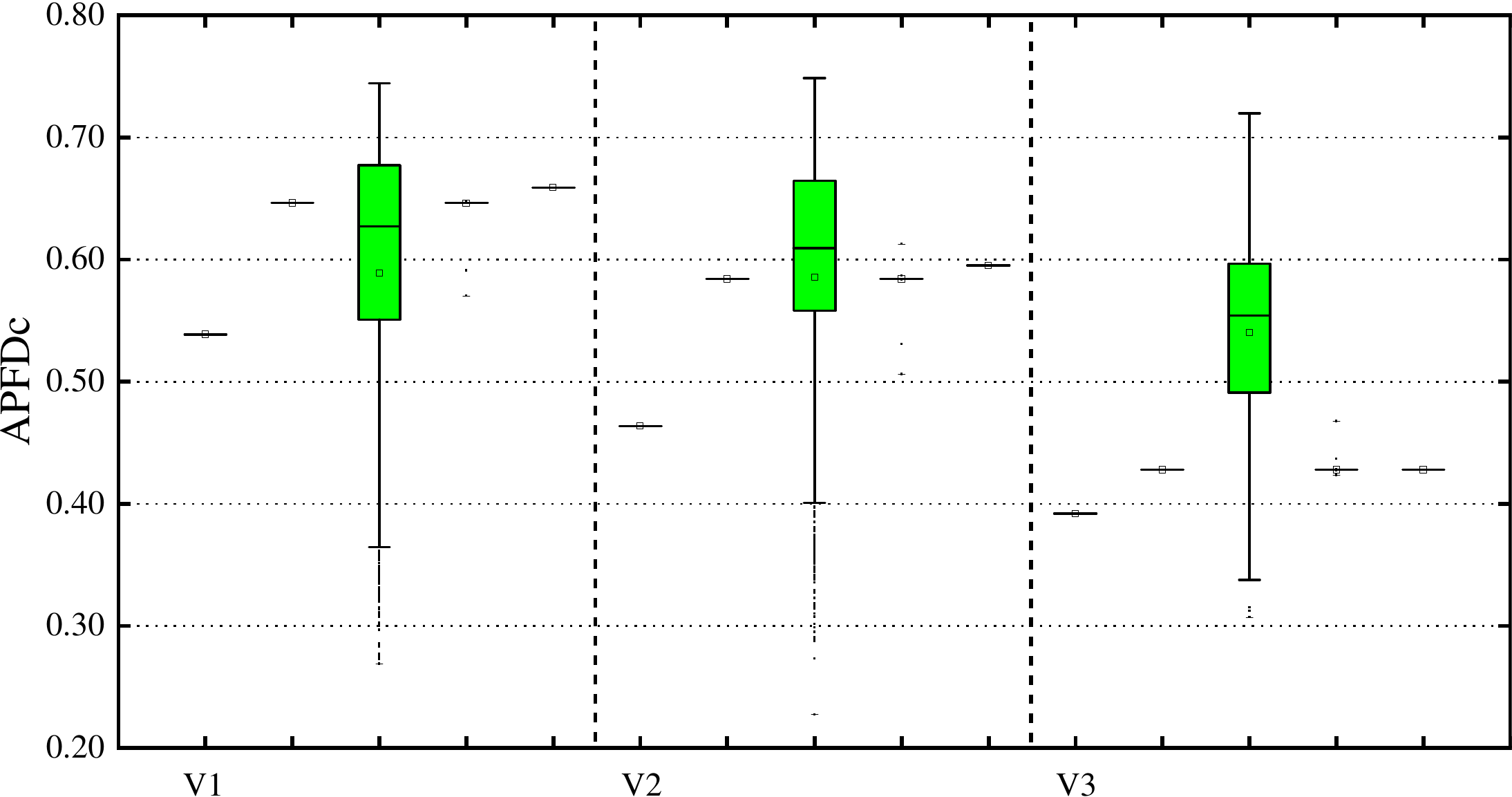}
        \label{apfdc_jtopas_bc}
    }
        \subfigure[$jtopas$-method coverage]
    {
        \includegraphics[width=0.319\textwidth]{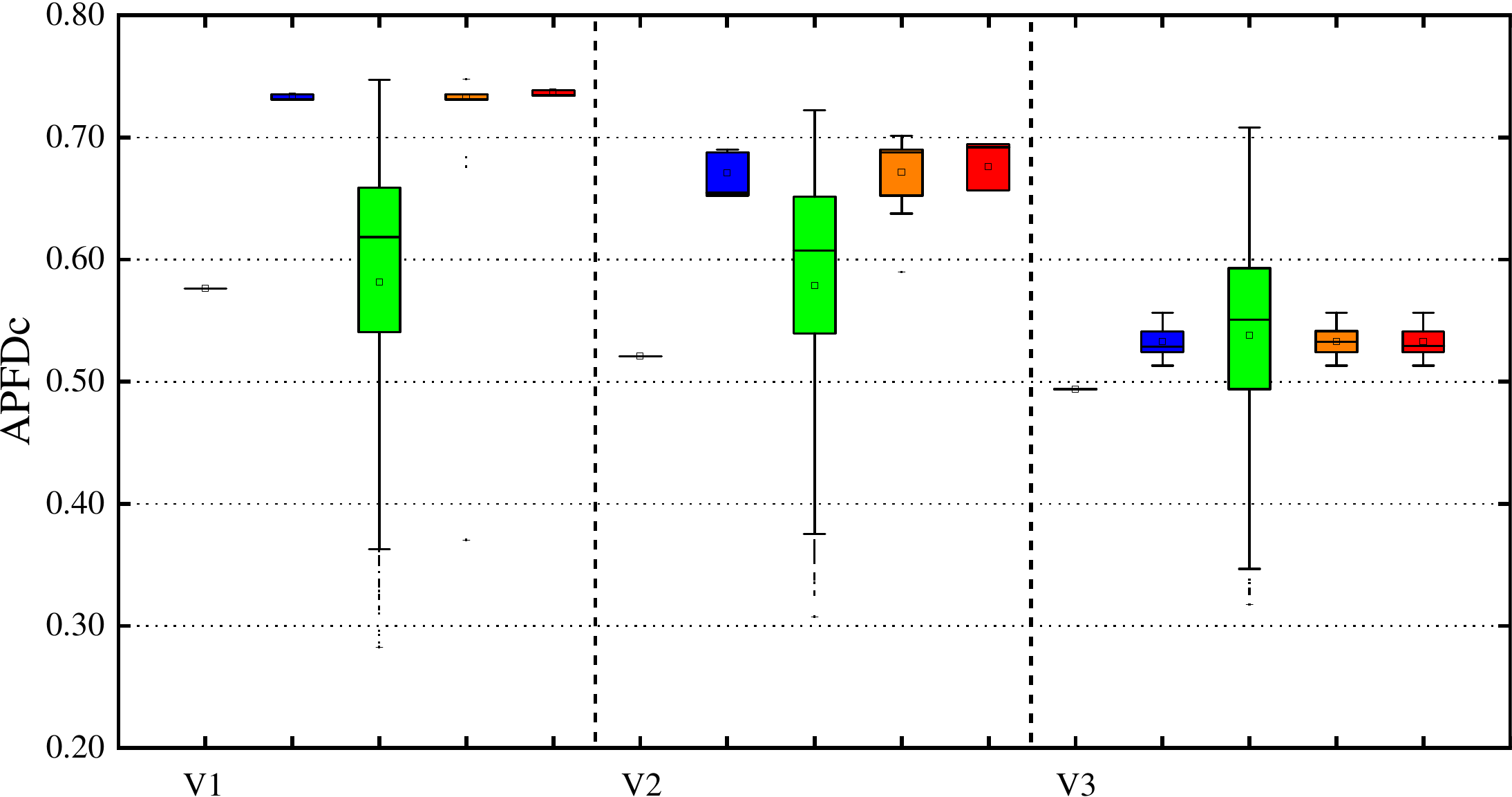}
        \label{apfdc_jtopas_mc}
    }
        \subfigure[$xmlsec$-statement coverage]
    {
        \includegraphics[width=0.319\textwidth]{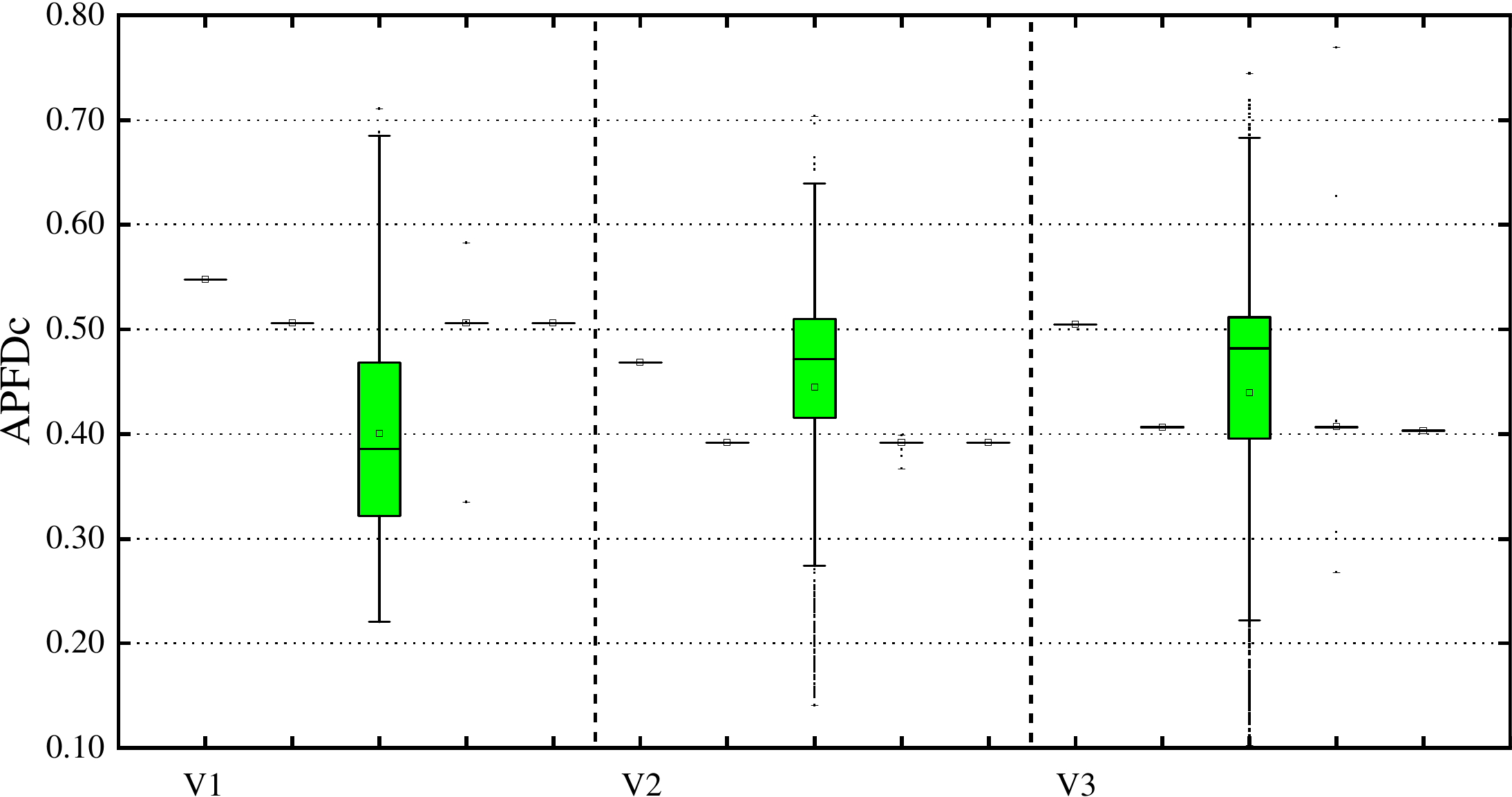}
        \label{apfdc_xmlsec_sc}
    }
        \subfigure[$xmlsec$-branch coverage]
    {
        \includegraphics[width=0.319\textwidth]{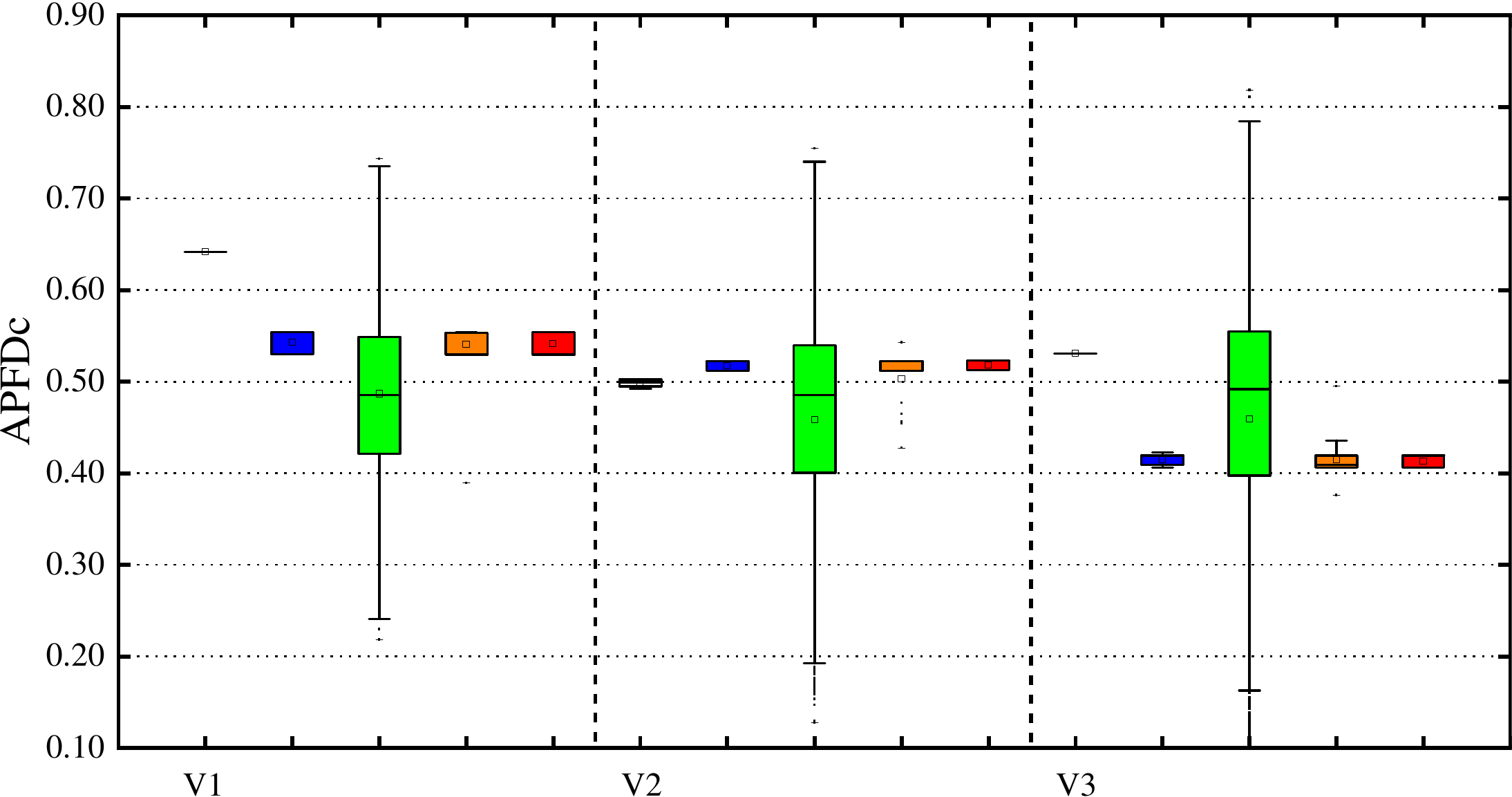}
        \label{apfdc_xmlsec_bc}
    }
        \subfigure[$xmlsec$-method coverage]
    {
        \includegraphics[width=0.319\textwidth]{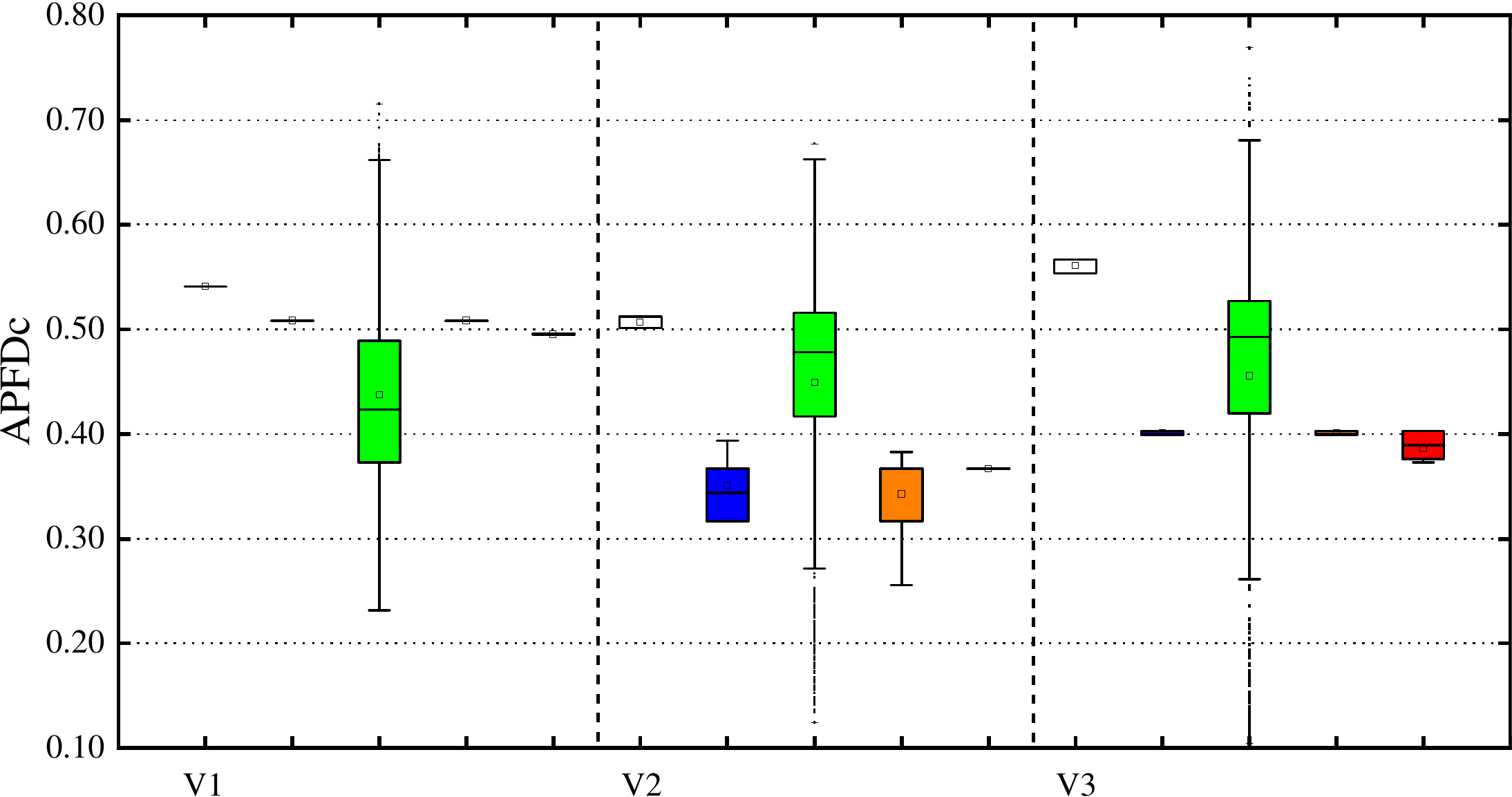}
        \label{apfdc_xmlsec_mc}
    }
    \caption{\textbf{Effectiveness: }\textbf{APFD}$_\textbf{c}$  results for \textbf{Java programs} at the \textbf{test-class level}}
    \label{FIG:apfdc-java-c}
\end{figure*}

\section{Results and Analysis}
\label{res}

This section presents the experimental results to answer the research questions.

To answer RQ1 to RQ4, Figures \ref{FIG:apfd-java-c} to \ref{FIG:SBM} present box plots of the distribution of the APFD or APFD$_\textrm{c}$ values (averaged over 1000 iterations).
Each box plot shows the mean (square in the box), median (line in the box), and upper and lower quartiles (25th and 75th percentile) for the APFD or APFD$_\textrm{c}$ values for the RTCP techniques. Statistical analyses are also provided in Tables \ref{TAB:apfd-java-c} to \ref{TAB:SBM}
for each pairwise APFD or APFD$_\textrm{c}$ comparison between CCCP and the other RTCP techniques.
For example, for a comparison between two methods
$\textit{TCP}_\textit{ccc}$ and $M$,
where
$M \in \{\textit{TCP}_\textit{tot}, \textit{TCP}_\textit{add}, \textit{TCP}_\textit{art}, \textit{TCP}_\textit{search}\}$,
the symbol \ding{52} means that $\textit{TCP}_\textit{ccc}$ is better
($p$-value is less than 0.05, and the effect size $\hat{\textrm{A}}_{12}(\textit{TCP}_\textit{ccc},M)$ is greater than 0.50);
the symbol \ding{54} means that $M$ is better
(the $p$-value is less than 0.05, and $\hat{\textrm{A}}_{12}(\textit{TCP}_\textit{ccc},M)$ is less than 0.50);
and the symbol \ding{109} means that there is no statistically significant difference between them (the $p$-value is greater than 0.05).

To answer RQ5, Table \ref{TAB:time} provides comparisons of the execution times for the different RTCP techniques.
To answer RQ6, Figure \ref{FIG:2wise} shows the APFD and APFD$_\textrm{c}$ results for CCCP with $\lambda=2$, and Table \ref{TAB:2wise} presents the corresponding statistical analysis.

\subsection{\textbf{RQ1: Effectiveness of CCCP Measured by APFD}}
Here, we provide the APFD results for CCCP for  different code coverage and test case granularities.
Figures \ref{FIG:apfd-java-c} to \ref{FIG:apfd-c} show the APFD results for the Java programs at the test-class level;
at the test-method level; and
the C programs, respectively.
Each sub-figure in these figures has the program versions across the $x$-axis, and the APFD values for the five RTCP techniques on the $y$-axis.
Tables \ref{TAB:apfd-java-c} to \ref{TAB:apfd-c} present the corresponding statistical comparisons.


\subsubsection{APFD Results: Java Programs (Test-Class Level)}

Based on Figure \ref{FIG:apfd-java-c} and Table \ref{TAB:apfd-java-c}, we have the following observations:

(1)
Compared with the total test prioritization technique, CCCP achieves better performances for the program $xmlsec$, irrespective of code coverage granularity, with differences between the mean and median APFD values reaching about 3\%.
For the other programs ($ant$, $jmeter$, and $jtopas$), however, they have very similar APFD results.

(2)
CCCP performs much better than adaptive random test prioritization, regardless of subject program and code coverage granularity, with the maximum mean and median APFD differences reaching about 12\%.

(3)
CCCP has very similar performance to the additional and search-based test prioritization techniques, with the mean and median APFD differences approximately equal to 1\%.

(4)
There is a statistically significant difference between $\textit{TCP}_\textit{ccc}$ and $\textit{TCP}_\textit{art}$, which supports the above observations.
However, none of the other three techniques ($\textit{TCP}_\textit{tot}$, $\textit{TCP}_\textit{add}$, or $\textit{TCP}_\textit{search}$) is either always better or always worse than $\textit{TCP}_\textit{ccc}$,  with $\textit{TCP}_\textit{ccc}$ sometimes performing better for some programs, and sometimes worse.

(5)
Considering all Java programs:
Overall, because all $p$-values are less than 0.05, and the relevant effect size $\hat{\textrm{A}}_{12}$ ranges from 0.58 to 0.98,
$\textit{TCP}_\textit{ccc}$ performs better than $\textit{TCP}_\textit{tot}$ and $\textit{TCP}_\textit{art}$.
However, CCCP has very similar (or slightly worse) performance to $\textit{TCP}_\textit{add}$ and $\textit{TCP}_\textit{search}$,
with $\hat{\textrm{A}}_{12}$ values of either 0.49 or 0.50.

\subsubsection{APFD Results: Java Programs (Test-Method Level)}

Based on Figure \ref{FIG:apfd-java-m} and Table \ref{TAB:apfd-java-m}, we have the following observations:

(1)
Our proposed method achieves much higher mean and median APFD values than $\textit{TCP}_\textit{tot}$ for all programs with all code coverage granularities, with the maximum differences reaching approximately 30\%.

(2)
CCCP has very similar performance to $\textit{TCP}_\textit{add}$, with their mean and median APFD differences at around 1\%.

(3)
Other than for some versions of $jtopas$, CCCP has much better performance than $\textit{TCP}_\textit{art}$.

(4)
Other than for a few cases (such as $jtopas\_v2$ with method coverage, and $xmlsec\_v1$ with  statement coverage),
CCCP usually has better performance than $\textit{TCP}_\textit{search}$.

(5)
Overall, the statistical analysis supports the above box plots observations.
Looking at all Java  programs together,
CCCP is,  on the whole, better than $\textit{TCP}_\textit{tot}$, $\textit{TCP}_\textit{art}$, and $\textit{TCP}_\textit{search}$:
The $p$-values are all less than 0.05, indicating that their differences are significant;
and the effect size $\hat{\textrm{A}}_{12}$ values range from 0.54 to 0.97, which means that $\textit{TCP}_\textit{ccc}$ is better than the other three RTCP techniques.
Finally, while the $p$-values for comparisons between $\textit{TCP}_\textit{ccc}$ and $\textit{TCP}_\textit{add}$ are less than 0.05 (which means that the differences are significant), the $\hat{\textrm{A}}_{12}$ values range from 0.49 to 0.51, indicating that they are very similar.

\subsubsection{APFD Results: C Subject Programs}

Based on Figure \ref{FIG:apfd-c} and Table \ref{TAB:apfd-c}, we have the following observations:

(1)
Our proposed CCCP approach has much better performance than $\textit{TCP}_\textit{tot}$ and $\textit{TCP}_\textit{add}$,
for all programs and code coverage granularities,
except for $gzip$ with method coverage (for which $\textit{TCP}_\textit{ccc}$ has very similar, or better performance).
The maximum difference in mean and median APFD values between $\textit{TCP}_\textit{ccc}$ and $\textit{TCP}_\textit{tot}$ is more than 40\%;
while between $\textit{TCP}_\textit{ccc}$ and $\textit{TCP}_\textit{add}$, it is about 10\%.

(2)
$\textit{TCP}_\textit{ccc}$ has similar or better APFD performance than $\textit{TCP}_\textit{art}$ and $\textit{TCP}_\textit{search}$ for some subject programs
(such as $flex$ and $gzip$, with all code coverage granularities),
but also has slightly worse performance for some others
(such as $grep$ with method coverage and $sed$ with statement coverage).
However, the difference in mean and median APFD values between $\textit{TCP}_\textit{ccc}$ and $\textit{TCP}_\textit{art}$ or $\textit{TCP}_\textit{search}$
is less than 5\%.

(3)
Overall, the statistical results support the box plot observations.
All $p$ values for the comparisons between $\textit{TCP}_\textit{ccc}$ and $\textit{TCP}_\textit{tot}$ or $\textit{TCP}_\textit{add}$ are less than 0.05,
indicating that their APFD scores are significantly different.
The $\hat{\textrm{A}}_{12}$ values are also much greater than 0.50,
ranging from 0.56 to 1.00 (except for the programs $gzip\_v3$, $gzip\_v4$, and $gzip\_v5$, with method coverage).
However, although all $p$ values for the comparisons between $\textit{TCP}_\textit{ccc}$ and $\textit{TCP}_\textit{art}$ or $\textit{TCP}_\textit{search}$
are also less than 0.05,
their  $\hat{\textrm{A}}_{12}$ values are much greater than 0.50 in some cases, but also  much less than 0.50 in others.
Nevertheless, considering all the C programs,
not only does $\textit{TCP}_\textit{ccc}$ have significantly different APFD values to the other four RTCP techniques,
but it also has better performances overall (except for $\textit{TCP}_\textit{art}$ with method coverage).

\begin{table*}[!t]
\scriptsize
\centering
 \caption{Statistical \textbf{effectiveness}  comparisons of  \textbf{APFD}$_\textbf{c}$  for \textbf{Java programs} at the \textbf{test-class level}.
For a comparison between two methods
$\textit{TCP}_\textit{ccc}$ and $M$,
where
$M \in \{\textit{TCP}_\textit{tot}, \textit{TCP}_\textit{add}, \textit{TCP}_\textit{art}, \textit{TCP}_\textit{search}\}$,
the symbol \ding{52} means that $\textit{TCP}_\textit{ccc}$ is better
($p$-value is less than 0.05, and the effect size $\hat{\textrm{A}}_{12}(\textit{TCP}_\textit{ccc},M)$ is greater than 0.50);
the symbol \ding{54} means that $M$ is better
(the $p$-value is less than 0.05, and $\hat{\textrm{A}}_{12}(\textit{TCP}_\textit{ccc},M)$ is less than 0.50);
and the symbol \ding{109} means that there is no statistically significant difference between them (the $p$-value is greater than 0.05).
}
  \label{TAB:apfdc-java-c}
    \setlength{\tabcolsep}{1.8mm}{
    \begin{tabular}{c|cccc|cccc|cccc}
     \hline
       \multirow{2}*{\textbf{Program Name}} &\multicolumn{4}{c|}{\textbf{Statement Coverage}} &\multicolumn{4}{c|}{\textbf{Branch Coverage}} &\multicolumn{4}{c}{\textbf{Method Coverage}} \\
        \cline{2-13}

         &$\textit{TCP}_\textit{tot}$ &$\textit{TCP}_\textit{add}$ &$\textit{TCP}_\textit{art}$ &$\textit{TCP}_\textit{search}$  & $\textit{TCP}_\textit{tot}$ &$\textit{TCP}_\textit{add}$ &$\textit{TCP}_\textit{art}$ &$\textit{TCP}_\textit{search}$ &$\textit{TCP}_\textit{tot}$ & $\textit{TCP}_\textit{add}$ &$\textit{TCP}_\textit{art}$ &$\textit{TCP}_\textit{search}$\\
            \hline
$ant\_v1$	&\ding{52}(1.00)	&\ding{52}(1.00)	&\ding{52}(1.00)	&\ding{52}(0.97)	&\ding{52}(1.00)	&\ding{52}(0.59)	&\ding{52}(1.00)	&\ding{52}(0.58)	&\ding{52}(1.00)	&\ding{52}(0.59)	&\ding{52}(1.00)	&\ding{52}(0.56)	\\
$ant\_v2$	&\ding{52}(1.00)	&\ding{52}(0.75)	&\ding{52}(0.59)	&\ding{52}(0.70)	&\ding{52}(1.00)	&\ding{52}(1.00)	&\ding{54}(0.21)	&\ding{52}(0.96)	&\ding{52}(1.00)	&\ding{109}(0.50)	&\ding{54}(0.32)	&\ding{52}(0.53)	\\
$ant\_v3$	&\ding{52}(1.00)	&\ding{52}(0.65)	&\ding{52}(0.98)	&\ding{52}(0.66)	&\ding{52}(1.00)	&\ding{52}(1.00)	&\ding{54}(0.45)	&\ding{52}(0.85)	&\ding{52}(1.00)	&\ding{109}(0.50)	&\ding{54}(0.44)	&\ding{52}(0.53)	\\\hline
$ant$	&\ding{52}(1.00)	&\ding{52}(0.60)	&\ding{52}(0.85)	&\ding{52}(0.59)	&\ding{52}(0.83)	&\ding{52}(0.62)	&\ding{52}(0.61)	&\ding{52}(0.60)	&\ding{52}(0.98)	&\ding{109}(0.51)	&\ding{52}(0.59)	&\ding{52}(0.52)	\\\hline
$jmeter\_v1$	&\ding{54}(0.00)	&\ding{54}(0.00)	&\ding{52}(1.00)	&\ding{54}(0.00)	&\ding{54}(0.00)	&\ding{52}(0.56)	&\ding{52}(0.96)	&\ding{109}(0.52)	&\ding{54}(0.00)	&\ding{52}(0.67)	&\ding{52}(1.00)	&\ding{52}(0.64)	\\
$jmeter\_v2$	&\ding{54}(0.00)	&\ding{54}(0.14)	&\ding{52}(0.99)	&\ding{54}(0.12)	&\ding{54}(0.00)	&\ding{109}(0.49)	&\ding{52}(0.86)	&\ding{54}(0.47)	&\ding{54}(0.00)	&\ding{52}(1.00)	&\ding{52}(1.00)	&\ding{52}(1.00)	\\
$jmeter\_v3$	&\ding{54}(0.00)	&\ding{52}(0.66)	&\ding{52}(0.69)	&\ding{52}(0.67)	&\ding{54}(0.00)	&\ding{52}(0.54)	&\ding{52}(0.86)	&\ding{109}(0.48)	&\ding{54}(0.00)	&\ding{54}(0.47)	&\ding{52}(0.63)	&\ding{54}(0.46)	\\
$jmeter\_v4$	&\ding{109}(0.49)	&\ding{52}(0.73)	&\ding{52}(0.78)	&\ding{52}(0.75)	&\ding{54}(0.00)	&\ding{52}(0.53)	&\ding{52}(0.81)	&\ding{54}(0.46)	&\ding{54}(0.00)	&\ding{52}(0.70)	&\ding{52}(0.58)	&\ding{52}(0.72)	\\
$jmeter\_v5$	&\ding{52}(1.00)	&\ding{52}(0.61)	&\ding{52}(0.92)	&\ding{52}(0.64)	&\ding{52}(0.92)	&\ding{52}(0.62)	&\ding{52}(0.92)	&\ding{52}(0.55)	&\ding{52}(1.00)	&\ding{52}(0.57)	&\ding{52}(0.98)	&\ding{52}(0.62)	\\\hline
$jmeter$	&\ding{54}(0.33)	&\ding{109}(0.49)	&\ding{52}(0.90)	&\ding{109}(0.50)	&\ding{54}(0.18)	&\ding{109}(0.51)	&\ding{52}(0.86)	&\ding{109}(0.50)	&\ding{54}(0.36)	&\ding{52}(0.54)	&\ding{52}(0.84)	&\ding{52}(0.54)	\\\hline
$jtopas\_v1$	&\ding{52}(1.00)	&\ding{52}(1.00)	&\ding{52}(0.69)	&\ding{52}(1.00)	&\ding{52}(1.00)	&\ding{52}(1.00)	&\ding{52}(0.62)	&\ding{52}(1.00)	&\ding{52}(1.00)	&\ding{52}(0.74)	&\ding{52}(1.00)	&\ding{52}(0.74)	\\
$jtopas\_v2$	&\ding{52}(1.00)	&\ding{52}(1.00)	&\ding{109}(0.49)	&\ding{52}(1.00)	&\ding{52}(1.00)	&\ding{52}(1.00)	&\ding{54}(0.44)	&\ding{52}(1.00)	&\ding{52}(1.00)	&\ding{52}(0.76)	&\ding{52}(0.88)	&\ding{52}(0.74)	\\
$jtopas\_v3$	&\ding{52}(1.00)	&\ding{109}(0.50)	&\ding{54}(0.12)	&\ding{109}(0.50)	&\ding{52}(1.00)	&\ding{109}(0.50)	&\ding{54}(0.10)	&\ding{109}(0.50)	&\ding{52}(1.00)	&\ding{109}(0.50)	&\ding{54}(0.40)	&\ding{109}(0.49)	\\\hline
$jtopas$	&\ding{52}(0.78)	&\ding{52}(0.61)	&\ding{109}(0.50)	&\ding{52}(0.61)	&\ding{52}(0.78)	&\ding{52}(0.61)	&\ding{54}(0.47)	&\ding{52}(0.61)	&\ding{52}(0.87)	&\ding{52}(0.56)	&\ding{52}(0.73)	&\ding{52}(0.55)	\\\hline
$xmlsec\_v1$	&\ding{54}(0.00)	&\ding{109}(0.50)	&\ding{52}(0.87)	&\ding{109}(0.50)	&\ding{54}(0.00)	&\ding{54}(0.36)	&\ding{52}(0.72)	&\ding{109}(0.52)	&\ding{54}(0.00)	&\ding{54}(0.00)	&\ding{52}(0.76)	&\ding{54}(0.00)	\\
$xmlsec\_v2$	&\ding{54}(0.00)	&\ding{109}(0.50)	&\ding{54}(0.22)	&\ding{109}(0.50)	&\ding{52}(1.00)	&\ding{52}(0.75)	&\ding{52}(0.64)	&\ding{52}(0.81)	&\ding{54}(0.00)	&\ding{52}(0.66)	&\ding{54}(0.16)	&\ding{52}(0.76)	\\
$xmlsec\_v3$	&\ding{54}(0.00)	&\ding{54}(0.00)	&\ding{54}(0.27)	&\ding{54}(0.00)	&\ding{54}(0.00)	&\ding{54}(0.27)	&\ding{54}(0.28)	&\ding{54}(0.27)	&\ding{54}(0.00)	&\ding{54}(0.20)	&\ding{54}(0.19)	&\ding{54}(0.20)	\\\hline
$xmlsec$	&\ding{54}(0.22)	&\ding{54}(0.44)	&\ding{54}(0.48)	&\ding{54}(0.44)	&\ding{54}(0.28)	&\ding{109}(0.49)	&\ding{52}(0.54)	&\ding{109}(0.51)	&\ding{54}(0.00)	&\ding{54}(0.41)	&\ding{54}(0.35)	&\ding{54}(0.44)	\\\hline
\textit{\textbf{All Java Programs}}	&\ding{52}(0.58)	&\ding{52}(0.52)	&\ding{52}(0.68)	&\ding{52}(0.53)	&\ding{52}(0.53)	&\ding{52}(0.52)	&\ding{52}(0.65)	&\ding{52}(0.52)	&\ding{52}(0.58)	&\ding{109}(0.50)	&\ding{52}(0.67)	&\ding{52}(0.51)	\\
\hline
  \end{tabular}}
\end{table*}

Table \ref{TAB:APFDsummary} summarizes the statistical results, presenting the total number of times $\textit{TCP}_\textit{ccc}$
is better (\ding{52}), worse (\ding{54}), or not statistically different (\ding{109}), compared to the other RTCP techniques.
Based on this table, we can answer RQ1 as follows:
\begin{enumerate}
    \item
        When prioritizing Java test suites at the test-class level,
        $\textit{TCP}_\textit{ccc}$ performs much better than $\textit{TCP}_\textit{art}$;
        similarly to $\textit{TCP}_\textit{tot}$ and $\textit{TCP}_\textit{seaerch} $; and
        slightly worse than $\textit{TCP}_\textit{add}$.

    \item
        When prioritizing Java test suites at the test-method level,
        $\textit{TCP}_\textit{ccc}$ performs much better than $\textit{TCP}_\textit{tot}$, $\textit{TCP}_\textit{art}$, and $\textit{TCP}_\textit{search}$; and
        similarly to $\textit{TCP}_\textit{add}$.

    \item
        When prioritizing C test suites,
        $\textit{TCP}_\textit{ccc}$ performs much better than $\textit{TCP}_\textit{tot}$, $\textit{TCP}_\textit{add}$, and $\textit{TCP}_\textit{search}$; and
        slightly better than $\textit{TCP}_\textit{art}$.

\end{enumerate}
In other words, as will be discussed in detail later (Sections \ref{SEC:impactCCG} and \ref{SEC:impactTCG}), code coverage granularity and test case granularity may  impact on the effectiveness of CCCP, in terms of APFD.
Nevertheless,
the ratios of $\textit{TCP}_\textit{ccc}$ performing better (\ding{52}) rather than worse (\ding{54}) than
$\textit{TCP}_\textit{tot}$, $\textit{TCP}_\textit{add}$, $\textit{TCP}_\textit{art}$, and $\textit{TCP}_\textit{search}$, are:
6.89 (131/19), 4.26 (81/19), 3.05 (113/37), and 1.93 (81/42), respectively.
In conclusion, overall, the proposed CCCP approach is more effective than the other four RTCP techniques, in terms of testing effectiveness, as measured by APFD.

\subsection{\textbf{RQ2: Effectiveness of CCCP Measured by APFD$_\textbf{c}$}}

\begin{figure*}[!t]
\graphicspath{{graphs/}}
\centering
    \subfigure[$ant$-statement coverage]
    {
        \includegraphics[width=0.319\textwidth]{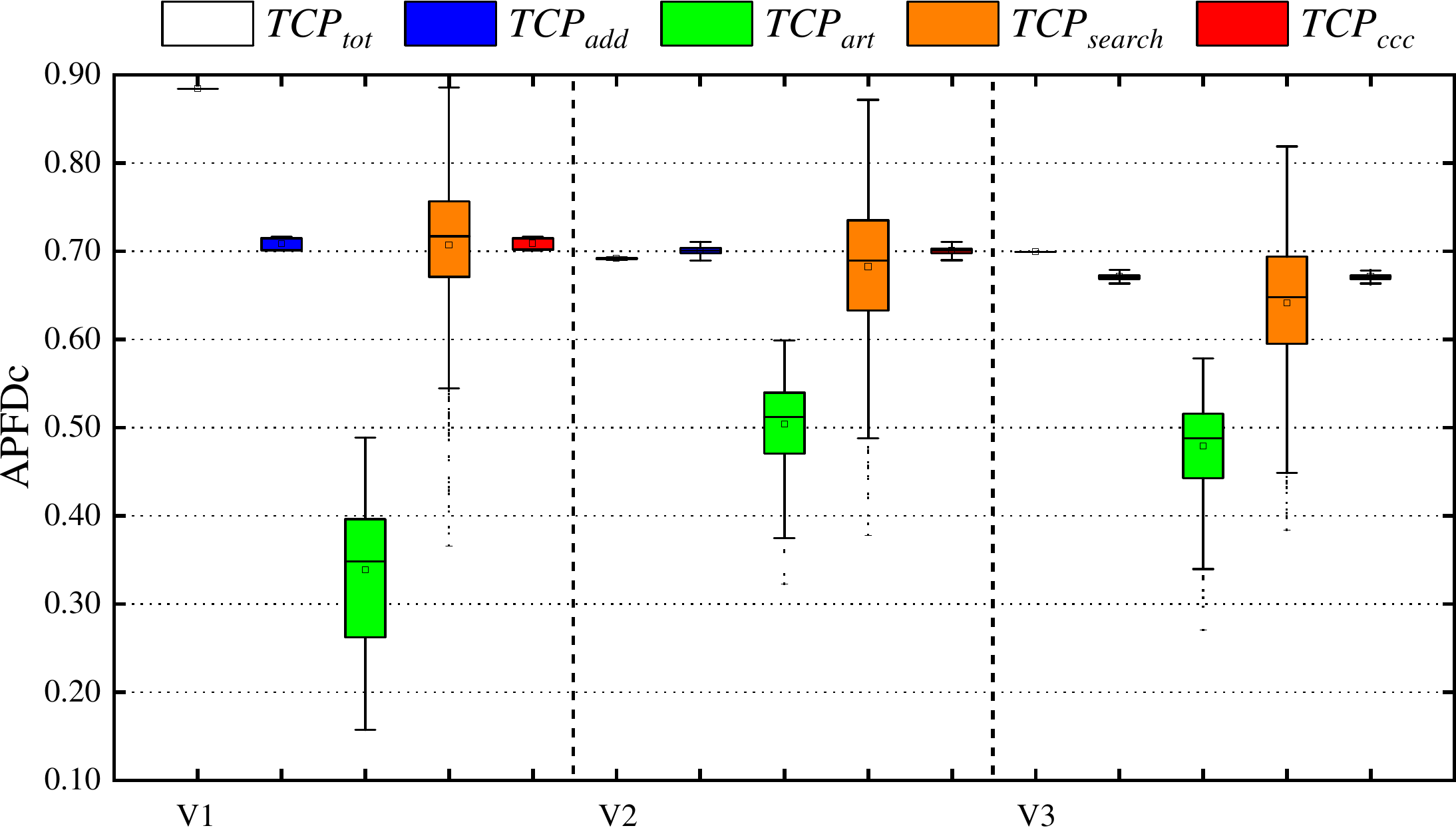}
        \label{apfdc_ant_sc}
    }
    \subfigure[$ant$-branch coverage]
    {
        \includegraphics[width=0.319\textwidth]{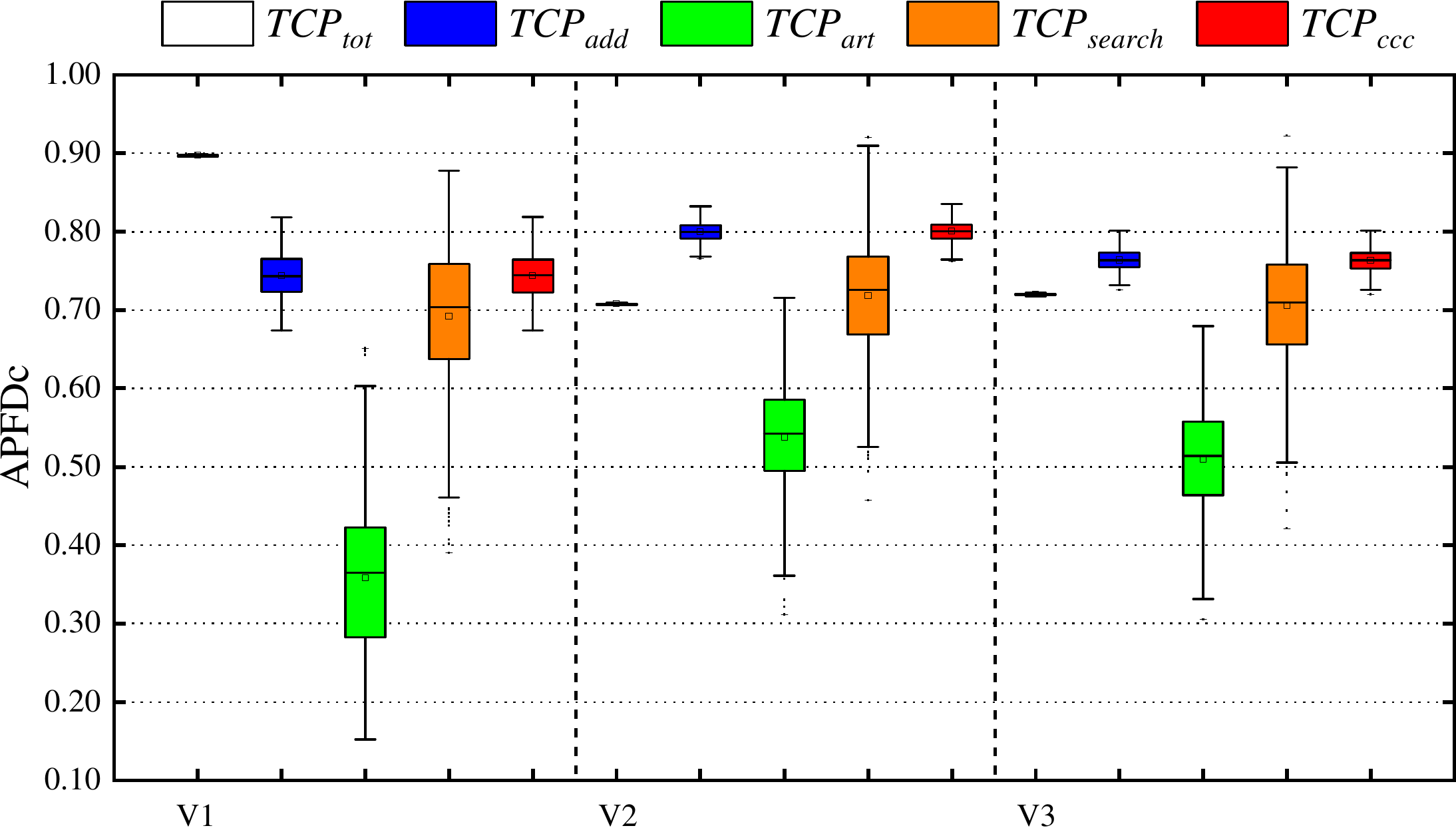}
        \label{apfdc_ant_bc}
    }
        \subfigure[$ant$-method coverage]
    {
        \includegraphics[width=0.319\textwidth]{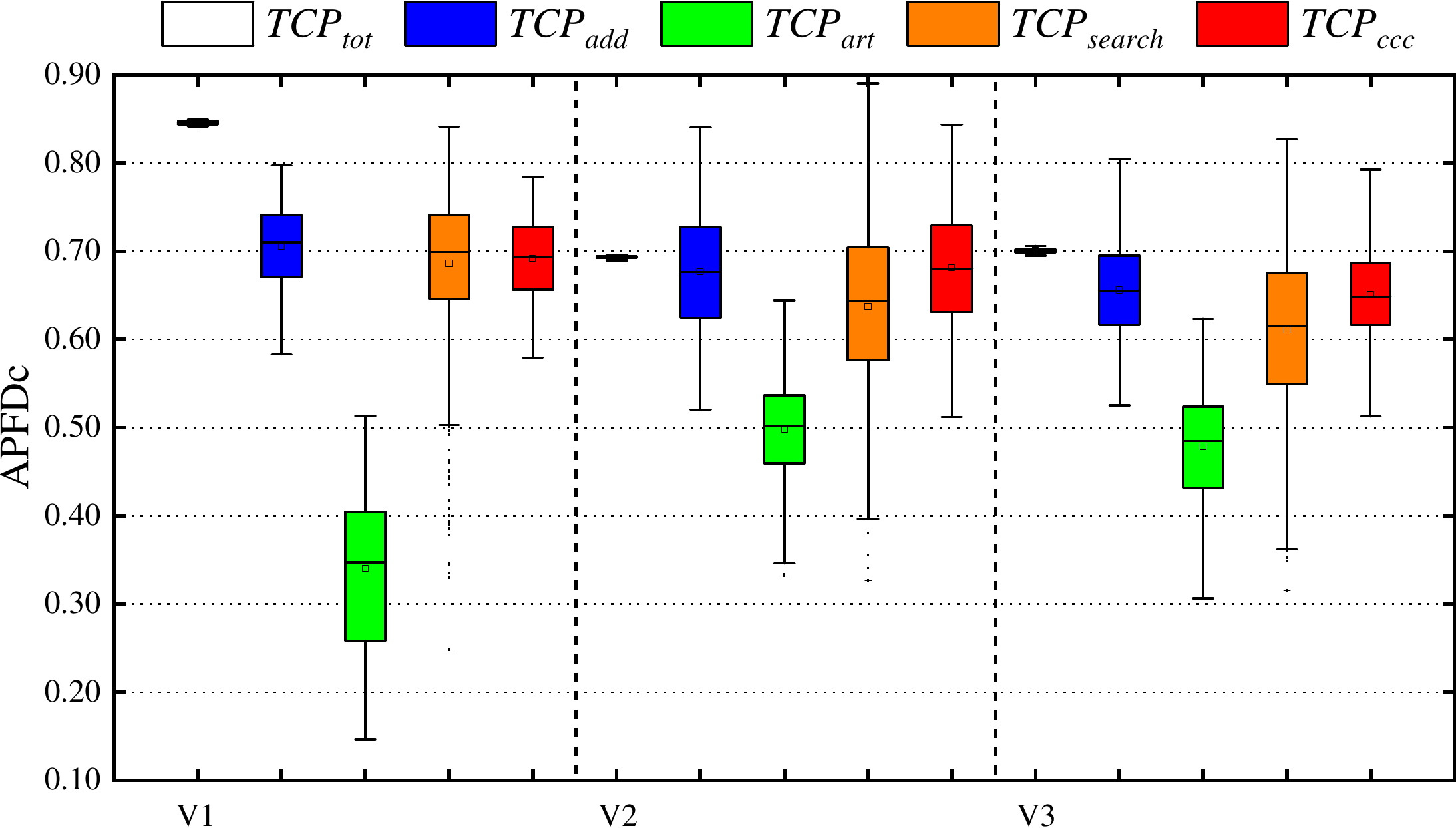}
        \label{apfdc_ant_mc}
    }
    \subfigure[$jmeter$-statement coverage]
    {
        \includegraphics[width=0.319\textwidth]{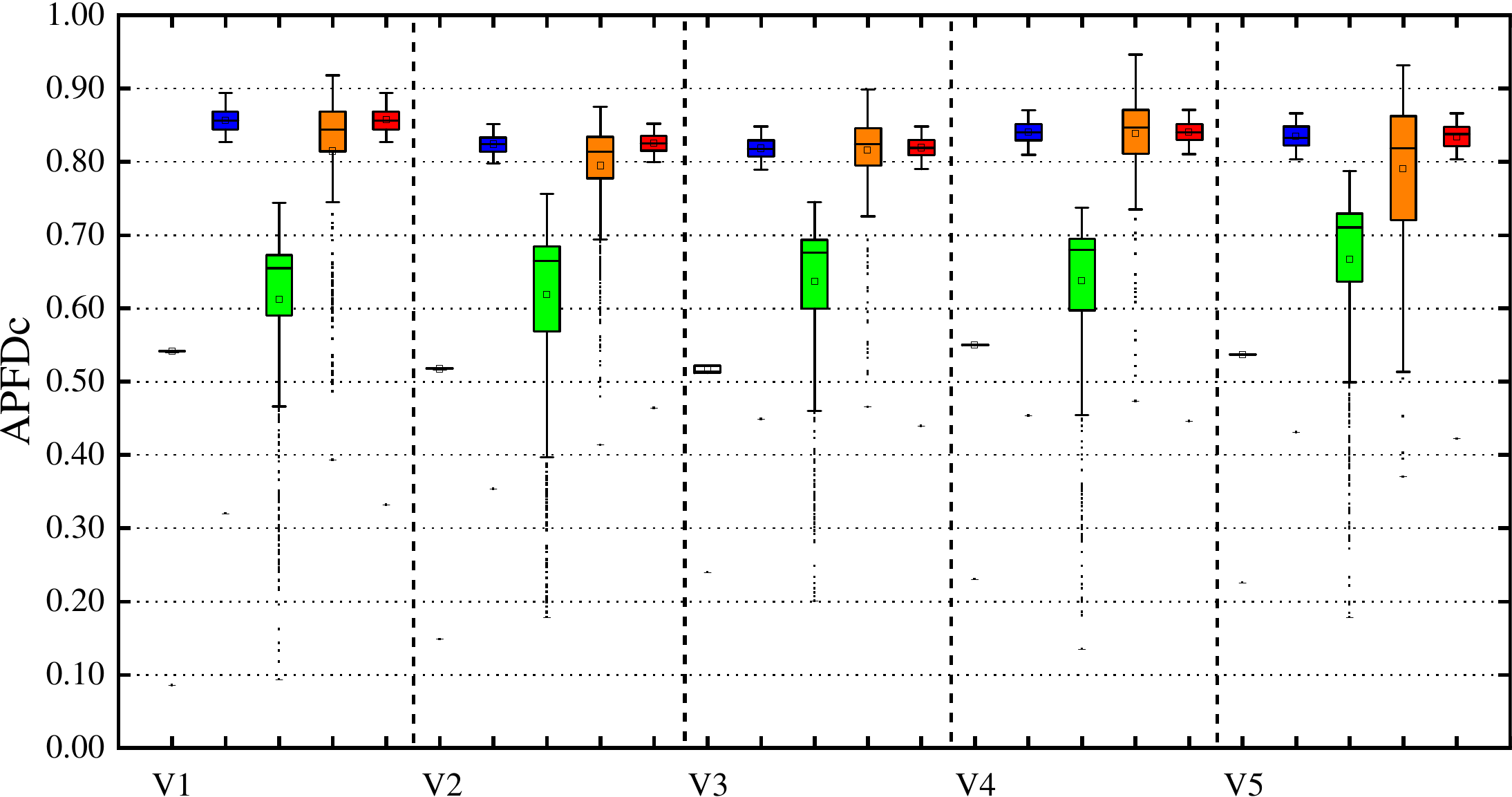}
        \label{apfdc_grep_sc}
    }
        \subfigure[$jmeter$-branch coverage]
    {
        \includegraphics[width=0.319\textwidth]{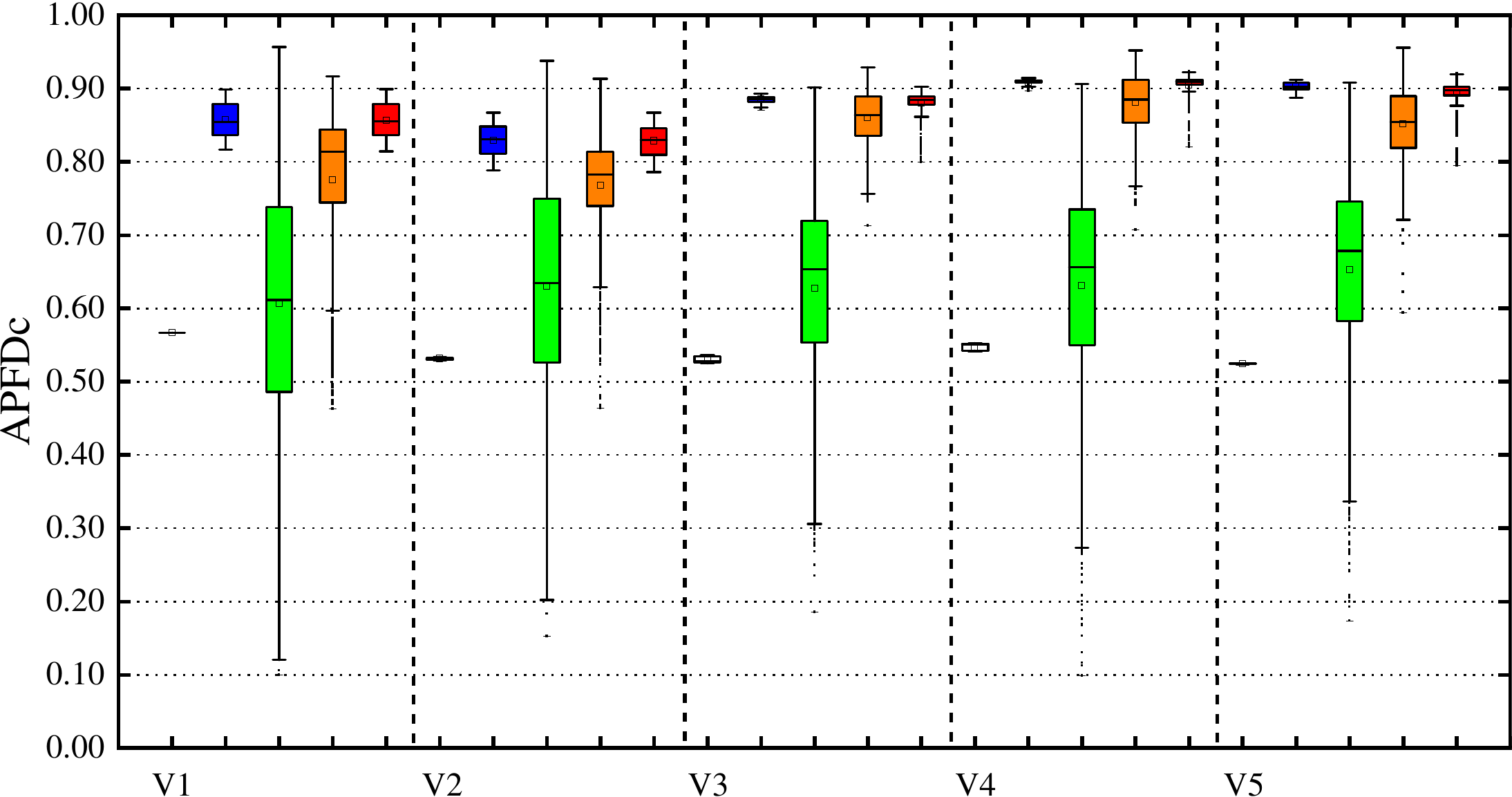}
        \label{apfdc_jmeter_bc}
    }
        \subfigure[$jmeter$-method coverage]
    {
        \includegraphics[width=0.319\textwidth]{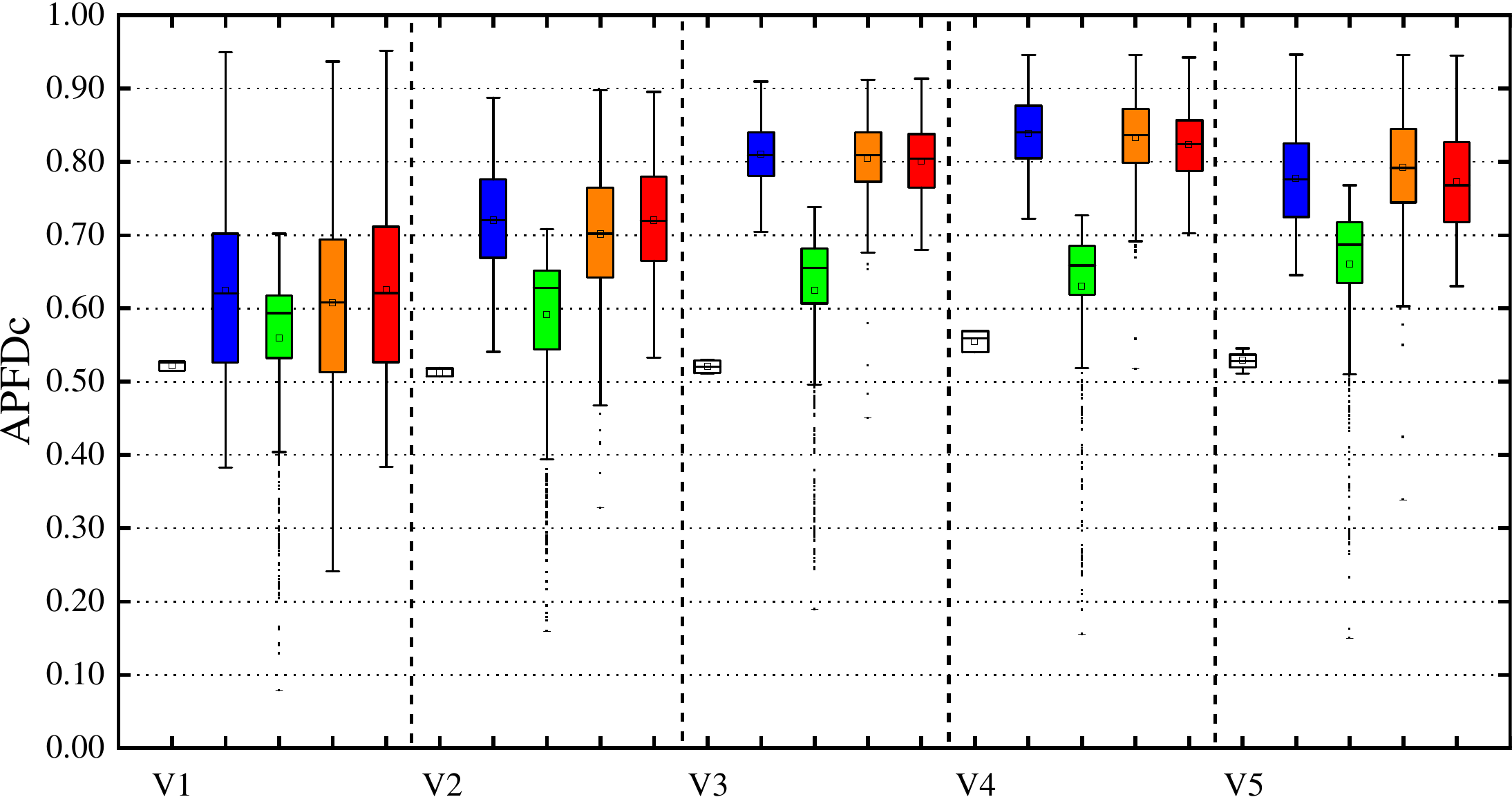}
        \label{apfdc_jmeter_mc}
    }
        \subfigure[$jtopas$-statement coverage]
    {
        \includegraphics[width=0.319\textwidth]{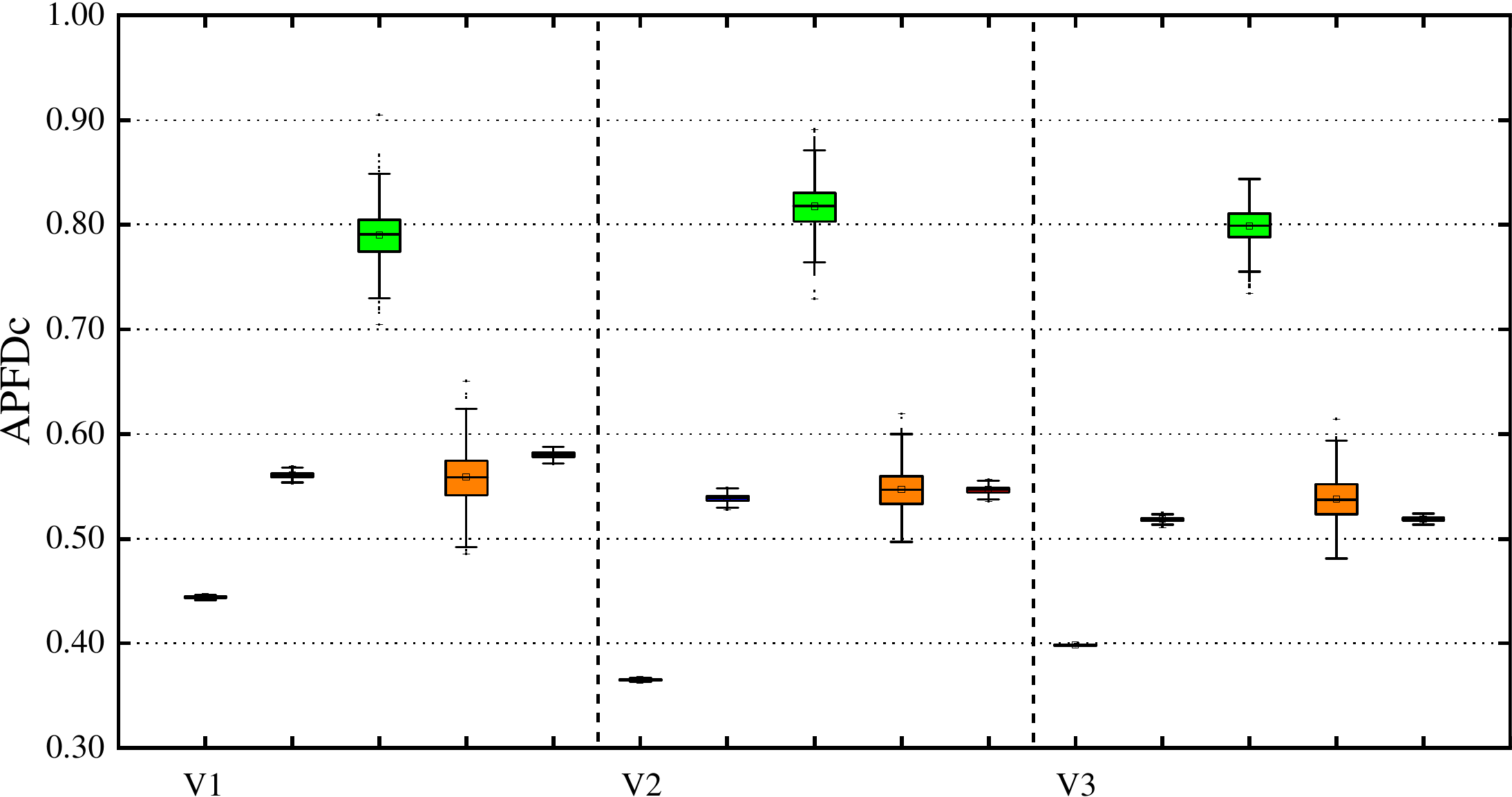}
        \label{apfdc_jtopas_sc}
    }
        \subfigure[$jtopas$-branch coverage]
    {
        \includegraphics[width=0.319\textwidth]{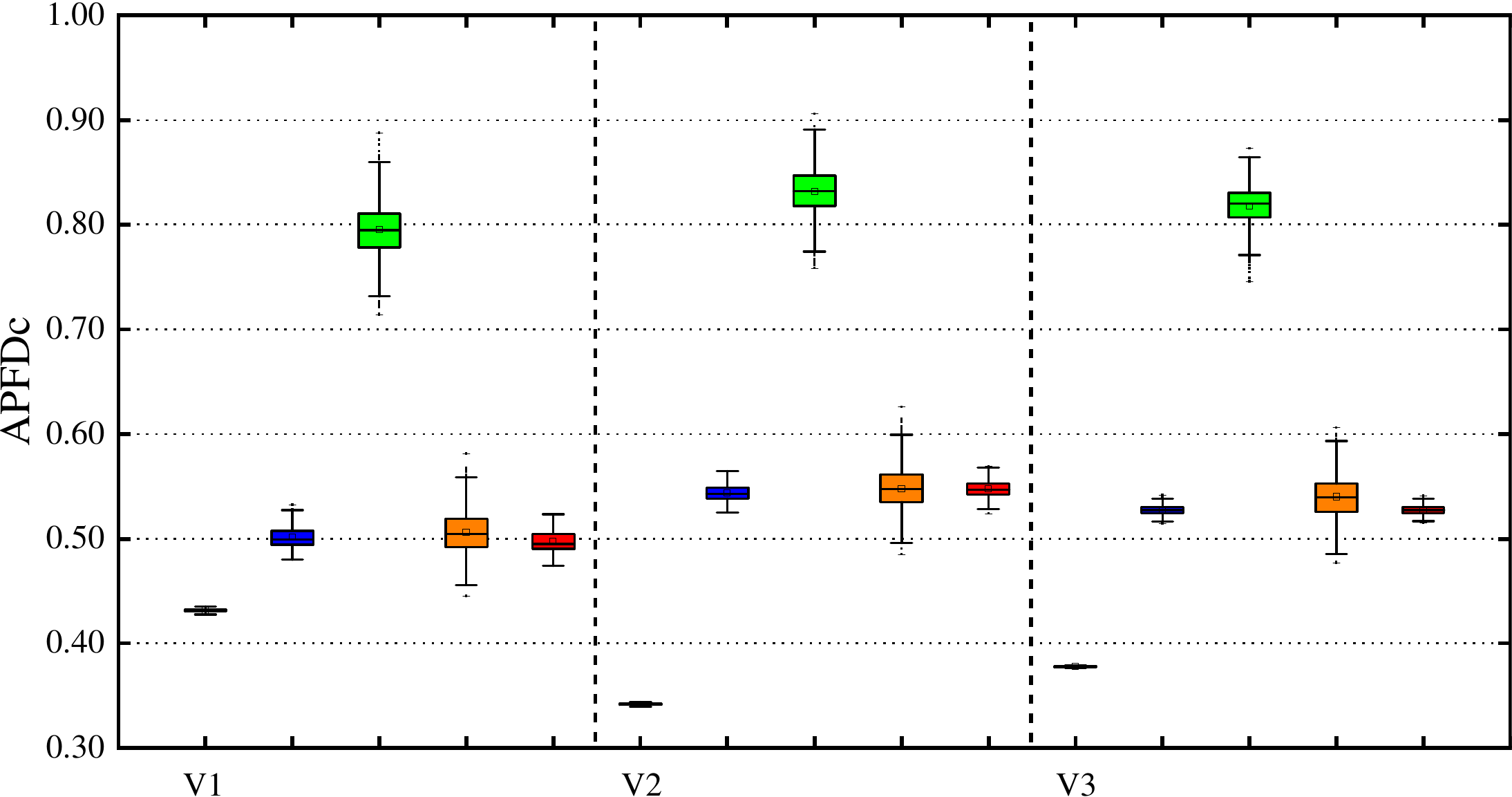}
        \label{apfdc_jtopas_bc}
    }
        \subfigure[$jtopas$-method coverage]
    {
        \includegraphics[width=0.319\textwidth]{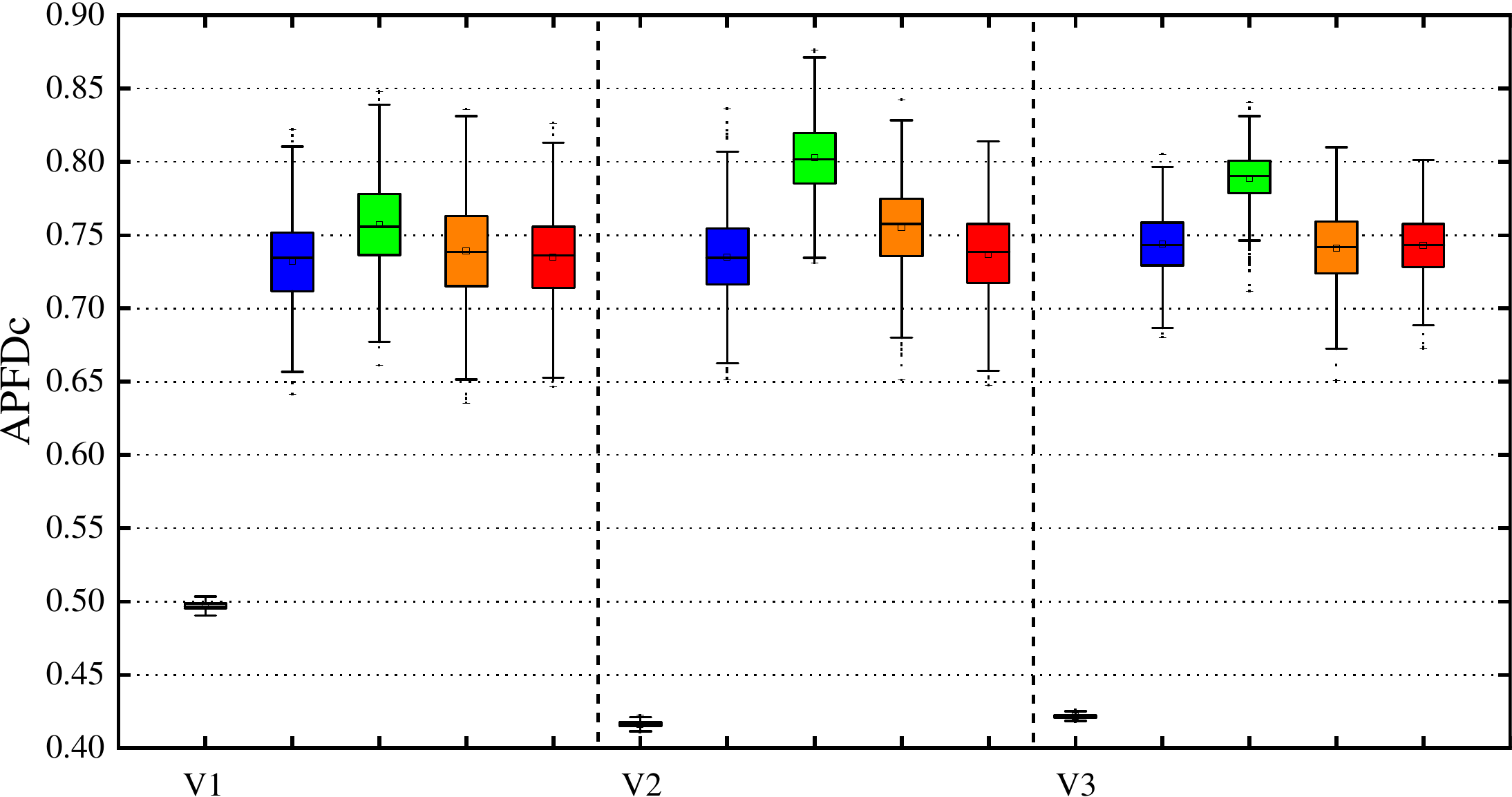}
        \label{apfdc_jtopas_mc}
    }
        \subfigure[$xmlsec$-statement coverage]
    {
        \includegraphics[width=0.319\textwidth]{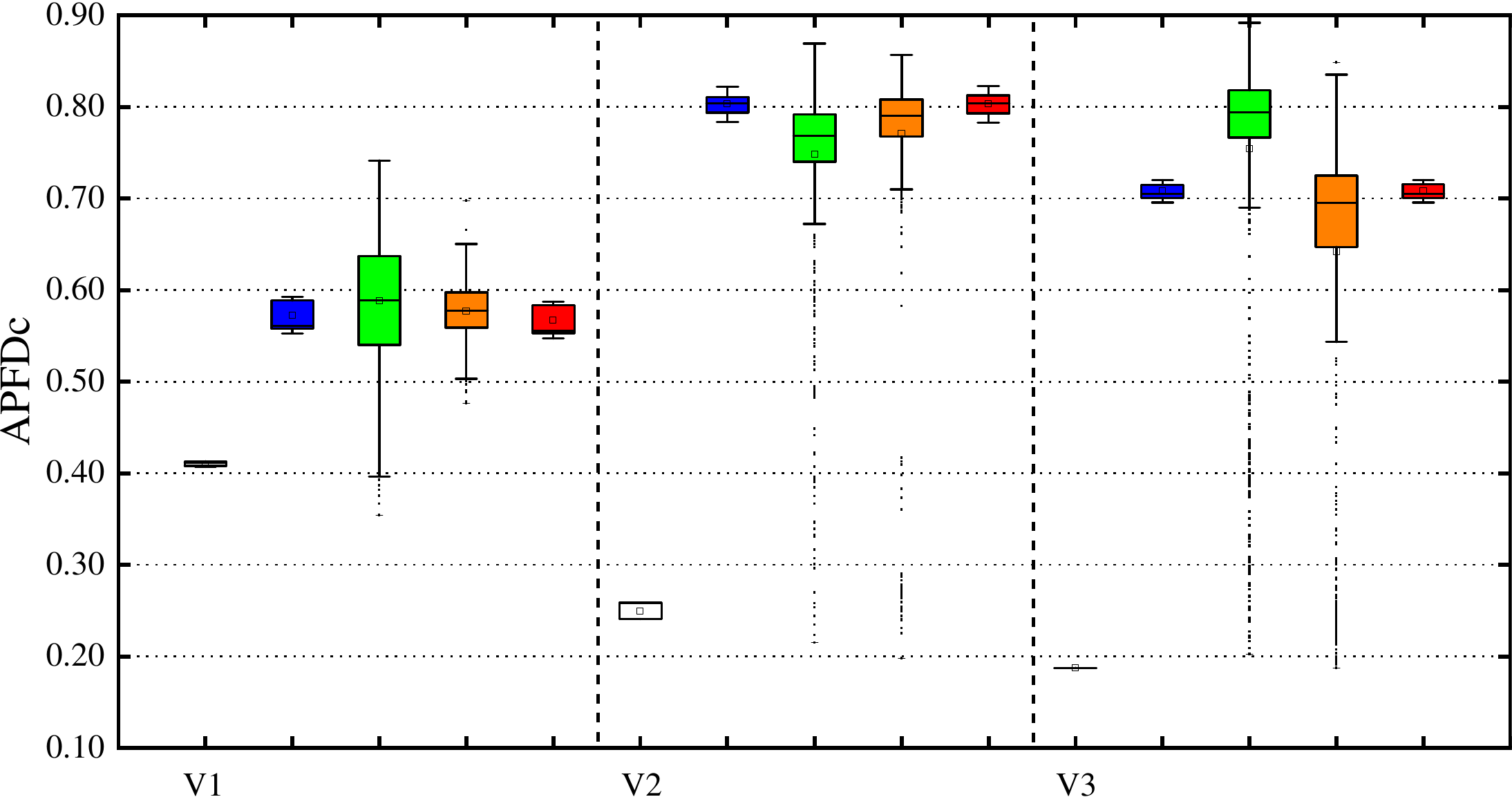}
        \label{apfdc_xmlsec_sc}
    }
        \subfigure[$xmlsec$-branch coverage]
    {
        \includegraphics[width=0.319\textwidth]{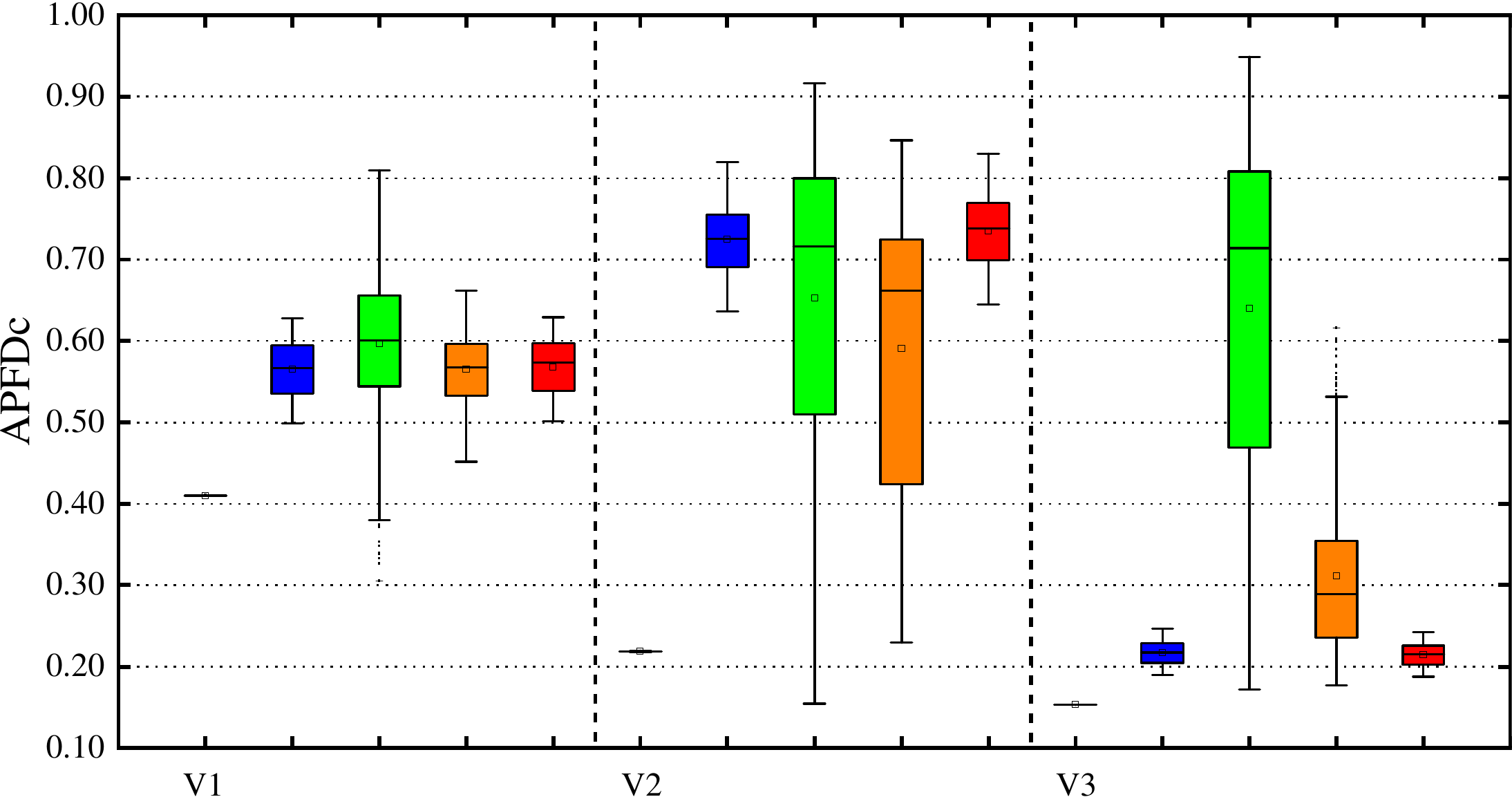}
        \label{apfdc_xmlsec_bc}
    }
        \subfigure[$xmlsec$-method coverage]
    {
        \includegraphics[width=0.319\textwidth]{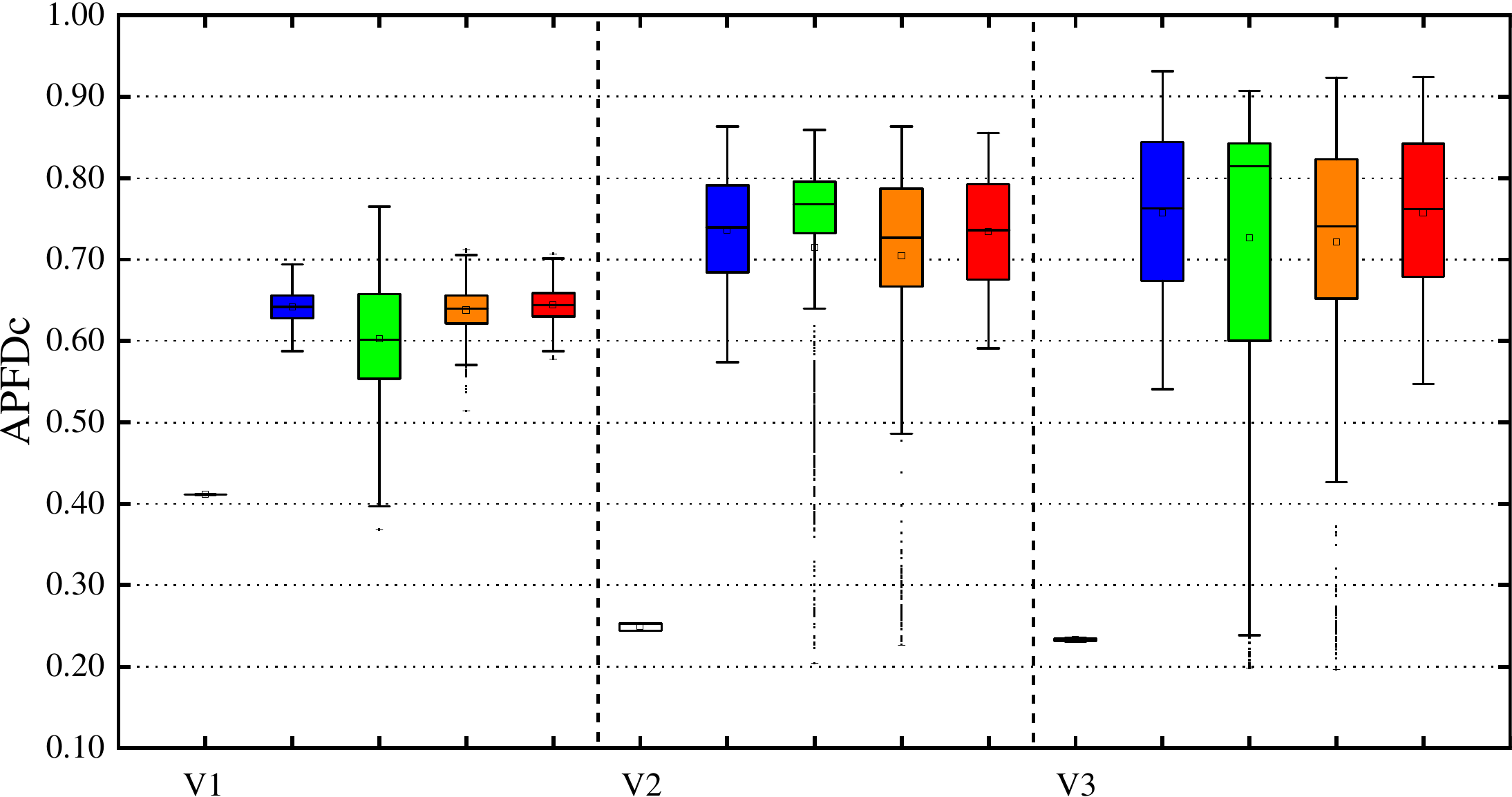}
        \label{apfdc_xmlsec_mc}
    }
    \caption{\textbf{Effectiveness: }\textbf{APFD}$_\textbf{c}$  results for \textbf{Java programs} at the \textbf{test-method level}}
    \label{FIG:apfdc-java-m}
\end{figure*}

\begin{figure*}[!t]
\graphicspath{{graphs/}}
\centering
        \subfigure[$flex$-statement coverage]
    {
        \includegraphics[width=0.319\textwidth]{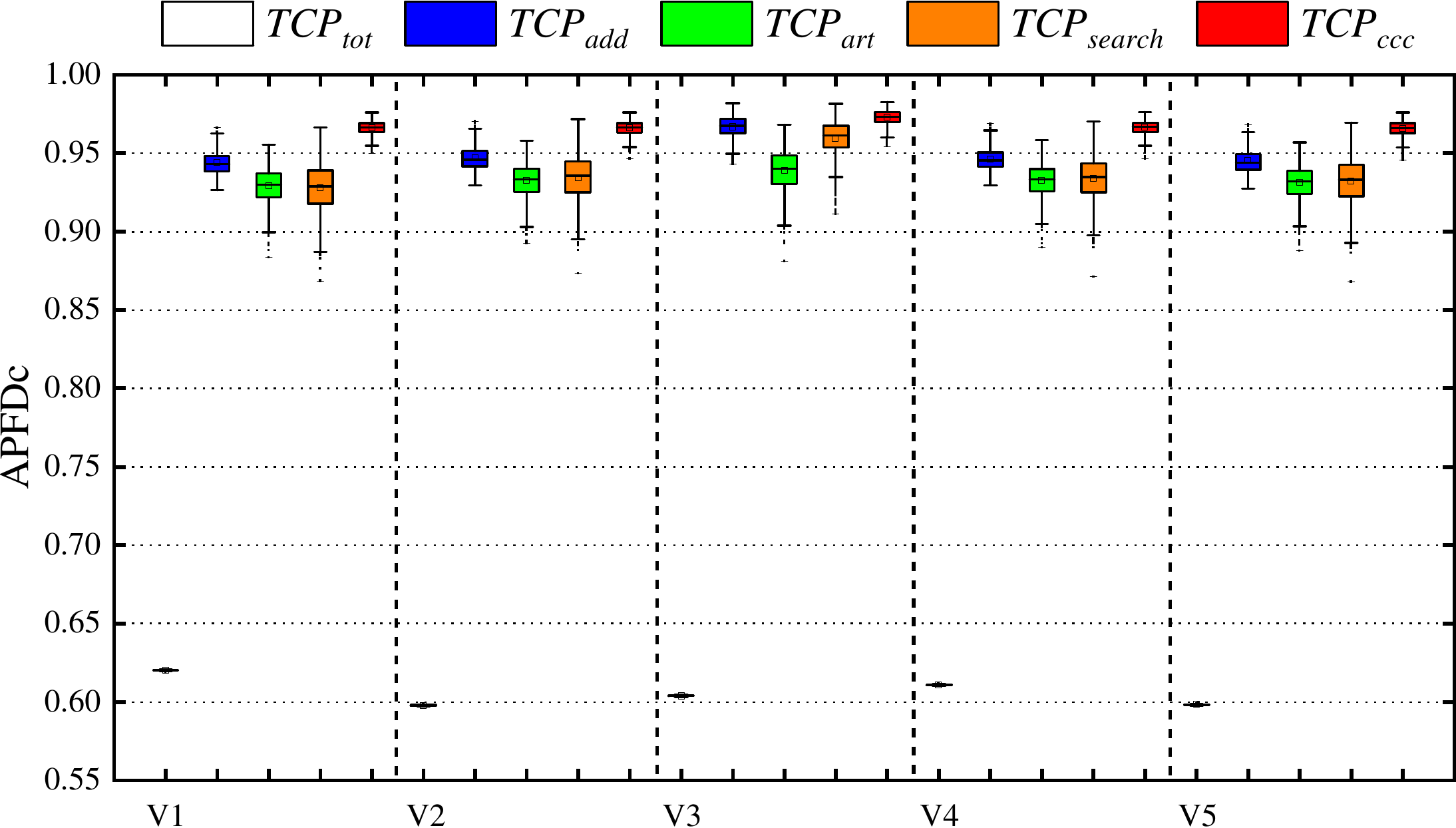}
        \label{apfdc_flex_sc}
    }
        \subfigure[$flex$-branch coverage]
    {
        \includegraphics[width=0.319\textwidth]{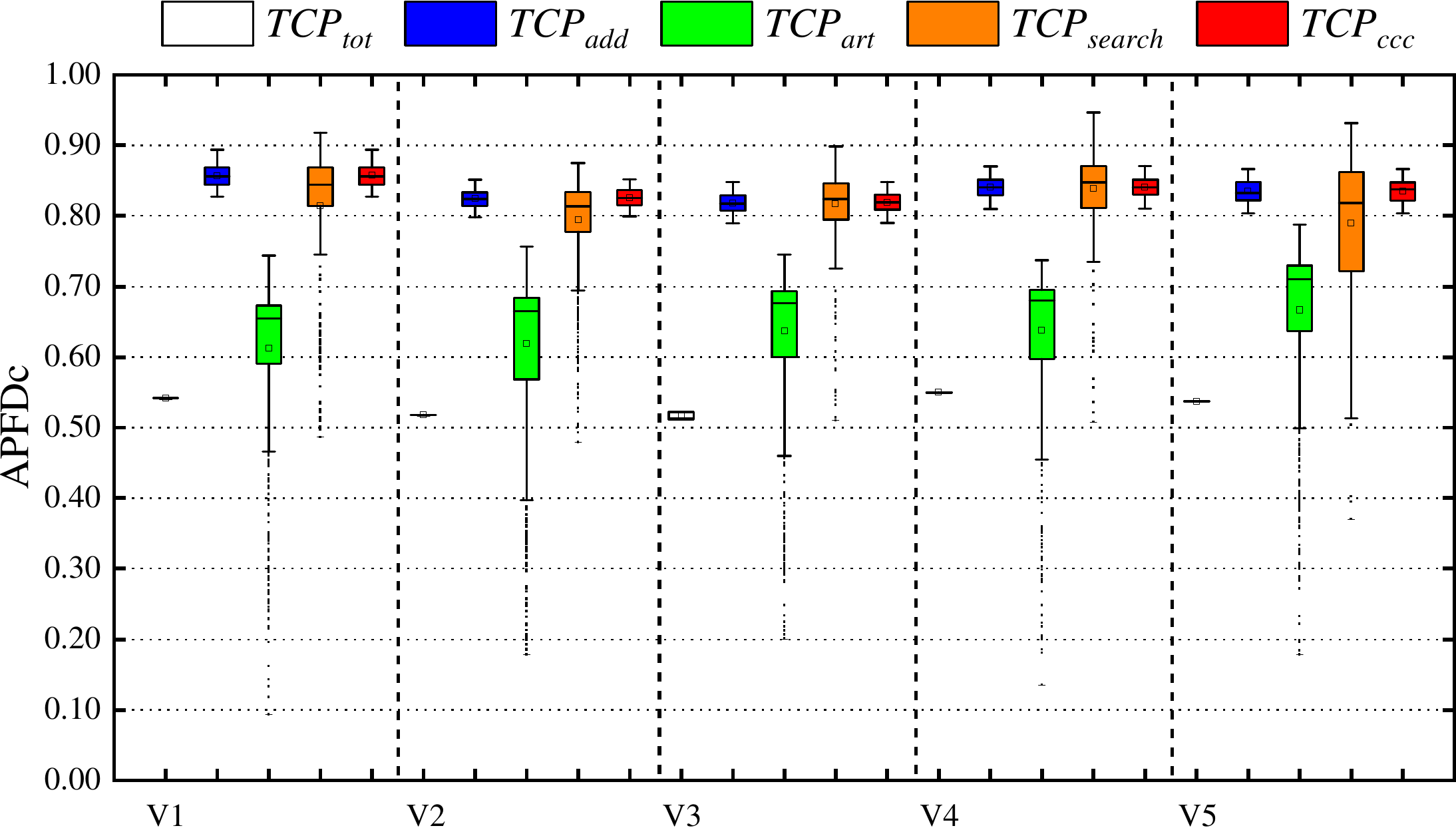}
        \label{apfdc_flex_bc}
    }
        \subfigure[$flex$-method coverage]
    {
        \includegraphics[width=0.319\textwidth]{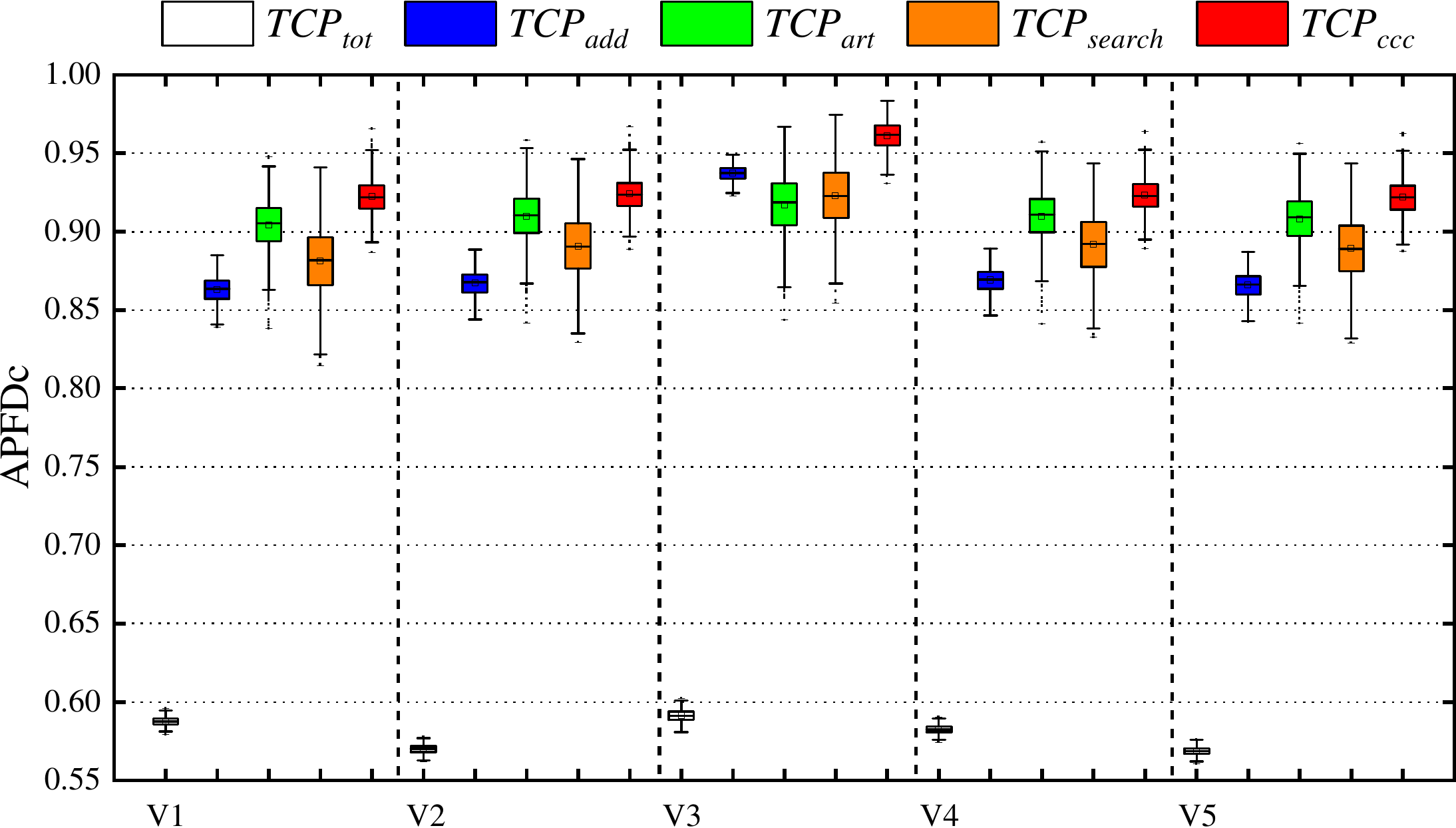}
        \label{apfdc_flex_mc}
    }
        \subfigure[$grep$-statement coverage]
    {
        \includegraphics[width=0.319\textwidth]{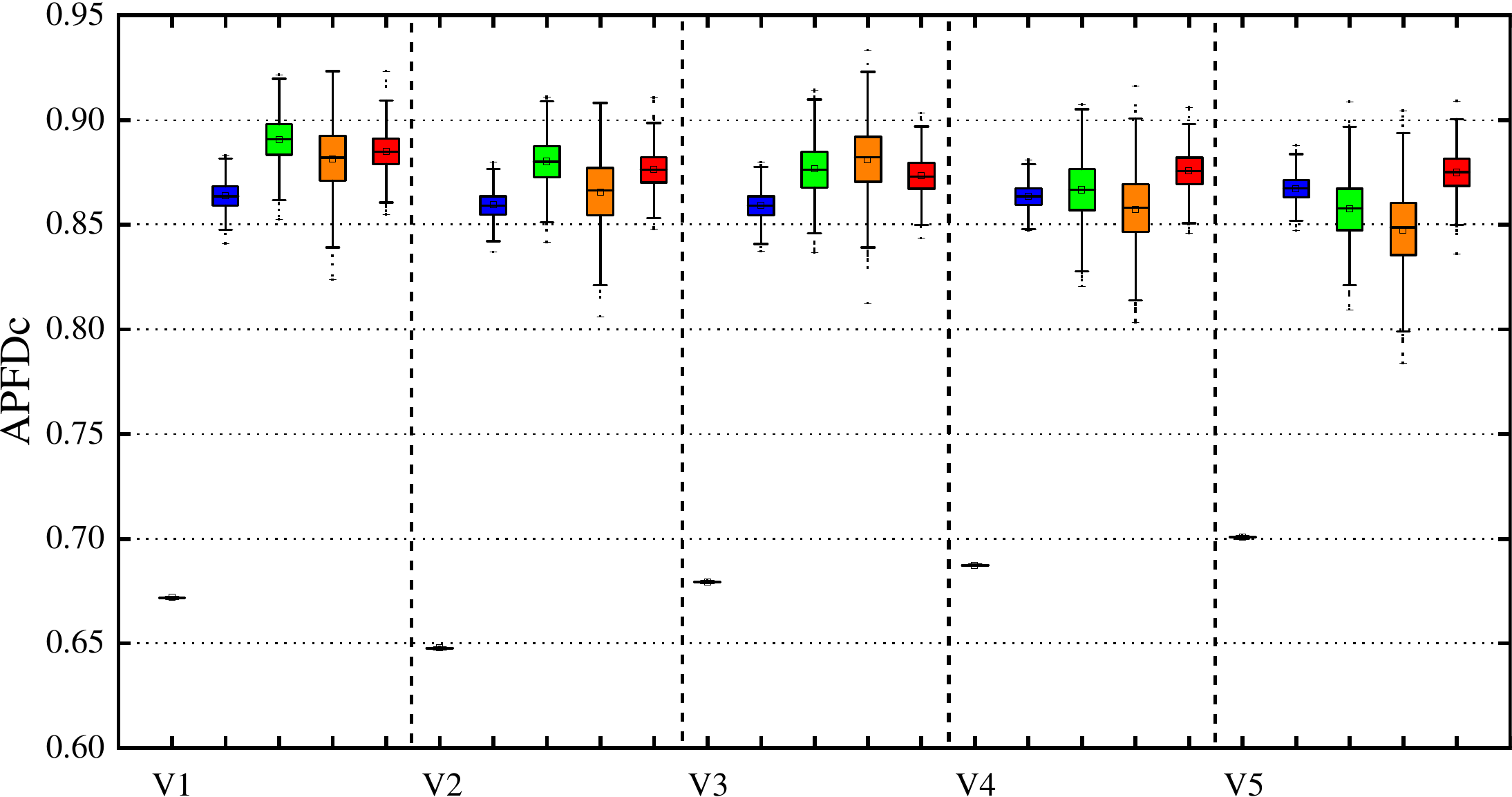}
        \label{apfdc_grep_sc}
    }
        \subfigure[$grep$-branch coverage]
    {
        \includegraphics[width=0.319\textwidth]{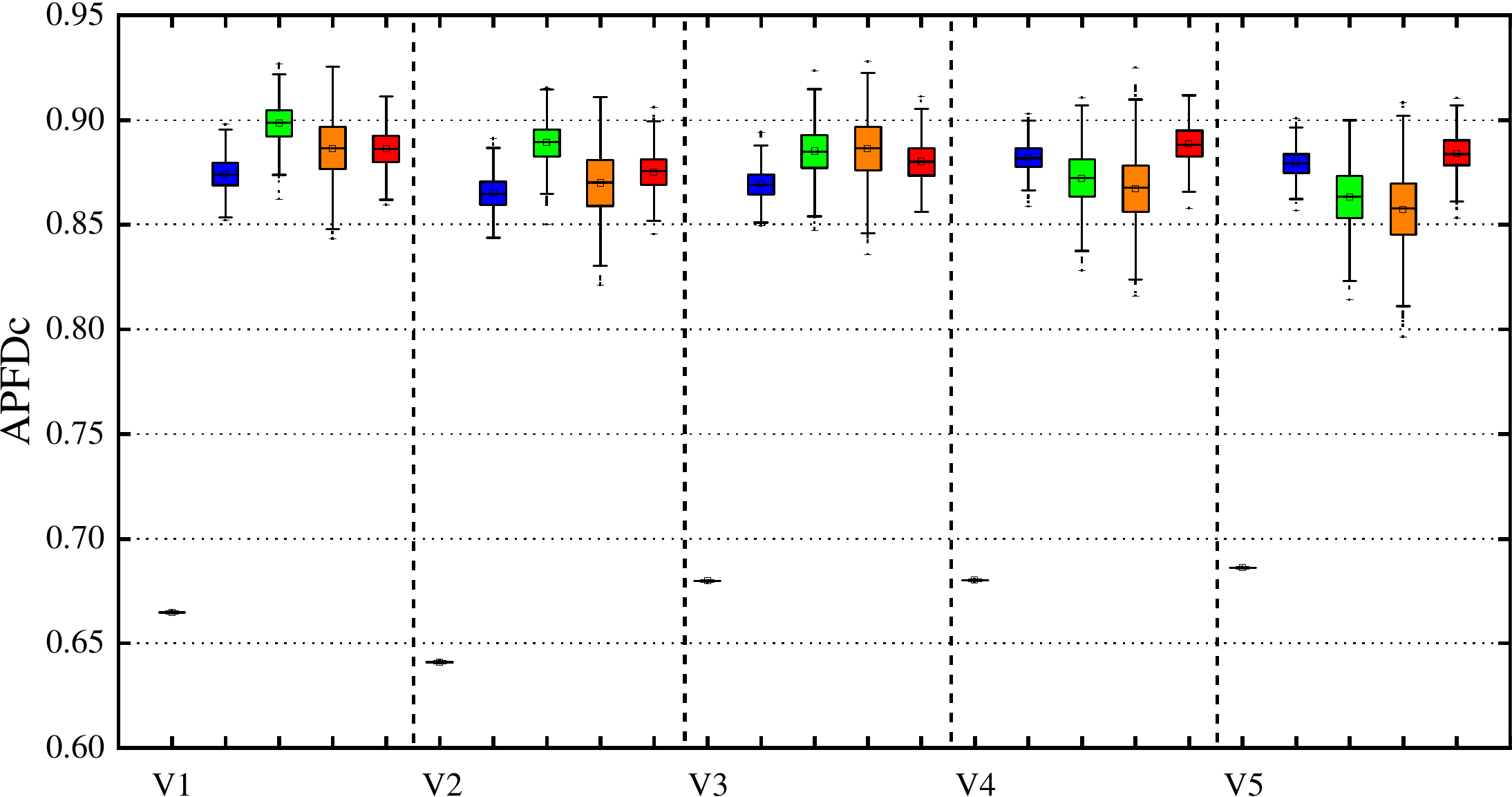}
        \label{apfdc_grep_bc}
    }
        \subfigure[$grep$-method coverage]
    {
        \includegraphics[width=0.319\textwidth]{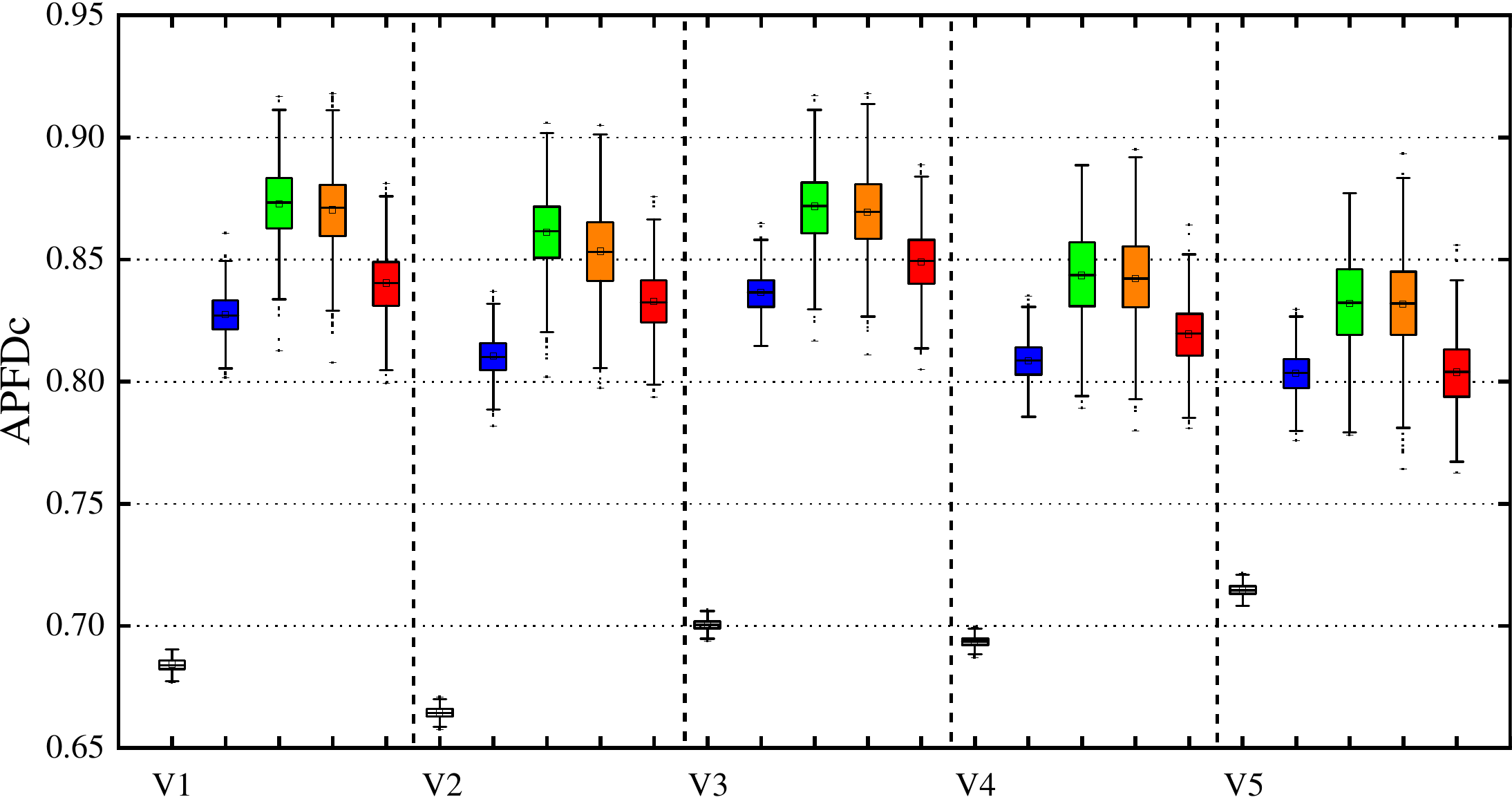}
        \label{apfdc_grep_mc}
    }
        \subfigure[$gzip$-statement coverage]
    {
        \includegraphics[width=0.319\textwidth]{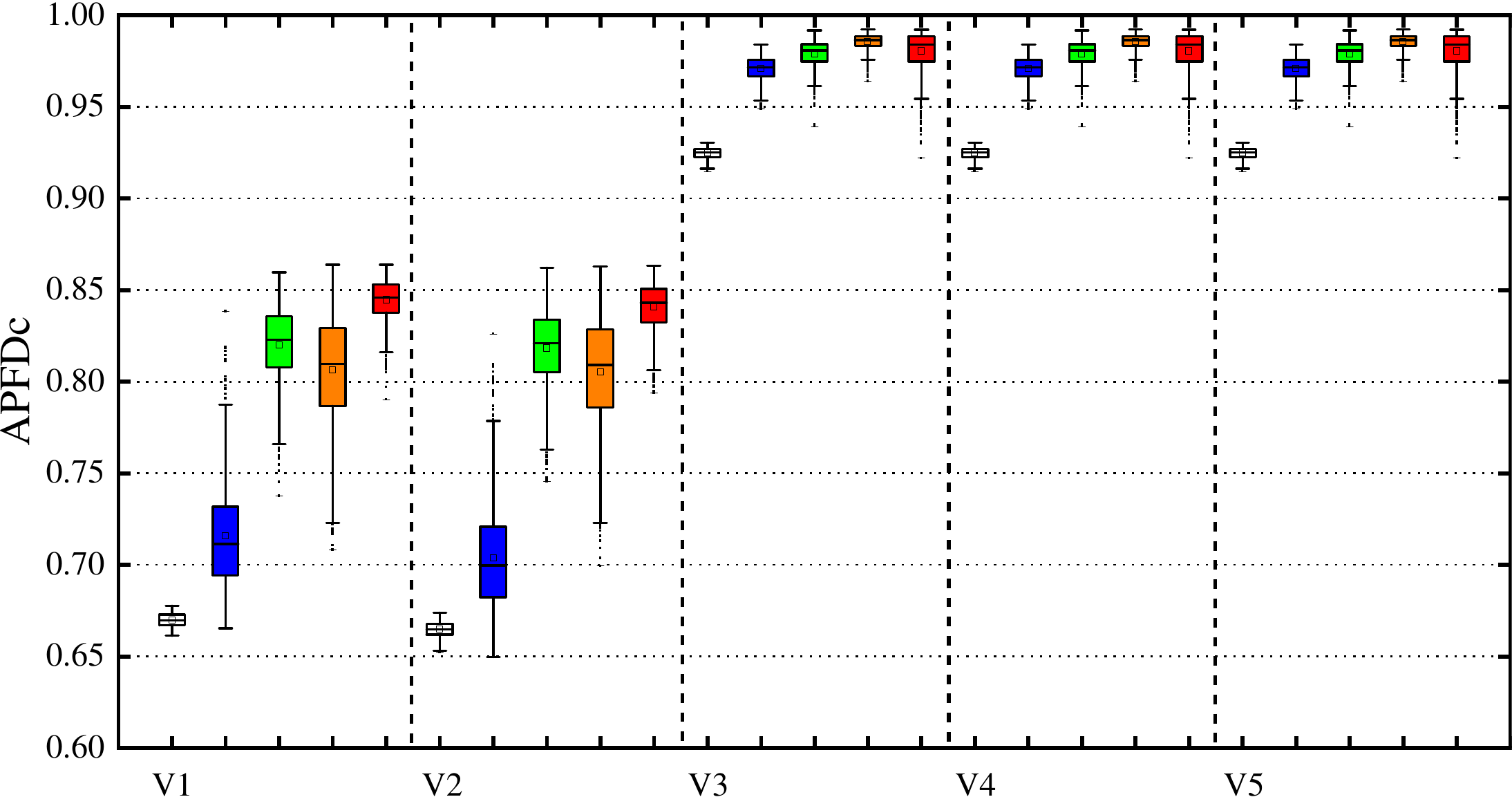}
        \label{apfdc_gzip_sc}
    }
        \subfigure[$gzip$-branch coverage]
    {
        \includegraphics[width=0.319\textwidth]{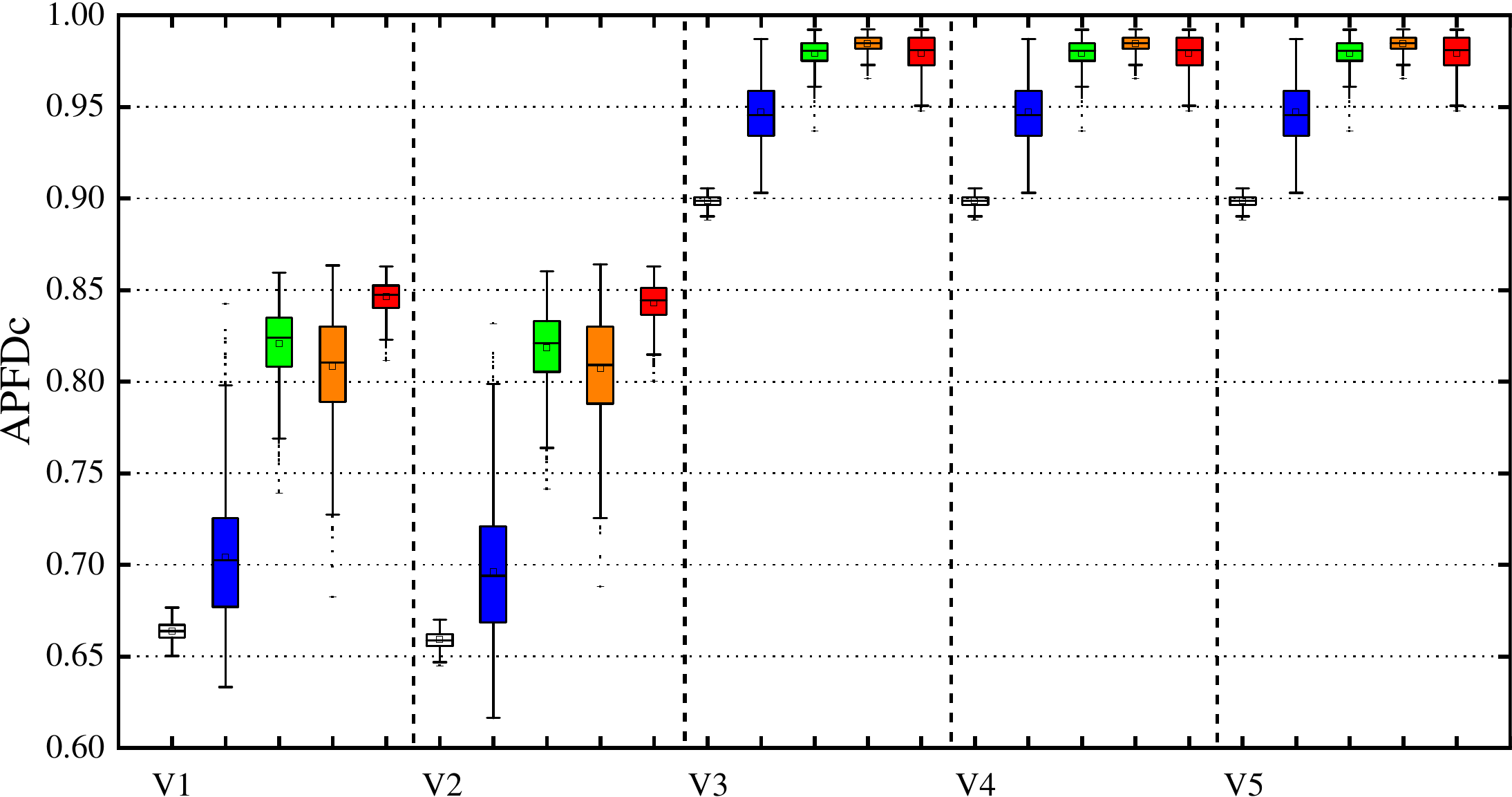}
        \label{apfdc_gzip_bc}
    }
        \subfigure[$gzip$-method coverage]
    {
        \includegraphics[width=0.319\textwidth]{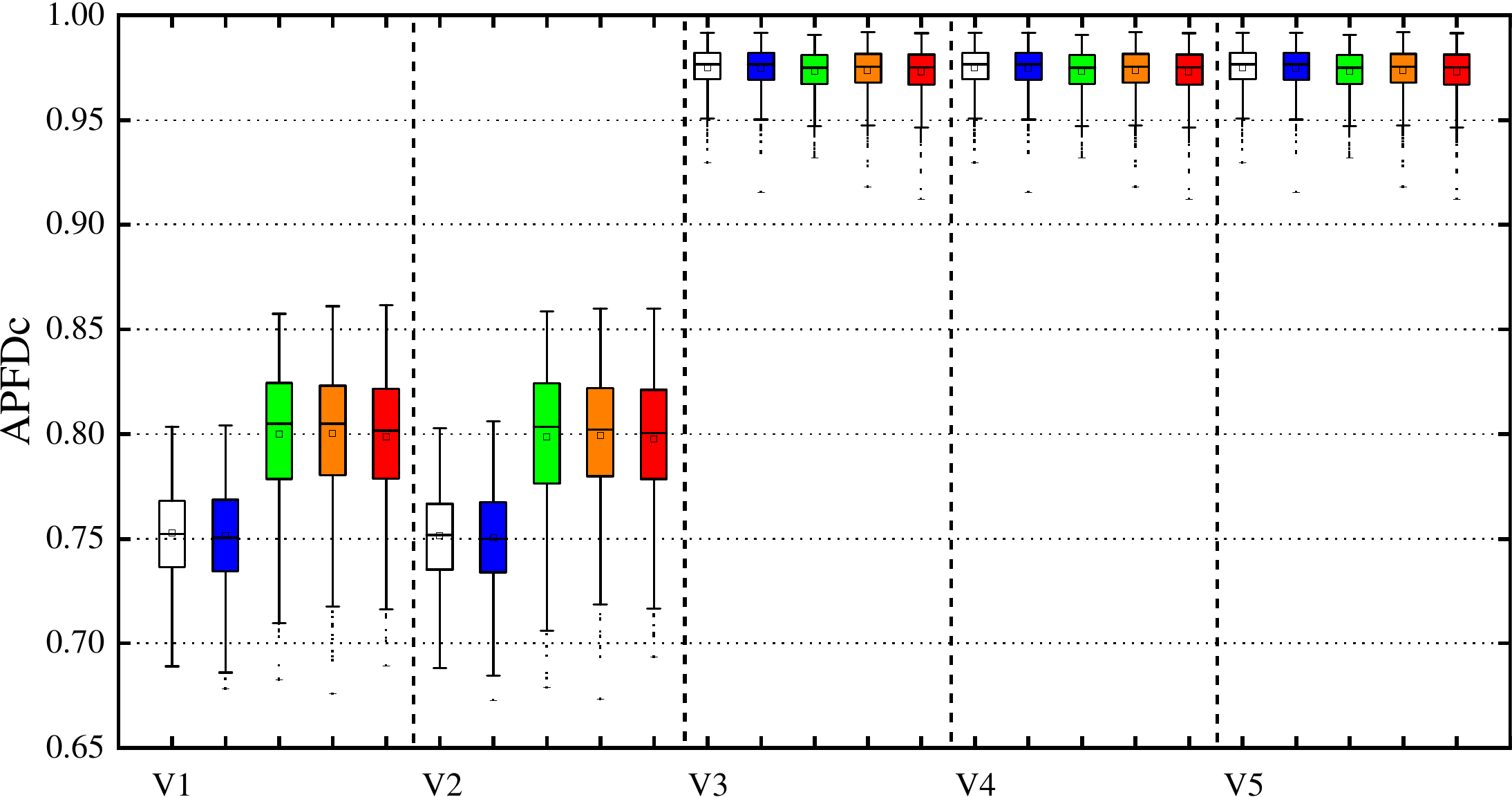}
        \label{apfdc_gzip_mc}
    }
        \subfigure[$make$-statement coverage]
    {
        \includegraphics[width=0.319\textwidth]{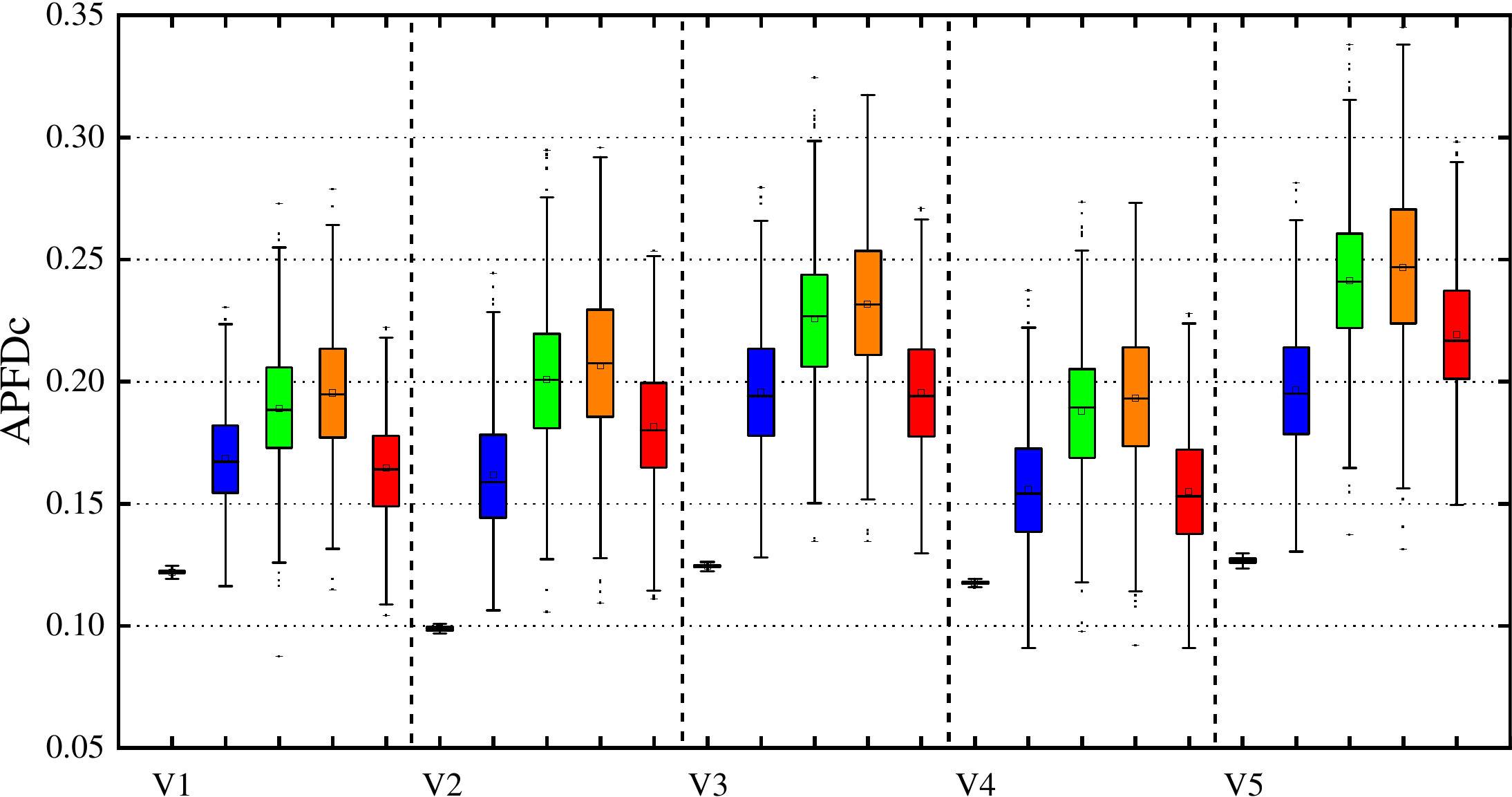}
        \label{apfdc_make_sc}
    }
        \subfigure[$make$-branch coverage]
    {
        \includegraphics[width=0.319\textwidth]{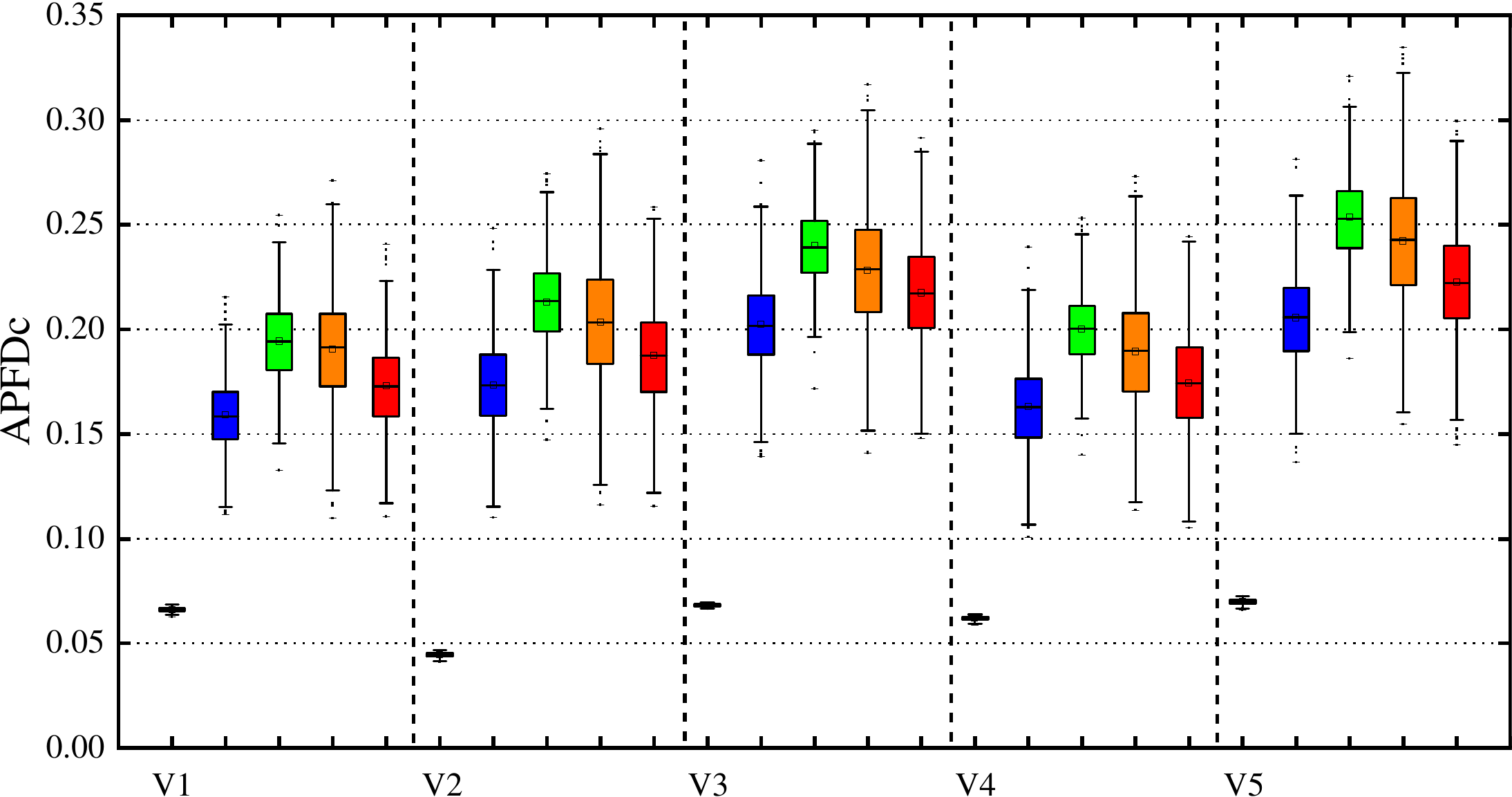}
        \label{apfdc_make_bc}
    }
        \subfigure[$make$-method coverage]
    {
        \includegraphics[width=0.319\textwidth]{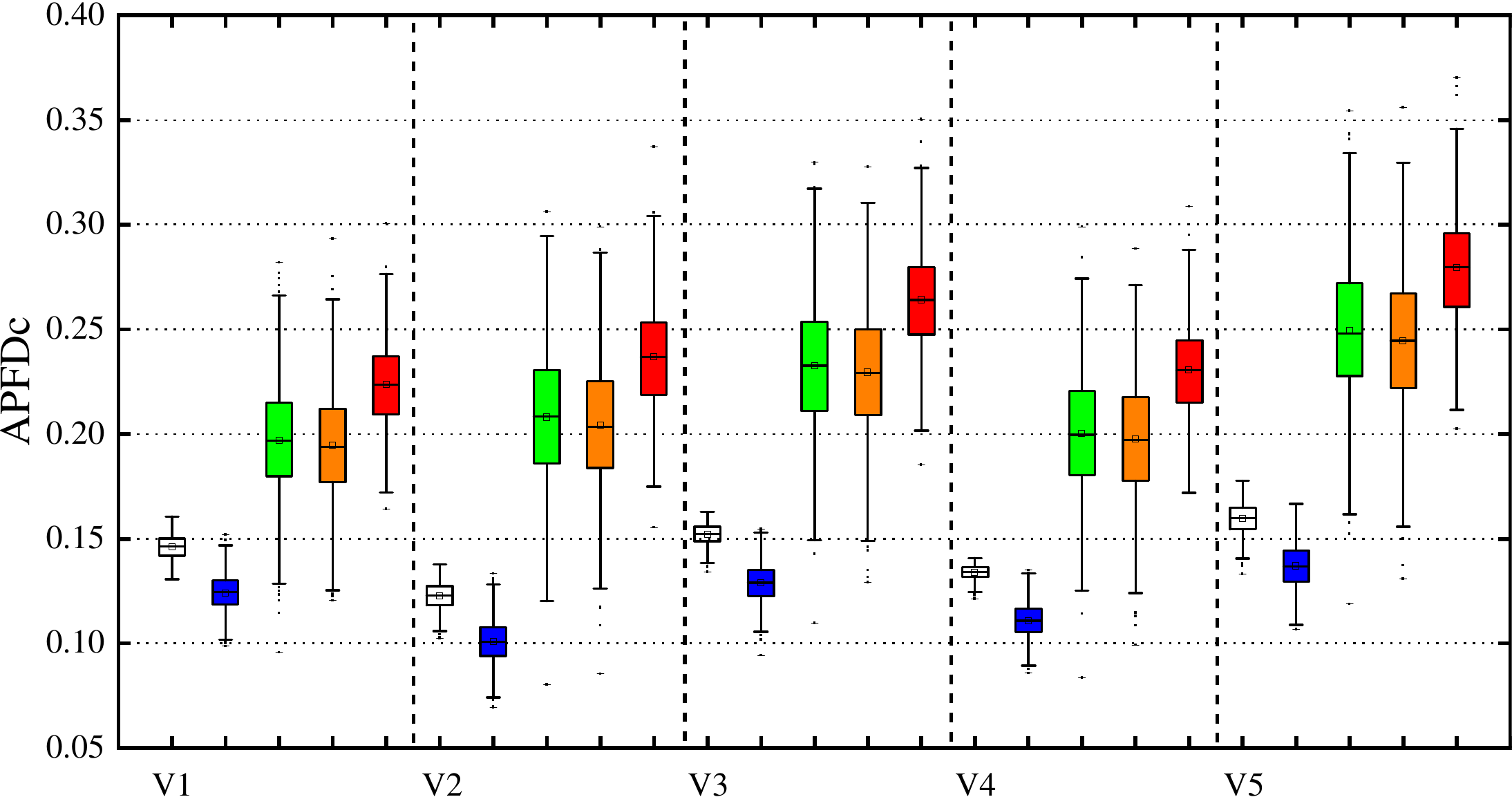}
        \label{apfdc_make_mc}
    }
        \subfigure[$sed$-statement coverage]
    {
        \includegraphics[width=0.319\textwidth]{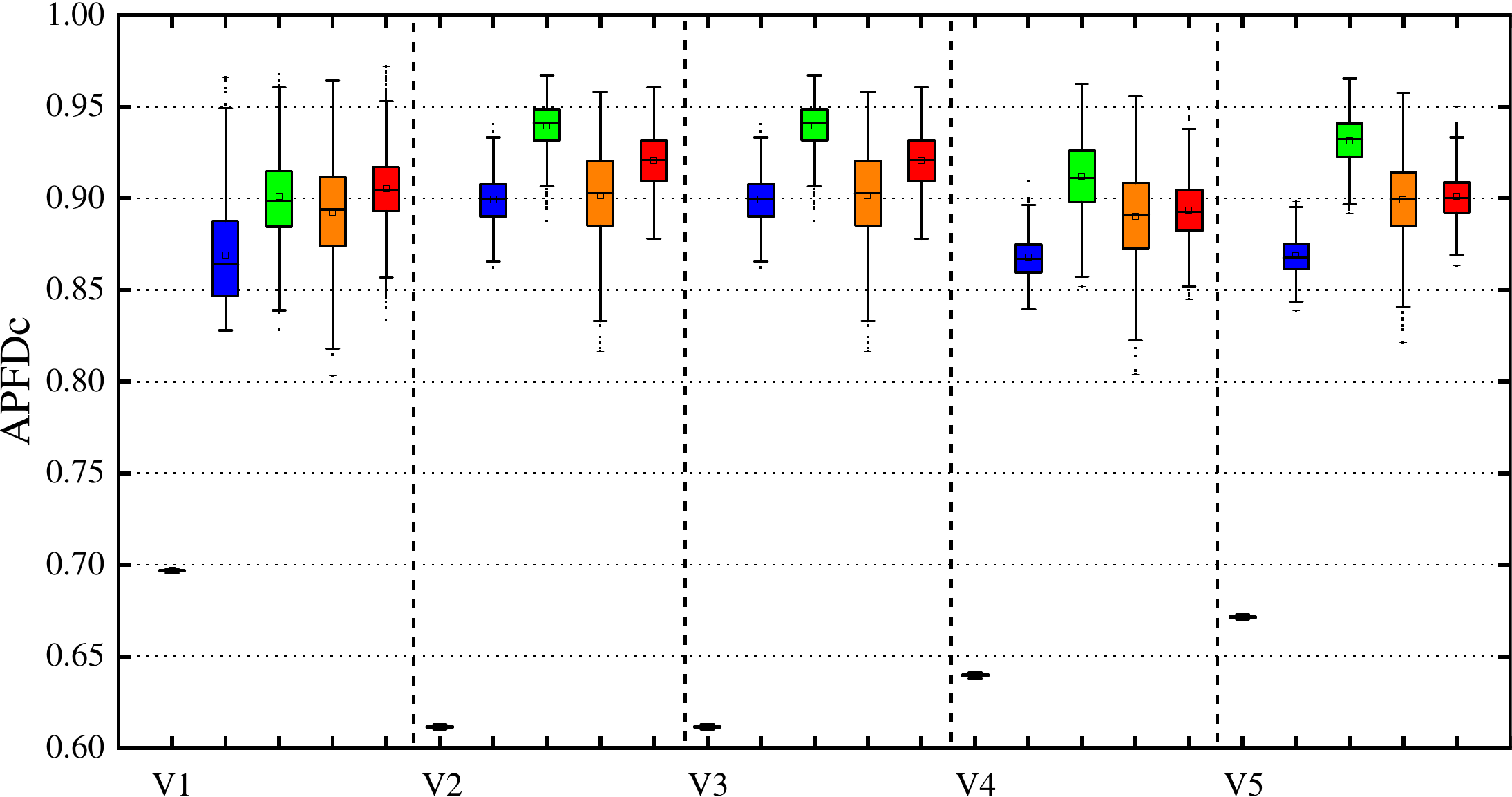}
        \label{apfdc_sed_sc}
    }
        \subfigure[$sed$-branch coverage]
    {
        \includegraphics[width=0.319\textwidth]{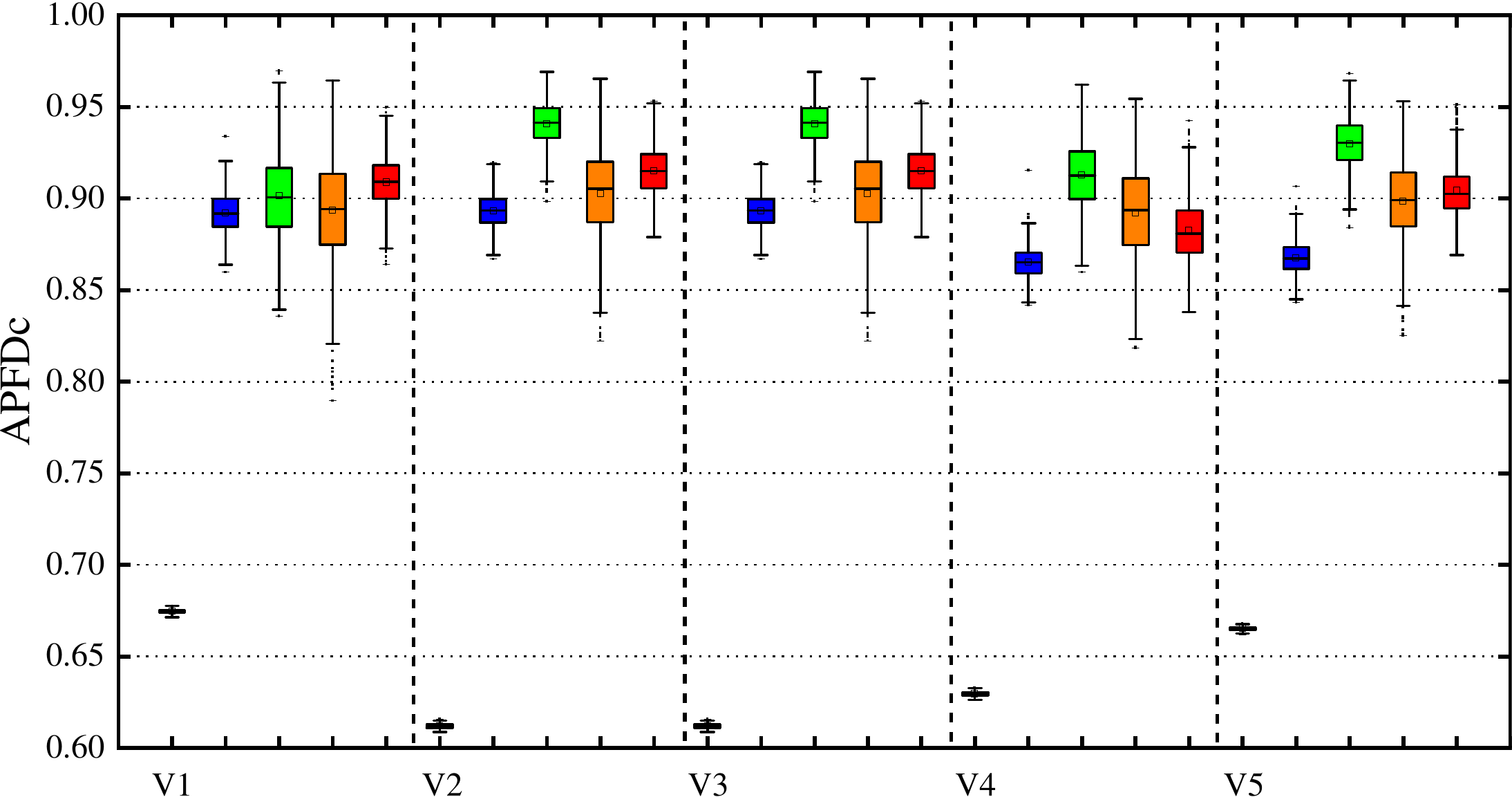}
        \label{apfdc_sed_bc}
    }
        \subfigure[$sed$-method coverage]
    {
        \includegraphics[width=0.319\textwidth]{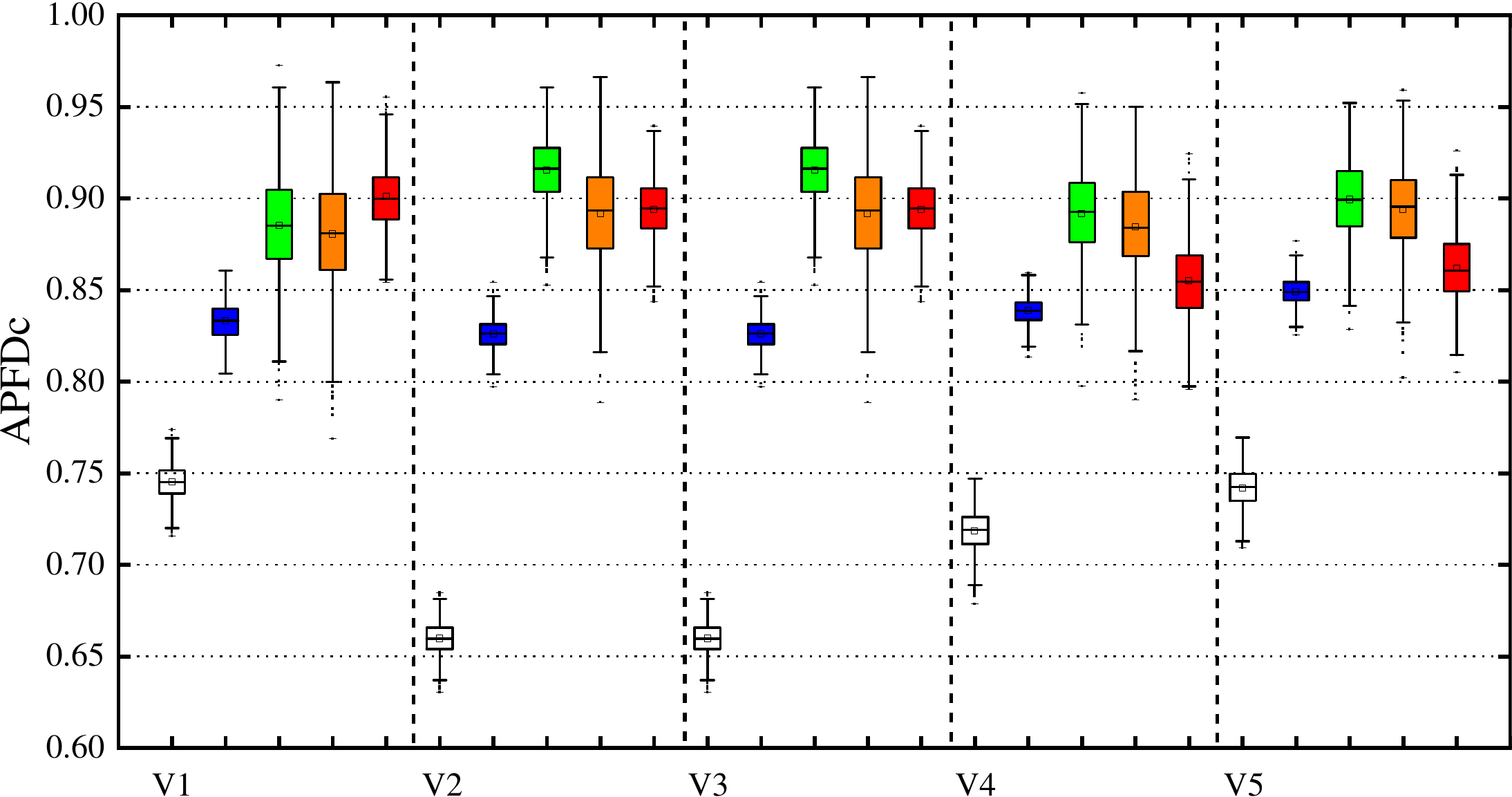}
        \label{apfdc_sed_mc}
    }
    \caption{\textbf{Effectiveness: }\textbf{APFD}$_\textbf{c}$ results for \textbf{C programs}}
    \label{FIG:apfdc-c}
\end{figure*}

Next, we provide the APFD$_\textrm{c}$ results for CCCP for different code coverage and test case granularities.
Figures \ref{FIG:apfdc-java-c} to \ref{FIG:apfdc-c} show the APFD$_\textrm{c}$ results for the Java programs at the test-class level;
at the test-method level;
and the C programs, respectively.
Each sub-figure in these figures has the program versions across the $x$-axis, and the APFD$_\textrm{c}$ values for the five RTCP techniques on the y-axis.
Tables \ref{TAB:apfdc-java-c} to \ref{TAB:apfdc-c} present the corresponding statistical comparisons.


\begin{table*}[!t]
\scriptsize
\centering
 \caption{Statistical \textbf{effectiveness}  comparisons of  \textbf{APFD}$_\textbf{c}$ for \textbf{Java programs} at the \textbf{test-method level}.
For a comparison between two methods
$\textit{TCP}_\textit{ccc}$ and $M$,
where
$M \in \{\textit{TCP}_\textit{tot}, \textit{TCP}_\textit{add}, \textit{TCP}_\textit{art}, \textit{TCP}_\textit{search}\}$,
the symbol \ding{52} means that $\textit{TCP}_\textit{ccc}$ is better
($p$-value is less than 0.05, and the effect size $\hat{\textrm{A}}_{12}(\textit{TCP}_\textit{ccc},M)$ is greater than 0.50);
the symbol \ding{54} means that $M$ is better
(the $p$-value is less than 0.05, and $\hat{\textrm{A}}_{12}(\textit{TCP}_\textit{ccc},M)$ is less than 0.50);
and the symbol \ding{109} means that there is no statistically significant difference between them (the $p$-value is greater than 0.05).
}
  \label{TAB:apfdc-java-m}
    \setlength{\tabcolsep}{1.8mm}{
    \begin{tabular}{c|cccc|cccc|cccc}
     \hline
       \multirow{2}*{\textbf{Program Name}} &\multicolumn{4}{c|}{\textbf{Statement Coverage}} &\multicolumn{4}{c|}{\textbf{Branch Coverage}} &\multicolumn{4}{c}{\textbf{Method Coverage}} \\
        \cline{2-13}

         &$\textit{TCP}_\textit{tot}$ &$\textit{TCP}_\textit{add}$ &$\textit{TCP}_\textit{art}$ &$\textit{TCP}_\textit{search}$  & $\textit{TCP}_\textit{tot}$ &$\textit{TCP}_\textit{add}$ &$\textit{TCP}_\textit{art}$ &$\textit{TCP}_\textit{search}$ &$\textit{TCP}_\textit{tot}$ & $\textit{TCP}_\textit{add}$ &$\textit{TCP}_\textit{art}$ &$\textit{TCP}_\textit{search}$\\
            \hline
$ant\_v1$	&\ding{54}(0.00)	&\ding{52}(0.54)	&\ding{52}(1.00)	&\ding{54}(0.45)	&\ding{54}(0.00)	&\ding{109}(0.50)	&\ding{52}(1.00)	&\ding{52}(0.68)	&\ding{54}(0.00)	&\ding{54}(0.42)	&\ding{52}(1.00)	&\ding{109}(0.48)	\\
$ant\_v2$	&\ding{52}(0.99)	&\ding{109}(0.50)	&\ding{52}(1.00)	&\ding{52}(0.56)	&\ding{52}(1.00)	&\ding{109}(0.52)	&\ding{52}(1.00)	&\ding{52}(0.87)	&\ding{54}(0.44)	&\ding{109}(0.52)	&\ding{52}(0.99)	&\ding{52}(0.64)	\\
$ant\_v3$	&\ding{54}(0.00)	&\ding{109}(0.49)	&\ding{52}(1.00)	&\ding{52}(0.63)	&\ding{52}(1.00)	&\ding{109}(0.48)	&\ding{52}(1.00)	&\ding{52}(0.77)	&\ding{54}(0.17)	&\ding{109}(0.47)	&\ding{52}(0.99)	&\ding{52}(0.64)	\\\hline
$ant$	&\ding{54}(0.40)	&\ding{109}(0.51)	&\ding{52}(1.00)	&\ding{52}(0.54)	&\ding{52}(0.63)	&\ding{109}(0.50)	&\ding{52}(1.00)	&\ding{52}(0.76)	&\ding{54}(0.24)	&\ding{54}(0.47)	&\ding{52}(0.99)	&\ding{52}(0.59)	\\\hline
$jmeter\_v1$	&\ding{52}(1.00)	&\ding{52}(0.53)	&\ding{52}(1.00)	&\ding{52}(0.63)	&\ding{52}(1.00)	&\ding{109}(0.50)	&\ding{52}(0.93)	&\ding{52}(0.80)	&\ding{52}(0.76)	&\ding{109}(0.50)	&\ding{52}(0.63)	&\ding{52}(0.53)	\\
$jmeter\_v2$	&\ding{52}(1.00)	&\ding{52}(0.53)	&\ding{52}(1.00)	&\ding{52}(0.65)	&\ding{52}(1.00)	&\ding{109}(0.49)	&\ding{52}(0.90)	&\ding{52}(0.82)	&\ding{52}(1.00)	&\ding{109}(0.50)	&\ding{52}(0.88)	&\ding{52}(0.56)	\\
$jmeter\_v3$	&\ding{52}(1.00)	&\ding{109}(0.52)	&\ding{52}(1.00)	&\ding{54}(0.47)	&\ding{52}(1.00)	&\ding{54}(0.44)	&\ding{52}(0.99)	&\ding{52}(0.66)	&\ding{52}(1.00)	&\ding{54}(0.45)	&\ding{52}(1.00)	&\ding{54}(0.47)	\\
$jmeter\_v4$	&\ding{52}(1.00)	&\ding{109}(0.51)	&\ding{52}(1.00)	&\ding{109}(0.48)	&\ding{52}(1.00)	&\ding{109}(0.49)	&\ding{52}(1.00)	&\ding{52}(0.69)	&\ding{52}(1.00)	&\ding{54}(0.41)	&\ding{52}(1.00)	&\ding{54}(0.44)	\\
$jmeter\_v5$	&\ding{52}(1.00)	&\ding{109}(0.50)	&\ding{52}(1.00)	&\ding{52}(0.61)	&\ding{52}(1.00)	&\ding{54}(0.30)	&\ding{52}(1.00)	&\ding{52}(0.77)	&\ding{52}(1.00)	&\ding{109}(0.48)	&\ding{52}(0.86)	&\ding{54}(0.41)	\\\hline
$jmeter$	&\ding{52}(1.00)	&\ding{109}(0.51)	&\ding{52}(1.00)	&\ding{52}(0.57)	&\ding{52}(1.00)	&\ding{54}(0.46)	&\ding{52}(0.97)	&\ding{52}(0.69)	&\ding{52}(0.95)	&\ding{54}(0.48)	&\ding{52}(0.85)	&\ding{109}(0.49)	\\\hline
$jtopas\_v1$	&\ding{52}(1.00)	&\ding{52}(1.00)	&\ding{54}(0.00)	&\ding{52}(0.81)	&\ding{52}(1.00)	&\ding{54}(0.37)	&\ding{54}(0.00)	&\ding{54}(0.36)	&\ding{52}(1.00)	&\ding{52}(0.53)	&\ding{54}(0.31)	&\ding{54}(0.47)	\\
$jtopas\_v2$	&\ding{52}(1.00)	&\ding{52}(0.95)	&\ding{54}(0.00)	&\ding{109}(0.51)	&\ding{52}(1.00)	&\ding{52}(0.64)	&\ding{54}(0.00)	&\ding{109}(0.50)	&\ding{52}(1.00)	&\ding{52}(0.53)	&\ding{54}(0.04)	&\ding{54}(0.33)	\\
$jtopas\_v3$	&\ding{52}(1.00)	&\ding{109}(0.51)	&\ding{54}(0.00)	&\ding{54}(0.18)	&\ding{52}(1.00)	&\ding{109}(0.50)	&\ding{54}(0.00)	&\ding{54}(0.28)	&\ding{52}(1.00)	&\ding{109}(0.49)	&\ding{54}(0.05)	&\ding{109}(0.51)	\\\hline
$jtopas$	&\ding{52}(1.00)	&\ding{52}(0.61)	&\ding{54}(0.00)	&\ding{109}(0.50)	&\ding{52}(1.00)	&\ding{109}(0.50)	&\ding{54}(0.00)	&\ding{54}(0.42)	&\ding{52}(1.00)	&\ding{52}(0.52)	&\ding{54}(0.14)	&\ding{54}(0.43)	\\\hline
$xmlsec\_v1$	&\ding{52}(1.00)	&\ding{54}(0.27)	&\ding{54}(0.38)	&\ding{54}(0.38)	&\ding{52}(1.00)	&\ding{52}(0.53)	&\ding{54}(0.36)	&\ding{109}(0.52)	&\ding{52}(1.00)	&\ding{52}(0.53)	&\ding{52}(0.69)	&\ding{52}(0.56)	\\
$xmlsec\_v2$	&\ding{52}(1.00)	&\ding{109}(0.50)	&\ding{52}(0.83)	&\ding{52}(0.67)	&\ding{52}(1.00)	&\ding{52}(0.57)	&\ding{52}(0.56)	&\ding{52}(0.78)	&\ding{52}(1.00)	&\ding{109}(0.49)	&\ding{54}(0.46)	&\ding{52}(0.54)	\\
$xmlsec\_v3$	&\ding{52}(1.00)	&\ding{109}(0.50)	&\ding{54}(0.12)	&\ding{52}(0.61)	&\ding{52}(1.00)	&\ding{54}(0.45)	&\ding{54}(0.01)	&\ding{54}(0.14)	&\ding{52}(1.00)	&\ding{109}(0.50)	&\ding{109}(0.48)	&\ding{52}(0.55)	\\\hline
$xmlsec$	&\ding{52}(1.00)	&\ding{54}(0.47)	&\ding{54}(0.46)	&\ding{52}(0.54)	&\ding{52}(0.83)	&\ding{109}(0.51)	&\ding{54}(0.34)	&\ding{52}(0.53)	&\ding{52}(1.00)	&\ding{109}(0.50)	&\ding{52}(0.53)	&\ding{52}(0.54)	\\\hline
\textit{\textbf{All Java Programs}}	&\ding{52}(0.85)	&\ding{109}(0.51)	&\ding{52}(0.64)	&\ding{52}(0.54)	&\ding{52}(0.77)	&\ding{109}(0.50)	&\ding{52}(0.62)	&\ding{52}(0.57)	&\ding{52}(0.87)	&\ding{54}(0.49)	&\ding{52}(0.67)	&\ding{52}(0.51)	\\
\hline
  \end{tabular}}
\end{table*}

\begin{table*}[!t]
\scriptsize
\centering
 \caption{Statistical \textbf{effectiveness}  comparisons of \textbf{APFD}$_\textbf{c}$  for \textbf{C programs}.
For a comparison between two methods
$\textit{TCP}_\textit{ccc}$ and $M$,
where
$M \in \{\textit{TCP}_\textit{tot}, \textit{TCP}_\textit{add}, \textit{TCP}_\textit{art}, \textit{TCP}_\textit{search}\}$,
the symbol \ding{52} means that $\textit{TCP}_\textit{ccc}$ is better
($p$-value is less than 0.05, and the effect size $\hat{\textrm{A}}_{12}(\textit{TCP}_\textit{ccc},M)$ is greater than 0.50);
the symbol \ding{54} means that $M$ is better
(the $p$-value is less than 0.05, and $\hat{\textrm{A}}_{12}(\textit{TCP}_\textit{ccc},M)$ is less than 0.50);
and the symbol \ding{109} means that there is no statistically significant difference between them (the $p$-value is greater than 0.05).
}
  \label{TAB:apfdc-c}
    \setlength{\tabcolsep}{1.8mm}{
    \begin{tabular}{c|cccc|cccc|cccc}
     \hline
       \multirow{2}*{\textbf{Program Name}} &\multicolumn{4}{c|}{\textbf{Statement Coverage}} &\multicolumn{4}{c|}{\textbf{Branch Coverage}} &\multicolumn{4}{c}{\textbf{Method Coverage}} \\
        \cline{2-13}

         &$\textit{TCP}_\textit{tot}$ &$\textit{TCP}_\textit{add}$ &$\textit{TCP}_\textit{art}$ &$\textit{TCP}_\textit{search}$  & $\textit{TCP}_\textit{tot}$ &$\textit{TCP}_\textit{add}$ &$\textit{TCP}_\textit{art}$ &$\textit{TCP}_\textit{search}$ &$\textit{TCP}_\textit{tot}$ & $\textit{TCP}_\textit{add}$ &$\textit{TCP}_\textit{art}$ &$\textit{TCP}_\textit{search}$\\
            \hline

$flex\_v1$ &\ding{52}(1.00) &\ding{52}(0.99) &\ding{52}(1.00) &\ding{52}(1.00) &\ding{52}(1.00) &\ding{52}(1.00) &\ding{52}(1.00) &\ding{52}(1.00) &\ding{52}(1.00) &\ding{52}(1.00) &\ding{52}(0.82) &\ding{52}(0.96) \\
$flex\_v2$ &\ding{52}(1.00) &\ding{52}(0.98) &\ding{52}(1.00) &\ding{52}(0.99) &\ding{52}(1.00) &\ding{52}(1.00) &\ding{52}(1.00) &\ding{52}(0.99) &\ding{52}(1.00) &\ding{52}(1.00) &\ding{52}(0.76) &\ding{52}(0.93) \\
$flex\_v3$ &\ding{52}(1.00) &\ding{52}(0.75) &\ding{52}(1.00) &\ding{52}(0.88) &\ding{52}(1.00) &\ding{52}(0.71) &\ding{52}(1.00) &\ding{52}(0.94) &\ding{52}(1.00) &\ding{52}(0.99) &\ding{52}(0.99) &\ding{52}(0.96) \\
$flex\_v4$ &\ding{52}(1.00) &\ding{52}(0.99) &\ding{52}(1.00) &\ding{52}(0.99) &\ding{52}(1.00) &\ding{52}(1.00) &\ding{52}(1.00) &\ding{52}(1.00) &\ding{52}(1.00) &\ding{52}(1.00) &\ding{52}(0.75) &\ding{52}(0.92) \\
$flex\_v5$ &\ding{52}(1.00) &\ding{52}(0.98) &\ding{52}(1.00) &\ding{52}(0.99) &\ding{52}(1.00) &\ding{52}(1.00) &\ding{52}(1.00) &\ding{52}(1.00) &\ding{52}(1.00) &\ding{52}(1.00) &\ding{52}(0.75) &\ding{52}(0.92) \\\hline
$flex$ &\ding{52}(1.00) &\ding{52}(0.89) &\ding{52}(1.00) &\ding{52}(0.94) &\ding{52}(1.00) &\ding{52}(0.86) &\ding{52}(1.00) &\ding{52}(0.95) &\ding{52}(1.00) &\ding{52}(0.86) &\ding{52}(0.79) &\ding{52}(0.87) \\\hline

$grep\_v1$ &\ding{52}(1.00) &\ding{52}(0.97) &\ding{54}(0.34) &\ding{52}(0.57) &\ding{52}(1.00) &\ding{52}(0.84) &\ding{54}(0.17) &\ding{109}(0.49) &\ding{52}(1.00) &\ding{52}(0.79) &\ding{54}(0.06) &\ding{54}(0.08) \\
$grep\_v2$ &\ding{52}(1.00) &\ding{52}(0.93) &\ding{54}(0.39) &\ding{52}(0.71) &\ding{52}(1.00) &\ding{52}(0.80) &\ding{54}(0.14) &\ding{52}(0.61) &\ding{52}(1.00) &\ding{52}(0.92) &\ding{54}(0.09) &\ding{54}(0.18) \\
$grep\_v3$ &\ding{52}(1.00) &\ding{52}(0.90) &\ding{54}(0.42) &\ding{54}(0.33) &\ding{52}(1.00) &\ding{52}(0.82) &\ding{54}(0.37) &\ding{54}(0.36) &\ding{52}(1.00) &\ding{52}(0.79) &\ding{54}(0.13) &\ding{54}(0.17) \\
$grep\_v4$ &\ding{52}(1.00) &\ding{52}(0.87) &\ding{52}(0.70) &\ding{52}(0.82) &\ding{52}(1.00) &\ding{52}(0.73) &\ding{52}(0.85) &\ding{52}(0.88) &\ding{52}(1.00) &\ding{52}(0.75) &\ding{54}(0.14) &\ding{54}(0.16) \\
$grep\_v5$ &\ding{52}(1.00) &\ding{52}(0.75) &\ding{52}(0.84) &\ding{52}(0.91) &\ding{52}(1.00) &\ding{52}(0.67) &\ding{52}(0.90) &\ding{52}(0.91) &\ding{52}(1.00) &\ding{109}(0.51) &\ding{54}(0.13) &\ding{54}(0.13) \\\hline
$grep$ &\ding{52}(1.00) &\ding{52}(0.88) &\ding{52}(0.54) &\ding{52}(0.66) &\ding{52}(1.00) &\ding{52}(0.74) &\ding{109}(0.50) &\ding{52}(0.66) &\ding{52}(1.00) &\ding{52}(0.68) &\ding{54}(0.19) &\ding{54}(0.22) \\\hline

$gzip\_v1$ &\ding{52}(1.00) &\ding{52}(1.00) &\ding{52}(0.86) &\ding{52}(0.89) &\ding{52}(1.00) &\ding{52}(1.00) &\ding{52}(0.89) &\ding{52}(0.90) &\ding{52}(0.88) &\ding{52}(0.88) &\ding{109}(0.48) &\ding{109}(0.48) \\
$gzip\_v2$ &\ding{52}(1.00) &\ding{52}(1.00) &\ding{52}(0.82) &\ding{52}(0.86) &\ding{52}(1.00) &\ding{52}(1.00) &\ding{52}(0.87) &\ding{52}(0.88) &\ding{52}(0.88) &\ding{52}(0.88) &\ding{109}(0.48) &\ding{109}(0.49) \\
$gzip\_v3$ &\ding{52}(1.00) &\ding{52}(0.80) &\ding{52}(0.61) &\ding{54}(0.40) &\ding{52}(1.00) &\ding{52}(0.95) &\ding{52}(0.53) &\ding{54}(0.36) &\ding{54}(0.46) &\ding{54}(0.46) &\ding{109}(0.50) &\ding{109}(0.49) \\
$gzip\_v4$ &\ding{52}(1.00) &\ding{52}(0.80) &\ding{52}(0.61) &\ding{54}(0.40) &\ding{52}(1.00) &\ding{52}(0.95) &\ding{52}(0.53) &\ding{54}(0.36) &\ding{54}(0.46) &\ding{54}(0.46) &\ding{109}(0.50) &\ding{109}(0.49) \\
$gzip\_v5$ &\ding{52}(1.00) &\ding{52}(0.80) &\ding{52}(0.61) &\ding{54}(0.40) &\ding{52}(1.00) &\ding{52}(0.95) &\ding{52}(0.53) &\ding{54}(0.36) &\ding{54}(0.46) &\ding{54}(0.46) &\ding{109}(0.50) &\ding{109}(0.49) \\\hline
$gzip$ &\ding{52}(0.76) &\ding{52}(0.69) &\ding{52}(0.59) &\ding{52}(0.52) &\ding{52}(0.76) &\ding{52}(0.74) &\ding{52}(0.57) &\ding{52}(0.51) &\ding{52}(0.55) &\ding{52}(0.55) &\ding{109}(0.50) &\ding{109}(0.49) \\\hline

$make\_v1$ &\ding{52}(0.98) &\ding{54}(0.44) &\ding{54}(0.23) &\ding{54}(0.19) &\ding{52}(1.00) &\ding{52}(0.70) &\ding{54}(0.23) &\ding{54}(0.30) &\ding{52}(1.00) &\ding{52}(1.00) &\ding{52}(0.79) &\ding{52}(0.80) \\
$make\_v2$ &\ding{52}(1.00) &\ding{52}(0.71) &\ding{54}(0.31) &\ding{54}(0.27) &\ding{52}(1.00) &\ding{52}(0.67) &\ding{54}(0.21) &\ding{54}(0.34) &\ding{52}(1.00) &\ding{52}(1.00) &\ding{52}(0.76) &\ding{52}(0.79) \\
$make\_v3$ &\ding{52}(1.00) &\ding{109}(0.50) &\ding{54}(0.22) &\ding{54}(0.19) &\ding{52}(1.00) &\ding{52}(0.68) &\ding{54}(0.23) &\ding{54}(0.39) &\ding{52}(1.00) &\ding{52}(1.00) &\ding{52}(0.79) &\ding{52}(0.81) \\
$make\_v4$ &\ding{52}(0.94) &\ding{109}(0.49) &\ding{54}(0.19) &\ding{54}(0.16) &\ding{52}(1.00) &\ding{52}(0.64) &\ding{54}(0.20) &\ding{54}(0.35) &\ding{52}(1.00) &\ding{52}(1.00) &\ding{52}(0.79) &\ding{52}(0.81) \\
$make\_v5$ &\ding{52}(1.00) &\ding{52}(0.73) &\ding{54}(0.29) &\ding{54}(0.26) &\ding{52}(1.00) &\ding{52}(0.69) &\ding{54}(0.18) &\ding{54}(0.32) &\ding{52}(1.00) &\ding{52}(1.00) &\ding{52}(0.76) &\ding{52}(0.79) \\\hline
$make$ &\ding{52}(0.98) &\ding{52}(0.56) &\ding{54}(0.30) &\ding{54}(0.27) &\ding{52}(1.00) &\ding{52}(0.63) &\ding{54}(0.29) &\ding{54}(0.37) &\ding{52}(1.00) &\ding{52}(1.00) &\ding{52}(0.73) &\ding{52}(0.75) \\\hline

$sed\_v1$ &\ding{52}(1.00) &\ding{52}(0.85) &\ding{52}(0.57) &\ding{52}(0.64) &\ding{52}(1.00) &\ding{52}(0.83) &\ding{52}(0.61) &\ding{52}(0.68) &\ding{52}(1.00) &\ding{52}(1.00) &\ding{52}(0.68) &\ding{52}(0.71) \\
$sed\_v2$ &\ding{52}(1.00) &\ding{52}(0.86) &\ding{54}(0.18) &\ding{52}(0.73) &\ding{52}(1.00) &\ding{52}(0.91) &\ding{54}(0.08) &\ding{52}(0.65) &\ding{52}(1.00) &\ding{52}(1.00) &\ding{54}(0.19) &\ding{109}(0.52) \\
$sed\_v3$ &\ding{52}(1.00) &\ding{52}(0.86) &\ding{54}(0.18) &\ding{52}(0.73) &\ding{52}(1.00) &\ding{52}(0.91) &\ding{54}(0.08) &\ding{52}(0.65) &\ding{52}(1.00) &\ding{52}(1.00) &\ding{54}(0.19) &\ding{109}(0.52) \\
$sed\_v4$ &\ding{52}(1.00) &\ding{52}(0.90) &\ding{54}(0.24) &\ding{52}(0.53) &\ding{52}(1.00) &\ding{52}(0.82) &\ding{54}(0.12) &\ding{54}(0.37) &\ding{52}(1.00) &\ding{52}(0.76) &\ding{54}(0.13) &\ding{54}(0.19) \\
$sed\_v5$ &\ding{52}(1.00) &\ding{52}(0.98) &\ding{54}(0.06) &\ding{109}(0.51) &\ding{52}(1.00) &\ding{52}(0.99) &\ding{54}(0.10) &\ding{52}(0.57) &\ding{52}(1.00) &\ding{52}(0.73) &\ding{54}(0.10) &\ding{54}(0.15) \\\hline
$sed$ &\ding{52}(1.00) &\ding{52}(0.82) &\ding{54}(0.29) &\ding{52}(0.63) &\ding{52}(1.00) &\ding{52}(0.82) &\ding{54}(0.24) &\ding{52}(0.59) &\ding{52}(1.00) &\ding{52}(0.94) &\ding{54}(0.29) &\ding{54}(0.43) \\\hline

\textit{\textbf{All C Programs}} &\ding{52}(0.79) &\ding{52}(0.58) &\ding{52}(0.52) &\ding{52}(0.54) &\ding{52}(0.80) &\ding{52}(0.59) &\ding{52}(0.51) &\ding{52}(0.53) &\ding{52}(0.75) &\ding{52}(0.61) &\ding{54}(0.49) &\ding{52}(0.51) \\

\hline

  \end{tabular}}
\end{table*}

\subsubsection{APFD$_\textrm{c}$ Results: Java Programs (Test-Class Level)}

Based on Figure \ref{FIG:apfdc-java-c} and Table \ref{TAB:apfdc-java-c}, we have the following observations:

(1)
Compared with $\textit{TCP}_\textit{tot}$, $\textit{TCP}_\textit{ccc}$ has much better APFD$_\textrm{c}$ results for the programs $ant$ and $jtopas$,
irrespective of program version and code coverage granularity, with the maximum difference between the mean and median values being up to about 30\%.
For the programs $jmeter$ and $xmlsec$, however, $\textit{TCP}_\textit{ccc}$ performs worse than $\textit{TCP}_\textit{tot}$,
with a maximum APFD$_\textrm{c}$ difference of about 15\%.

(2)
Although $\textit{TCP}_\textit{ccc}$ performs better than $\textit{TCP}_\textit{art}$ in many cases
(for example, with $jmeter$, for all code coverage granularities),
it also sometimes performs worse  (including with $ant\_v2$ and $ant\_v3$ for the branch and method coverage levels).
Nevertheless, the $\textit{TCP}_\textit{ccc}$ APFD$_\textrm{c}$ values have much lower variation than $\textit{TCP}_\textit{art}$.

(3)
$\textit{TCP}_\textit{ccc}$ has very similar performance to $\textit{TCP}_\textit{add}$ and $\textit{TCP}_\textit{search}$,
sometimes performing slightly better (for example, with $jtopas$, using statement coverage) or worse (such as with $xmlsec\_v3$, for method coverage).
The differences among the mean and median APFD$_\textrm{c}$ values for the three RTCP techniques are very small, at most, about 5\%.

(4)
Overall, the statistical analyses support the box plot observations.
Looking at all Java  programs together,
$\textit{TCP}_\textit{ccc}$ is better than $\textit{TCP}_\textit{tot}$ and $\textit{TCP}_\textit{art}$, for all code coverage granularities:
The $p$-values are much less than $0.05$;
and the $\hat{\textrm{A}}_{12}$ values range from 0.53 to 0.68.
Furthermore, $\textit{TCP}_\textit{ccc}$ performs similarly, or slightly better, compared with $\textit{TCP}_\textit{add}$ and $\textit{TCP}_\textit{search}$,
with the $\hat{\textrm{A}}_{12}$ values ranging from 0.50 to 0.53.

\subsubsection{APFD$_\textrm{c}$ Results: Java Programs (Test-Method Level)}

Based on Figure \ref{FIG:apfdc-java-m} and Table \ref{TAB:apfdc-java-m}, we have the following observations:

(1)
Apart from some cases with the program $ant$ (for example, $ant\_v1$),
$\textit{TCP}_\textit{ccc}$ has much better APFD$_\textrm{c}$ performance than $\textit{TCP}_\textit{tot}$,
with the maximum mean and median differences reaching about 50\%.

(2)
$\textit{TCP}_\textit{ccc}$ performs much better than $\textit{TCP}_\textit{art}$ for the programs $ant$ and $jmeter$,
with the maximum  mean and median APFD$_\textrm{c}$ differences being about 30\%.
In contrast, $\textit{TCP}_\textit{art}$ performs much better than $\textit{TCP}_\textit{ccc}$ for the program $ jtopas$.
For the program $xmlsec$, however, neither $\textit{TCP}_\textit{art}$ nor $\textit{TCP}_\textit{ccc}$ is always better:
At branch coverage level, for example,
$\textit{TCP}_\textit{ccc}$ is better for version $v2$;
$\textit{TCP}_\textit{art}$ is better for version $v3$; and
they have similar performance for version $v1$.

(3)
The $\textit{TCP}_\textit{ccc}$ and $\textit{TCP}_\textit{add}$
APFD$_\textrm{c}$ distributions are very similar,
in most cases,
which means that, on the whole both techniques have very similar effectiveness (according to APFD$_\textrm{c}$).

(4)
Although there are some large performance differences between $\textit{TCP}_\textit{ccc}$ and $\textit{TCP}_\textit{search}$
(such as for $ant\_v2$ and $xmlsec$, with branch coverage), overall, they have similar mean and median APFD$_\textrm{c}$ values.
In most cases, $\textit{TCP}_\textit{ccc}$ has lower variation in APFD$_\textrm{c}$ values than $\textit{TCP}_\textit{search}$.

(5)
Overall, the statistical analyses support the box plot observations.
Looking at all Java programs together,
$\textit{TCP}_\textit{ccc}$ is better than $\textit{TCP}_\textit{tot}$ and $\textit{TCP}_\textit{art}$,
with $p$-values much less than 0.05,
and the $\hat{\textrm{A}}_{12}$ values ranging from 0.62 to 0.87.
$\textit{TCP}_\textit{ccc}$ is better than $\textit{TCP}_\textit{search}$,
with the $p$-values less than 0.05, and the $\hat{\textrm{A}}_{12}$ values ranging from 0.51 to 0.57.
Finally, $\textit{TCP}_\textit{ccc}$ and $\textit{TCP}_\textit{add}$ have very similar performance:
The $\hat{\textrm{A}}_{12}$ values only range between 0.49 and 0.51; and
the $p$-values are greater than 0.05 for statement and branch coverage, but less than 0.05 for method coverage.

\begin{table*}[!t]
\scriptsize
\centering
 \caption{An analysis of statistical \textbf{effectiveness} results of \textbf{APFD}$_\textbf{c}$.
Each cell represents the total times of (\ding{52}), worse (\ding{54}), and (\ding{109}) for corresponding prioritization scenarios described in Tables \ref{TAB:apfdc-java-c} to \ref{TAB:apfdc-c}.
 }
  \label{TAB:APFDcsummary}
  \setlength{\tabcolsep}{0.5mm}{
    \begin{tabular}{c|c|cccc|cccc|cccc|cccc}
     \hline
   \multirow{2}*{\textbf{Language}} &\multirow{2}*{\textbf{Status}}&\multicolumn{4}{c|}{\textbf{Statement Coverage}} &\multicolumn{4}{c|}{\textbf{Branch Coverage}} &\multicolumn{4}{c|}{\textbf{Method Coverage}} &\multicolumn{4}{c}{\textit{\textbf{Sum }}$\sum$ }\\
        \cline{3-18}

         &&$\textit{TCP}_\textit{tot}$ &$\textit{TCP}_\textit{add}$ &$\textit{TCP}_\textit{art}$ &$\textit{TCP}_\textit{search}$  & $\textit{TCP}_\textit{tot}$ &$\textit{TCP}_\textit{add}$ &$\textit{TCP}_\textit{art}$ &$\textit{TCP}_\textit{search}$ &$\textit{TCP}_\textit{tot}$ & $\textit{TCP}_\textit{add}$ &$\textit{TCP}_\textit{art}$ &$\textit{TCP}_\textit{search}$ &$\textit{TCP}_\textit{tot}$ & $\textit{TCP}_\textit{add}$ &$\textit{TCP}_\textit{art}$ &$\textit{TCP}_\textit{search}$\\


            \hline

            \multirow{3}*{Java (test-class)}
                             &\ding{52} &7   &8  &10 &8  &8  &10 &9  &7  &7  &8   &9  &10 &22 &26 &28 &25\\
                             &\ding{54} &6   &3  &3  &3  &6  &2  &5  &3  &7  &3   &5  &3 &19 &8 &13 &9\\
                             &\ding{109}&1   &3  &1  &3  &0  &2  &0  &4  &0  &3   &0  &1 &1 &8 &1 &8\\\hline
            \multirow{3}*{Java (test-method)}
                             &\ding{52} &12  &5   &9  &8  &13 &3  &9  &9  &11 &3  &9  &7 &36 &11 &27 &24\\
                             &\ding{54} &2   &1   &5  &4  &1  &4  &5  &3  &3  &3  &4  &5 &6 &8 &14 &12\\
                             &\ding{109}&0   &8   &0  &2  &0  &7  &0  &2  &0  &8  &1  &2 &0 &23 &1 &6\\\hline
            \multirow{3}*{C}  &\ding{52} &25  &22  &13 &15 &25 &25 &13 &14 &22 &21 &11 &11 &72 &68 &37 &40\\
                             &\ding{54} &0   &1   &12 &9  &0  &0  &12 &10 &3  &3  &9  &7 &3 &4 &33 &26\\
                             &\ding{109}&0   &2   &0  &1  &0  &0  &0  &1  &0  &1  &5  &7 &0 &3 &5 &9\\\hline
           \multirow{3}*{\textit{\textbf{Sum}} $\sum$}
                             &\ding{52} &44  &35  &32 &31  &45 &46 &38 &31 &40 &32 &29 &28 &129 &113 &99 &90\\
                             &\ding{54} &8   &5   &20 &16  &4  &7  &6  &22 &13 &9  &18 &15 &25 &21 &44 &53\\
                             &\ding{109}&1   &13  &1  &6   &4  &0  &9  &0  &0  &12 &6  &10 &5 &25 &16 &16\\\hline
  \end{tabular}}
\end{table*}

\subsubsection{APFD$_\textrm{c}$ Results: C Programs}

Based on  Figure \ref{FIG:apfdc-c} and Table \ref{TAB:apfdc-c}, we have the following observations:

(1)
Except for some very few cases
(such as $gzip\_v3$, $gzip\_v4$, and $gzip\_v5$, with  method coverage),
$\textit{TCP}_\textit{ccc}$ has much higher APFD$_\textrm{c}$ values than $\textit{TCP}_\textit{tot}$ and $\textit{TCP}_\textit{add}$,
with the maximum mean and median APFD$_\textrm{c}$ differences between
$\textit{TCP}_\textit{ccc}$ and $\textit{TCP}_\textit{tot}$ reaching more than 35\%;
and between $\textit{TCP}_\textit{ccc}$ and $\textit{TCP}_\textit{add}$ being about 15\%.

(2)
$\textit{TCP}_\textit{ccc}$ performs differently compared with
$\textit{TCP}_\textit{art}$ and $\textit{TCP}_\textit{search}$
for different programs and different code coverage granularities:
With the program $flex$, for example, for all versions and code coverage granularities, $\textit{TCP}_\textit{ccc}$ is more effective;
however, with $make$, for both statement and branch coverage,
$\textit{TCP}_\textit{art}$ and $\textit{TCP}_\textit{search}$ are more effective.

(3)
Overall, the statistical results support the box plot observations.
Considering all C programs, the $p$ values for all comparisons between
$\textit{TCP}_\textit{ccc}$ and
$\textit{TCP}_\textit{tot}$, $\textit{TCP}_\textit{add}$, $\textit{TCP}_\textit{art}$, and $\textit{TCP}_\textit{search}$
are less than 0.05,
indicating that the APFD$_\textrm{c}$ scores are all significantly different.
According to the effect size $\hat{\textrm{A}}_{12}$ values,
$\textit{TCP}_\textit{ccc}$ outperforms $\textit{TCP}_\textit{tot}$ and $\textit{TCP}_\textit{add}$;
and performs slightly better than $\textit{TCP}_\textit{search}$ and $\textit{TCP}_\textit{art}$ (except at the method coverage level).

Table \ref{TAB:APFDcsummary}
summarizes the statistical results, presenting the total number of times $\textit{TCP}_\textit{ccc}$
is better (\ding{52}), worse (\ding{54}), or not statistically different (\ding{109}), compared to the other RTCP techniques.
Based on this table, we can answer RQ2 as follows:
\begin{enumerate}
    \item
        When prioritizing Java test suites at the test-class level,
        $\textit{TCP}_\textit{ccc}$ performs much better than $\textit{TCP}_\textit{add}$, $\textit{TCP}_\textit{art}$ and $\textit{TCP}_\textit{search} $; and
        slightly better than $\textit{TCP}_\textit{tot}$.

    \item
        When prioritizing Java test suites at the test-method level,
        $\textit{TCP}_\textit{ccc}$ performs much better than $\textit{TCP}_\textit{tot}$, $\textit{TCP}_\textit{art}$, and $\textit{TCP}_\textit{search}$; and
        slightly better than $\textit{TCP}_\textit{add}$.

    \item
        When prioritizing C test suites,
        $\textit{TCP}_\textit{ccc}$ performs much better than $\textit{TCP}_\textit{tot}$, $\textit{TCP}_\textit{add}$, and $\textit{TCP}_\textit{search}$; and
        slightly better than $\textit{TCP}_\textit{art}$.

\end{enumerate}
Similar to the APFD results,
the CCCP APFD$_\textrm{c}$ performance is influenced by different factors, including the type of test suite, and the code coverage granularity
---
both of which will be discussed in detail in the following two sections (Sections \ref{SEC:impactCCG} and \ref{SEC:impactTCG}).
Nevertheless, the ratios of $\textit{TCP}_\textit{ccc}$ performing better (\ding{52}) rather than worse (\ding{54}) than
$\textit{TCP}_\textit{tot}$, $\textit{TCP}_\textit{add}$, $\textit{TCP}_\textit{art}$, and $\textit{TCP}_\textit{search}$, are:
5.16 (129/25), 5.38 (113/21), 2.25 (99/44), and 1.70 (90/53), respectively.
In conclusion, overall, the proposed CCCP approach is more effective than the four other RTCP techniques, in terms of testing effectiveness, as measured by APFD$_\textrm{c}$.

\subsection{\textbf{RQ3: Impact of Code Coverage Granularity}}
\label{SEC:impactCCG}

In this study, we examined  three types of code coverage (statement, branch, and method).
According to the APFD results (Figures \ref{FIG:apfd-java-c} to \ref{FIG:apfd-c}) and APFD$_\textrm{c}$ results (Figures \ref{FIG:apfdc-java-c} to \ref{FIG:apfdc-c}),
in spite of some cases where the three types of code coverage provide very different APFD or APFD$_\textrm{c}$ results for CCCP,
they do, overall, deliver comparable performance.
This means that the choice of code coverage granularity may have little overall impact on the effectiveness of CCCP.

\begin{figure}[!b]
\graphicspath{{graphs/}}
\centering
    \includegraphics[width=0.47\textwidth]{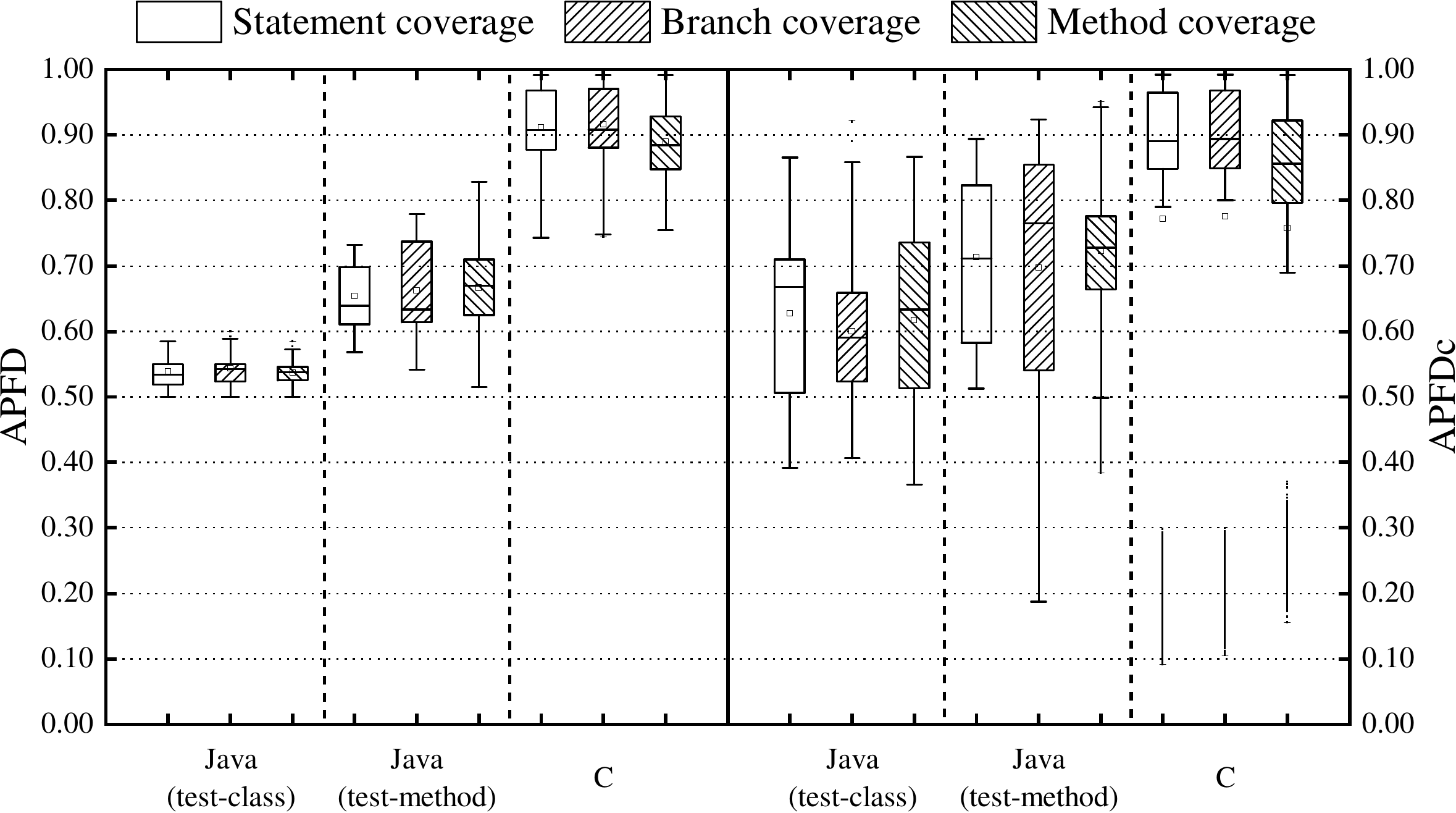}
    \caption{\textbf{Effectiveness:} CCCP \textbf{APFD} and \textbf{APFD}$_\textbf{c}$ results  with different code coverage and test case granularities for \textbf{all programs}}
    \label{FIG:SBM}
\end{figure}

\begin{table*}[!t]
\scriptsize
\centering
 \caption{Statistical \textbf{effectiveness} comparisons of \textbf{APFD} and \textbf{APFD}$_\textbf{c}$ between different granularities for CCCP.
 Each cell in the \textbf{Mean}, \textbf{Median}, and \textbf{Comparison} columns represents the mean APFD or APFD$_\textrm{c}$ value, the median value, and the $p$-values/effect size $\hat{\textrm{A}}_{12}$ for the different code coverage granularity comparisons, respectively.
 }
  \label{TAB:SBM}
  \setlength{\tabcolsep}{1.3mm}{
    \begin{tabular}{c|c|rrr|rrr|rrr}
     \hline
       \multirow{2}*{\textbf{Metric}} &\multirow{2}*{\textbf{Language}} &\multicolumn{3}{c|}{\textbf{Mean}} &\multicolumn{3}{c|}{\textbf{Median}} &\multicolumn{3}{c}{\textbf{Comparison}}\\
        \cline{3-11}

 & &Statement &Branch &Method &Statement &Branch &Method &Statement \textit{vs} Branch &Statement \textit{vs} Method &Branch \textit{vs} Method \\
            \hline
 \multirow{3}*{APFD}
&Java (test-class)	&0.5385 	&0.5430 	&0.5369 	&0.5342 	&0.5420 	&0.5379 	&2.5E-60/0.44	&6.1E-07/0.52	&9.9E-106/0.58	\\\cline{2-11}
&Java (test-method)	&0.6542 	&0.6626 	&0.6664 	&0.6390 	&0.6339 	&0.6698 	&1.83E-40/0.45	&1.4E-75/0.44	&2.5E-07/0.48	\\\cline{2-11}
&C	                 &0.9115 	&0.9156 	&0.8901 	&0.9076 	&0.9077 	&0.8842 	&4.4E-18/0.48	&0/0.62         &0/0.64	\\\hline

 \multirow{3}*{APFD$_c$}
&Java (test-class)	&0.6270 	&0.6002 	&0.6171 	&0.6679 	&0.5904 	&0.6336 	&2.3E-247/0.62	&7.6E-08/0.52	&1.1E-29/0.46	\\\cline{2-11}
&Java (test-method)	&0.7130 	&0.6972 	&0.7227 	&0.7112 	&0.7651 	&0.7279 	&2.3E-08/0.48	&4.9E-02/0.49	&5.0E-25/0.54	\\\cline{2-11}
&C	&0.7722 	&0.7758 	&0.7581 	&0.8909 	&0.8937 	&0.8565 	&1.1E-11/0.48	&7.9E-222/0.58	&1.4E-252/0.59	\\\hline
  \end{tabular}}
\end{table*}

Figure \ref{FIG:SBM} presents the APFD and APFD$_\textrm{c}$ results for the three types of code coverage,
according to the subject programs' language or test suites
(the language or test case granularity is shown on the $x$-axis;
the APFD scores on the left $y$-axis; and
the APFD$_\textrm{c}$ on the right $y$-axis). It can be observed that for C programs, statement and branch coverage are very considerable, but are more effective than method coverage. For Java programs, however, there is no best one among them, because they have similar APFD and APFD$_\textrm{c}$ values.

Table \ref{TAB:SBM} presents a comparison of the mean and median APFD and APFD$_\textrm{c}$ values,
and also shows the $p$-values/effect size $\hat{\textrm{A}}_{12}$ for the different code coverage granularity comparisons.
It can be seen from the table that
the APFD and APFD$_\textrm{c}$ values are similar,
with the maximum mean and median value differences being less than 3\%, and less than 8\%, respectively.
According to the statistical comparisons,
there is no single best code coverage type for CCCP, with each type sometimes achieving the best results.
Nevertheless, branch coverage appears slightly more effective than statement and method coverage for CCCP.

In conclusion, the code coverage granularity may only provide a small impact on CCCP testing effectiveness,
with branch coverage possibly performing slightly better than statement and method coverage.

\subsection{\textbf{RQ4: Impact of Test Case Granularity}}
\label{SEC:impactTCG}

To answer RQ4, we focus on the Java programs, each of which had two levels of test cases (the test-class and test-method levels).
As can be seen in the comparisons between Figures \ref{FIG:apfd-java-c} and \ref{FIG:apfd-java-m},
and between Figures \ref{FIG:apfdc-java-c} and \ref{FIG:apfdc-java-m},
CCCP usually has  significantly lower average APFD and APFD$_\textrm{c}$ values for prioritizing test cases at the test-class level than at the test-method level.

Considering all the Java programs, as can be seen in Table \ref{TAB:SBM},
the mean and median APFD and APFD$_\textrm{c}$ values at the test-method level are much higher than at the test-class level,
regardless of code coverage granularity.
One possible explanation for this is:
Because a test case at the test-class level consists of a number of test cases at the test-method level,
prioritization at the test-method level may be more flexible,
giving better fault detection effectiveness \cite{Zhang2013}.

In conclusion,
CCCP has better fault detection effectiveness when prioritizing test cases at the test-method level than at the test-class level.

\renewcommand\arraystretch{2.0}
\begin{sidewaystable*}[!tp]
\tiny
\centering
 \caption{\textbf{Efficiency: }Comparisons of execution costs in milliseconds for different RTCP techniques.
The ``\textbf{Comp.}"~column presents the compilation times of the subject programs, and the ``\textbf{Instr.}"~column presents the instrumentation time (to collect the information of statement, branch, and method coverage).
Apart from the first four columns, each cell in the table shows the prioritization time using each RTCP technique, for each program, presented as $\mu/\sigma$ (where $\mu$ is the mean time and $\sigma$ is the standard deviation over the 1000 independent runs).
 }
  \label{TAB:time}
  \setlength{\tabcolsep}{0.6mm}{
    \begin{tabular}{@{}c|c|r|r|rrrrr|rrrrr|rrrrr|rrrrr@{}}
     \hline
\multirow{2}*{\textbf{Language}} &\multirow{2}*{\textbf{Program}} &\multicolumn{2}{c|}{\textbf{Time}}  &\multicolumn{5}{c|}{\textbf{Statement Coverage}} &\multicolumn{5}{c|}{\textbf{Branch Coverage}} &\multicolumn{5}{c|}{\textbf{Method Coverage}} &\multicolumn{5}{c}{\textit{\textbf{Sum }}$\sum$ }\\
        \cline{3-24}

         &&\textbf{Comp.} &\textbf{Instr.} &$\textit{TCP}_\textit{tot}$ &$\textit{TCP}_\textit{add}$ &$\textit{TCP}_\textit{art}$ &$\textit{TCP}_\textit{search}$ &$\textit{TCP}_\textit{ccc}$  & $\textit{TCP}_\textit{tot}$ &$\textit{TCP}_\textit{add}$ &$\textit{TCP}_\textit{art}$ &$\textit{TCP}_\textit{search}$ &$\textit{TCP}_\textit{ccc}$ &$\textit{TCP}_\textit{tot}$ & $\textit{TCP}_\textit{add}$ &$\textit{TCP}_\textit{art}$ &$\textit{TCP}_\textit{search}$ &$\textit{TCP}_\textit{ccc}$ &$\textit{TCP}_\textit{tot}$ & $\textit{TCP}_\textit{add}$ &$\textit{TCP}_\textit{art}$ &$\textit{TCP}_\textit{search}$ &$\textit{TCP}_\textit{ccc}$\\
            \hline

&$ant\_v1$	&10,386	&62,671	&0.3/0.5	&6.4/0.5	&149.6/13.2	&8,629.9/100.4	&5.0/0.7	&0.1/0.3	&1.3/0.5	&40.9/5.2	&1,899.6/59.1	&1.4/0.6	&0.0/0.2	&1.0/0.2	&29.1/4.3	&1,491.3/94.5	&0.9/0.3	&0.4/--	&8.7/--	&219.6/--	&12,020.8/--	&7.3/--	\\
&$ant\_v2$	&14,123	&115,818	&0.8/0.9	&19.2/1.4	&739.9/54.0	&21,885.1/799.4	&19.2/2.5	&0.2/0.7	&4.9/1.6	&206.7/15.9	&5,999.2/210.5	&5.3/1.2	&0.1/0.3	&3.1/0.2	&131.9/9.7	&3,832.8/150.5	&3.4/0.7	&1.1/--	&27.2/--	&1,078.5/--	&31,717.1/--	&27.9/--	\\
&$ant\_v3$	&36,126	&116,404	&0.8/0.5	&19.2/1.9	&740.2/52.3	&22,076.9/2,134.4	&19.2/2.3	&0.2/0.4	&4.9/1.1	&204.3/16.4	&6,120.5/234.1	&5.3/0.7	&0.1/0.4	&3.1/0.9	&132.5/10.4	&3,903.5/172.4	&3.5/0.6	&1.1/--	&27.2/--	&1,077.0/--	&32,100.9/--	&28.0/--	\\
&$jmeter\_v1$	&4,212	&38,406	&0.1/0.3	&0.9/0.3	&6.6/1.1	&7,343.6/71.2	&0.9/1.7	&0.1/0.3	&0.3/0.5	&2.5/0.6	&1,842.0/52.1	&0.3/0.5	&0.0/0.1	&0.1/0.2	&0.6/0.5	&276.2/11.2	&0.1/0.2	&0.2/--	&1.3/--	&9.7/--	&9,461.8/--	&1.3/--	\\
&$jmeter\_v2$	&4,737	&40,469	&0.2/0.7	&1.1/0.2	&8.8/1.4	&7,740.5/76.5	&1.0/0.1	&0.1/0.2	&0.4/0.5	&3.3/0.8	&1,992.8/49.5	&0.4/0.5	&0.0/0.1	&0.1/0.3	&0.8/0.5	&316.4/7.5	&0.1/0.2	&0.3/--	&1.6/--	&12.9/--	&10,049.7/--	&1.5/--	\\
&$jmeter\_v3$	&6,290	&45,950	&0.3/0.5	&2.9/0.3	&38.2/5.7	&17,497.2/220.2	&3.2/0.4	&0.1/0.3	&1.1/0.2	&13.8/2.0	&6,032.3/80.6	&1.2/0.4	&0.0/0.1	&0.3/0.5	&3.4/0.8	&763.6/25.8	&0.2/0.4	&0.4/--	&4.3/--	&55.4/--	&24,293.1/--	&4.6/--	\\
Java &$jmeter\_v4$	&13,395	&41,501	&0.1/0.3	&1.1/0.6	&21.0/2.6	&4,865.3/57.4	&1.3/0.5	&0.0/0.1	&0.2/0.4	&4.5/0.7	&613.8/20.1	&0.3/0.8	&0.0/0.1	&0.2/0.4	&4.1/0.7	&516.8/14.7	&0.2/0.4	&0.1/--	&1.5/--	&29.6/--	&5,995.9/--	&1.8/--	\\
(test-class)&$jmeter\_v5$	&13,070	&41,397	&0.2/0.4	&2.0/1.2	&50.0/4.8	&7,661.5/84.7	&3.1/1.1	&0.0/0.2	&0.3/0.5	&7.7/1.2	&784.0/22.5	&0.4/0.5	&0.0/0.2	&0.4/0.5	&8.9/1.2	&761.4/22.7	&0.5/0.5	&0.2/--	&2.7/--	&66.6/--	&9,206.9/--	&4.0/--	\\
&$jtopas\_v1$	&15,327	&22,317	&0.2/0.5	&3.1/1.1	&83.5/7.2	&9,028.7/129.9	&4.0/0.3	&0.0/0.2	&0.4/0.5	&11.9/1.5	&867.4/25.9	&0.6/0.7	&0.1/0.3	&0.7/0.5	&17.7/1.8	&1,050.6/33.9	&0.8/0.4	&0.3/--	&4.2/--	&113.1/--	&10,946.7/--	&5.4/--	\\
&$jtopas\_v2$	&15,028	&24,571	&0.2/0.4	&3.1/1.3	&85.5/7.2	&9,156.8/141.8	&4.2/0.6	&0.0/0.2	&0.4/0.5	&12.2/1.5	&929.8/27.3	&0.6/0.6	&0.0/0.2	&0.7/0.5	&18.3/2.0	&1,106.2/35.4	&0.9/1.4	&0.2/--	&4.2/--	&116.0/--	&11,192.8/--	&5.7/--	\\
&$jtopas\_v3$	&16,879	&35,646	&0.2/0.4	&4.0/0.8	&122.3/9.9	&9,996.7/258.5	&5.4/2.2	&0.0/0.2	&0.6/0.5	&14.9/1.7	&1,021.1/37.7	&0.7/0.4	&0.1/0.2	&0.9/0.4	&25.4/2.4	&1,249.2/51.8	&1.0/0.1	&0.3/--	&5.5/--	&162.6/--	&12,267.0/--	&7.1/--	\\
&$xmlsec\_v1$	&9,722	&12,570	&0.2/0.4	&1.6/0.5	&21.0/2.3	&6,781.8/74.3	&1.3/0.5	&0.0/0.2	&0.5/0.5	&4.9/1.0	&716.8/17.3	&0.3/0.5	&0.0/0.2	&0.2/0.4	&3.2/0.7	&536.7/12.6	&0.2/0.4	&0.2/--	&2.3/--	&29.1/--	&8,035.3/--	&1.8/--	\\
&$xmlsec\_v2$	&10,352	&25,234	&0.2/0.4	&1.7/0.5	&21.8/2.7	&6,781.6/728.6	&1.4/0.5	&0.1/0.2	&0.4/0.5	&5.0/1.2	&808.6/18.2	&0.3/0.5	&0.0/0.2	&0.2/0.4	&3.1/0.7	&512.6/12.6	&0.2/0.4	&0.3/--	&2.3/--	&29.9/--	&8,102.8/--	&1.9/--	\\
&$xmlsec\_v3$	&10,276	&31,700	&0.2/0.7	&1.3/0.4	&14.5/1.9	&4,926.8/986.1	&1.1/0.8	&0.0/0.2	&0.4/0.5	&3.4/0.9	&643.7/16.3	&0.3/0.4	&0.0/0.2	&0.2/0.4	&2.0/0.5	&409.8/10.8	&0.1/0.4	&0.2/--	&1.9/--	&19.9/--	&5,980.3/--	&1.5/--	\\

\hline

&$ant\_v1$	&--	&--	&1.1/0.6	&82.7/3.2	&7,895.2/363.0	&25,552.6/685.0	&67.1/4.5	&0.4/0.8	&17.9/2.4	&723.8/42.6	&5,950.3/400.1	&19.3/3.1	&0.3/1.0	&15.1/1.9	&1,456.9/70.7	&5,055.7/339.3	&13.0/2.0	&1.8/--	&115.7/--	&10,075.9/--	&36,558.6/--	&99.4/--	\\
&$ant\_v2$	&--	&--	&2.7/1.3	&308.3/5.6	&46,943.6/2,484.2	&56,464.7/12,834.1	&287.1/10.0	&0.9/1.6	&77.3/1.5	&2,307.4/151.0	&17,620.8/1,007.5	&77.9/5.9	&0.5/1.1	&56.6/2.5	&8,740.3/309.3	&13,052.6/755.0	&52.7/3.9	&4.1/--	&442.2/--	&57,991.3/--	&87,138.1/--	&417.7/--	\\
&$ant\_v3$	&--	&--	&2.8/1.5	&303.8/5.2	&46,487.4/1,950.7	&56,055.4/12,828.2	&284.0/8.9	&0.8/1.0	&77.0/1.4	&2,284.0/146.0	&11,463.7/2,429.0	&76.8/2.7	&0.5/1.0	&55.0/0.9	&8,591.8/308.4	&9,456.1/1,669.7	&52.0/2.3	&4.1/--	&435.8/--	&57,363.2/--	&76,975.2/--	&412.8/--	\\
&$jmeter\_v1$	&--	&--	&1.0/1.0	&63.4/1.8	&10,945.7/3,639.1	&48,223.8/3,045.9	&71.9/3.3	&0.4/0.7	&23.7/1.1	&2,912.9/109.8	&16,169.4/880.2	&26.5/1.7	&0.1/0.3	&6.3/0.5	&727.7/29.2	&2,697.8/355.8	&5.5/0.5	&1.5/--	&93.4/--	&14,586.3/--	&67,091.0/--	&103.9/--	\\
&$jmeter\_v2$	&--	&--	&1.0/0.5	&66.7/1.8	&11,800.5/3,838.7	&50,776.5/3,564.1	&75.3/3.0	&0.4/0.5	&25.0/1.5	&3,061.5/128.3	&17,139.4/1,003.3	&28.2/1.6	&0.1/0.3	&6.6/0.5	&792.7/31.6	&3,122.9/341.7	&6.0/0.3	&1.5/--	&98.3/--	&15,654.7/--	&71,038.8/--	&109.5/--	\\
&$jmeter\_v3$	&--	&--	&2.7/2.1	&263.0/4.6	&67,417.5/24,856.1	&115,587.2/31,331.9	&360.8/10.2	&1.0/0.6	&98.2/2.7	&35,013.6/7,080.4	&47,254.9/5,475.1	&132.7/4.4	&0.3/0.4	&28.6/0.7	&9,556.6/494.8	&15,477.5/1,017.4	&27.9/2.0	&4.0/--	&389.8/--	&111,987.7/--	&178,319.6/--	&521.4/--	\\
Java &$jmeter\_v4$	&--	&--	&0.2/0.4	&7.8/0.4	&854.6/94.2	&18,956.3/3,208.7	&9.3/1.7	&0.1/0.2	&1.7/0.5	&62.3/16.8	&3,388.9/805.1	&2.1/0.2	&0.0/0.2	&1.4/0.5	&168.0/33.4	&4,024.7/564.2	&1.7/0.5	&0.3/--	&10.9/--	&1,084.9/--	&26,369.9/--	&13.1/--	\\
(test-method)&$jmeter\_v5$	&--	&--	&0.5/1.3	&18.4/1.4	&2,631.4/200.2	&11,459.0/2,003.0	&22.6/1.8	&0.1/0.3	&2.8/0.4	&114.2/26.7	&5,524.5/539.1	&3.5/0.8	&0.1/0.3	&3.1/0.4	&455.9/64.2	&1,829.2/57.9	&3.8/1.1	&0.7/--	&24.3/--	&3,201.5/--	&18,812.7/--	&29.9/--	\\
&$jtopas\_v1$	&--	&--	&0.6/0.5	&23.5/0.6	&3,408.0/273.4	&17,396.6/702.3	&27.3/2.7	&0.1/0.3	&3.3/0.5	&144.5/34.2	&1,717.5/73.2	&4.2/0.6	&0.1/0.4	&4.9/0.7	&721.3/91.1	&2,489.7/110.4	&5.4/1.7	&0.8/--	&31.7/--	&4,273.8/--	&21,603.8/--	&36.9/--	\\
&$jtopas\_v2$	&--	&--	&0.7/0.7	&24.4/2.2	&3,519.3/289.1	&17,671.1/705.3	&27.9/2.6	&0.1/0.3	&3.3/0.8	&148.8/34.6	&1,910.7/84.7	&4.3/1.4	&0.2/0.4	&4.9/0.4	&746.8/96.4	&2,774.4/108.2	&5.4/0.8	&1.0/--	&32.6/--	&4,414.9/--	&22,356.2/--	&37.6/--	\\
&$jtopas\_v3$	&--	&--	&0.7/1.0	&28.9/0.7	&4,476.6/352.7	&20,455.8/577.5	&34.0/3.2	&0.1/0.3	&3.9/1.0	&182.5/37.9	&2,091.8/100.0	&5.0/0.9	&0.2/0.9	&5.9/0.9	&922.1/112.4	&3,105.3/129.6	&6.5/0.9	&1.0/--	&38.7/--	&5,581.2/--	&25,652.9/--	&45.5/--	\\
&$xmlsec\_v1$	&--	&--	&0.9/1.0	&36.8/2.1	&5,623.0/413.0	&16,075.6/468.9	&34.5/3.3	&0.2/0.5	&8.8/0.7	&282.4/55.3	&2,172.4/111.3	&8.4/1.4	&0.1/0.3	&5.6/1.2	&828.4/102.2	&2,235.5/78.4	&5.3/1.0	&1.2/--	&51.2/--	&6,733.8/--	&20,483.5/--	&48.2/--	\\
&$xmlsec\_v2$	&--	&--	&0.9/0.6	&41.2/2.3	&6,438.4/443.9	&13,324.8/3,148.0	&38.1/2.9	&0.3/0.6	&10.9/0.6	&356.0/60.1	&2,581.1/122.9	&9.8/1.6	&0.1/0.3	&6.1/0.9	&903.5/105.9	&2,149.4/73.5	&5.8/1.5	&1.3/--	&58.2/--	&7,697.9/--	&18,055.3/--	&53.7/--	\\
&$xmlsec\_v3$	&--	&--	&0.9/1.3	&33.7/2.2	&4,257.9/734.8	&10,055.2/2,334.9	&29.7/2.2	&0.3/0.4	&9.0/1.3	&265.0/50.0	&2,113.3/107.5	&8.0/1.6	&0.1/0.6	&4.5/0.5	&626.2/78.5	&1,322.8/202.3	&4.2/1.1	&1.3/--	&47.2/--	&5,149.1/--	&13,491.3/--	&41.9/--	\\
\hline
       \multirow{5}*{C}
&$flex$	&401	&10,075	&7.8/4.0	&482.9/14.4	&6,746.2/166.5	&4,308.3/155.7	&503.6/11.4	&4.6/3.7	&304.7/12.4	&5,805.9/165.7	&3,855.0/160.9	&260.2/8.0	&1.2/3.4	&72.9/2.8	&402.0/23.9	&2,871.9/63.9	&28.9/2.6	&13.6/--	&860.5/--	&12,954.1/--	&11,035.2/--	&792.7/--	\\
&$grep$	&1,833	&8,555	&4.4/3.6	&235.3/6.7	&3,547.3/123.2	&2,874.9/63.0	&216.2/7.5	&3.7/3.6	&217.0/6.7	&4,527.4/122.5	&2,966.2/78.3	&172.6/6.8	&0.8/0.8	&48.8/1.1	&226.6/9.1	&2,255.3/37.9	&17.6/1.9	&8.9/--	&501.1/--	&8,301.3/--	&8,096.4/--	&406.4/--	\\
&$gzip$	&308	&1,882	&0.9/0.7	&13.6/0.6	&70.2/4.4	&517.1/10.0	&16.5/1.9	&0.6/0.7	&9.4/0.5	&50.2/3.6	&460.9/10.1	&10.2/1.4	&0.3/0.7	&3.5/0.5	&7.4/1.2	&368.5/9.5	&2.3/0.8	&1.8/--	&26.5/--	&127.8/--	&1,346.5/--	&29.0/--	\\
&$make$	&1,483	&8,520	&1.8/0.9	&21.8/0.7	&116.3/8.1	&783.3/16.8	&28.9/2.2	&1.3/1.0	&14.9/0.3	&120.2/7.9	&604.9/12.6	&18.5/1.9	&0.3/0.8	&2.2/0.4	&5.9/1.1	&228.6/6.5	&2.5/0.6	&3.4/--	&38.9/--	&242.4/--	&1,616.8/--	&49.9/--	\\
&$sed$	&837	&2,956	&1.7/1.1	&68.2/1.2	&1,049.5/36.5	&1,597.8/32.0	&60.8/3.7	&1.1/1.0	&48.7/1.6	&849.0/32.7	&1,499.7/28.5	&31.6/2.6	&0.5/0.9	&20.6/0.6	&87.8/4.4	&1,324.7/27.5	&7.3/1.1	&3.3/--	&137.5/--	&1,986.3/--	&4,422.2/--	&99.7/--	\\

\hline

\multicolumn{4}{c|}{\textbf{\textit{Sum}} $\sum$} &37.3/--	&2,192.0/--	&236,331.5/--	&632,508.4/--	&2,265.9/--	&17.4/--	&973.6/--	&59,747.6/--	&176,757.0/--	&917.2/--	&6.2/--	&363.8/--	&36,348.9/--	&92,569.7/--	&265.9/--	&60.9 /--	&3,529.4 /--	&332,428.0 /--	&901,835.1 /--	&3,449.0 /--	\\
\hline

  \end{tabular}}
\end{sidewaystable*}

\subsection{\textbf{RQ5: CCCP Efficiency}}

Table \ref{TAB:time} presents the time overheads, in milliseconds, for the five RTCP techniques.
The ``\textbf{Comp.}"~column presents the compilation times of the subject programs, and the ``\textbf{Instr.}"~column presents the instrumentation time (to collect the information of statement, branch, and method coverage).
Apart from the first four columns, each cell in the table shows the prioritization time using each RTCP technique, for each program, presented as $\mu/\sigma$ (where $\mu$ is the mean time and $\sigma$ is the standard deviation over the 1000 independent runs).

The Java programs had each version individually adapted to collect the code coverage information, with different versions using different test cases.
Because of this, the execution time was collected for each Java program version. In contrast, each $P_{V0}$ version of the C programs was compiled and instrumented to collect the code coverage information for each test case, and all program versions used the same test cases.
Because of this, each C program version has the same compilation and instrumentation time.
Furthermore, because all the studied RTCP techniques prioritized test cases after the coverage information was collected,
they were all deemed to have the same compilation and instrumentation time for each version of each program.

Based on the time overheads, we have the following observations:

(1)
As expected,
the time overheads for all RTCP techniques (including CCCP) were lowest with method coverage,
followed by branch,
and then statement coverage,
irrespective of test case type.
The reason for this, as shown in Table \ref{TAB:programs},
is that the number of methods is much lower than the number of branches, which in turn is lower than the number of statements;
the related converted test cases are thus shorter, requiring less time to prioritize.

(2)
It was also expected that (for the Java programs)  prioritization at the test-method level would take longer than at the test-class level,
regardless of code coverage granularity.
The reason for this, again,
relates to the number of test cases to be prioritized at the test-method level being more than at the test-class level.

(3)
$\textit{TCP}_\textit{ccc}$ requires much less time to prioritize test cases than
$\textit{TCP}_\textit{art}$ and $\textit{TCP}_\textit{search}$, and very similar time to $\textit{TCP}_\textit{add}$,
irrespective of subject program, and code coverage and test case granularities.
Also, because $\textit{TCP}_\textit{tot}$ does not use feedback information during the prioritization process,
it has much faster prioritization speeds than $\textit{TCP}_\textit{ccc}$.

In conclusion,
$\textit{TCP}_\textit{ccc}$ prioritizes test cases faster than
$\textit{TCP}_\textit{art}$ and $\textit{TCP}_\textit{search}$;
has similar speed to $\textit{TCP}_\textit{add}$;
but performs slower than $\textit{TCP}_\textit{tot}$.

\begin{figure}[!b]
\graphicspath{{graphs/}}
\centering
        \subfigure[APFD]
    {
        \includegraphics[width=0.47\textwidth]{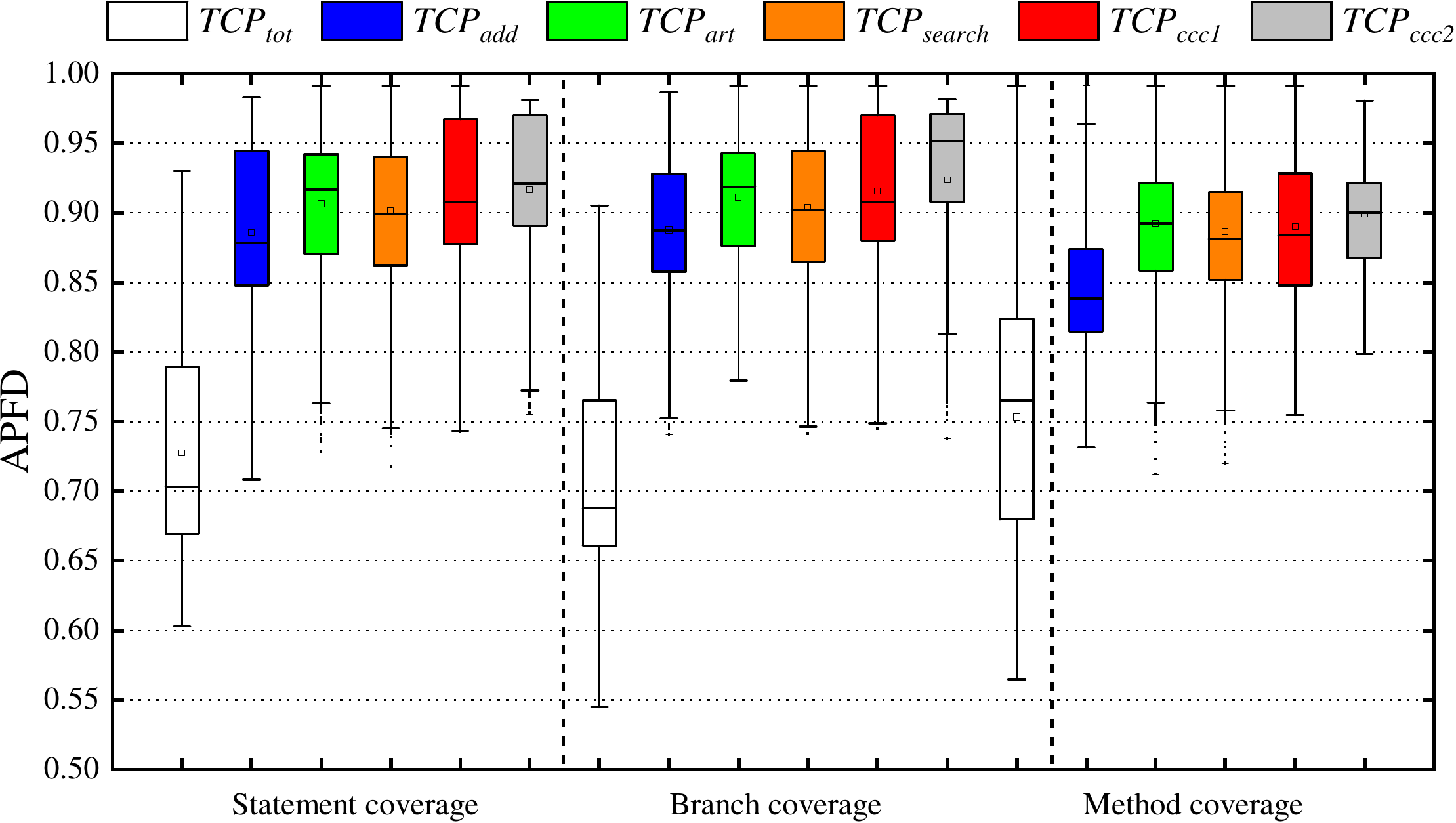}
        \label{apfdc_flex_sc}
    }
        \subfigure[APFD$_\textrm{c}$]
    {
        \includegraphics[width=0.47\textwidth]{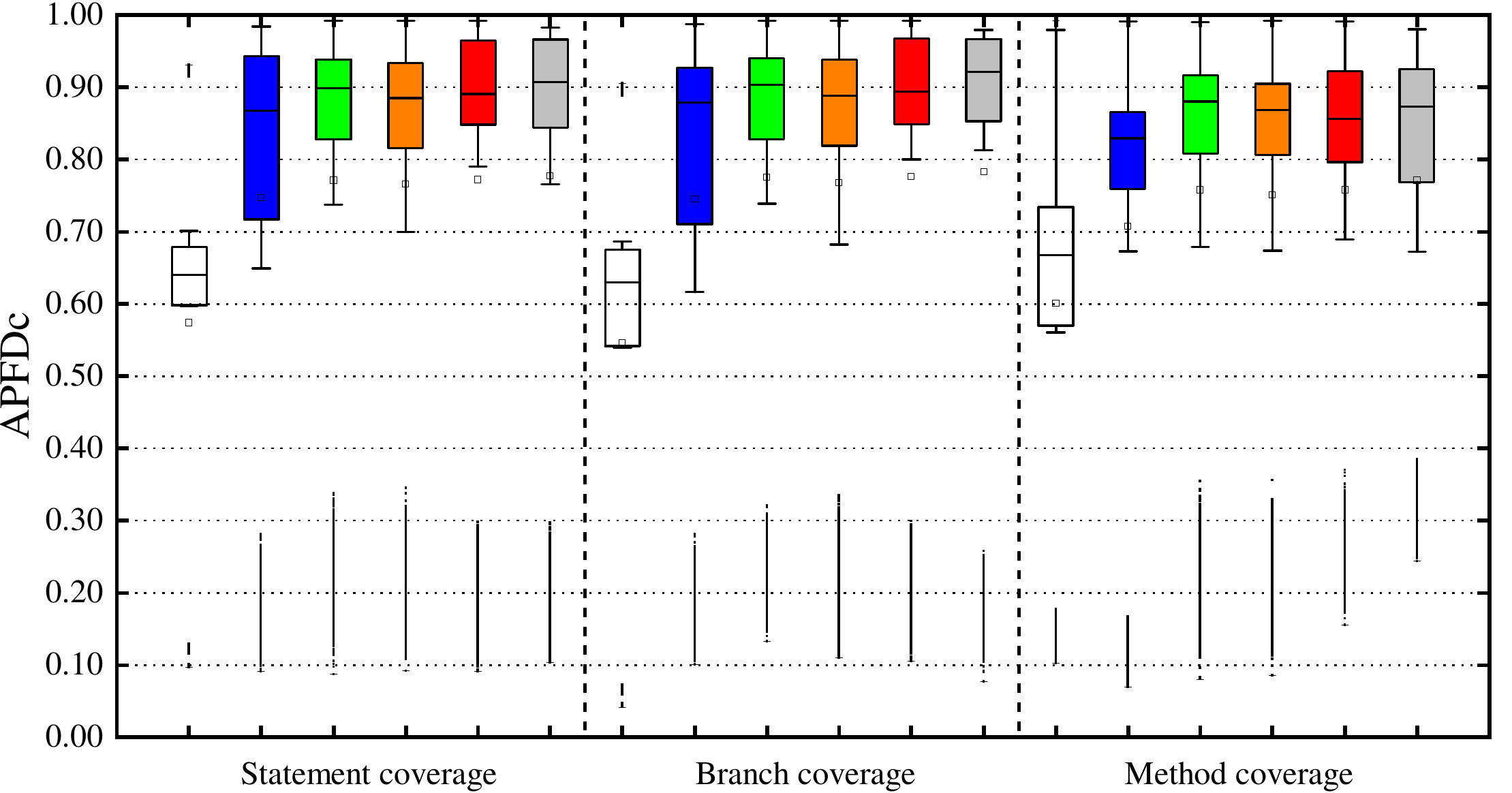}
        \label{apfdc_flex_bc}
    }

    \caption{\textbf{Effectiveness: }\textbf{APFD} and \textbf{APFD}$_\textbf{c}$ results for \textbf{all C programs}}
    \label{FIG:2wise}
\end{figure}

\subsection{\textbf{RQ6: CCCP Effectiveness with $\lambda=2$}}

To answer RQ6, this section briefly discusses the effectiveness of CCCP when $\lambda =2$.
Figure \ref{FIG:2wise} shows the detailed APFD and APFD$\textrm{c}$ results for the C programs, with the code coverage granularity on the $x$-axis, and the $y$-axis giving the APFD or APFD$\textrm{c}$ scores.
For ease of presentation, $\textit{TCP}_\textit{ccc1}$ and $\textit{TCP}_\textit{ccc2}$ denote CCCP with $\lambda =1$ and $\lambda =2$, respectively.
Table \ref{TAB:2wise} presents the statistical comparisons of the $\textit{TCP}_\textit{ccc2}$ APFD and APFD$\textrm{c}$ scores with those of the other five RTCP techniques:
Each data cell shows the $p$-value/effect size $\hat{\textrm{A}}_{12}$ value.

Based on the experimental data, we have the following observations:

(1)
$\textit{TCP}_\textit{ccc2}$ has the higher mean and median APFD and APFD$_\textrm{c}$ values than
 $\textit{TCP}_\textit{tot}$ and $\textit{TCP}_\textit{add}$,
 and better or similar to $\textit{TCP}_\textit{art}$, $\textit{TCP}_\textit{search}$, and $\textit{TCP}_\textit{ccc1}$,
 regardless of code coverage granularity.

(2)
The statistical results confirm the box plot observations.
All $p$-values are much less than 0.05,
indicating a statistically significant difference between the
$\textit{TCP}_\textit{ccc2}$
and each of other five RTCP techniques, regardless of APFD and APFD$_\textrm{c}$ values.
The $\hat{\textrm{A}}_{12}$ results also show
$\textit{TCP}_\textit{ccc2}$ to outperform $\textit{TCP}_\textit{tot}$, $\textit{TCP}_\textit{add}$, $\textit{TCP}_\textit{art}$,
$\textit{TCP}_\textit{search}$, and $\textit{TCP}_\textit{ccc1}$,
with probabilities ranging
from 74\% to 97\%,
62\% to 77\%,
52\% to 60\%,
54\% to 62\%,
and 53\% to 56\%, respectively.

\renewcommand\arraystretch{1.0}
\begin{table}[!t]
\centering
\scriptsize
 \caption{Statistical \textbf{effectiveness} comparisons of \textbf{APFD} and \textbf{APFD}$_\textbf{c}$ between CCCP with $\lambda=2$ and the other five RTCP techniques for \textbf{all C programs}}
  \label{TAB:2wise}
  \setlength{\tabcolsep}{1.6mm}{
    \begin{tabular}{clrrr}
     \hline
        \textbf{Metric} &\textbf{Comparison} &\textbf{Statement}  &\textbf{Branch} &\textbf{Method} \\
        \hline
        \multirow{5}*{APFD}
        &vs $\textit{TCP}_\textit{tot}$ &0/0.94 &0/0.97 &0/0.85  \\
        &vs $\textit{TCP}_\textit{add}$ &0/0.66 &0/0.71 &0/0.77  \\
        &vs $\textit{TCP}_\textit{art}$ &2.3E-149/0.57 &0/0.60 &6.3E-55/0.54  \\
        &vs $\textit{TCP}_\textit{search}$ &8.4E-262/0.58 &0/0.62 &2.2E-253/0.59  \\
        &vs $\textit{TCP}_\textit{ccc1}$ &3.0E-262/0.53 &2.0E-80/0.55 &4.8E-103/0.56  \\
        \hline

        \multirow{5}*{APFD$_\textrm{c}$}
        &vs $\textit{TCP}_\textit{tot}$ &0/0.79 &0/0.82 &0/0.74  \\
        &vs $\textit{TCP}_\textit{add}$ &0/0.60 &0/0.65 &0/0.62  \\
        &vs $\textit{TCP}_\textit{art}$ &5.0E-68/0.55 &4.0E-158/0.57 &1.4E-10/0.52  \\
        &vs $\textit{TCP}_\textit{search}$ &2.5E-135/0.56 &4.8E-237/0.58 &1.4E-59/0.54  \\
        &vs $\textit{TCP}_\textit{ccc1}$ &1.4E-40/0.53 &1.5E-66/0.54 &1.4E-22/0.53  \\
        \hline
  \end{tabular}}
\end{table}

These observations partly confirm our hypothesis about the performance of CCCP:
As the $\lambda$ for unit combination increases, the testing information for guiding prioritization is greater, which may will result in improved performance.

Finally, regarding the prioritization time:
$\textit{TCP}_\textit{ccc2}$ requires about 351073.3, 159881.4, and 501.8 milliseconds for the C programs when using statement, branch and method coverage, respectively.
This low prioritization time of 501.8 milliseconds for $\textit{TCP}_\textit{ccc2}$ with method coverage is less than the prioritization time for $\textit{TCP}_\textit{add}$ using statement or branch coverage (2192.0 and 973.6 milliseconds, respectively).
As shown in Figure \ref{FIG:2wise}, $\textit{TCP}_\textit{ccc2}$ with method coverage has comparable fault detection effectiveness to $\textit{TCP}_\textit{add}$ with statement or branch coverage.
Because method coverage is usually much less expensive to achieve than statement or branch coverage,
$\textit{TCP}_\textit{ccc2}$ with method coverage should be a better choice than $\textit{TCP}_\textit{add}$ with statement or branch coverage.
Furthermore, method-level coverage
---
which is the most natural for projects that are large in scale and high in complexity
---
has greater potential practical application than statement and branch criteria,
making $\textit{TCP}_\textit{ccc2}$ with method coverage more feasible
(2-wise code combinations coverage may incur much more complex calculations for statement and branch coverage than for method coverage).

\subsection{Practical Guidelines}

Here, we present some practical guidelines for how to choose the combination strength and code-coverage level for CCCP, under different testing scenarios:

(1)
    When testing resources are limited, it is (obviously) recommended that the lowest combination strength ($\lambda=1$) be chosen for CCCP.
    This not only achieves better testing effectiveness than other prioritization techniques,
    but also has comparable testing speed to the \textit{additional} test prioritization technique.

(2)
    When there are sufficient testing resources available, $\lambda=2$ is recommended for CCCP, because of the higher fault detection rates it can deliver.

(3)
    If the system under test is large in scale and high in complexity, method coverage is recommended to be used for CCCP.

\subsection{Threats to Validity}
\label{threats}
To facilitate the investigation of potential threats and to support the replication of experiments, we have made the relevant materials (including source code, subject programs, test suites, and mutants) available on our project website: \href{https://github.com/huangrubing/CCCP/}{https://github.com/huangrubing/CCCP/}.
Despite that, our study still face some threats to validity, listed as follows.

\subsubsection{Internal Validity}
The main threat to internal validity lies in the implementation of our experiment.
First, the randomized computations may affect the performance of CCCP:
To address this, we repeated the prioritization process 1000 times and used statistical tests to assess the strategies.
Second, the data structures used in the prioritization algorithms, and the faults in the source code, may introduce noise when evaluating the effectiveness and efficiency:
To minimize these threats, we used data structures that were as similar as possible, and carefully reviewed all source code before conducting the experiment.
Third, although we used the APFD and APFD$_\textrm{c}$ metrics, which have been extensively adopted to assess the performance of RTCP techniques,
APFD only reflects the rate at which faults are detected, ignoring the time and space costs, and
APFD$_\textrm{c}$ assumes that all faults have the same fault severity.
To address this threat, our future work will involve additional metrics that can measure other practical performance aspects of prioritization strategies.

\subsubsection{External Validity}

All the programs used in the experiment were medium-sized, and written in C or Java,
which means that the results may not be generalizable to programs written in other languages (such as C++ and C\#) and of different sizes.
To reduce this threat, other relevant programs will be adopted to evaluate the CCCP performance.
Mutation testing has been argued to be an appropriate approach for assessing fault detection performance~\cite{Andrews2005,Do2005,Just2014}.
Mutation testing has also been used in recent RTCP research studies \cite{Luo2016,Luo2019,Henard2016,Lu2016}.
However, Luo et al. \cite{Luo2018} has highlighted the differences between real faults and mutants, explaining that the relative performances of RTCP techniques on mutants may not translate to similar relative performances with real faults.
To address this threat, additional studies will be conducted to investigate the performance of RTCP on programs with real regression faults in the future.



\section{Related Work}
\label{related}

A considerable amount of research has been conducted into regression testing techniques with a goal of improving the testing performance.
This includes test case prioritization~\cite{Rothermel1999,2018Miranda}, reduction~\cite{2017Chen,2015Shi} and selection~\cite{2018Zhang,2015Gligoric-selection}.
This Related Work section focuses on test case prioritization, which aims to detect faults as early as possible through the reordering of regression test cases~\cite{2012yoo,2018khatibsyarbini}.

\textbf{Prioritization Strategies}.
The most widely investigated prioritization strategies are the \textit{total} and \textit{additional} techniques~\cite{Rothermel1999}.
Because existing greedy strategies may produce suboptimal results, Li et al.~\cite{Li2007} translated the RTCP problem into a search problem and proposed several search-based algorithms, including a hill-climbing and genetic one.
Motivated by random tie-breaking, Jiang et al.~\cite{Jiang2009} applied adaptive random testing to RTCP and proposed a family of adaptive random test cases prioritization techniques that aim to select a test case with the greatest distance from  already selected ones.

More recently, as the \textit{total} strategy and the \textit{additional} strategy can be seen as two extreme instances, Zhang et al.~\cite{Zhang2013} proposed a basic and an extended model to unify the two strategies.
Saha et al.~\cite{2015Saha} proposed an RTCP approach based on information retrieval without dynamic profiling or static analysis.
Many existing RTCP approaches use code coverage to schedule the test cases, but  do not consider the likely distribution of faults.
To address this limitation, instead of traditional code coverage, Wang et al.~\cite{2017Wang} used the quality-aware code coverage calculated by code inspection techniques to guide prioritization process.

\textbf{Coverage criteria}.
In terms of coverage criteria, structural coverage has been widely adopted in test case prioritization.
In addition to statement~\cite{Rothermel1999}, branch~\cite{Jiang2009}, method~\cite{Zhang2013,2017Wang}, block~\cite{Li2007} and modified condition/decision coverage~\cite{2003Jones}, Elbaum et al.~\cite{Elbaum2000} proposed a fault-exposing-potential (FEP) criterion based on the probability of the test case detecting a fault.
Recently, Chi et al.~\cite{2018Chi} used function call sequences, arguing that basic structural coverage may not be optimal for dynamic prioritization.

\textbf{Empirical studies}.
A large number of empirical studies have been performed aiming to offer practical guidelines for using RTCP techniques.

In addition to studies on traditional dynamic test prioritization~\cite{Rothermel1999,Elbaum2000,2006Do,2010Do}, recently, Lu et al.~\cite{Lu2016} were the first to investigate how real-world software evolution impacts on the performance of prioritization strategies:
They reported that source code changes have a low impact on the effectiveness of traditional dynamic techniques, but that the opposite was true when considering new tests in the process of evolution.

Citing a lack of comprehensive studies comparing static and dynamic test prioritization techniques, Luo et al.~\cite{Luo2016,Luo2019} compared static RTCP techniques with dynamic ones. Henard et al.~\cite{Henard2016} compared white-box and back-box RTCP techniques.

\section{Conclusions and Future work}
\label{conclude}
In this paper, we have introduced a new coverage criterion that combines the concepts of code and combination coverage.
Based on this, we proposed a new prioritization technique, \textit{code combinations coverage based prioritization} (CCCP).
Results from our empirical studies have demonstrated that CCCP with the lowest combination strength ($\lambda=1$) can achieve better fault detection rates than four well-known, popular prioritization techniques (\textit{total}, \textit{additional}, \textit{adaptive random}, and \textit{search-based} test prioritization).
CCCP was also found to have comparable testing efficiency to the \textit{additional} test prioritization technique, while requiring much less time to prioritize test cases than the \textit{adaptive random} and \textit{search-based} techniques.
The results also show that CCCP with a higher combination strength ($\lambda=2$) can be more effective than all other prioritization techniques, in terms of both APFD and APFD$_\textrm{c}$.


Our future work will include examining more real-life programs to further investigate the performance of CCCP, including the impact of combination strengths.
In this paper, we have only applied the concept of code combinations coverage to the traditional greedy prioritization strategy.
It will be very interesting to examine new prioritization techniques based on code combinations coverage adopting other prioritization strategies such as search-based strategy.





\section*{Acknowledgements}
We would like to thank the anonymous reviewers for their many constructive comments.
We would also like to thank Christopher Henard for providing us the fault data for the five C subject programs.
This work is supported by the National Natural Science Foundation of China under grant nos.~61502205, 61872167, and U1836116, the project funded by China Postdoctoral Science Foundation under grant no. 2019T120396, and the Senior Personnel Scientific Research Foundation of Jiangsu University under grant no.~14JDG039. This work is also in part supported by the Young Backbone Teacher Cultivation Project of Jiangsu University, and the Postgraduate Research \& Practice Innovation Program of Jiangsu Province under grant no.~KYCX19\_1614.

\section*{References}
\bibliography{CCCP}
\bibliographystyle{elsarticle-num}

\vbox{}
\vbox{}
\scriptsize
\noindent
\textbf{Rubing Huang}
received the Ph.D. degree in computer science and technology from the Huazhong University of Science and Technology, Wuhan, China, in 2013. From 2016 to 2018, he was a visiting scholar at Swinburne University of Technology and at Monash University, Australia. He is an associate professor in the Department of Software Engineering, School of Computer Science and Communication Engineering, Jiangsu University, Zhenjiang, China. His current research interests include software testing (including adaptive random testing, random testing, combinatorial testing, and regression testing), debugging, and maintenance. He has more than 50 publications in journals and proceedings, including in IEEE Transactions on Software Engineering, IEEE Transactions on Reliability, Journal of Systems and Software, Information and Software Technology, IET Software, The Computer Journal, International Journal of Software Engineering and Knowledge Engineering, ICSE, ICST, COMPSAC, QRS, SEKE, and SAC. He is a senior member of the China Computer Federation, and a member of the IEEE and the ACM. More about him and his work is available online at https://huangrubing.github.io/.

\vbox{}
\noindent
\textbf{Quanjun Zhang}
received the B.Eng. degree in computer science and technology in 2017 from Jiangsu University, Zhenjiang, China, where he is currently working toward the M.Eng. degree with the School of Computer Science and Communication Engineering.
His current research interests include software testing and software maintenance.

\vbox{}
\noindent
\textbf{Dave Towey}
received the B.A. and M.A. degrees in computer science, linguistics, and languages from the University of Dublin, Trinity College, Ireland; the M.Ed. degree in education leadership from the University of Bristol, U.K.; and the Ph.D. degree in computer science from The University of Hong Kong, China.
He is an associate professor at University of Nottingham Ningbo China (UNNC), in Zhejiang, China, where he serves as the director of
teaching and learning, and deputy head of school, for the School of Computer Science. He is also the deputy director of the International Doctoral
Innovation Centre at UNNC. He is a member of the UNNC Artificial Intelligence and Optimization research group. His current research interests include software testing (especially adaptive random testing, for which he was amongst the earliest researchers who established the field, and metamorphic testing), computer security, and technology-enhanced education.
He co-founded the ICSE International Workshop on Metamorphic Testing in 2016.
He is a member of both the IEEE and the ACM.

\vbox{}
\noindent
\textbf{Weifeng Sun}
received the B.Eng. degree in computer science and technology in 2018 from Jiangsu University, Zhenjiang, China, where he is currently working toward the M.Eng. degree with the School of Computer Science and Communication Engineering.
His current research interests include software testing and software debugging. His work has been published in journals and proceedings, including in IEEE Transactions on Software Engineering, IEEE Transactions on Reliability, and the IEEE International Conference on Software Testing, Verification and Validation (ICST).
He is a student member of the China Computer Federation and the ACM.

\vbox{}
\noindent
\textbf{Jinfu Chen}
received the BE degree in 2004 from Nanchang Hangkong University, Nanchang, China and the PhD degree in 2009 from Huazhong University of Science and Technology, Wuhan, China, both in computer science. He is currently a full professor in the School of Computer Science and Communication Engineering, Jiangsu University, Zhenjiang, China. His major research interests include software testing, software analysis, and trusted software.

\end{document}